\documentstyle[prd,aps]{revtex}

\newcommand{\bea}{\begin{eqnarray}}
\newcommand{\eea}{\end{eqnarray}}

\begin{document}
\draft

\title{Second-order Perturbations of the Friedmann World Model}
\author{Hyerim Noh${}^{(a)}$ and Jai-chan Hwang${}^{(b)}$}
\address{${}^{(a)}$ Korea Astronomy Observatory, Daejon, Korea \\
         ${}^{(b)}$ Department of Astronomy and Atmospheric Sciences, 
                    Kyungpook National University, Taegu, Korea}
\date{\today}
\maketitle

\begin{abstract}

We consider instability of the Friedmann world model to the
second-order in perturbations.
We present the perturbed set of equations up to the second-order
in the Friedmann background world model with general spatial curvature
and the cosmological constant.
We consider systems with the completely general imperfect fluids,
the minimally coupled scalar fields, the electro-magnetic field,
and the generalized gravity theories.
We also present the case of null geodesic equations, 
and the one based on the relativistic Boltzmann equation.
In due stage a decomposition is made for the scalar-, vector- and tensor-type
perturbations which couple each other to the second-order.
Gauge issue is resolved to each order.
The basic equations are presented without imposing any gauge condition,
thus in a gauge-ready form so that we can use the full advantage of 
having the gauge freedom in analysing the problems.
As an application we show that to the second-order in perturbation
the relativistic pressureless ideal fluid of the scalar-type 
reproduces exactly the known Newtonian result.
As another application we rederive the large-scale conserved quantities
(of the pure scalar- and tensor-perturbations) to the second order, 
first shown by Salopek and Bond, now from the exact equations.
Several other applications are made as well.

\end{abstract}

\noindent

\begin{quote}
``{\sl the linear perturbations are so surprisingly simple that a perturbation 
       analysis accurate to second order may be feasible \dots}''

      \hskip 12.6cm Sachs and Wolfe (1967)
\end{quote}


\tableofcontents

\section{Introduction}

We consider instabilities of the spatially homogeneous and isotropic
comological spacetime to the second-order in perturbations.
The relativistic cosmological perturbation plays fundamental roles
in the modern theory of large-scale cosmic structure formation.
The original analysis of linear perturbations based on Einstein gravity with 
a hydrodynamic fluid
was made by Lifshitz in 1946 \cite{Lifshitz-1946} in an almost complete form.
Due to the extremely low level anisotropies of the cosmic microwave
background (CMB) radiation, the cosmological dynamics of the structures
in the large-scale and in the early universe are generally believed to be
small deviations from the homogeneous and isotropic background 
Friedmann world model \cite{Friedmann-1922}.
The conventional relativisitic cosmological perturbation analysis 
considers such deviations small enough so that one could treat
them as {\it linear}.
The linear perturbation theory works as the basic framework in handling 
the cosmological structure formation processes.
Recent observations of the CMB anisotropies in all sky
by the WMAP satellite and others \cite{WMAP}, for example, assure the 
validity of the basic assumptions used in the cosmological perturbation theory,
i.e., the linearity of the relevant cosmic structures 
\cite{Bond-Efstathiou-1987}.

Still, as the observed relatively small-scale structures
are apparently nonlinear, the gravitational instability
based on the pure linear theory is not enough for the complete picture.
It is agreed that such small-scale nonlinear structures could
be handled by Newtonian gravity often based on numerical simulations.
The current paradigm of large-scale structure generation and evolution 
processes is based on an underlying {\it assumption} that linear 
processes dominated until nonlinear processes take over on subhorizon 
scales in the Newtonian regime.
Thus, it seems this paradigm of our understanding of the
origin and evolution of the large-scale structures is rather satisfactory
within the linear theory concerning the regimes where the
relativistic gravity theory is needed.
It is well known that in the linear theory there can be no structure formation.
In fact, this ``no structure formation'' in the scenario is precisely
why we were successful in describing the structure generation and evolution
processes in a simple manner as we describe below.
In the standard scenario, the initial conditions (seed fluctuations) generated 
from quantum fluctuations are imprinted into ripples in the spacetime, 
and its spatial structures are preserved as the raw large-scale structures 
(i.e. are as yet unaffected by nonlinear processes).
This is a trait which can be traced to the linearity assumption we adopt.
It gives a simple, but fictitious system.
We could, perhaps, fairly describe the current situation as:
the linear paradigm is {\it not inconsistent} with observations, 
especially with the low level of observed anisotropy in the CMB \cite{WMAP}.
However, we should remember that the actual equations
we are dealing with, both in the gravity and the quantum, are highly nonlinear.
It forms an intrinsically complex system.

We can decompose the perturbations into three different types:
the scalar-type (associated with the density condensation),
the vector-type (rotation) and the tensor-type (gravitational wave)
perturbations.
To the linear-order perturbation, due to the high symmetry of the
background space, these three types of perturbations decouple
from each other and evolve independently.
Both (the linearity and the homogeneous-isotropic background)
conditions are necessary to have natural descriptions of the
three types of perturbations independently.
We will see how couplings occur to the second-order perturbations;
for the couplings in a simplest spatially homogeneous but anisotropic 
spacetime, see \cite{Bianchi-pert}.

Another aspect of the simple nature of the linear processes 
in relativistic gravity theories is characterized by conservations 
(in expanding phase) of certain amplitudes in the super-horizon scale 
where we naively anticipate the independent evolution of causally 
disconnected regions.
On superhorizon scales this conserved character is presented by an equation
of the form
\bea
   & & \Phi ({\bf x},t) = C ({\bf x}),
   \label{Conservations}
\eea
which applies for both the scalar- and tensor-type perturbations;
for the vector-type perturbation the angular momentum is conserved 
in all scales.
$C({\bf x})$ is an integration constant of the integral form general solutions 
available in the large-scale limit,
 see eqs. (\ref{Phi-sol},\ref{rotation-sol},\ref{GW-sol-LS}).
The coefficient $C ({\bf x})$ contains information about spatial structure
which will eventually grow to the large-scale structure and the
gravitational wave background.
It can be considered as an initial condition for each perturbation variable
which is preserved during the linear evolution.
Whether a similar conserved variable can exist even in
a nonlinear analysis is apparently an interesting question:
for the presence of such variables to the second-order in perturbation
see \S \ref{sec:pure-scalar}-\ref{sec:pure-tensor}.
In the analyses of the large-scale structures in the linear stage, 
the simple behavior of the conserved variables is practically important. 
In fact, if we know $C$ the behavior of all the other variables 
can be determined through linear algebra.
Using  the conserved quantity one can trivially relate the currently observable
(or deducible) linear structure directly to the initial state of the
structure at the early universe; probably, just after the scale effectively
becomes the large-scale during the hypothetical early acceleration
(inflation) stage.
Of course, the underlying assumption for all of these results is the
applicability of the linear analysis.
As long as this assumption is valid, the initial condition is
imprinted onto the large-scale structure and is preserved
until the nonlinear effects become important.
Although this is a big advantage, in a sense this is very consistent
with the fact that no structure formation occurs in the linearized system.

The linear perturbation theory is currently well developed; see 
\cite{Bardeen-1980,Bardeen-1988,Kodama-Sasaki-1984,MFB-1992,Peebles-1980,Liddle-Lyth-2000}.
Although the observations do not desparately demand to go beyond
the linear theory, the second-order perturbation theory is a natural next step
in the theoretical investigations.
The second-order perturbation theory, if well developed, will
have important implications on our understanding of the 
large-scale structure formation processes.
Not only the structures we discover are nonlinear, according to the
gravitational instability there should occur (perhaps smooth) transitions
from the linear to the nonlinear ones.
Even in the theoretical point of view, in order to know the
limit of the linear perturbation theory we need the behavior of
perturbations beyond the linear theory.
It is not possible to know the limit of linear theory within
the context of the linear theory.
It is yet unclear whether the second-order `perturbation theory'
will provide an answer to such a question, but we expect it could
provide a better perspective on the problem than the simple linear theory.
There will be more practical applications as well, like investigating
the non-Gaussian signature in the inflation generated seed fluctuations
which could have left a detectable signature in the CMB anisotropies
and the large-scale structures.
Other possible situations where the trans-linear analyses might be useful
are summarized in \S \ref{sec:Discussions}.

Now, we discuss the gauge issue present in the relativistic
perturbation theory briefly. 
Since the unperturbed background spacetime is spatially
homogeneous and isotropic, to the linear-order the ambiguity caused 
by spatial gauge (coordinate transformation) freedom does not play a role 
\cite{Bardeen-1988}.
Thus, to the linear-order it is appropriate to write the perturbed 
set of equations in terms of natural combinations of variables 
which are invariant under spatial gauge transformation.
However, the temporal gauge freedom can be conveniently used in analyzing
various aspects of perturbation problem in the Friedmann background.
There are infinitely many different ways of taking the temporal gauge
(choosing the spatial hypersurface) conditions, and we can identify 
several fundamental gauge choices \cite{Bardeen-1988,Hwang-1991}.  
Except for a widely used temporal synchronous gauge fixing condition
($\delta g_{00} \equiv 0$), each of the other gauge conditions 
fixes completely the temporal gauge freedom.
Thus each having its own corresponding gauge invariant formulation.
In our study of the linear theory we found that
some particular gauge invariant variables show the correct Newtonian behaviors.
A perturbed density variable in the comoving gauge and a perturbed potential 
variable and a perturbed velocity variable in the zero-shear gauge most 
closely resemble the behaviors of the corresponding Newtonian variables
\cite{Hwang-Noh-Newtonian}.
Also, the scalar field perturbation in the uniform-curvature gauge
most closely resembles the scalar field equation in the
quantum field in curved space \cite{Hwang-QFT,Hwang-MSF}.
Since the gauge conditions mentioned allow no room for remaining gauge mode
the variable in a given such gauges has the unique corresponding gauge 
invariant combination of variables.
Thus, the variable in such a gauge is equivalent to the
corresponding gauge invariant combination.

In the gauge theory it is well known that proper choice of the gauge
condition is often necessary for proper handling of the problem.
Either by fixing certain gauge conditions or by choosing certain
gauge-invariant combinations in the early calculation stage we are likely to
lose possible advantages available in the other gauge conditions.
In order to use the various gauge conditions as advantage in handling
cosmological perturbations we have proposed a {\it gauge-ready} method
which allows the flexible use of the various fundamental gauge conditions.
The strategy is that, in order to use the various available temporal
gauge conditions as advantage, we had better present the basic equations
without choosing any temporal gauge condition, and arrange the equations 
so that we could choose the fundamental gauge conditions conveniently.
In this work we will elaborate further the gauge-ready approach to
the second-order perturbations.
Our gauge-ready strategy, together with our notation for
indicating the gauge-invariant combinations, allows us to use 
the gauge freedom as the advantage in analysing various given problems.
We follow the wisdom suggested by Bardeen in 1988
``{\sl the moral is that one should work in the gauge that is mathematically 
most convenient for the problem at hand}'' \cite{Bardeen-1988}.

As long as we are taking perturbative approach the gauge issue 
in the higher-orders can be resolved similarly as in the linear theory. 
To the second-order we will identify two variables 
which can be used to fix the spatial gauge freedom.
One gauge condition completely removes the spatial gauge mode,
whereas the other condition does not; 
{\it i.e.}, in the latter case even after imposing the gauge condition 
there still remains a degree of freedom which is a gauge mode 
(a coordinate artifact).
We call this incomplete gauge condition the $B$-gauge 
($\tilde g_{0\alpha} \equiv 0$ where $\alpha$ is a spatial index)
whereas the other complete condition is called the $C$-gauge.
To the second-order we can identify the same several temporal gauge fixing
conditions.
Again, except for the synchronous gauge each of the other gauge
conditions removes completely the temporal gauge modes.

It is amusing to note that the classic study by Lifshitz \cite{Lifshitz-1946}
adopted the synchronous gauge condition which is a combination of
the temporal synchronous gauge condition and the spatial $B$-gauge condition,
thus failing to fix both the temporal and  the spatial gauge modes completely.
This has caused some prevalent errors in the literature based on 
the synchronous gauge: see the Appendix of \cite{Hwang-Noh-Newtonian}.
However, we note that these errors are simple algebraic ones probably 
caused by slightly more complicated algebra due to the presence 
of the gauge mode after the synchronous gauge fixing.
We would like to emphasize that the gauge condition should be
appropriately used according to the character of each problem at hand.
We have such a freedom because Einstein's gravity theory might be 
regarded as a gauge theory \cite{Gauge-theory}.
In this sense, although the temporal synchronous and the spatial
$B$-gauge conditions do not remove completely the gauge modes, 
often even these conditions could possibly turn out to be convenient 
in certain problems.
Since physically measurable quantities should be gauge invariant
we propose to use the gauge conditions in this pragmatic sense.

In a classic study of CMB anisotropy in 1967 Sachs and Wolfe have mentioned 
that 
``{\sl the linear perturbations are so surprisingly simple that a perturbation
       analysis accurate to second order may be feasible \dots}''
\cite{SW-1967}. 
In this work, we will present the basic formulation of the second-order
perturbation of the Friedmann world model in details.
We will present the basic equations needed to investigate
the second-order perturbation in a rather general context.
We will consider the most general Friedmann background with $K$ and $\Lambda$.
We will consider the most general imperfect fluid situation.
This includes multiple imperfect fluids with general interactions among them.
We will also include minimally coupled scalar fields,
a class of generalized gravity theories, the electromagnetic fields,
the null geodesic, and the relativistic Boltzmann equation.
In order to use the gauge fixing conditions optimally we will present 
the complete sets of perturbed equations in the gauge-ready form.
In this manner, as in the linear theory, we could easily apply the equations 
to any gauge conditions which make the mathematical analyses of given 
problems simplest.
Our formulation will be suitable to handle the nonlinear evolutions
in the perturbative manner.
Ours will be a useful complement to the other methods
suggested in the literature to investigate the trans-linear regimes.
In the Discussions we summarize the related studies, 
the different methods we could employ for further applications, 
and the cosmological situations where our formultion could be applied 
fruitfully.
Although we will present some trivial applications in the later part,
the main applications are left for future studies.

In \S \ref{sec:basic-eqs} we summarize the basic equations of the Einstein's 
gravity expressed in the ADM ($3+1$) formulation and in the 
covariant ($1+3$) formulation.
In \S \ref{sec:perturbed-quantities} we introduce our definition of the
metric and the energy-momentum tensor to the second-order
in perturbation and present some useful quantities appearing
in the ADM and the covariant formulations.
In \S \ref{sec:Equations} we present the complete sets of 
perturbed equations up to the second-order in the Friedmann world model.
We consider the general spatial curvature and the cosmological constant
in the background.
We consider systems with the completely general imperfect fluids.
Such a general formulation can be reinterpreted to include the
cases of the minimally coupled scalar fields, the electro-magnetic field,
and even the generalized gravity theories.
We present the complete sets of equations for these systems as well.
We also present the case of null geodesic equations, 
and the one based on the relativistic Boltzmann equation.
In \S \ref{sec:Decomposition} we introduce decomposition of perturbations
to three different types and show how these couple to each other 
to the second-order.
All equations up to this point are presented without introducing
any gauge condition.
Thus, the equations are presented in the most general forms,
and any suitable gauge conditions can be easily deployed to these equations.
In this sense, our set of equations is in a gauge-ready form.
In \S \ref{sec:Gauges} we address the gauge issue,
and show that the gauge issue can be resolved to each perturbation order,
just like the case in the linear perturbation.
We implement our gauge-ready strategy to the second-order in perturbations.
In \S \ref{sec:Applications} we make several applications.
In \S \ref{sec:Discussions} we summarize the main results and 
outline future applications of our work.

As a unit we set $c \equiv 1$.

\section{Basic equations}
                                                     \label{sec:basic-eqs}

\subsection{ADM ($3+1$) equations}
                                                     \label{sec:ADM-eqs}

The ADM (Arnowitt-Deser-Misner) equations \cite{ADM}
are based on splitting the spacetime into the spatial and the temporal parts 
using a normal vector field $\tilde n_a$. 
The metric is written as (we put tilde on the covariant variables):
\bea
   & & \tilde g_{00} \equiv - N^2 + N^\alpha N_\alpha, \quad
       \tilde g_{0\alpha} \equiv N_\alpha, \quad
       \tilde g_{\alpha\beta} \equiv h_{\alpha\beta},
   \nonumber \\
   & & \tilde g^{00} = - N^{-2}, \quad
       \tilde g^{0\alpha} = N^{-2} N^\alpha, \quad
       \tilde g^{\alpha\beta} = h^{\alpha\beta} - N^{-2} N^\alpha N^\beta,
   \label{ADM-metric-def}
\eea
where $N_\alpha$ is based on $h_{\alpha\beta}$ as the metric
and $h^{\alpha\beta}$ is an inverse metric of $h_{\alpha\beta}$.
Indices $a, b, \dots $ indicate the spacetime indices,
and $\alpha, \beta, \dots$ indicate the spatial ones.
The normal vector $\tilde n_a$ is introduced as:
\bea
   & & \tilde n_0 \equiv - N, \quad 
       \tilde n_\alpha \equiv 0, \quad
       \tilde n^0 = N^{-1}, \quad
       \tilde n^\alpha = - N^{-1} N^\alpha.
   \label{n_a-def}
\eea
The fluid quantities are defined as:
\bea
   E \equiv \tilde n_a \tilde n_b \tilde T^{ab}, \quad
       J_\alpha \equiv - \tilde n_b \tilde T^b_\alpha, \quad
       S_{\alpha\beta} \equiv \tilde T_{\alpha\beta}, \quad
       S \equiv h^{\alpha\beta} S_{\alpha\beta}, \quad
       \bar S_{\alpha\beta} \equiv S_{\alpha\beta}
       - {1\over 3} h_{\alpha\beta} S,
   \label{ADM-fluid-def}
\eea
where $J_\alpha$ and $S_{\alpha\beta}$ are based on $h_{\alpha\beta}$.
The extrinsic curvature is introduced as
\bea
   & & K_{\alpha\beta} \equiv {1\over 2N} \left( N_{\alpha:\beta} 
       + N_{\beta:\alpha} - h_{\alpha\beta,0} \right), \quad
       K \equiv h^{\alpha\beta} K_{\alpha\beta}, \quad
       \bar K_{\alpha\beta} \equiv K_{\alpha\beta} 
       - {1\over 3} h_{\alpha\beta} K,
   \label{extrinsic-curvature-def}
\eea
where $K_{\alpha\beta}$ is based on $h_{\alpha\beta}$.
A colon `$:$' denotes a covariant derivative based on $h_{\alpha\beta}$.
The connections become:
\bea
   & & \tilde \Gamma^0_{00} = {1\over N} \left( N_{,0} + N_{,\alpha} N^\alpha
       - K_{\alpha\beta} N^\alpha N^\beta \right), \quad
       \tilde \Gamma^0_{0\alpha} = {1\over N} \left( N_{,\alpha}
       - K_{\alpha\beta} N^\beta \right), \quad
       \tilde \Gamma^0_{\alpha\beta} = - {1\over N} K_{\alpha\beta},
   \nonumber \\
   & & \tilde \Gamma^\alpha_{00} = {1\over N} N^\alpha \left( - N_{,0}
       - N_{,\beta} N^\beta + K_{\beta\gamma} N^\beta N^\gamma \right)
       + N N^{,\alpha} + N^\alpha_{\;\; ,0} - 2 N K^{\alpha\beta} N_\beta
       + N^{\alpha:\beta} N_\beta,
   \nonumber \\
   & & \tilde \Gamma^\alpha_{0\beta} = - {1\over N} N_{,\beta} N^\alpha
       - N K^\alpha_\beta + N^\alpha_{\;\;\; :\beta}
       + {1\over N} N^\alpha N^\gamma K_{\beta\gamma}, \quad
       \tilde \Gamma^\alpha_{\beta\gamma}
       = \Gamma^{(h)\alpha}_{\;\;\;\;\beta\gamma}
       + {1\over N} N^\alpha K_{\beta\gamma}, 
   \label{connection-ADM}
\eea
where $\Gamma^{(h)\alpha}_{\;\;\;\;\beta\gamma}$ is the connection based on
$h_{\alpha\beta}$ as the metric,
$\Gamma^{(h)\alpha}_{\;\;\;\;\;\beta\gamma}
 \equiv {1\over 2} h^{\alpha\delta} \left(
 h_{\beta\delta,\gamma} + h_{\delta\gamma,\beta}
 - h_{\beta\gamma,\delta} \right)$.
The intrinsic curvatures are based on $h_{\alpha\beta}$ as the metric:
\bea
   & & R^{(h)\alpha}_{\;\;\;\;\;\;\;\beta\gamma\delta}
       \equiv
       \Gamma^{(h)\alpha}_{\;\;\;\;\;\beta\delta,\gamma}
       - \Gamma^{(h)\alpha}_{\;\;\;\;\;\beta\gamma,\delta}
       + \Gamma^{(h)\epsilon}_{\;\;\;\;\;\beta\delta}
       \Gamma^{(h)\alpha}_{\;\;\;\;\;\gamma\epsilon}
       - \Gamma^{(h)\epsilon}_{\;\;\;\;\;\beta\gamma}
       \Gamma^{(h)\alpha}_{\;\;\;\;\;\delta\epsilon},
   \nonumber \\
   & & R^{(h)}_{\alpha\beta} 
       \equiv R^{(h)\gamma}_{\;\;\;\;\;\;\;\alpha\gamma\beta}, \quad
       R^{(h)} \equiv h^{\alpha\beta} R^{(h)}_{\alpha\beta}, \quad
       \bar R^{(h)}_{\alpha\beta} \equiv R^{(h)}_{\alpha\beta}
       - {1\over 3} h_{\alpha\beta} R^{(h)}.
   \label{ADM-curvature}
\eea
A complete set of the ADM equations is the following \cite{Bardeen-1980}.

\noindent
Energy constraint equation:
\bea
   & & R^{(h)} = \bar K^{\alpha\beta} \bar K_{\alpha\beta}
       - {2 \over 3} K^2 + 16 \pi G E + 2 \Lambda,
   \label{E-constraint}
\eea
where $\Lambda$ is the cosmological constant.

\noindent
Momentum constraint equation:
\bea
   & & \bar K^\beta_{\alpha:\beta} - {2 \over 3} K_{,\alpha} 
       = 8 \pi G J_\alpha.
   \label{Mom-constraint}
\eea
Trace of ADM propagation equation:
\bea
   & & K_{,0} N^{-1} - K_{,\alpha} N^\alpha N^{-1}
       + N^{:\alpha}_{\;\;\;\;\alpha} N^{-1}
       - \bar K^{\alpha\beta} \bar K_{\alpha\beta}
       - {1\over 3} K^2 - 4 \pi G \left( E + S \right) + \Lambda = 0.
   \label{Trace-prop}
\eea
Tracefree ADM propagation equation:
\bea
   & & \bar K^\alpha_{\beta,0} N^{-1}
       - \bar K^\alpha_{\beta:\gamma} N^\gamma N^{-1}
       + \bar K_{\beta\gamma} N^{\alpha:\gamma} N^{-1}
       - \bar K^\alpha_\gamma N^\gamma_{\;\;\;:\beta} N^{-1}
   \nonumber \\
   & & \qquad = K \bar K^\alpha_\beta
       - \left( N^{:\alpha}_{\;\;\;\;\beta} 
       - {1\over 3} \delta^\alpha_\beta N^{:\gamma}_{\;\;\;\;\gamma} \right)
       N^{-1}
       + \bar R^{(h)\alpha}_{\;\;\;\;\;\;\beta} - 8 \pi G \bar S^\alpha_\beta.
   \label{Tracefree-prop}
\eea
Energy conservation equation:
\bea
   & & E_{,0} N^{-1} - E_{,\alpha} N^\alpha N^{-1} 
       - K \left( E + {1 \over 3} S \right)
       - \bar S^{\alpha\beta} \bar K_{\alpha\beta}
       + N^{-2} \left( N^2 J^\alpha \right)_{:\alpha} 
       = 0.
   \label{E-conservation}
\eea
Momentum conservation equation:
\bea
   & & J_{\alpha,0} N^{-1} - J_{\alpha:\beta} N^\beta N^{-1}
       - J_\beta N^\beta_{\;\;\;:\alpha} N^{-1} - K J_\alpha
       + E N_{,\alpha} N^{-1} 
       + S^\beta_{\alpha:\beta}
       + S_\alpha^\beta N_{,\beta} N^{-1} = 0.
   \label{Mom-conservation}
\eea

\subsection{Covariant ($1+3$) equations}
                                              \label{sec:covariant-eqs}

The covariant formulation of Einstein gravity was investigated in
\cite{Ehlers-1993,Ellis-1971}.
The $1+3$ covariant decomposition is based on the time-like normalized 
($\tilde u^a \tilde u_a \equiv - 1$) four-vector field $\tilde u_a$ 
introduced in all spacetime points.
The expansion ($\tilde \theta$), the acceleration ($\tilde a_a$), 
the rotation ($\tilde \omega_{ab}$), and the shear ($\tilde \sigma_{ab}$) are
kinematic quantities of the projected covariant derivative of 
flow vector $\tilde u_a$ introduced as
\bea
   & & \tilde h^c_a \tilde h^d_b \tilde u_{c;d} 
       = \tilde h^c_{[a} \tilde h^d_{b]} \tilde u_{c;d} 
       + \tilde h^c_{(a} \tilde h^d_{b)} \tilde u_{c;d} 
       \equiv \tilde \omega_{ab} + \tilde \theta_{ab} 
       = \tilde u_{a;b} + \tilde a_a \tilde u_b,
   \nonumber \\
   & & \tilde \sigma_{ab} \equiv \tilde \theta_{ab} 
       - {1 \over 3} \tilde \theta \tilde h_{ab}, \quad
       \tilde \theta \equiv \tilde u^a_{\;\; ;a}, \quad
       \tilde a_a \equiv {\tilde {\dot {\tilde u}}}_a
       \equiv \tilde u_{a;b} \tilde u^b,
   \label{kinematic-shear}
\eea
where $\tilde h_{ab} \equiv \tilde g_{ab} + \tilde u_a \tilde u_b$
is the projection tensor with $\tilde h_{ab} \tilde u^b = 0$
and $\tilde h^a_a = 3$.
An overdot with tilde $\tilde {\dot {}}$
indicates a covariant derivative along $\tilde u^a$.
We have
\bea
   & & \tilde u_{a;b} = \tilde \omega_{ab} + \tilde \sigma_{ab} 
       + {1 \over 3} \tilde \theta \tilde h_{ab} - \tilde a_a \tilde u_b.
   \label{kinematic-cov-2}
\eea
We introduce
\bea
   & & \tilde \omega^a \equiv {1 \over 2} \tilde \eta^{abcd} \tilde u_b 
       \tilde \omega_{cd}, \quad
       \tilde \omega_{ab} = \tilde \eta_{abcd} \tilde \omega^c \tilde u^d, \quad
       \tilde \omega^2 \equiv {1 \over 2} \tilde \omega^{ab} \tilde \omega_{ab}
       = \tilde \omega^a \tilde \omega_a, \quad
       \tilde \sigma^2 \equiv {1 \over 2} \tilde \sigma^{ab} \tilde \sigma_{ab},
   \label{kinematic-cov-3}
\eea
where $\tilde \omega^a$ is a {\it vorticity vector} which has the same 
information as the vorticity tensor $\tilde \omega_{ab}$.
We have $\tilde \eta^{abcd} = \tilde \eta^{[abcd]}$ with
$\tilde \eta^{1234} = 1/\sqrt{- \tilde g}$;
indices surrounded by $()$ and $[]$ are the symmetrization 
and anti-symmetrization symbols, respectively.

Our convention of the Riemann curvature and Einstein's equation are:
\bea
   & & \tilde u_{a;bc} - \tilde u_{a;cb} = \tilde u_d \tilde R^d_{\;\;abc}, 
   \\
   & & \tilde R_{ab} - {1 \over 2} \tilde R \tilde g_{ab} 
       = 8 \pi G \tilde T_{ab} - \Lambda \tilde g_{ab}.
\eea
The Weyl (conformal) curvature is introduced as
\bea
   & & \tilde C_{abcd} \equiv \tilde R_{abcd} - {1 \over 2} \left(
       \tilde g_{ac} \tilde R_{bd} + \tilde g_{bd} \tilde R_{ac} 
       - \tilde g_{bc} \tilde R_{ad} - \tilde g_{ad} \tilde R_{bc} \right)
       + {\tilde R \over 6} \left( \tilde g_{ac} \tilde g_{bd} 
       - \tilde g_{ad} \tilde g_{bc} \right).
   \label{Weyl}
\eea
The electric and magnetic parts of the Weyl curvature are introduced as:
\bea
   & & \tilde E_{ab} \equiv \tilde C_{acbd} \tilde u^c \tilde u^d, \quad
       \tilde H_{ab} \equiv {1 \over 2} \tilde \eta_{ac}^{\;\;\;\;ef} 
       \tilde C_{efbd} \tilde u^c \tilde u^d.
   \label{EM-Weyl}
\eea
The energy-momentum tensor is decomposed into fluid quantities
based on the four-vector field $\tilde u^a$ as
\bea
   & & \tilde T_{ab} \equiv \tilde \mu \tilde u_a \tilde u_b
       + \tilde p \left( \tilde g_{ab} + \tilde u_a \tilde u_b \right)
       + \tilde q_a \tilde u_b + \tilde q_b \tilde u_a
       + \tilde \pi_{ab},
   \label{Tab}
\eea
where
\bea
   & & \tilde u^a \tilde q_a = 0 = \tilde u^a \tilde \pi_{ab}, \quad
       \tilde \pi_{ab} = \tilde \pi_{ba}, \quad
       \tilde \pi^a_a = 0.
   \label{q-pi-relations}
\eea
The variables $\tilde \mu$, $\tilde p$, $\tilde q_a$ and $\tilde \pi_{ab}$
are the energy density, the isotropic pressure (including the entropic one),
the energy flux and the anisotropic pressure based on 
$\tilde u_a$-frame, respectively.
We have
\bea
   & & \tilde \mu \equiv \tilde T_{ab} \tilde u^a \tilde u^b, \quad
       \tilde p \equiv {1 \over 3} \tilde T_{ab} \tilde h^{ab}, \quad
       \tilde q_a \equiv - \tilde T_{cd} \tilde u^c \tilde h_a^d, \quad
       \tilde \pi_{ab} \equiv \tilde T_{cd} \tilde h_a^c \tilde h_b^d 
       - \tilde p \tilde h_{ab}.
   \label{fluid-Tab}
\eea

The specific entropy $\tilde S$ can be introduced by 
$\tilde T d \tilde S = d \tilde \varepsilon + \tilde p_T d \tilde v$
where $\tilde \varepsilon$ is specific internal energy density
with $\tilde \mu = \tilde \varrho ( 1 + \tilde \varepsilon )$,
$\tilde p_T$ the thermodynamic pressure, $\tilde v \equiv 1/\tilde \varrho$
the specific volume, and $\tilde T$ the temperature.
We have the isotropic pressure $\tilde p = \tilde p_T + \tilde \pi$
where $\tilde \pi$ is the entropic pressure.
Using eq. (\ref{cov-E-conserv}) below we can show
\bea
   & & \tilde \varrho \tilde T \tilde {\dot {\tilde S}}
       = - \left( \tilde \pi \tilde \theta
       + \tilde \pi^{ab} \tilde \sigma_{ab}
       + \tilde q^a_{\;\;;a} + \tilde q^a \tilde a_a \right).
   \label{entropy}
\eea
Thus, we notice that $\tilde \pi$, $\tilde \pi^{ab}$ and $\tilde q^a$
generate the entropy.
Using a four-vector $\tilde S^a \equiv \tilde \varrho \tilde u^a \tilde S
+ {1 \over \tilde T} \tilde q^a$ which is termed the entropy flow density
\cite{Ehlers-1993} we can derive 
\bea
   & & \tilde S^a_{\;\; ;a}
       = - {1 \over \tilde T^2} \left( \tilde T_{,a} 
       + \tilde T \tilde a_a \right) \tilde q^a
       - {1 \over \tilde T} \left( \tilde \pi \tilde \theta
       + \tilde \pi^{ab} \tilde \sigma_{ab} \right).
\eea

The covariant formulation provides a useful complement to the ADM formulation.
We summarize the covariant $(1+3)$ set of equations in the following.
{}For details, see \cite{Ehlers-1993,Ellis-1971} 
and the Appendix in \cite{HV-1990}.

\noindent
The energy and the momentum conservation equations:
\bea
   & & \tilde {\dot {\tilde \mu}} 
       + \left( \tilde \mu + \tilde p \right) \tilde \theta 
       + \tilde \pi^{ab} \tilde \sigma_{ab} 
       + \tilde q^a_{\;\;;a} + \tilde q^a \tilde a_a = 0, 
   \label{cov-E-conserv} \\
   & & \left( \tilde \mu + \tilde p \right) \tilde a_a 
       + \tilde h^b_a \left( \tilde p_{,b} + \tilde \pi^c_{b;c} 
       + \tilde {\dot {\tilde q}}_b \right) 
       + \left( \tilde \omega_{ab} + \tilde \sigma_{ab} 
       + {4 \over 3} \tilde \theta \tilde h_{ab} \right) \tilde q^b = 0.
   \label{cov-Mom-conserv} 
\eea
Raychaudhuri equation:
\bea
   & & \tilde {\dot {\tilde \theta}} + {1 \over 3} \tilde \theta^2 
       - \tilde a^a_{\;\; ;a} 
       + 2 \left( \tilde \sigma^2 - \tilde \omega^2 \right)
       + 4 \pi G \left( \tilde \mu + 3 \tilde p \right) - \Lambda = 0.
   \label{Raychaudhury-eq} 
\eea
Vorticity propagation:
\bea
   & & \tilde h^a_b \tilde {\dot {\tilde \omega}}{}^b 
       + {2 \over 3} \tilde \theta \tilde \omega^a 
       = \tilde \sigma^a_b \tilde \omega^b
       + {1 \over 2} \tilde \eta^{abcd} \tilde u_b \tilde a_{c;d}.
   \label{cov-vorticity-prop} 
\eea
Shear propagation:
\bea
   & & \tilde h_a^c \tilde h_b^d \left( \tilde {\dot {\tilde \sigma}}_{cd} 
       - \tilde a_{(c;d)} \right)
       - \tilde a_a \tilde a_b + \tilde \omega_a \tilde \omega_b 
       + \tilde \sigma_{ac} \tilde \sigma^c_b 
       + {2 \over 3} \tilde \theta \tilde \sigma_{ab}
       - {1 \over 3} \tilde h_{ab} \left( \tilde \omega^2 
       + 2 \tilde \sigma^2
       - \tilde a^c_{\;\; ;c} \right) + \tilde E_{ab}
       - 4 \pi G \tilde \pi_{ab} = 0.
   \label{cov-shear-prop} 
\eea
Three constraint equations:
\bea
   & & \tilde h_{ab} \left( \tilde \omega^{bc}_{\;\;\; ;c} 
       - \tilde \sigma^{bc}_{\;\;\; ;c} 
       + {2 \over 3} \tilde \theta^{;b} \right)
       + \left( \tilde \omega_{ab} + \tilde \sigma_{ab} \right) \tilde a^b 
       = 8 \pi G \tilde q_a,
   \label{cov-constr-1} \\
   & & \tilde \omega^a_{\;\; ;a} = 2 \tilde \omega^b \tilde a_b,
   \label{cov-constr-2} \\
   & & \tilde H_{ab} 
       = 2 \tilde a_{(a} \tilde \omega_{b)} 
       - \tilde h_a^c \tilde h_b^d \left( \tilde \omega_{(c}^{\;\;\; e;f}
       + \tilde \sigma_{(c}^{\;\;\; e;f} \right) \tilde \eta_{d)gef} \tilde u^g.
   \label{cov-constr-3} 
\eea
{}Four quasi-Maxwellian equations:
\bea
   & & \tilde h^a_b \tilde h^c_d \tilde E^{bd}_{\;\;\;\; ;c} 
       - \tilde \eta^{abcd} \tilde u_b \tilde \sigma_c^e \tilde H_{de}
       + 3 \tilde H^a_b \tilde \omega^b 
       = 4 \pi G \left( {2 \over 3} \tilde h^{ab} \tilde \mu_{,b} 
       - \tilde h^a_b \tilde \pi^{bc}_{\;\;\;\; ;c}
       - 3 \tilde \omega^a_{\;\;b} \tilde q^b 
       + \tilde \sigma^a_b \tilde q^b
       + \tilde \pi^a_b \tilde a^b 
       - {2 \over 3} \tilde \theta \tilde q^a \right),
   \label{cov-Maxwell-1} \\
   & & \tilde h^a_b \tilde h^c_d \tilde H^{bd}_{\;\;\;\; ;c} 
       + \tilde \eta^{abcd} \tilde u_b \tilde \sigma_c^e \tilde E_{de}
       - 3 \tilde E^a_b \tilde \omega^b 
       = 4 \pi G \left\{ 2 \left( \tilde \mu 
       + \tilde p \right) \tilde \omega^a 
       + \tilde \eta^{abcd} \tilde u_b \left[ \tilde q_{c;d}
       + \tilde \pi_{ce} \left( \tilde \omega^e_{\;\; d} 
       + \tilde \sigma^e_{\;\; d} \right) \right] \right\},
   \label{cov-Maxwell-2} \\
   & &  \tilde h^a_c \tilde h^b_d \tilde {\dot {\tilde E}} {}^{cd} 
        + \left( \tilde H^f_{d;e} \tilde h_f^{(a} 
        - 2 \tilde a_d \tilde H_e^{(a} \right) \tilde \eta^{b)cde} \tilde u_c 
        + \tilde h^{ab} \tilde \sigma^{cd} \tilde E_{cd}
        + \tilde \theta \tilde E^{ab} 
        - \tilde E_c^{(a} \left( 3 \tilde \sigma^{b)c} 
        + \tilde \omega^{b)c} \right)
       = 4 \pi G \Big[ 
       - \left( \tilde \mu + \tilde p \right) \tilde \sigma^{ab} 
   \nonumber \\
   & & \qquad
       - 2 \tilde a^{(a} \tilde q^{b)}
       - \tilde h^{(a}_c \tilde h^{b)}_d \left( \tilde q^{c;d} 
       + \tilde {\dot {\tilde \pi}} {}^{cd} \right)
       - \left( \tilde \omega_c^{\;\;(a} 
       + \tilde \sigma^{(a}_c \right) \tilde \pi^{b)c} 
       - {1 \over 3} \tilde \theta \tilde \pi^{ab} 
       + {1 \over 3} \left( \tilde q^c_{\;\; ;c} 
       + \tilde a_c \tilde q^c
       + \tilde \pi^{cd} \tilde \sigma_{cd} \right) \tilde h^{ab} \Big],
   \label{cov-Maxwell-3} \\
   & & \tilde h^a_c \tilde h^b_d \tilde {\dot {\tilde H}} {}^{cd} 
       - \left( \tilde E^f_{d;e} \tilde h_f^{(a} 
       - 2 \tilde a_d \tilde E_e^{(a} \right) \tilde \eta^{b)cde} \tilde u_c 
       + \tilde h^{ab} \tilde \sigma^{cd} \tilde H_{cd}
       + \tilde \theta \tilde H^{ab} 
       - \tilde H_c^{(a} \left( 3 \tilde \sigma^{b)c} 
       + \tilde \omega^{b)c} \right)
   \nonumber \\
   & & \qquad
       = 4 \pi G \left[ 
       \left( \tilde q_e \tilde \sigma^{(a}_d 
       - \tilde \pi^f_{d;e} \tilde h_f^{(a} \right)
       \tilde \eta^{b)cde} \tilde u_c 
       + \tilde h^{ab} \tilde \omega_c \tilde q^c 
       - 3 \tilde \omega^{(a} \tilde q^{b)} \right].
   \label{cov-Maxwell-4} 
\eea
Evaluated in the normal-frame eqs. 
(\ref{cov-E-conserv},\ref{cov-Mom-conserv},\ref{Raychaudhury-eq},\ref{cov-shear-prop},\ref{cov-constr-1})
reproduce eqs. 
(\ref{E-conservation},\ref{Mom-conservation},\ref{Trace-prop},\ref{Tracefree-prop},\ref{Mom-constraint}) 
in the ADM formulation.

Now, we take the normal-frame vector, thus $\tilde u_a = \tilde n_a$
with $\tilde n_\alpha \equiv 0$, thus $\tilde \omega_{ab} = 0$.
The trace and tracefree parts of the Gauss equation give \cite{HV-1990}:
\bea
   & & \tilde R^{(3)} = 2 \left( - {1 \over 3} \tilde \theta^2 
       + \tilde \sigma^2 + 8 \pi G \tilde \mu + \Lambda \right),
   \label{cov-Gauss-trace} \\
   & & \tilde R^{(3)}_{ab} - {1 \over 3} \tilde R^{(3)} \tilde h_{ab}
       = \tilde h^c_a \tilde h^d_b \left(
       - \tilde {\dot {\tilde \sigma}}_{cd} 
       - \tilde \theta \tilde \sigma_{cd} + \tilde a_{(c;d)} \right)
       + \tilde a_a \tilde a_b 
       - {1 \over 3} \tilde h_{ab} \tilde a^c_{\;\; ;c} 
       + 8 \pi G \tilde \pi_{ab},
   \label{cov-Gauss-tracefree} 
\eea
where $\tilde R^{(3)}_{ab}$ and $\tilde R^{(3)}$ are the Ricci and
scalar curvatures of the hypersurface normal to $\tilde n_a$;
for an arbitrary vector $\tilde V_a$ we have
\bea
   & & \tilde R^{(3)}_{abcd} \tilde V^b
       \equiv 2 \tilde \nabla^{(3)}_{[c} \tilde \nabla^{(3)}_{d]} \tilde V_a
       \equiv 2 \tilde h_c^e \tilde h_d^f \tilde h_a^g \tilde \nabla_{[e}
       \left( \tilde h^h_{f]} \tilde h^i_g \tilde \nabla_h \tilde V_i \right), 
       \quad
       \tilde R^{(3)}_{ab} \equiv \tilde h^{cd} \tilde R^{(3)}_{cadb}, \quad
       \tilde R^{(3)} \equiv \tilde h^{ab} \tilde R^{(3)}_{ab}.
\eea
{}From this we have
\bea
   & & \tilde R^{(3)}_{abcd} 
       = \tilde h_a^e \tilde h_b^f \tilde h_c^g \tilde h_d^h \tilde R_{efgh} 
       - \tilde \theta_{ca} \tilde \theta_{db}
       + \tilde \theta_{bc} \tilde \theta_{ad},
\eea
which is the Gauss equation.
We can show that
$\tilde R^{(3)}_{\alpha\beta\gamma\delta} = R^{(h)}_{\alpha\beta\gamma\delta}$.
Equation (\ref{cov-Gauss-tracefree}) follows from eq. (\ref{cov-shear-prop})
evaluated in the normal-frame.
Using eqs. (\ref{Weyl},\ref{EM-Weyl}) we can show that 
eq. (\ref{cov-Gauss-tracefree})
reproduces eq. (\ref{Tracefree-prop}) in the ADM formulation.
Equation (\ref{cov-Gauss-trace}) gives eq. (\ref{E-constraint}) 
in the ADM formulation.

Compared with the ADM equations in (\ref{E-constraint}-\ref{Mom-conservation})
part of the covariant equations in 
(\ref{cov-vorticity-prop},\ref{cov-constr-2},\ref{cov-constr-3},\ref{cov-Maxwell-1}-\ref{cov-Maxwell-4})
look new.
In the normal-frame eqs. (\ref{cov-vorticity-prop},\ref{cov-constr-2}) are
identically satisfied; using eqs.
(\ref{kinematic-shear},\ref{connection-ADM},\ref{n_a-def}) we can show
\bea
   & & \tilde a_\alpha = \left( \ln{N} \right)_{,\alpha},
\eea
thus $\tilde a_{[\beta;\gamma]} = 0$.
Still, eqs. (\ref{E-constraint}-\ref{Mom-conservation}) provide a complete set.
These additional equations in the covariant forms should be regarded as 
complementary equations which could possibly show certain aspects 
of the system better.
In our perturbation analyses we will use parts of these equations as complement
ones.
Although the covariant set of equations is based on the general frame vector,
this does not add any new physics which is not covered by the normal-frame 
taken in the ADM formulation, see \S \ref{sec:Frame}.

The covariant equations for the scalar fields, generalized gravity, 
electromagnetic field, null geodesic, and Boltzmann equation 
will be introduced individually in the corresponding sections later.

\subsection{Multi-component situation}

In the multi-component situation we have
\bea
   & & \tilde T_{ab} = \sum_l \tilde T_{(l)ab}; \quad
       \tilde T_{(i)a;b}^{\;\;\;\; b} \equiv \tilde I_{(i)a}, \quad
       \sum_l \tilde I_{(l)a} = 0.
   \label{Tab-i}
\eea
Based on the normal-frame vector, we have
\bea
   & & \tilde \mu = \sum_l \tilde \mu_{(l)}, \quad
       \tilde p = \sum_l \tilde p_{(l)}, \quad
       \tilde q_a = \sum_l \tilde q_{(l)a}, \quad
       \tilde \pi_{ab} = \sum_l \tilde \pi_{(l)ab}.
   \label{fluids-sum}
\eea

The ADM formulation is based on the normal-frame vector 
$\tilde u_a = \tilde n_a$.
The ADM fluid quantities in eq. (\ref{ADM-fluid-def}) correspond
to the fluid quantities based on the normal-frame vector as
\bea
   & & E = \tilde \mu, \quad
       S = 3 \tilde p, \quad
       J_\alpha = \tilde q_\alpha, \quad
       \bar S_{\alpha\beta} = \tilde \pi_{\alpha\beta}.
   \label{ADM-normal-fluid}
\eea
{}From eq. (\ref{ADM-fluid-def}) or eq. (\ref{ADM-normal-fluid}) we have
\bea
   & & E = \sum_l E_{(l)}, \quad
       S = \sum_l S_{(l)}, \quad
       J_\alpha = \sum_l J_{(l)\alpha}, \quad
       S_{\alpha\beta} = \sum_l S_{(l)\alpha\beta}. 
   \label{ADM-fluid-sum}
\eea
Equation (\ref{Tab-i}) gives
\bea
   & & E_{(i),0} N^{-1} - E_{(i),\alpha} N^\alpha N^{-1} 
       - K \left( E_{(i)} + {1 \over 3} S_{(i)} \right)
       - \bar S^{\alpha\beta}_{(i)} \bar K_{\alpha\beta}
       + N^{-2} \left( N^2 J^\alpha_{(i)} \right)_{:\alpha} 
       = - {1 \over N} \left( \tilde I_{(i)0} 
       - \tilde I_{(i)\alpha} N^\alpha \right),
   \label{E-conservation-i} \\
   & & J_{(i)\alpha,0} N^{-1} - J_{(i)\alpha:\beta} N^\beta N^{-1}
       - J_{(i)\beta} N^\beta_{\;\;\;:\alpha} N^{-1} - K J_{(i)\alpha}
       + E_{(i)} N_{,\alpha} N^{-1} 
       + S^{\;\;\;\;\beta}_{(i)\alpha:\beta}
       + S_{(i)\alpha}^{\;\;\;\;\beta} N_{,\beta} N^{-1} 
       = \tilde I_{(i)\alpha}.
   \label{Mom-conservation-i}
\eea
The ADM equations in eqs. (\ref{E-constraint}-\ref{Mom-conservation}) 
remain valid, with the above additional equations of motion 
for the individual component.
Thus, in the multi-component situation eqs. 
(\ref{E-constraint}-\ref{Mom-conservation},\ref{ADM-fluid-sum}-\ref{Mom-conservation-i})
provide a complete set.

\section{Perturbed quantities}
                                            \label{sec:perturbed-quantities}

\subsection{Metric and connections}

We use the following convention for the metric variables:
\bea
   & & \tilde g_{00} \equiv - a^2 \left( 1 + 2 A \right), \quad
       \tilde g_{0\alpha} \equiv - a^2 B_\alpha, \quad
       \tilde g_{\alpha\beta} \equiv a^2 \left( g^{(3)}_{\alpha\beta}
       + 2 C_{\alpha\beta} \right),
   \label{metric-def}
\eea
where $A$, $B_\alpha$ and $C_{\alpha\beta}$ are perturbed order variables
and are {\it assumed} to be based on $g^{(3)}_{\alpha\beta}$ as the metric.
To the second-order, we can write the perturbation variables explicitly as:
\bea
   & & A \equiv A^{(1)} + A^{(2)}, \quad
       B_\alpha \equiv B_\alpha^{(1)} + B_\alpha^{(2)}, \quad
       C_{\alpha\beta} \equiv C_{\alpha\beta}^{(1)} + C_{\alpha\beta}^{(2)}.
   \label{metric-expansion}
\eea
As we are interested in the perturbation to the second-order,
as our ansatz, we include up to second-order (quadratic) terms
in the deviation from the Friedmann background.
This can be extended to any higher-order perturbation as long as we take
the perturbative approach where the lower-order solutions
drive (work as sources for) the next higher-order variables.
Thus, in this work we ignore the terms higher than quadratic 
(second-order) combination 
of the perturbed metric ($A$, $B_\alpha$, $C_{\alpha\beta}$), the perturbed
fluid quantities ($\delta \mu$, $\delta p$, $Q_\alpha$, $\Pi_{\alpha\beta}$)
to be introduced in eq. (\ref{fluid-def}), 
the perturbed field ($\delta \phi$)
to be introduced in eq. (\ref{field-perturbation}), etc.

The inverse metric expanded to the second-order in perturbation variables is:
\bea
   \tilde g^{00} 
   &=& {1 \over a^2} \left( - 1 + 2 A - 4 A^2 + B_\alpha B^\alpha \right), 
   \nonumber \\
   \tilde g^{0\alpha} 
   &=& {1 \over a^2} \left( - B^\alpha + 2 A B^\alpha 
       + 2 B_\beta C^{\alpha\beta} \right), 
   \nonumber \\
   \tilde g^{\alpha\beta} 
   &=& {1 \over a^2} \left( g^{(3)\alpha\beta} - 2 C^{\alpha\beta} 
       - B^\alpha B^\beta
       + 4 C^\alpha_\gamma C^{\beta\gamma} \right).
   \label{inverse-metric}
\eea    
The connections are:
\bea
   \tilde \Gamma^0_{00} 
   &=& {a^\prime \over a} + A^\prime - 2 A A^\prime
       - A_{,\alpha} B^\alpha + B_\alpha \left( B^{\alpha\prime}
       + {a^\prime \over a} B^\alpha \right),
   \nonumber \\
   \tilde \Gamma^0_{0\alpha} 
   &=& A_{,\alpha} - {a^\prime \over a} B_\alpha
       - 2 A A_{,\alpha} + 2 {a^\prime \over a} A B_\alpha
       - B_\beta C^{\beta\prime}_\alpha + B^\beta B_{[\beta|\alpha]} ,
   \nonumber \\
   \tilde \Gamma^\alpha_{00} 
   &=& A^{|\alpha} - B^{\alpha\prime} - {a^\prime \over a} B^\alpha 
       + A^\prime B^\alpha
       - 2 A_{,\beta} C^{\alpha\beta} + 2 C^\alpha_\beta
       \left( B^{\beta\prime} + {a^\prime \over a} B^\beta \right),
   \nonumber \\
   \tilde \Gamma^0_{\alpha\beta} 
   &=& {a^\prime \over a} g^{(3)}_{\alpha\beta}
       - 2 {a^\prime \over a} g^{(3)}_{\alpha\beta} A
       + B_{(\alpha|\beta)}
       + C^\prime_{\alpha\beta} + 2 {a^\prime \over a} C_{\alpha\beta} 
   \nonumber \\
   & & + {a^\prime \over a } g^{(3)}_{\alpha\beta}
       \left( 4 A^2 - B_\gamma B^\gamma \right)
       - 2 A \left( B_{(\alpha|\beta)} + C^\prime_{\alpha\beta}
       + 2 {a^\prime \over a} C_{\alpha\beta} \right)
       - B_\gamma \left( 2 C^\gamma_{(\alpha|\beta)}
       - C_{\alpha\beta}^{\;\;\;\;|\gamma} \right),
   \nonumber \\
   \tilde \Gamma^\alpha_{0\beta} 
   &=& {a^\prime \over a} \delta^\alpha_\beta
       + {1\over 2} \left( B_\beta^{\;\;|\alpha}
       - B^\alpha_{\;\;|\beta} \right) + C^{\alpha\prime}_\beta
       + B^\alpha \left( A_{,\beta} - {a^\prime \over a} B_\beta \right)
       + 2 C^{\alpha\gamma} \left( B_{[\gamma|\beta]}
       - C_{\gamma \beta}^\prime \right),
   \nonumber \\
   \tilde \Gamma^\alpha_{\beta\gamma} 
   &=& \Gamma^{(3)\alpha}_{\;\;\;\;\;\beta\gamma}
       + {a^\prime \over a} g^{(3)}_{\beta\gamma} B^\alpha
       + 2 C^\alpha_{(\beta|\gamma)}
       - C_{\beta\gamma}^{\;\;\;\;|\alpha}
       - 2 C^\alpha_\delta \left( 2 C^\delta_{(\beta|\gamma)}
       - C_{\beta\gamma}^{\;\;\;\;|\delta} \right)
   \nonumber \\
   & & - 2 {a^\prime \over a} g^{(3)}_{\gamma\beta}
       \left( A B^\alpha + B^\delta C^\alpha_\delta \right)
       + B^\alpha \left( B_{(\beta|\gamma)}
       + C_{\beta\gamma}^\prime
       + 2 {a^\prime \over a} C_{\beta\gamma} \right),
   \label{connections}
\eea
where a vertical bar indicates a covariant derivative based on
$g^{(3)}_{\alpha\beta}$.
An index $0$ indicates the conformal time $\eta$,
and a prime indicates a time derivative with respect to $\eta$. 
The components of the frame four-vector $\tilde u_a$ are introduced as:
\bea
   & & \tilde u^0 \equiv {1 \over a} \left( 1 - A + {3\over 2} A^2 + {1\over 2}
       V^\alpha V_\alpha - V^\alpha B_\alpha \right), \quad
       \tilde u^\alpha \equiv {1 \over a} V^\alpha,
   \nonumber \\
   & & \tilde u_0 = - a \left( 1 + A - {1\over 2} A^2
       + {1\over 2} V^\alpha V_\alpha \right), \quad
       \tilde u_\alpha = a \left( V_\alpha - B_\alpha + A B_\alpha
       + 2 V^\beta C_{\alpha\beta} \right),
   \label{u-def}
\eea
where $V^\alpha$ is based on $g^{(3)}_{\alpha\beta}$.

\subsection{Normal-frame quantities}
                                   \label{sec:normal-frame-quantities}

The normal-frame vector $\tilde n_a$ has a property $\tilde n_\alpha \equiv 0$.
Thus we have
\bea
   & & \tilde n^0 \equiv {1 \over a} \left( 1 - A + {3\over 2} A^2 - {1\over 2}
       B^\alpha B_\alpha \right), \quad
       \tilde n^\alpha \equiv {1 \over a} \left( B^\alpha - A B^\alpha
       - 2 B^\beta C^\alpha_\beta \right), 
   \nonumber \\
   & & \tilde n_0 = - a \left( 1 + A - {1\over 2} A^2 
       + {1\over 2} B^\alpha B_\alpha \right), \quad
       \tilde n_\alpha = 0.
   \label{normal-vector}
\eea
{}Using eqs. (\ref{ADM-metric-def},\ref{n_a-def}) 
the ADM metric variables become:
\bea
   & & N = a \left( 1 + A - {1\over 2} A^2 
       + {1\over 2} B^\alpha B_\alpha \right), \quad
       N_\alpha = - a^2 B_\alpha, \quad
       N^\alpha = - B^\alpha + 2 B^\beta C^\alpha_\beta, 
   \nonumber \\
   & & h_{\alpha\beta} = a^2 \left( g^{(3)}_{\alpha\beta}
       + 2 C_{\alpha\beta} \right), \quad
       h^{\alpha\beta} = {1 \over a^2} \left( g^{(3)\alpha\beta}
       - 2 C^{\alpha\beta} + 4 C^\alpha_\gamma C^{\beta\gamma} \right).
   \label{ADM-metric-pert}
\eea
The connection becomes
\bea 
   & & \Gamma^{(h)\gamma}_{\;\;\;\;\;\alpha\beta}
       = \Gamma^{(3)\gamma}_{\;\;\;\;\;\alpha\beta} + \left( g^{(3)\gamma\delta}
       - 2 C^{\gamma\delta} \right) \left( C_{\delta\alpha|\beta}
       + C_{\delta\beta|\alpha} - C_{\alpha\beta|\delta} \right). 
\eea

\noindent
The extrinsic curvature in eq. (\ref{extrinsic-curvature-def}) gives:
\bea
   K_{\alpha\beta} 
   &=& 
       - a \Bigg[ \left( {a^\prime \over a} 
       g^{(3)}_{\alpha\beta} + B_{(\alpha|\beta)} + C^\prime_{\alpha\beta} 
       + 2 {a^\prime \over a} C_{\alpha\beta} \right) (1 - A ) 
       + {1\over 2} {a^\prime \over a} g^{(3)}_{\alpha\beta} \left( 3 A^2
       - B_\gamma B^\gamma \right) 
       - B_\gamma \left( 2 C^\gamma_{(\alpha|\beta)} 
       - C_{\alpha\beta}^{\;\;\;\;|\gamma} \right) \Bigg], 
   \nonumber \\
   K 
   &=& 
       - {1 \over a} \Bigg[ \left( 3 {a^\prime \over a} 
       + B^\alpha_{\;\;|\alpha} + C^{\alpha\prime}_\alpha \right) ( 1 - A )
       + {3 \over 2} {a^\prime \over a} \left( 3 A^2 
       - B^\alpha B_{\alpha} \right) 
       - B^\beta \left( 2 C^\alpha_{\beta|\alpha} 
       - C^\alpha_{\alpha|\beta} 
       \right) - 2 C^{\alpha\beta} \left( C^\prime_{\alpha\beta} 
       + B_{\alpha|\beta} \right) \Bigg],
   \nonumber \\
   \bar K_{\alpha\beta} 
   &=& 
       - a \Bigg\{ 
       \left( B_{(\alpha|\beta)} + C^\prime_{\alpha\beta} \right) (1 - A ) 
       - B_\gamma \left( 2 C^\gamma_{(\alpha|\beta)} 
       - C_{\alpha\beta}^{\;\;\;\;|\gamma} \right) 
       - {2 \over 3} C_{\alpha\beta} \left( B^\gamma_{\;\;|\gamma}
       + C^{\gamma\prime}_\gamma \right)
   \nonumber \\
   & & 
       - {1 \over 3} g^{(3)}_{\alpha\beta} 
       \left[ \left( B^\gamma_{\;\;|\gamma} + C^{\gamma\prime}_\gamma \right)
       \left( 1 - A \right)
       - B^\gamma \left( 2 C^\delta_{\gamma|\delta} 
       - C^\delta_{\delta|\gamma} \right)
       - 2 C^{\gamma\delta} \left( C^\prime_{\gamma\delta}
       + B_{\gamma|\delta} \right) \right]
       \Bigg\}.
   \label{Extrinsic-curvature}
\eea
The intrinsic curvature in eq. (\ref{ADM-curvature}) becomes:
\bea
   R^{(h)}_{\alpha\beta} 
   &=& 
       R^{(3)}_{\alpha\beta} + \left( g^{(3)\gamma\delta} 
       - 2 C^{\gamma\delta}
       \right) \left( C_{\delta\alpha|\beta\gamma} 
       + C_{\delta\beta|\alpha\gamma}
       - C_{\alpha\beta|\delta\gamma} - C_{\delta\gamma|\alpha\beta} \right) 
       + 2 C^{\gamma\delta}_{\;\;\;\;|\beta} C_{\gamma\delta|\alpha}
   \nonumber \\
   & & 
       - \left( 2 C^\gamma_{\delta|\gamma} - C^\gamma_{\gamma|\delta} \right)
       \left( C^\delta_{\alpha|\beta} + C^\delta_{\beta|\alpha}
       - C_{\alpha\beta}^{\;\;\;\;|\delta} \right) 
       - \left( C^\delta_{\alpha|\gamma} + C^\delta_{\gamma|\alpha}
       - C_{\alpha\gamma}^{\;\;\;\;|\delta} \right)
       \left( C^\gamma_{\beta|\delta} + C^\gamma_{\delta|\beta}
       - C_{\beta\delta}^{\;\;\;\;|\gamma} \right),
   \nonumber \\
   R^{(h)} 
   &=& 
       {1 \over a^2} \Bigg[ R^{(3)} - 2 C^{\alpha\beta} 
       R^{(3)}_{\alpha\beta} + 2 C^{\beta|\alpha}_{\alpha\;\;\;\beta}
       - 2 C^{\alpha|\beta}_{\alpha\;\;\;\beta} 
       + 4 C^\alpha_\gamma C^{\beta\gamma} R^{(3)}_{\alpha\beta} 
       + 4 C^{\alpha\beta} \left( - C^\gamma_{\alpha|\beta\gamma}
       - C^{\gamma}_{\alpha|\gamma\beta}
       + C^{\;\;\;\;|\gamma}_{\alpha\beta\;\;\;\gamma}
       + C^\gamma_{\gamma|\alpha\beta} \right) 
   \nonumber \\
   & & 
       - \left( 2 C^\gamma_{\beta|\gamma} - C^\gamma_{\gamma|\beta} \right)
       \left( 2 C^{\alpha\beta}_{\;\;\;\;|\alpha} - C^{\alpha|\beta}_\alpha 
       \right) 
       + C^{\alpha\beta|\gamma} \left( 3 C_{\alpha\beta|\gamma}
       - 2 C_{\alpha\gamma|\beta} \right)
       \Bigg],
   \label{Intrinsic-curvature}
\eea
where
\bea
   & & R^{(3)\alpha}_{\;\;\;\;\;\;\;\beta\gamma\delta}
       = {1 \over 6} R^{(3)} \left( \delta^\alpha_\gamma g^{(3)}_{\beta\delta}
       - \delta^\alpha_\delta g^{(3)}_{\beta\gamma} \right), \quad
       R^{(3)}_{\alpha\beta} = {1 \over 3} R^{(3)} g^{(3)}_{\alpha\beta}, \quad
       R^{(3)} = 6 K,
\eea
with a normalized $K (= 0, \pm 1)$, the sign of the background three-space 
curvature.
Thus,
\bea
   \bar R^{(h)\alpha}_{\;\;\;\;\;\beta}
   &=& 
       {1 \over a^2} \Bigg\{
       C^{\alpha\gamma}_{\;\;\;\;\;|\beta\gamma}
       + C^{\gamma|\alpha}_{\beta\;\;\;\;\gamma}
       - C^{\alpha|\gamma}_{\beta\;\;\;\;\gamma} 
       - C^{\gamma|\alpha}_{\gamma\;\;\;\;\beta} 
       - {2 \over 3} R^{(3)} C^\alpha_\beta
   \nonumber \\
   & & 
       - 2 C^{\gamma\delta} 
       \left( C^\alpha_{\delta|\beta\gamma}
       + C^{\;\;\;\;|\alpha}_{\delta\beta\;\;\;\;\gamma}
       - C^\alpha_{\beta|\delta\gamma} 
       - C^{\;\;\;\;|\alpha}_{\delta\gamma\;\;\;\;\beta} \right)
       - 2 C^{\alpha\gamma} \left( 
       C^\delta_{\gamma|\beta\delta}
       + C^\delta_{\beta|\gamma\delta}
       - C^{\;\;\;\;|\delta}_{\beta\gamma\;\;\;\delta}
       - C^\delta_{\delta|\gamma\beta} \right)
       + {4\over 3} R^{(3)} C^\alpha_\gamma C^\gamma_\beta
   \nonumber \\
   & & 
       - \left( 2 C^\gamma_{\delta|\gamma}
       - C^\gamma_{\gamma|\delta} \right)
       \left( C^{\alpha\delta}_{\;\;\;\;|\beta}
       + C^{\delta|\alpha}_\beta
       - C^{\alpha|\delta}_\beta \right) 
       + C_{\gamma\delta|\beta} C^{\gamma\delta|\alpha}
       + 2 C^{\alpha\gamma|\delta} \left( C_{\beta\gamma|\delta}
       - C_{\beta\delta|\gamma} \right) 
   \nonumber \\
   & & 
       - {1\over 3} \delta^\alpha_\beta 
       \Bigg[ - {2 \over 3} R^{(3)} C^\gamma_\gamma 
       + 2 C^{\delta|\gamma}_{\gamma\;\;\;\;\delta}
       - 2 C^{\gamma|\delta}_{\gamma\;\;\;\;\delta} 
       + 4 C^{\gamma\delta} \left( - C^\epsilon_{\gamma|\delta\epsilon}
       - C^{\epsilon}_{\gamma|\epsilon\delta}
       + C^{\;\;\;\;|\epsilon}_{\gamma\delta\;\;\;\;\epsilon}
       + C^\epsilon_{\epsilon|\gamma\delta} \right) 
   \nonumber \\
   & & 
       + {4 \over 3} R^{(3)} C^\delta_\gamma C^\gamma_\delta 
       - \left( 2 C^\epsilon_{\delta|\epsilon} 
       - C^\epsilon_{\epsilon|\delta} \right)
       \left( 2 C^{\gamma\delta}_{\;\;\;\;|\gamma} 
       - C^{\gamma|\delta}_\gamma 
       \right) 
       + C^{\gamma\delta|\epsilon} \left( 3 C_{\gamma\delta|\epsilon}
       - 2 C_{\gamma\epsilon|\delta} \right)
       \Bigg] \Bigg\}.
   \label{Ricci}
\eea
It is convenient to have
\bea
   & & B^\alpha_{\;\;\;|\beta\gamma} = B^\alpha_{\;\;\;|\gamma\beta}
       - R^{(3)\alpha}_{\;\;\;\;\;\;\; \delta \beta \gamma} B^\delta, \quad
       B_{\alpha|\beta\gamma} = B_{\alpha|\gamma\beta}
       + R^{(3)\delta}_{\;\;\;\;\;\;\; \alpha \beta \gamma} B_\delta.
   \label{three-curvature}
\eea

\subsection{General ($\tilde u_a$)-frame quantities}

To the second-order perturbation,
using eqs. (\ref{metric-def},\ref{inverse-metric},\ref{u-def}),
the kinematic quantities in eq. (\ref{kinematic-shear}) become:
\bea
   \tilde h_{\alpha\beta} 
   &=& a^2 \left[ g^{(3)}_{\alpha\beta}
       + 2 C_{\alpha\beta} + ( V_\alpha - B_\alpha ) ( V_\beta - B_\beta )
       \right], \quad
       \tilde h_{0\alpha} 
       = - a^2 \left( V_\alpha + A V_\alpha
       + 2 C_{\alpha\beta} V^\beta \right), \quad
       \tilde h_{00} = a^2 V^\alpha V_\alpha, 
   \label{projection-pert} \\
   \tilde \theta 
   &=& a^{-1} \Bigg[ 3 {a^\prime \over a} (1 - A)
       + {V^\alpha}_{|\alpha} + C^{\alpha\prime}_\alpha
       + {9\over 2} {a^\prime \over a} A^2 + B^\alpha B_\alpha^\prime
       - 2 C^{\alpha\beta} C^\prime_{\alpha\beta} - A C^{\alpha\prime}_\alpha
   \nonumber \\
   & & 
       - V^{\alpha\prime} B_\alpha + V^\alpha \left( V^\prime_\alpha
       + {3\over 2} {a^\prime \over a} V_\alpha - B^\prime_\alpha
       - 3 {a^\prime \over a} B_\alpha + A_{,\alpha} + C^\beta_{\beta|\alpha}
       \right) \Bigg],
   \\
   \tilde a_\alpha 
   &=& A_{,\alpha} + V_\alpha^\prime - B_\alpha^\prime
       + {a^\prime \over a} \left(V_\alpha - B_\alpha \right)
       + A^\prime B_\alpha + A \left( -2 A_{,\alpha} + 2 B^\prime_\alpha
       + 2 {a^\prime \over a} B_\alpha - V^\prime_\alpha
       - {a^\prime \over a} V_\alpha \right)
   \nonumber \\
   & & 
       + V^\beta \left( V_{\alpha|\beta} + B_{\beta|\alpha}
       - B_{\alpha|\beta} \right) + 2 C_{\alpha\beta}
       \left( V^{\beta\prime} + {a^\prime \over a} V^\beta \right)
       + 2 C^\prime_{\alpha\beta} V^\beta, \quad
       \tilde a_0 
       = - V^\alpha \tilde a_\alpha,
   \\
   \tilde \omega_{\alpha\beta} 
   &=& a \Big( V_{[\alpha|\beta]}
       - B_{[\alpha|\beta]} + A B_{[\alpha|\beta]}
       - V_{[\alpha} A_{,\beta]} + 2 B_{[\alpha} A_{,\beta]} - B_{[\alpha}
       B^\prime_{\beta]}
   \nonumber \\
   & & 
       - V_{[\alpha} V^\prime_{\beta]} + B_{[\alpha} V^\prime_{\beta]}
       + V_{[\alpha} B^\prime_{\beta]} + 2 V_\gamma C^\gamma_{[\alpha|\beta]}
       + 2 C_{\gamma[\alpha} {V^\gamma}_{|\beta]} \Big), \quad
       \tilde \omega_{0\alpha} 
       = V^\beta \tilde \omega_{\alpha\beta}, \quad
       \tilde \omega_{00} = 0,
   \\
   \tilde \sigma_{\alpha\beta} 
   &=& a \Bigg[ V_{(\alpha|\beta)}
       + C^\prime_{\alpha\beta} - {1\over 3} g^{(3)}_{\alpha\beta}
       \left( {V^\gamma}_{|\gamma} + C^{\gamma\prime}_\gamma \right)
       + V_{(\alpha} V^\prime_{\beta)} - V_{(\alpha} B^\prime_{\beta)}
       - V^\prime_{(\alpha} B_{\beta)} + B_{(\alpha} B^\prime_{\beta)}
   \nonumber \\
   & & 
       + V_{(\alpha} A_{,\beta)} + V_\gamma {C_{\alpha\beta}}^{|\gamma}
       - A C^\prime_{\alpha\beta}
       + 2 C_{\delta(\alpha} {V^\delta}_{|\beta)}
       - {2\over 3} C_{\alpha\beta} \left( {V^\gamma}_{|\gamma}
       + C^{\gamma\prime}_\gamma \right) 
   \nonumber \\
   & & 
       - {1 \over 3} g^{(3)}_{\alpha\beta} \left( V^\gamma V^\prime_\gamma
       - V^\gamma B^\prime_\gamma - V^{\gamma\prime} B_\gamma + B^\gamma
       B^\prime_\gamma + V^\gamma A_{,\gamma} + V_{\delta}
       C_\gamma^{\gamma|\delta} - A C_\gamma^{\gamma\prime} 
       - 2 C^{\delta\gamma} C^\prime_{\delta\gamma} \right)
       \Bigg],
   \nonumber \\
   \tilde \sigma_{0\alpha} 
   &=& - V^\beta \tilde \sigma_{\alpha\beta}, \quad
       \tilde \sigma_{00} = 0.
   \label{shear-pert}
\eea
In the normal-frame we have $\tilde u_\alpha \equiv 0$, thus 
$V_\alpha = B_\alpha - A B_\alpha - 2 B^\beta C_{\alpha\beta}$.
In this frame we have
\bea
   & & \tilde h_{\alpha\beta} = a^2 \left( g^{(3)}_{\alpha\beta}
       + 2 C_{\alpha\beta} \right), \quad
       \tilde h_{0\alpha} = - a^2 B_\alpha, \quad
       \tilde h_{00} = a^2 B^\alpha B_\alpha, 
   \\
   & & \tilde \theta = - K,
   \\
   & & \tilde a_\alpha = \left( \ln{N} \right)_{,\alpha}
       = \left( A - A^2 + {1 \over 2} B^\beta B_\beta \right)_{,\alpha}, 
       \quad
       \tilde a_0 = - B^\alpha A_{,\alpha},
   \label{a-normal} \\
   & & \tilde \sigma_{\alpha\beta} = - \bar K_{\alpha\beta}, \quad
       \tilde \sigma_{0\alpha} = B^\beta \bar K_{\alpha\beta}, \quad
       \tilde \sigma_{00} = 0,
   \\
   & & \tilde \omega_{ab} = 0.
   \label{shear-pert-normal}
\eea
In this frame we have $\tilde \theta = - K$ and 
$\tilde \sigma_{\alpha\beta} = - \bar K_{\alpha\beta}$.
These are natural because $K$ and $\bar K_{\alpha\beta}$ are 
the same as negatives of the expansion scalar and the
shear, respectively, of the normal-frame vector field.

\subsection{Fluid quantities}

To the perturbed order we decompose the fluid quantities as:
\bea
   & & \tilde \mu \equiv \mu + \delta \mu, \quad
       \tilde p \equiv p + \delta p, \quad
       \tilde q_\alpha \equiv a Q_\alpha, \quad
       \tilde \pi_{\alpha\beta} \equiv a^2 \Pi_{\alpha\beta},
   \label{fluid-def}
\eea
where $Q_\alpha$ and $\Pi_{\alpha\beta}$ are based on $g^{(3)}_{\alpha\beta}$.
In the Friedmann world model we have $\tilde \mu = \mu$ and $\tilde p = p$
and zeros for the other fluid quantities.
We have
\bea
   & & \Pi^\alpha_\alpha - 2 C^{\alpha\beta} \Pi_{\alpha\beta} = 0,
\eea
which follows from $\tilde \pi^a_a = 0$ or $\bar S^\alpha_\alpha = 0$.
The perturbed order fluid quantities $\delta \mu$, $\delta p$, 
$Q_\alpha$ and $\Pi_{\alpha\beta}$ in eq. (\ref{fluid-def}) can be 
expanded similarly as in equation (\ref{metric-expansion})
\bea
   & & \delta \mu = \delta \mu^{(1)} + \delta \mu^{(2)}, \quad
       \delta p = \delta p^{(1)} + \delta p^{(2)}, \quad
       Q_\alpha = Q_\alpha^{(1)} + Q_\alpha^{(2)}, \quad
       \Pi_{\alpha\beta} = \Pi_{\alpha\beta}^{(1)} + \Pi_{\alpha\beta}^{(2)}.
   \label{fluids-second-order}
\eea
In the multi-component situation, from eqs. (\ref{fluids-sum},\ref{fluid-def}) 
we set
\bea
   & & \mu = \sum_l \mu_{(l)}, \quad
       p = \sum_l p_{(l)}, 
   \nonumber \\
   & & \delta \mu = \sum_l \delta \mu_{(l)}, \quad
       \delta p = \sum_l \delta p_{(l)}, \quad
       Q_\alpha = \sum_l Q_{(l)\alpha}, \quad
       \Pi_{\alpha\beta} = \sum_l \Pi_{(l)\alpha\beta}.
   \label{fluid-sum}
\eea
Thus, from eq. (\ref{ADM-normal-fluid}) the ADM fluid variables become:
\bea
   & & E \equiv \mu + \delta \mu, 
   \nonumber \\
   & & S \equiv 3 ( p + \delta p ), 
   \nonumber \\
   & & J_\alpha \equiv a Q_\alpha, \quad 
       J^\alpha = {1 \over a} \left( Q^\alpha
       - 2 C^{\alpha\beta} Q_\beta \right), 
   \nonumber \\
   & & \bar S_{\alpha\beta} \equiv a^2 \Pi_{\alpha\beta}, \quad
       \bar S^\alpha_\beta=\Pi^\alpha_\beta 
       - 2 C^{\alpha\gamma} \Pi_{\beta\gamma}, 
       \quad \bar S^{\alpha\beta} = {1 \over a^2} \left( \Pi^{\alpha\beta}
       - 4 C^{\gamma (\alpha} \Pi^{\beta)}_\gamma \right).
   \label{ADM-fluid-pert}
\eea
{}From eqs. (\ref{q-pi-relations},\ref{fluid-def}) we have
\bea
   & & \tilde q_\alpha \equiv a Q_\alpha, \quad
       \tilde q_0 = - a Q_\alpha B^\alpha,
   \nonumber \\
   & & \tilde \pi_{\alpha\beta} \equiv a^2 \Pi_{\alpha\beta}, \quad
       \tilde \pi_{0\alpha} = - a^2 \Pi_{\alpha\beta} B^\beta, \quad
       \tilde \pi_{00} = 0.
\eea
{}From eqs. (\ref{Tab},\ref{fluid-def},\ref{normal-vector}) we have
\bea
   & & \tilde T_{00} = a^2 \left[ \mu + \delta \mu + 2 \mu A 
       + 2 \delta \mu A + \left( \mu + p \right) B^\alpha B_\alpha
       + 2 Q_\alpha B^\alpha \right],
   \nonumber \\
   & & \tilde T_{0\alpha} = - a^2 \left( Q_\alpha + p B_\alpha
       + \delta p B_\alpha + A Q_\alpha + \Pi_{\alpha\beta} B^\beta \right),
   \nonumber \\
   & & \tilde T_{\alpha\beta} = a^2 \left( p g^{(3)}_{\alpha\beta}
       + \delta p g^{(3)}_{\alpha\beta} + \Pi_{\alpha\beta}
       + 2 p C_{\alpha\beta} + 2 \delta p C_{\alpha\beta} \right).
   \label{Tab-fluid-1}
\eea
{}From eqs. (\ref{ADM-fluid-def},\ref{ADM-fluid-pert}) we have
\bea
   & & \mu + \delta \mu = \tilde T_{ab} \tilde n^a \tilde n^b, \quad
       p + \delta p = {1 \over 3} h^{\alpha\beta} \tilde T_{\alpha\beta}, \quad
       Q_\alpha = - {1 \over a} \tilde T_{\alpha b} \tilde n^b, \quad
       \Pi_{\alpha\beta} = { 1\over a^2} \left( \tilde T_{\alpha\beta}
       - h_{\alpha\beta} \tilde p \right).
   \label{Tab-fluid}
\eea
Equation (\ref{entropy}) gives
\bea
   & & \tilde \varrho \tilde T \tilde {\dot {\tilde S}}
       = \tilde \pi K 
       + {1 \over a^2} \Pi^{\alpha\beta} \bar K_{\alpha\beta}
       - 2 {1 \over a} Q^\alpha A_{,\alpha}
       - {1 \over a} Q^\alpha_{\;\;|\alpha}
       + 2 {1 \over a} \left( C^{\alpha\beta} Q_\beta \right)_{|\alpha}
       - {1 \over a} C^\alpha_{\alpha|\beta} Q^\beta.
\eea

{}For the interaction terms in eq. (\ref{Tab-i}) we set
\bea
   & & \tilde I_{(i)0} \equiv I_{(i)0} + \delta I_{(i)0}, \quad
       \tilde I_{(i)\alpha} \equiv \delta I_{(i)\alpha},
\eea
where $\delta I_{(i)\alpha}$ is based in $g^{(3)}_{\alpha\beta}$.

\subsection{Frame choice}
                                                        \label{sec:Frame}

The energy-momentum tensor in the general ($\tilde u_a$) frame follows from 
eqs. (\ref{Tab},\ref{fluid-def},\ref{u-def}):
\bea
   \tilde T^0_0
   &=& 
       - \mu - \delta \mu - \left( \mu + p \right) V^\alpha 
       \left( V_\alpha - B_\alpha \right)
       - Q^\alpha \left( 2 V_\alpha - B_\alpha \right),
   \nonumber \\
   \tilde T^0_\alpha
   &=&
       \left( 1 - A \right) \left[ Q_\alpha + \left( \mu + p \right)
       \left( V_\alpha - B_\alpha \right) \right]
       + \left( \mu + p \right) \left( A B_\alpha
       + 2 V^\beta C_{\alpha \beta} \right)
       + \left( \delta \mu + \delta p \right) 
       \left( V_\alpha - B_\alpha \right)
       + \left( V^\beta - B^\beta \right) \Pi_{\alpha\beta},
   \nonumber \\
   \tilde T^\alpha_\beta
   &=&
       \left( p + \delta p \right) \delta^\alpha_\beta
       + \Pi^\alpha_\beta 
       + V^\alpha \left[ Q_\beta + \left( \mu + p \right)
       \left( V_\beta - B_\beta \right) \right]
       + Q^\alpha \left( V_\beta - B_\beta \right)
       - 2 C^{\alpha\gamma} \Pi_{\beta\gamma}.
   \label{Tab-pert-general}
\eea 
In the energy-frame we set $Q_\alpha \equiv 0$, thus $\tilde q_a = 0$.
In the normal-frame we have $\tilde u_\alpha \equiv 0$, thus from
eq. (\ref{u-def}) we have
\bea
   & & {\rm Energy\!-\!frame}: \quad 
       Q_\alpha \equiv 0,
   \nonumber \\
   & & {\rm Normal\!-\!frame}: \quad 
       V_\alpha - B_\alpha + A B_\alpha + 2 B^\beta C_{\alpha\beta} \equiv 0.
   \label{frames}
\eea
Although we can take infinitely many different combination of the two frames,
the energy- and the normal-frames are the ones often used in the literature
\cite{Israel-1976}.
By choosing a frame (which is a decision about $Q_\alpha$ and $V_\alpha$)
we lose no generality.
This is because we have $10$ independent informations in $\tilde T_{ab}$
which can be allocated to the energy density $\tilde \mu$ (one),
the pressure $\tilde p$ (one), the anisotropic stress $\Pi_{\alpha\beta}$
(five, because it is tracefree).
The remaining (three) informations can be assigned to either the velocity
$V_\alpha$ (three) or the flux $Q_\alpha$ (three); or some combinations
of $V_\alpha$ and $Q_\alpha$ with total three informations.

Thus, in the normal-frame (indicated by a superscript $N$) we have
\bea
   \tilde T^0_0
   &=& 
       - \mu - \delta \mu^N - Q^{N\alpha} B_\alpha,
   \nonumber \\
   \tilde T^0_\alpha
   &=&
       \left( 1 - A \right) Q^N_\alpha,
   \nonumber \\
   \tilde T^\alpha_\beta
   &=&
       \left( p + \delta p^N \right) \delta^\alpha_\beta
       + \Pi^{N\alpha}_{\;\;\;\beta} + B^\alpha Q^N_\beta
       - 2 C^{\alpha\gamma} \Pi^N_{\beta\gamma},
   \label{Tab-pert-normal}
\eea 
which is consistent with eq. (\ref{Tab-fluid-1}).
To the linear-order we notice that
$\delta \mu$, $\delta p$, $\Pi_{\alpha\beta}$ are independent of
the frame choice, and $Q_\alpha + ( \mu + p ) ( V_\alpha - B_\alpha )$ 
is a frame-invariant combination \cite{HV-1990}.
However, to the second-order we no longer have such a luxury.
As the fluid quantities are defined based on the frame vector as in
eq. (\ref{fluid-Tab}) the values of $\delta \mu$, $\delta p$ 
and $\Pi_{\alpha\beta}$ are dependent on the frame.

By comparing the energy-momentum tensor in the normal-frame in 
eq. (\ref{Tab-pert-normal}) with the one in the general frame in
eq. (\ref{Tab-pert-general}),  
we find that by replacing the normal-frame fluid quantities to 
\bea
   \delta \mu^N 
   &\equiv& \delta \mu + \left( V^\alpha - B^\alpha \right)
       \left[ \left( \mu + p \right) \left( V_\alpha - B_\alpha \right)
       + 2 Q_\alpha \right],
   \nonumber \\
   \delta p^N 
   &\equiv& \delta p + {1 \over 3} \left( V^\alpha - B^\alpha \right)
       \left[ \left( \mu + p \right) \left( V_\alpha - B_\alpha \right)
       + 2 Q_\alpha \right],
   \nonumber \\
   Q^N_\alpha 
   &\equiv& 
       Q_\alpha + \left( \mu + p \right) \left( V_\alpha - B_\alpha \right)
       + \left( \mu + p \right) \left( A B_\alpha 
       + 2 V^\beta C_{\alpha\beta} \right)
       + \left( \delta \mu + \delta p \right) 
       \left( V_\alpha - B_\alpha \right)
       + \left( V^\beta - B^\beta \right) \Pi_{\alpha\beta},
   \nonumber \\
   \Pi^N_{\alpha\beta} 
   &\equiv& \Pi_{\alpha\beta}
       + \left( \mu + p \right) \left( V_\alpha - B_\alpha \right)
       \left( V_\beta - B_\beta \right)
       + 2 Q_{(\alpha} \left( V_{\beta)} - B_{\beta)} \right)
       - {1 \over 3} g^{(3)}_{\alpha\beta} \left( V^\gamma - B^\gamma \right)
       \left[ \left( \mu + p \right) \left( V_\gamma - B_\gamma \right)
       + 2 Q_\gamma \right],
   \label{frame-inv-def-N} 
\eea
we recover the general frame energy-momentum tensor.
Thus, by imposing the energy-frame condition ($Q_\alpha = 0$)
we recover the fluid quantities in the energy-frame.

Now, similarly, by comparing eq. (\ref{Tab-pert-general})
with the same one evaluated in the energy-frame,
we find that by replacing the energy-frame fluid quantities (indicated
by a superscript $E$) to 
\bea
   \delta \mu^E
   &\equiv& \delta \mu - {1 \over \mu + p} Q^\alpha Q_\alpha,
   \nonumber \\
   \delta p^E
   &\equiv& \delta p - {1 \over 3} {1 \over \mu + p} Q^\alpha Q_\alpha,
   \nonumber \\
   \left( \mu + p \right) \left( V^E_\alpha - B_\alpha \right)
   &\equiv&
       \left( \mu + p \right) \left( V_\alpha - B_\alpha \right)
       + Q_\alpha 
       - 2 Q^\beta C_{\alpha\beta} 
       - {\delta \mu + \delta p \over \mu + p} Q_\alpha
       - Q^\beta {\Pi_{\alpha\beta} \over \mu + p},
   \nonumber \\
   \Pi^E_{\alpha\beta} 
   &\equiv& \Pi_{\alpha\beta}
       - {1 \over \mu + p} \left( Q_\alpha Q_\beta
       - {1 \over 3} g^{(3)}_{\alpha\beta} Q^\gamma Q_\gamma \right),
   \label{frame-inv-def-E}
\eea
we recover the general frame energy-momentum tensor.
By imposing the normal-frame condition in eq. (\ref{frames})
we recover the fluid quantities in the normal-frame.
Thus, using eqs. (\ref{frame-inv-def-N},\ref{frame-inv-def-E})
we can transform the fluid quantities in one frame to the other:
\bea
   & & \delta \mu^N 
       = \delta \mu^E 
       + \left( \mu + p \right) \left( V^{E\alpha} - B^\alpha \right)
       \left( V^E_\alpha - B_\alpha \right), 
   \nonumber \\
   & & \delta p^N 
       =\delta p^E 
       + {1 \over 3} \left( \mu + p \right) 
       \left( V^{E\alpha} - B^\alpha \right)
       \left( V^E_\alpha - B_\alpha \right),
   \nonumber \\
   & & Q^N_\alpha 
       = \left( \mu + p \right) \left( V^E_\alpha - B_\alpha \right)
       + \left( \mu + p \right) \left( A B_\alpha 
       + 2 V^{E\beta} C_{\alpha\beta} \right)
       + \left( \delta \mu^E + \delta p^E \right) 
       \left( V^E_\alpha - B_\alpha \right)
       + \left( V^{E\beta} - B^\beta \right) \Pi^E_{\alpha\beta},
   \nonumber \\
   & & \Pi^N_{\alpha\beta} 
       = \Pi^E_{\alpha\beta}
       + \left( \mu + p \right) \left( V^E_\alpha - B_\alpha \right)
       \left( V^E_\beta - B_\beta \right)
       - {1 \over 3} g^{(3)}_{\alpha\beta} \left( \mu + p \right) 
       \left( V^{E\gamma} - B^\gamma \right)
       \left( V^E_\gamma - B_\gamma \right);
   \label{frame-N-E} \\
   & & \delta \mu^E = \delta \mu^N - {1 \over \mu + p} Q^{N\alpha} Q^N_\alpha,
   \nonumber \\
   & & \delta p^E
       = \delta p^N - {1 \over 3} {1 \over \mu + p} Q^{N\alpha} Q^N_\alpha,
   \nonumber \\
   & & \left( \mu + p \right) \left( V^E_\alpha - B_\alpha \right)
       = 
       Q^N_\alpha 
       - 2 Q^{N\beta} C_{\alpha\beta} 
       - {\delta \mu^N + \delta p^N \over \mu + p} Q^N_\alpha
       - Q^{N\beta} {\Pi^N_{\alpha\beta} \over \mu + p}
       - \left( \mu + p \right) 
       \left( A B_\alpha + 2 B^\beta C_{\alpha\beta} \right),
   \nonumber \\
   & & \Pi^E_{\alpha\beta} 
       = \Pi^N_{\alpha\beta}
       - {1 \over \mu + p} \left( Q^N_\alpha Q^N_\beta
       - {1 \over 3} g^{(3)}_{\alpha\beta} Q^{N\gamma} Q^N_\gamma \right).
   \label{frame-E-N}
\eea

\subsection{Spacetime curvatures to the linear-order}

Although straightforward, it is not an easy task to derive the spacetime 
curvature to the second-order. 
{}For our purpose, fortunately, it is not necessary to have the forms.
Still, it is convenient to have the curvatures to the linear-order
and we present them in the following.
These follow from 
eqs. (\ref{metric-def},\ref{inverse-metric},\ref{connections}).

\noindent
Curvature tensors are:
\bea
   \tilde R^a_{\;\;b00} 
   &=& 0, \quad
       \tilde R^0_{\;\;00\alpha}
       = - \left( { a^\prime \over a} \right)^\prime B_\alpha, \quad
       \tilde R^0_{\;\;0\alpha\beta} = 0,
   \nonumber \\
   \tilde R^0_{\;\;\alpha 0 \beta}
   &=& 
       \left( {a^\prime \over a} \right)^\prime g^{(3)}_{\alpha\beta}
       - \left[ {a^\prime \over a} A^\prime
       + 2 \left( {a^\prime \over a} \right)^\prime A \right]
       g^{(3)}_{\alpha\beta}
       - A_{,\alpha|\beta}
       + B_{(\alpha|\beta)}^\prime
       + {a^\prime \over a} B_{(\alpha|\beta)}
       + C_{\alpha\beta}^{\prime\prime}
       + {a^\prime \over a} C_{\alpha\beta}^\prime
       + 2 \left( {a^\prime \over a} \right)^\prime C_{\alpha\beta},
   \nonumber \\
   \tilde R^0_{\;\;\alpha\beta\gamma}
   &=& 
       2 {a^\prime \over a} g^{(3)}_{\alpha[\beta} A_{,\gamma]}
       - B_{\alpha|[\beta\gamma]}
       + {1 \over 2} ( B_{\gamma|\alpha\beta} - B_{\beta|\alpha\gamma} )
       - 2 C^\prime_{\alpha[\beta|\gamma]},
   \nonumber \\
   \tilde R^\alpha_{\;\;00\beta}
   &=& 
       \left( {a^\prime \over a} \right)^\prime \delta^\alpha_\beta
       - {a^\prime \over a} A^\prime \delta^\alpha_\beta
       - A^{|\alpha}_{\;\;\;\beta}
       + {1 \over 2} \left( B_\beta^{\;\;|\alpha} + B^\alpha_{\;\;|\beta}
       \right)^\prime
       + {1 \over 2} {a^\prime \over a} \left( B_\beta^{\;\;|\alpha}
       + B^\alpha_{\;\;|\beta} \right)
       + C^{\alpha\prime\prime}_\beta
       + {a^\prime \over a} C^{\alpha\prime}_\beta,
   \nonumber \\
   \tilde R^\alpha_{\;\;0\beta\gamma}
   &=& 
       2 {a^\prime \over a} \delta^\alpha_{[\beta} A_{,\gamma]}
       - B_{[\beta\;\;\;\gamma]}^{\;\;\;|\alpha}
       + B^\alpha_{\;\;|[\beta\gamma]}
       - 2 \left( {a^\prime \over a} \right)^2 \delta^\alpha_{[\beta}
       B_{\gamma]}
       - 2 C^{\alpha\prime}_{[\beta|\gamma]}
   \nonumber \\
   \tilde R^\alpha_{\;\;\beta 0\gamma}
   &=& 
       {a^\prime \over a} \left( g^{(3)}_{\beta\gamma} A^{,\alpha}
       - \delta^\alpha_\gamma A_{,\beta} \right)
       + \left( {a^\prime \over a} \right)^\prime g^{(3)}_{\beta\gamma}
       B^\alpha
       - \left( {a^\prime \over a} \right)^2
       \left( g^{(3)}_{\beta\gamma} B^{\alpha}
       - \delta^\alpha_\gamma B_{\beta} \right)
       - {1 \over 2} \left( B_\beta^{\;\;|\alpha} - B^\alpha_{\;\;|\beta}
       \right)_{|\gamma}
       + C^{\alpha\prime}_{\gamma|\beta}
       - C_{\beta\gamma}^{\prime\;\;\;|\alpha},
   \nonumber \\
   \tilde R^\alpha_{\;\;\beta\gamma\delta}
   &=& 
       R^{(3)\alpha}_{\;\;\;\;\;\;\;\beta\gamma\delta}
       + \left( {a^\prime \over a} \right)^2
       \left( \delta^\alpha_\gamma g^{(3)}_{\beta\delta}
       - \delta^\alpha_\delta g^{(3)}_{\beta\gamma} \right) ( 1 - 2 A )
   \nonumber \\
   & &
       + {1 \over 2} {a^\prime \over a}
       \left[ g^{(3)}_{\beta\delta} \left( B_\gamma^{\;\;|\alpha}
       + B^\alpha_{\;\;|\gamma} \right)
       - g^{(3)}_{\beta\gamma} \left( B_\delta^{\;\;|\alpha}
       + B^\alpha_{\;\;|\delta} \right)
       + 2 \delta^\alpha_\gamma B_{(\beta|\delta)}
       - 2 \delta^\alpha_\delta B_{(\beta|\gamma)} \right]
   \nonumber \\
   & & 
       + {a^\prime \over a} \left[
       g^{(3)}_{\beta\delta} C^{\alpha\prime}_\gamma
       - g^{(3)}_{\beta\gamma} C^{\alpha\prime}_\delta
       + \delta^\alpha_\gamma C^\prime_{\beta\delta}
       - \delta^\alpha_\delta C^\prime_{\beta\gamma}
       + 2 {a^\prime \over a} \left(
       \delta^\alpha_\gamma C_{\beta\delta}
       - \delta^\alpha_\delta C_{\beta\gamma} \right) \right]
   \nonumber \\
   & &
       + 2 C^\alpha_{(\beta|\delta)\gamma}
       - 2 C^\alpha_{(\beta|\gamma)\delta}
       + C^{\;\;\;\;|\alpha}_{\beta\gamma\;\;\;\delta}
       - C^{\;\;\;\;|\alpha}_{\beta\delta\;\;\;\gamma},
   \label{Riemann-curvature-linear} \\
   \tilde R_{00}
   &=& 
       - 3 \left( {a^\prime \over a} \right)^\prime
       + 3 {a^\prime \over a} A^\prime
       + \Delta A
       - B^{\alpha\prime}_{\;\;|\alpha}
       - {a^\prime \over a} B^\alpha_{\;\;|\alpha}
       - C^{\alpha\prime\prime}_\alpha
       - {a^\prime \over a} C^{\alpha\prime}_\alpha,
   \nonumber \\
   \tilde R_{0\alpha}
   &=& 
       2 {a^\prime \over a} A_{,\alpha}
       - \left( {a^\prime \over a} \right)^\prime B_\alpha
       - 2 \left( {a^\prime \over a} \right)^2 B_\alpha
       + {1 \over 2} \Delta B_\alpha
       - {1 \over 2} B^\beta_{\;\;|\alpha\beta}
       - C^{\beta\prime}_{\beta|\alpha}
       + C^{\prime \;\;\;|\beta}_{\alpha\beta},
   \nonumber \\
   \tilde R_{\alpha\beta}
   &=& 
       2 K g^{(3)}_{\alpha\beta}
       + \left[ \left( {a^\prime \over a} \right)^\prime
       + 2 \left( {a^\prime \over a} \right)^2 \right]
       g^{(3)}_{\alpha\beta} ( 1 - 2 A )
       - {a^\prime \over a} A^\prime g^{(3)}_{\alpha\beta}
       - A_{,\alpha|\beta}
       + B^\prime_{(\alpha|\beta)}
       + 2 {a^\prime \over a} B_{(\alpha|\beta)}
       + {a^\prime \over a} g^{(3)}_{\alpha\beta} B^\gamma_{\;\;|\gamma}
   \nonumber \\
   & & 
       + C^{\prime\prime}_{\alpha\beta}
       + 2 {a^\prime \over a} C^{\prime}_{\alpha\beta}
       + 2 \left[ \left( {a^\prime \over a} \right)^\prime
       + 2 \left( {a^\prime \over a} \right)^2 \right] C_{\alpha\beta}
       + {a^\prime \over a} g^{(3)}_{\alpha\beta} C^{\gamma\prime}_\gamma
       + 2 C^\gamma_{(\alpha|\beta)\gamma}
       - C^\gamma_{\gamma|\alpha\beta}
       - \Delta C_{\alpha\beta},
   \\
   \tilde R 
   &=& 
       {1 \over a^2} \Bigg\{ 6 \left[
       \left( {a^\prime \over a} \right)^\prime
       + \left( {a^\prime \over a} \right)^2 + K \right]
       - 6 {a^\prime \over a} A^\prime
       - 12 \left[ \left( {a^\prime \over a} \right)^\prime
       + \left( {a^\prime \over a} \right)^2 \right] A
       - 2 \Delta A
   \nonumber \\
   & &
       + 2 B^{\alpha\prime}_{\;\; |\alpha}
       + 6 {a^\prime \over a} B^\alpha_{\;\;|\alpha}
       + 2 C^{\alpha\prime\prime}_\alpha
       + 6 {a^\prime \over a} C^{\alpha\prime}_\alpha
       - 4 K C^\alpha_\alpha
       - 2 \Delta C^\alpha_\alpha
       + 2 C^{\alpha\beta}_{\;\;\;\;|\alpha\beta} \Bigg\}.
   \label{Scalar-curvature-linear}
\eea
The nonvanishing components of the electric and magnetic parts of 
the Weyl curvature in eq. (\ref{EM-Weyl}) are:
\bea
   \tilde E_{\alpha\beta}
   &=& - \tilde C^0_{\;\;\alpha 0\beta}
   \nonumber \\
   &=& 
       {1 \over 2} A_{,\alpha|\beta}
       - {1 \over 2} B^\prime_{(\alpha|\beta)}
       - {1 \over 2} C^{\prime\prime}_{\alpha\beta}
       - {1 \over 2} \Delta C_{\alpha\beta} - 2 K C_{\alpha\beta}
       + C^\gamma_{(\alpha|\beta)\gamma}
       - {1 \over 2} C^\gamma_{\gamma|\alpha\beta}
   \nonumber \\
   & & 
       - {1 \over 3} g^{(3)}_{\alpha\beta} \left(
       {1 \over 2} \Delta A
       - {1 \over 2} B^{\gamma\prime}_{\;\;|\gamma}
       - {1 \over 2} C^{\gamma\prime\prime}_\gamma
       - \Delta C^\gamma_\gamma - 2 K C^\gamma_\gamma
       + C^{\gamma\delta}_{\;\;\;\;|\gamma\delta} \right)
   \nonumber \\
   &=& 
       {1 \over 2} \left( \nabla_\alpha \nabla_\beta
       - {1 \over 3} g^{(3)}_{\alpha\beta} \Delta \right)
       \left( \alpha - \varphi - {1 \over a} \chi^\prime
       + {a^\prime \over a^2} \chi \right)
       - {1 \over 2} \Psi^{(v)\prime}_{(\alpha|\beta)}
       - {1 \over 2} \left[ C^{(t)\prime\prime}_{\alpha\beta}
       + \left( \Delta - 2 K \right) C^{(t)}_{\alpha\beta} \right],
   \label{Electric-linear} \\
   \tilde H_{\alpha\beta}
   &=& 
       - {1 \over 2} \tilde \eta_{0(\alpha}^{\;\;\;\;\;\;\gamma\delta}
       \tilde C^0_{\;\;\beta)\gamma\delta}
   \nonumber \\
   &=& 
       - \eta_{(\alpha}^{\;\;\;\;\gamma\delta}
       \left( {1 \over 2} B_{\gamma|\beta)\delta}
       + C^\prime_{\beta)\gamma|\delta} \right)
       = - \eta_{(\alpha}^{\;\;\;\;\gamma\delta}
       \left( {1 \over 2} \Psi^{(v)}_{\gamma|\beta)\delta}
       + C^{(t)\prime}_{\beta)\gamma|\delta} \right),
   \label{Magnetic-linear} 
\eea
where the symmetrization is only over $\alpha$ and $\beta$ indices.
The last steps are evaluated in decomposed forms which will be
introduced in \S \ref{sec:Decomposition}.
We introduced $\eta^{\alpha\beta\gamma}$ which is based on 
$g^{(3)}_{\alpha\beta}$ with
$\eta^{\alpha\beta\gamma} \equiv \eta^{[\alpha\beta\gamma]}$ and
$\eta^{123} \equiv 1/\sqrt{g^{(3)}}$.
We have
\bea
   & & \tilde \eta^{0\alpha\beta\gamma} 
       = {1 \over a^4 \sqrt{1 + D}} \eta^{\alpha\beta\gamma},
   \nonumber \\
   & & \tilde g = - a^8 (1 + D ) g^{(3)}, \quad
       D \equiv 2 A + 2 C^\alpha_\alpha + 4 A C^\alpha_\alpha 
       + B_\alpha B^\alpha + 2 C^\alpha_\alpha C^\beta_\beta
       - 2 C^\alpha_\beta C^\beta_\alpha,
   \label{eta-relation}
\eea
which is valid to the second-order in perturbation.

Deriving the the electric and magnetic parts of Weyl tensor to the 
second-order using eqs. (\ref{Weyl},\ref{EM-Weyl}) requires 
quite lengthy algebra.
Instead, using eqs. (\ref{cov-shear-prop},\ref{cov-constr-3}) 
we can derive them easily.
Evaluated in the normal-frame we have
\bea
   \tilde E_{\alpha\beta}
   &=&
       {1 \over a} \left( 1 - A \right) \left( \bar K_{\alpha\beta}^\prime
       - 2 {a^\prime \over a} \bar K_{\alpha\beta} \right)
       - {1 \over a} \left( B_{(\alpha}^{\;\;\;|\gamma}
       - B^\gamma_{\;\;|(\alpha} 
       + 2 C^{\gamma\prime}_{(\alpha} \right) \bar K_{\beta)\gamma}
       + {1 \over a} \bar K_{\alpha\beta|\gamma} B^\gamma
       - {1 \over a^2} g^{(3)\gamma\delta} \bar K_{\alpha\gamma} 
       \bar K_{\beta\delta} 
       - {2 \over 3} K \bar K_{\alpha\beta}
   \nonumber \\
   & & 
       + \left( A - A^2 + {1 \over 2} B^\gamma B_\gamma \right)_{,\alpha|\beta}
       - \left( 2 C^\gamma_{(\alpha|\beta)}
       - C_{\alpha\beta}^{\;\;\;\;|\gamma} \right) A_{,\gamma}
       + A_{,\alpha} A_{,\beta}
       - {2 \over 3} C_{\alpha\beta} \Delta A
       + 4 \pi G a^2 \Pi_{\alpha\beta}
   \nonumber \\
   & & 
       - {1 \over 3} g^{(3)}_{\alpha\beta} \Bigg[ 
       - {1 \over a^2} g^{(3)\alpha\gamma} g^{(3)\beta\delta}
       \bar K_{\alpha\beta} \bar K_{\gamma\delta}
       + A_{,\gamma} \left( A^{,\gamma}
       - 2 C^{\gamma\delta}_{\;\;\;\;|\delta} 
       + C^{\delta|\gamma}_\delta \right)
       + \Delta \left( A - A^2 + {1 \over 2} B^\gamma B_\gamma \right)
       - 2 C^{\gamma\delta} A_{,\gamma|\delta} \Bigg],
   \nonumber \\
   \\
   \tilde H_{\alpha\beta}
   &=& \left\{ \left[ g^{(3)}_{\delta(\beta} \left( 1- C^\nu_\nu \right)
       + 2 C_{\delta(\beta} \right] \bar K_{\alpha)\gamma|\mu}
       - \left( C^\nu_{(\alpha|\mu} + C^\nu_{\mu|(\alpha}
       - C_{\mu(\alpha}^{\;\;\;\; |\nu} \right)
       g^{(3)}_{\beta)\delta} \bar K_{\gamma\nu}
       \right\} {1 \over a} \eta^{\delta\gamma\mu},
\eea
where $K$ and $\bar K_{\alpha\beta}$ are given 
in eq. (\ref{Extrinsic-curvature});
the symmetrization is only over $\alpha$ and $\beta$ indices.
We have $\tilde E_{ab} \tilde u^b = 0 = \tilde H_{ab} \tilde u^b$,
thus $\tilde E_{\alpha0} = - \tilde E_{\alpha\beta} B^\beta$,
$\tilde E_{00} = 0$, and similarly for $\tilde H_{ab}$.

{}For later use, it is convenient to have the spacetime scalar curvature
expanded to the second-order.
In terms of the ADM notation, using eq. (\ref{connection-ADM}), we have
\bea
   & & \tilde R = R^{(h)} + K^{\alpha\beta} K_{\alpha\beta} + K^2
       + {2 \over N} \left( - K_{,0}
       + K_{,\alpha} N^\alpha - N^{:\alpha}_{\;\;\;\alpha} \right).
   \label{R-ADM}
\eea
To the second-order in perturbation, using quantities in 
\S \ref{sec:normal-frame-quantities} we have
\bea
   \tilde R
   &\equiv& R + \delta R
   \nonumber \\
   &=& 6 \left( {K \over a^2} + \dot H + 2 H^2 \right)
   \nonumber \\
   & &
       - 6 H \dot A - 12 \left( \dot H + 2 H^2 \right) A
       - 2 {\Delta \over a^2} A
       + 2 \left( {1 \over a} B^\alpha_{\;\;|\alpha}
       + \dot C^\alpha_\alpha \right)^\cdot
       + 8 H \left( {1 \over a} B^\alpha_{\;\;|\alpha}
       + \dot C^\alpha_\alpha \right)
       + 2 {1 \over a^2} \left[ C^{\beta|\alpha}_{\alpha \;\;\; \beta}
       - \left( \Delta + 2 K \right) C^\alpha_\alpha \right]
   \nonumber \\
   & &
       + 24 H A \dot A
       - 4 A \left( {1 \over a} B^\alpha_{\;\;|\alpha}
       + \dot C^\alpha_\alpha \right)^\cdot
       - 2 \left( \dot A + 8 H A \right)
       \left( {1 \over a} B^\alpha_{\;\;|\alpha} + \dot C^\alpha_\alpha \right)
       + 24 \left( \dot H + 2 H^2 \right) A^2
       + 4 A {\Delta \over a^2} A
       + 2 {1 \over a^2} A^{,\alpha} A_{,\alpha}
   \nonumber \\
   & &
       - 6 H {1 \over a} A_{,\alpha} B^\alpha
       + 4 {1 \over a^2} A_{,\alpha|\beta} C^{\alpha\beta}
       + 2 { 1\over a^2} A_{,\beta} \left( 2 C^{\alpha\beta}_{\;\;\;\;\;|\alpha}
       - C^{\alpha|\beta}_\alpha \right)
       + {1 \over a^2} \left[
       B_{(\alpha|\beta)} B^{\alpha|\beta}
       + B^\alpha_{\;\;|\alpha} B^\beta_{\;\;|\beta} 
       - 2 \left( B_\alpha B^{\alpha|\beta} \right)_{|\beta} \right]
   \nonumber \\
   & & 
       - 6 H B^\alpha \dot B_\alpha
       - 6 \left( \dot H + 2 H^2 \right) B^\alpha B_\alpha
       - 2 {1 \over a} B^\alpha \left( 2 C^\beta_{\alpha|\beta}
       - C^\beta_{\beta|\alpha} \right)^\cdot
       + 2 {1 \over a} B^\alpha_{\;\;|\alpha} \dot C^\beta_\beta
   \nonumber \\
   & &
       - 2 {1 \over a} \left( 2 C^\beta_{\alpha|\beta}
       - C^\beta_{\beta|\alpha} \right)
       \left( \dot B^\alpha + 3 H B^\alpha \right)
       - 4 {1 \over a} C^{\alpha\beta} \dot B_{\alpha|\beta}
       - 2 { 1\over a} \left( \dot C^{\alpha\beta}
       + 6 H C^{\alpha\beta} \right) B_{\alpha|\beta}
       + 2 {1 \over a} B^\alpha \left( {1 \over a} B^\beta_{\;\;|\beta}
       + \dot C^\beta_\beta \right)_{|\alpha}
   \nonumber \\
   & &
       + \dot C^\alpha_\alpha \dot C^\beta_\beta
       - 3 \dot C^{\alpha\beta} \dot C_{\alpha\beta}
       - 4 C^{\alpha\beta} \left( \ddot C_{\alpha\beta}
       + 4 H \dot C_{\alpha\beta}
       - 2 {K \over a^2} C_{\alpha\beta} \right)
       + 4 {1 \over a^2} C^{\alpha\beta} \left(
       - C^\gamma_{\alpha|\beta\gamma}
       - C^\gamma_{\alpha|\gamma\beta}
       + \Delta C_{\alpha\beta}
       + C^\gamma_{\gamma|\alpha\beta} \right)
   \nonumber \\
   & & - {1 \over a^2} \left( 2 C^\gamma_{\beta|\gamma}
       - C^\gamma_{\gamma|\beta} \right)
       \left( 2 C^{\alpha\beta}_{\;\;\;\;|\alpha}
       - C^{\alpha|\beta}_\alpha \right)
       + {1 \over a^2} C^{\alpha\beta|\gamma}
       \left( 3 C_{\alpha\beta|\gamma} - 2 C_{\alpha\gamma|\beta} \right).
   \label{R}
\eea
An overdot indicates a time derivative with respect to $t$, 
with $dt = a d\eta$.

\section{Perturbed equations}
                                            \label{sec:Equations}

\subsection{Basic equations with general fluids}
                                            \label{sec:Basic-equations}

In the following we present complete sets of equations valid up to 
second-order in the perturbation without fixing the gauge conditions. 
As the basic set we consider eqs. 
(\ref{extrinsic-curvature-def},\ref{E-constraint}-\ref{Mom-conservation},\ref{E-conservation-i},\ref{Mom-conservation-i}) 
in the ADM formulation.

\noindent
Definition of $\delta K$:
\bea
   & & \bar K + 3 H
   \nonumber \\
   & & \qquad
       + \delta K - 3 H A + \dot C^\alpha_\alpha 
       + {1\over a} B^\alpha_{\;\;|\alpha}
   \nonumber \\
   & & \qquad \qquad
       = - A \left( {9 \over 2} H A - \dot C^\alpha_\alpha 
       - {1 \over a} B^\alpha_{\;\;|\alpha} \right)
       + {3 \over 2} H B^\alpha B_\alpha
       + {1\over a} B^\alpha \left( 2 C^\beta_{\alpha|\beta}
       - C^\beta_{\beta|\alpha} \right)
       + 2 C^{\alpha\beta} \left( \dot C_{\alpha\beta} 
       + {1\over a} B_{\alpha|\beta} \right)
   \nonumber \\
   & & \qquad \qquad
       \equiv N_0,
   \label{pert-eq0}
\eea
where $K \equiv \bar K + \delta K$ and $K$ is read from 
eq. (\ref{Extrinsic-curvature}).

\noindent
Energy constraint equation:
\bea
   & & 16 \pi G \mu + 2 \Lambda - 6 H^2 - {1\over a^2} R^{(3)}
   \nonumber \\
   & & \qquad
       + 16 \pi G \delta \mu + 4 H \delta K
       - {1\over a^2} \left(
       2 C^{\beta|\alpha}_{\alpha\;\;\;\;\beta}
       - 2 C^{\alpha|\beta}_{\alpha\;\;\;\;\beta} 
       - {2 \over 3} R^{(3)} C^\alpha_\alpha \right)
   \nonumber \\
   & & \qquad \qquad
       = {2 \over 3} \delta K^2
       - \left( \dot C_{\alpha\beta} + {1\over a} B_{(\alpha|\beta)} \right)
       \left( \dot C^{\alpha\beta} + {1\over a} B^{\alpha|\beta} \right)
       + {1\over 3} \left( \dot C^\alpha_\alpha 
       + {1\over a} B^\alpha_{\;\;|\alpha} \right)^2 
   \nonumber \\
   & & \qquad \qquad \qquad
       + {1 \over a^2} \Bigg[ 
       4 C^{\alpha\beta} \left( - C^\gamma_{\alpha|\beta\gamma}
       - C^{\gamma}_{\alpha|\gamma\beta}
       + C^{\;\;\;\;|\gamma}_{\alpha\beta\;\;\;\gamma}
       + C^\gamma_{\gamma|\alpha\beta} \right) 
       + {4 \over 3} R^{(3)} C^\alpha_\gamma C^\gamma_\alpha 
   \nonumber \\
   & & \qquad \qquad \qquad
       - \left( 2 C^\gamma_{\beta|\gamma} - C^\gamma_{\gamma|\beta} \right)
       \left( 2 C^{\alpha\beta}_{\;\;\;\;|\alpha} - C^{\alpha|\beta}_\alpha 
       \right) 
       + C^{\alpha\beta|\gamma} \left( 3 C_{\alpha\beta|\gamma}
       - 2 C_{\alpha\gamma|\beta} \right)
       \Bigg]
   \nonumber \\
   & & \qquad \qquad
       \equiv N_1.
   \label{pert-eq1}
\eea
Momentum constraint equation:
\bea
   & & \left[ \dot C^\beta_\alpha + {1\over 2a} \left( B^\beta_{\;\;|\alpha}
       + B_\alpha^{\;\;|\beta} \right) \right]_{|\beta}
       - {1\over 3} \left( \dot C^\gamma_\gamma 
       + {1\over a} B^\gamma_{\;\;|\gamma} \right)_{,\alpha}
       + {2\over 3} \delta K_{,\alpha}
       + 8 \pi G a Q_\alpha
   \nonumber \\
   & & \qquad
       = A \left( - {2\over 3} \delta K_{,\alpha} - 8 \pi G a Q_\alpha \right)
       + A_{,\beta} \left[ \dot C^\beta_\alpha
       + {1\over 2a} \left( B^\beta_{\;\;|\alpha}
       + B_\alpha^{\;\;|\beta} \right) \right]
   \nonumber \\
   & & \qquad \qquad
       + \left( 2 C^\beta_{\gamma|\beta}
       - C^\beta_{\beta|\gamma} \right)
       \left[ \dot C^\gamma_\alpha 
       + {1\over 2a} \left( B_\alpha^{\;\;|\gamma}
       + B^\gamma_{\;\;|\alpha} \right) \right]
       + 2 C^{\beta\gamma} \left( \dot C_{\alpha\gamma}
       + {1\over a} B_{(\alpha|\gamma)} \right)_{|\beta}
   \nonumber \\
   & & \qquad \qquad
       + {1\over a} \left[ B_\gamma \left( C^{\beta\gamma}_{\;\;\;\;\;|\alpha}
       + C^{\gamma|\beta}_\alpha
       - C^{\beta|\gamma}_\alpha \right) \right]_{|\beta}
       + {1\over 3} C^\gamma_{\beta|\alpha} \left(
       \dot C^\beta_\gamma + {1\over a} B^\beta_{\;\;|\gamma} \right)
   \nonumber \\
   & & \qquad \qquad
       - {1\over 3} \left\{ 
       A_{,\alpha} \left( \dot C^\gamma_\gamma
       + {1\over a} B^\gamma_{\;\;|\gamma} \right)
       + 2 C^{\gamma\delta} \left( \dot C_{\gamma\delta}
       + {1\over a} B_{\gamma|\delta} \right)_{|\alpha}
       + {1\over a}\left[ B^\delta \left( 2 C^\gamma_{\delta|\gamma}
       - C^\gamma_{\gamma|\delta} \right) \right]_{|\alpha} \right\}
   \nonumber \\
   & & \qquad
       \equiv N_{2\alpha}.
   \label{pert-eq2}
\eea
Trace of the ADM propagation equation:
\bea
   & & - \left[ 3 \dot H + 3 H^2 + 4 \pi G \left( \mu + 3 p \right) - \Lambda
       \right]
   \nonumber \\
   & & \qquad
       + \delta \dot K + 2 H \delta K 
       - 4 \pi G \left( \delta \mu + 3 \delta p \right) 
       + {1\over a^2} A^{|\alpha}_{\;\;\;\;\alpha}
       + 3 \dot H A 
   \nonumber \\
   & & \qquad \qquad
       = A \delta \dot K - {1\over a} \delta K_{,\alpha} B^\alpha
       + {1\over 3} \delta K^2
       + {3\over 2} \dot H \left( 3 A^2 - B^\alpha B_\alpha \right)
   \nonumber \\
   & & \qquad \qquad \qquad
       + {1 \over a^2} \left[ 2 A A^{|\alpha}_{\;\;\;\;\alpha}
       + A_{,\alpha} A^{,\alpha}
       - B^\beta B^{\;\;|\alpha}_{\beta\;\;\;\;\alpha}
       - B^{\beta|\alpha} B_{\beta|\alpha}
       + A^{,\alpha} \left( 2 C^\beta_{\alpha|\beta}
       - C^\beta_{\beta|\alpha} \right)
       + 2 C^{\alpha\beta} A_{,\alpha|\beta} \right]
   \nonumber \\
   & & \qquad \qquad \qquad
       + \left( \dot C_{\alpha\beta} + {1\over a} B_{(\alpha|\beta)} \right)
       \left( \dot C^{\alpha\beta} + {1\over a} B^{\alpha|\beta} \right)
       - {1\over 3} \left( \dot C^\alpha_\alpha 
       + {1\over a} B^\alpha_{\;\;|\alpha} \right)^2
   \nonumber \\
   & & \qquad \qquad
       \equiv N_3.
   \label{pert-eq3}
\eea
Tracefree ADM propagation equation:
\bea
   & & 
       \left[ \dot C^\alpha_\beta + {1\over 2a} \left( B^\alpha_{\;\;|\beta}
       + B_\beta^{\;\;|\alpha} \right) \right]^\cdot
       + 3 H 
       \left[ \dot C^\alpha_\beta + {1\over 2a} \left( B^\alpha_{\;\;|\beta}
       + B_\beta^{\;\;|\alpha} \right) \right]
       - {1\over a^2} A^{|\alpha}_{\;\;\;\;\beta}
   \nonumber \\
   & & 
       - {1\over 3} \delta^\alpha_\beta \left[ 
       \left( \dot C^\gamma_\gamma + {1\over a} B^\gamma_{\;\;|\gamma} 
       \right)^\cdot
       + 3 H 
       \left( \dot C^\gamma_\gamma + {1\over a} B^\gamma_{\;\;|\gamma} \right)
       - {1\over a^2} A^{|\gamma}_{\;\;\;\;\gamma} \right]
   \nonumber \\
   & & 
       + {1 \over a^2} \left[
       C^{\alpha\gamma}_{\;\;\;\;\;|\beta\gamma}
       + C^{\gamma|\alpha}_{\beta\;\;\;\;\gamma}
       - C^{\alpha|\gamma}_{\beta\;\;\;\;\gamma} 
       - C^{\gamma|\alpha}_{\gamma\;\;\;\;\beta} 
       - {2 \over 3} R^{(3)} C^\alpha_\beta
       - {1\over 3} \delta^\alpha_\beta \left(
       2 C^{\delta|\gamma}_{\gamma\;\;\;\;\delta}
       - 2 C^{\gamma|\delta}_{\gamma\;\;\;\;\delta} 
       - {2 \over 3} R^{(3)} C^\gamma_\gamma \right) \right]
       - 8 \pi G \Pi^\alpha_\beta
   \nonumber \\
   & & \qquad
       = 
       \left\{ 
       \left[ \dot C^\alpha_\beta + {1\over 2a} \left( B^\alpha_{\;\;|\beta}
       + B_\beta^{\;\;|\alpha} \right) \right] A
       + 2 C^{\alpha\gamma} \left( \dot C_{\beta\gamma}
       + {1\over a} B_{(\beta|\gamma)} \right)
       + {1\over a}B_\gamma \left( C^{\alpha\gamma}_{\;\;\;\;\;|\beta}
       + C^{\gamma|\alpha}_\beta - C^{\alpha|\gamma}_\beta \right)
       \right\}^\cdot
   \nonumber \\
   & & \qquad \qquad
       + 3 H \left\{
       \left[ \dot C^\alpha_\beta + {1\over 2a} \left( B^\alpha_{\;\;|\beta}
       + B_\beta^{\;\;|\alpha} \right) \right] A
       + 2 C^{\alpha\gamma} \left( \dot C_{\beta\gamma}
       + {1\over a} B_{(\beta|\gamma)} \right)
       + {1\over a}B_\gamma \left( C^{\alpha\gamma}_{\;\;\;\;\;|\beta}
       + C^{\gamma|\alpha}_\beta - C^{\alpha|\gamma}_\beta \right)
       \right\}
   \nonumber \\
   & & \qquad \qquad
       + \left[ \dot C^\alpha_\beta + {1\over 2a} \left( B^\alpha_{\;\;|\beta}
       + B_\beta^{\;\;|\alpha} \right) \right]^\cdot A
       - {1\over a} \left[ \dot C^\alpha_\beta + {1\over 2a}
       \left( B^\alpha_{\;\;\;|\beta} + B^{\;\;|\alpha}_\beta \right)
       \right]_{|\gamma} B^\gamma
       + \delta K 
       \left[ \dot C^\alpha_\beta + {1\over 2a} \left( B^\alpha_{\;\;|\beta}
       + B_\beta^{\;\;|\alpha} \right) \right]
   \nonumber \\
   & & \qquad \qquad
       + {1 \over a^2} \left[ - A A^{|\alpha}_{\;\;\;\;\beta}
       + {1 \over 2} \left( - A^2 + B^\gamma B_\gamma 
       \right)^{|\alpha}_{\;\;\;\;\beta}
       - 2 C^{\alpha\gamma} A_{,\beta|\gamma}
       - \left( C^{\alpha\gamma}_{\;\;\;\;\;|\beta}
       + C^{\gamma|\alpha}_\beta - C^{\alpha|\gamma}_\beta \right)
       A_{,\gamma} \right]
   \nonumber \\
   & & \qquad \qquad
       - {1 \over 3} \delta^\alpha_\beta \Bigg\{
       \left[ 
       \left( \dot C^\gamma_\gamma + {1\over a} B^\gamma_{\;\;|\gamma}
       \right) A
       + 2 C^{\gamma\delta} \left( \dot C_{\gamma\delta}
       + {1\over a} B_{\gamma|\delta} \right)
       + {1\over a} B^\delta \left( 2 C^\gamma_{\delta|\gamma}
       - C^\gamma_{\gamma|\delta} \right) \right]^\cdot
   \nonumber \\
   & & \qquad \qquad
       + 3 H \left[ 
       \left( \dot C^\gamma_\gamma 
       + {1\over a} B^\gamma_{\;\;|\gamma} \right) A
       + 2 C^{\gamma\delta} \left( \dot C_{\gamma\delta}
       + {1\over a} B_{\gamma|\delta} \right)
       + {1\over a} B^\delta \left( 2 C^\gamma_{\delta|\gamma}
       - C^\gamma_{\gamma|\delta} \right) \right]
   \nonumber \\
   & & \qquad \qquad
       + \left( \dot C^\gamma_\gamma + {1\over a} B^\gamma_{\;\;|\gamma}
       \right)^\cdot A
       - {1\over a} \left( \dot C^\gamma_\gamma 
       + {1\over a} B^\gamma_{\;\;|\gamma} \right)_{|\delta} B^\delta
       + \delta K \left( \dot C^\gamma_\gamma 
       + {1\over a} B^\gamma_{\;\;|\gamma} \right)
   \nonumber \\
   & & \qquad \qquad
       + {1 \over a^2} \left[ - A A^{|\gamma}_{\;\;\;\;\gamma}
       + {1 \over 2} \left( - A^2 + B^\delta B_\delta 
       \right)^{|\gamma}_{\;\;\;\;\gamma}
       - 2 C^{\gamma\delta} A_{,\gamma|\delta}
       - \left( 2 C^{\gamma\delta}_{\;\;\;\;|\gamma}
       - C^{\gamma|\delta}_\gamma \right) A_{,\delta} \right]
       \Bigg\}
   \nonumber \\
   & & \qquad \qquad
       + {1 \over a} B^\alpha_{\;\;|\gamma}
       \left[ \dot C^\gamma_\beta + {1\over 2a} \left( B^\gamma_{\;\;|\beta}
       + B_\beta^{\;\;|\gamma} \right) \right]
       - {1 \over a} B^\gamma_{\;\;|\beta} \left[ \dot C^\alpha_\gamma
       + {1 \over 2a} \left( B^\alpha_{\;\;|\gamma}
       + B_\gamma^{\;\;|\alpha} \right) \right]
   \nonumber \\
   & & \qquad \qquad
       + {1 \over a^2} \Bigg\{
       2 C^{\gamma\delta} \left( C^\alpha_{\delta|\beta\gamma}
       + C^{\;\;\;\;|\alpha}_{\delta\beta\;\;\;\;\gamma}
       - C^\alpha_{\beta|\delta\gamma} 
       - C^{\;\;\;\;|\alpha}_{\delta\gamma\;\;\;\;\beta} \right)
       + 2 C^{\alpha\gamma} \left( 
       C^\delta_{\gamma|\beta\delta}
       + C^\delta_{\beta|\gamma\delta}
       - C^{\;\;\;\;|\delta}_{\beta\gamma\;\;\;\delta}
       - C^\delta_{\delta|\gamma\beta} \right)
       - {4 \over 3} R^{(3)} C^\alpha_\gamma C^\gamma_\beta
   \nonumber \\
   & & \qquad \qquad
       + \left( 2 C^\gamma_{\delta|\gamma}
       - C^\gamma_{\gamma|\delta} \right)
       \left( C^{\alpha\delta}_{\;\;\;\;\;|\beta}
       + C^{\delta|\alpha}_\beta
       - C^{\alpha|\delta}_\beta \right) 
       - C_{\gamma\delta|\beta} C^{\gamma\delta|\alpha}
       + 2 C^{\alpha\gamma|\delta} \left( C_{\beta\delta|\gamma} 
       - C_{\beta\gamma|\delta} \right) 
   \nonumber \\
   & & \qquad \qquad
       - {1 \over 3} \delta^\alpha_\beta \Bigg[ 
       4 C^{\gamma\delta} \left( C^\epsilon_{\gamma|\delta\epsilon}
       + C^{\epsilon}_{\gamma|\epsilon\delta}
       - C^{\;\;\;\;|\epsilon}_{\gamma\delta\;\;\;\;\epsilon}
       - C^\epsilon_{\epsilon|\gamma\delta} \right) 
       - {4 \over 3} R^{(3)} C^\delta_\gamma C^\gamma_\delta 
   \nonumber \\
   & & \qquad \qquad
       + \left( 2 C^\epsilon_{\delta|\epsilon} 
       - C^\epsilon_{\epsilon|\delta} \right)
       \left( 2 C^{\gamma\delta}_{\;\;\;\;|\gamma} 
       - C^{\gamma|\delta}_\gamma 
       \right) 
       + C^{\gamma\delta|\epsilon} \left( 
       2 C_{\gamma\epsilon|\delta} - 3 C_{\gamma\delta|\epsilon} \right)
       \Bigg] \Bigg\}
   \nonumber \\
   & & \qquad \qquad
       -16 \pi G C^{\alpha\gamma} \Pi_{\beta\gamma}
   \nonumber \\
   & & \qquad
       \equiv N_{4\beta}^{\;\alpha}.
   \label{pert-eq4}
\eea
Energy conservation equation:
\bea
   & & \dot \mu + 3 H \left( \mu + p \right)
   \nonumber \\
   & & \qquad
       + \delta \dot \mu + 3 H \left( \delta \mu + \delta p \right)
       - \left( \mu + p \right) \left( \delta K - 3 H A \right)
       + {1\over a} Q^\alpha_{\;\;\;|\alpha} 
   \nonumber \\
   & & \qquad \qquad
       = - {1\over a} \delta \mu_{,\alpha} B^\alpha
       + \left( \delta \mu + \delta p \right) \left( \delta K - 3 H A \right)
       + \left( \mu + p \right) \left[ A \delta K
       + {3 \over 2} H \left( A^2 - B^\alpha B_\alpha \right) \right]
   \nonumber \\
   & & \qquad \qquad \qquad
       - {1 \over a} \left[ A Q^\alpha_{\;\;\;|\alpha}
       + Q^\alpha \left( 2 A_{,\alpha}
       + C^\beta_{\beta|\alpha} - 2 C^\beta_{\alpha|\beta} \right)
       - 2 C^{\alpha\beta} Q_{\alpha|\beta} \right]
       - \Pi^{\alpha\beta} \left( \dot C_{\alpha\beta}
       + {1\over a} B_{\alpha|\beta} \right)
   \nonumber \\
   & & \qquad \qquad
       \equiv N_5.
   \label{pert-eq5}
\eea
Momentum conservation equation:
\bea
   & & \dot Q_\alpha +4 H Q_\alpha + {1\over a} \left[ \left( \mu + p \right)
       A_{,\alpha} + \delta p_{,\alpha} + \Pi^\beta_{\alpha|\beta} \right]
   \nonumber \\
   & & \qquad
       = Q_\alpha \left( \delta K - 3 H A \right)
       + {1\over a} \Bigg\{
       - Q_{\alpha|\beta} B^\beta 
       - Q_\beta B^\beta_{\;\;|\alpha}
       - \left( \delta \mu + \delta p \right) A_{,\alpha}
       + A \left[ \left( \mu + p \right) A_{,\alpha} 
       - \delta p_{,\alpha}
       - \Pi^\beta_{\alpha|\beta} \right]
   \nonumber \\
   & & \qquad \qquad
       - \left( \mu + p \right) B^\beta B_{\beta|\alpha}
       + 2 \left( C^{\gamma\beta} \Pi_{\alpha\gamma} \right)_{|\beta}
       - \Pi^\gamma_\alpha C^\beta_{\beta|\gamma}
       + \Pi^\beta_\gamma C^\gamma_{\beta|\alpha}
       - A_{,\beta} \Pi^\beta_\alpha 
       \Bigg\}
   \nonumber \\
   & & \qquad 
       \equiv N_{6\alpha}.
   \label{pert-eq6}
\eea        
In the multi-component situation we additionally have the energy and 
the momentum conservation of individual component in eqs. 
(\ref{E-conservation-i},\ref{Mom-conservation-i}).

\noindent
Energy conservation equation for  the $i$-th component:
\bea
   & & \dot \mu_{(i)} + 3 H \left( \mu_{(i)} + p_{(i)} \right)
       + {1 \over a} I_{(i)0}
   \nonumber \\
   & & \qquad
       + \delta \dot \mu_{(i)} 
       + 3 H \left( \delta \mu_{(i)} + \delta p_{(i)} \right)
       - \left( \mu_{(i)} + p_{(i)} \right) \left( \delta K - 3 H A \right)
       + {1 \over a} Q_{(i)|\alpha}^\alpha
       + {1 \over a} \delta I_{(i)0}
   \nonumber \\
   & & \qquad \qquad
       = - {1 \over a} \delta \mu_{(i),\alpha} B^\alpha
       + \left( \delta \mu_{(i)} + \delta p_{(i)} \right) 
       \left( \delta K - 3 H A \right)
       + \left( \mu_{(i)} + p_{(i)} \right) A \delta K
       + {3 \over 2} H \left( \mu_{(i)} + p_{(i)} \right)
       \left( A^2 - B^\alpha B_\alpha \right)
   \nonumber \\
   & & \qquad \qquad \qquad
       + {1 \over a} \left[ 
       - Q^\alpha_{(i)|\alpha} A
       + 2 \left( C^{\alpha\beta} Q_{(i)\beta} \right)_{|\alpha}
       - C_\alpha^{\alpha|\beta} Q_{(i)\beta}
       - 2 A_{,\alpha} Q_{(i)}^\alpha \right]
       - \Pi^{\alpha\beta}_{(i)} \left( {1 \over a} B_{\alpha|\beta}
       + \dot C_{\alpha\beta} \right)
       - {1 \over a} \delta I_{(i)\alpha} B^\alpha
   \nonumber \\
   & & \qquad \qquad
       \equiv N_{(i)5}.
   \label{pert-eq5i}
\eea
Momentum conservation equation for the $i$-th component:
\bea
   & & \dot Q_{(i)\alpha} + 4 H Q_{(i)\alpha}
       + {1 \over a} \left( \mu_{(i)} + p_{(i)} \right) A_{,\alpha}
       + {1 \over a} \left( \delta p_{(i),\alpha} 
       + \Pi^{\;\;\;\; \beta}_{(i)\alpha|\beta}
       - \delta I_{(i)\alpha} \right)
   \nonumber \\
   & & \qquad
       = {1 \over a} \Bigg\{ 
       - \left( \delta p_{(i),\alpha} + \Pi^{\;\;\;\; \beta}_{(i)\alpha|\beta}
       - \delta I_{(i)\alpha} \right) A
       - \left( \delta \mu_{(i)} + \delta p_{(i)} \right) A_{,\alpha}
       + \left( \mu_{(i)} + p_{(i)} \right) \left( A A_{,\alpha}
       - B^\beta B_{\beta|\alpha} \right)
   \nonumber \\
   & & \qquad \qquad
       - Q_{(i)\alpha|\beta} B^\beta
       - Q_{(i)\beta} B^\beta_{\;\; |\alpha}
       + a \left( \delta K - 3 H A \right) Q_{(i)\alpha}
       + 2 \left( C^{\beta\gamma} \Pi_{(i)\alpha\gamma} \right)_{|\beta}
       - C^\beta_{\beta|\gamma} \Pi^{\;\;\;\;\gamma}_{(i)\alpha}
       + C^\gamma_{\beta|\alpha} \Pi^{\;\;\;\;\beta}_{(i)\gamma}
       - A_{,\beta} \Pi^{\;\;\;\;\beta}_{(i)\alpha}  
       \Bigg\}
   \nonumber \\
   & & \qquad
       \equiv N_{(i)6\alpha}.
   \label{pert-eq6i}
\eea
The collective fluid quantities are given in eq. (\ref{fluid-sum}).
The equations are presented with the quadratic combination of the
linear order terms located in the right-hand-side.
Still, notice that the equations are presented to the second-order
without separating the background order part.

Equations (\ref{pert-eq0}-\ref{pert-eq6i}) provide a
complete set valid for the Einstein's gravity with an imperfect fluid, 
thus the most general form of energy-momentum tensor.
We have not imposed any condition like the gauge condition.
In the following subsections we will consider the cases of
minimally coupled scalar fields, an electromagnetic field, 
and a broad class of generalized gravity theories.
We emphasize that even in these additional fields or
generalized gravity the above equations {\it remain valid}
with the fluid quantities reinterpreted to absorb
the contributions from the fields and the generalized gravity.

\subsection{Scalar field}

\subsubsection{Covariant equations}

The action for a minimally coupled scalar field is given as
\bea
   & & S = \int \sqrt{- \tilde g} \left[ {1 \over 16 \pi G} \tilde R
       - {1 \over 2} \tilde \phi^{,c} \tilde \phi_{,c} 
       - \tilde V (\tilde \phi) \right] d^4 x. 
   \label{action-MSF}
\eea
The equation of motion follows from the variation in $\tilde \phi$
\bea
   & & \tilde \phi^{;c}_{\;\;\; c} - \tilde V_{,\tilde \phi} = 0,
\eea
where 
$\tilde V_{,\tilde \phi} \equiv {\partial \tilde V \over \partial \tilde \phi}$.
{}From $\delta_{\tilde g_{ab}} {\cal L}_M \equiv {1\over 2} \sqrt{-\tilde g} 
\tilde T^{ab} \delta \tilde g_{ab}$ we have
the energy-momentum tensor
\bea
   & & \tilde T^{(\phi)}_{ab} = \tilde \phi_{,a} \tilde \phi_{,b} 
       - {1 \over 2} \tilde g_{ab} \tilde \phi^{,c} \tilde \phi_{,c}
       - \tilde g_{ab} \tilde V (\tilde \phi).
\eea

\subsubsection{Perturbations}

We decompose
\bea
   & & \tilde \phi \equiv \phi + \delta \phi. 
   \label{field-perturbation}
\eea
The equation of motion becomes
\bea
   & & \ddot \phi + 3 H \dot \phi + V_{,\phi}  
   \nonumber \\
   & & \qquad
       + \delta \ddot \phi + 3 H \delta \dot \phi
       - {\Delta \over a^2} \delta \phi + V_{,\phi\phi} \delta \phi 
       - 2 A \ddot \phi - \dot \phi \left( \dot A + 6 H A 
       - {1 \over a} B^\alpha_{\;\; |\alpha} - \dot C^\alpha_\alpha \right) 
   \nonumber \\
   & & \qquad \qquad
       = 2 A \left[ \delta \ddot \phi 
       + 3 H \delta \dot \phi
       - 2 A \ddot \phi 
       - \dot \phi \left( 2 \dot A + 6 H A 
       - {1 \over a} B^\alpha_{\;\;|\alpha} 
       - \dot C^\alpha_\alpha \right) \right]
       + \delta \dot \phi \left( 
       \dot A - {1 \over a} B^\alpha_{\;\;|\alpha} 
       - \dot C^\alpha_\alpha \right) 
       - 2 {1 \over a} B^\alpha \delta \dot \phi_{,\alpha} 
   \nonumber \\
   & & \qquad \qquad \qquad
       + {1 \over a} \delta \phi_{,\alpha} 
       \left( {1 \over a} A^{,\alpha} - \dot B^\alpha 
       - 2 H B^\alpha - 2 {1 \over a} C^{\alpha\beta}_{\;\;\;\; |\beta}
       + {1 \over a} C^{\beta|\alpha}_{\beta} \right) 
       - 2 {1 \over a^2} \delta \phi_{,\alpha|\beta} C^{\alpha\beta} 
       - {1 \over 2} V_{,\phi\phi\phi} \delta \phi^2 
   \nonumber \\
   & & \qquad \qquad \qquad
       + \ddot \phi B_\alpha B^\alpha 
       + \dot \phi \left[ 
       {1 \over a} A_{,\alpha} B^\alpha 
       + B_\alpha \dot B^\alpha 
       + 3 H B^\alpha B_\alpha 
       + {1 \over a} B_\beta \left( 2 C^{\alpha\beta}_{\;\;\;\; |\alpha} 
       - C^{\alpha|\beta}_\alpha \right) 
       + 2 C^{\alpha\beta} \left( {1 \over a} B_{\alpha|\beta} 
       + \dot C_{\alpha\beta} \right) \right]
   \nonumber \\
   & & \qquad \qquad
       \equiv N_\phi.
   \label{MSF-EOM-pert}
\eea
The energy-momentum tensor gives: 
\bea
   \tilde T^{(\phi)}_{00}
   &=& {1 \over 2} \phi^{\prime 2} + a^2 V
       + \phi^\prime \delta \phi^\prime 
       + a^2 \left( V_{,\phi} \delta \phi + 2 V A \right)
   \nonumber \\
   & & + {1 \over 2} \delta \phi^{\prime 2}
       + {1 \over 2} \delta \phi_{,\alpha} \delta \phi^{,\alpha}
       + {1 \over 2} a^2 V_{,\phi\phi} \delta \phi^2
       + 2 a^2 V_{,\phi} \delta \phi A
       - \phi^\prime \delta \phi_{,\alpha} B^\alpha
       + {1 \over 2} \phi^{\prime 2} B_\alpha B^\alpha,
   \nonumber \\
   \tilde T^{(\phi)}_{0\alpha}
   &=& \phi^\prime \delta \phi_{,\alpha}
       - \left( {1 \over 2} \phi^{\prime 2} - a^2 V \right) B_\alpha
       + \delta \phi^\prime \delta \phi_{,\alpha}
       + \left( - \phi^\prime \delta \phi^\prime 
       + a^2 V_{,\phi} \delta \phi 
       + \phi^{\prime 2} A \right) B_\alpha,
   \nonumber \\
   \tilde T^{(\phi)}_{\alpha\beta}
   &=& g^{(3)}_{\alpha\beta} \left( {1\over 2} \phi^{\prime 2} - a^2 V \right)
       + g^{(3)}_{\alpha\beta} \left(
       \phi^\prime \delta \phi^\prime
       - a^2 V_{,\phi} \delta \phi
       - \phi^{\prime 2} A \right)
       + \left( \phi^{\prime 2} - 2 a^2 V \right) C_{\alpha\beta}
   \nonumber \\
   & & + \delta \phi_{,\alpha} \delta \phi_{,\beta}
       + 2 \left[ \phi^\prime \delta \phi^\prime
       - a^2 V_{,\phi} \delta \phi
       - \phi^{\prime 2} A \right] C_{\alpha\beta}
   \nonumber \\
   & & - {1 \over 2} g^{(3)}_{\alpha\beta} \left[
       - \delta \phi^{\prime 2}
       + \delta \phi_{,\gamma} \delta \phi^{,\gamma}
       + a^2 V_{,\phi\phi} \delta \phi^2
       + 4 \phi^\prime \delta \phi^\prime A
       - 2 \phi^\prime \delta \phi_{,\gamma} B^\gamma
       + \phi^{\prime 2} \left( - 4 A^2 + B_\gamma B^\gamma \right) \right].
   \label{MSF-Tab-pert}
\eea
{}Fluid quantities can be read from eq. (\ref{Tab-fluid}) as:
\bea
   \mu^{(\phi)} + \delta \mu^{(\phi)} 
   &=& {1 \over 2} \dot \phi^2 + V
       + \dot \phi \delta \dot \phi - \dot \phi^2 A 
       + V_{,\phi} \delta \phi 
   \nonumber \\
   & & + {1 \over 2} \delta \dot \phi^2 
       + {1 \over 2 a^2} \delta \phi_{,\alpha} \delta \phi^{,\alpha} 
       + {1 \over 2} V_{,\phi\phi} \delta \phi^2
       - 2 \dot \phi \delta \dot \phi A 
       + {1 \over a} \dot \phi \delta \phi_{,\alpha} B^\alpha
       + 2 \dot \phi^2 A^2 
       - {1 \over 2} \dot \phi^2 B^\alpha B_\alpha, 
   \nonumber \\
   p^{(\phi)} + \delta p^{(\phi)} 
   &=& {1 \over 2} \dot \phi^2 - V
       + \dot \phi \delta \dot \phi - \dot \phi^2 A
       - V_{,\phi} \delta \phi 
   \nonumber \\
   & & + {1 \over 2} \delta \dot \phi^2
       - {1 \over 6 a^2} \delta \phi_{,\alpha} \delta \phi^{,\alpha} 
       - {1 \over 2} V_{,\phi\phi} \delta \phi^2
       - 2 \dot \phi \delta \dot \phi A
       + {1 \over a} \dot \phi \delta \phi_{,\alpha} B^\alpha 
       + 2 \dot \phi^2 A^2 
       - {1 \over 2} \dot \phi^2 B^\alpha B_\alpha,
   \nonumber \\
   Q^{(\phi)}_\alpha 
   &=& - {1 \over a} \left[ \dot \phi \delta \phi_{,\alpha} 
       + \delta \phi_{,\alpha} 
       \left( \delta \dot \phi - \dot \phi A \right) \right], 
   \nonumber \\
   \Pi^{(\phi)}_{\alpha\beta} 
   &=& {1 \over a^2} \left( \delta \phi_{,\alpha} \delta \phi_{,\beta} 
       - {1\over 3} g^{(3)}_{\alpha\beta} 
       \delta \phi_{,\gamma} \delta \phi^{,\gamma} \right). 
   \label{MSF-fluid}
\eea
We indicate the quadratic parts (the quadratic combinations of two
linear-order terms) as $\delta \mu^{(q)}$, $\delta p^{(q)}$, 
$Q^{(q)}_\alpha$, and $\Pi^{(q)}_{\alpha\beta}$.

\subsection{Scalar fields}

\subsubsection{Covariant equations}

The action for multiple components of minimally coupled scalar fields is
\bea
   & & S = \int \sqrt{- \tilde g} \left[ {1 \over 16 \pi G} \tilde R
       - {1 \over 2} \sum_{k} \tilde \phi_{(k)}^{\;\;\;\; ,c} 
       \tilde \phi_{(k),c} - \tilde V (\tilde \phi_{(l)}) \right] d^4 x, 
\eea
where $i, j, \dots = 1, 2 \dots n$ indicate the $n$ scalar fields.
The equation of motion for $i$-th component is
\bea
   & & \tilde \phi^{\;\;\; ;c}_{(i)\;\; c} 
       - \tilde V_{,\tilde \phi_{(i)}} = 0. 
\eea
The energy-momentum tensor is
\bea
   & & \tilde T^{(\phi)}_{ab} 
       = \sum_k \left( \tilde \phi_{(k),a} \tilde \phi_{(k),b} 
       - {1 \over 2} \tilde g_{ab} \tilde \phi_{(k)}^{\;\;\;\; ,c} 
       \tilde \phi_{(k),c} \right)
       - \tilde g_{ab} \tilde V (\tilde \phi_{(l)}).
\eea

\subsubsection{Perturbations}

We introduce
\bea
   & & \tilde \phi_{(i)} \equiv \phi_{(i)} + \delta \phi_{(i)}. 
\eea
The equation of motion for $i$-th component becomes
\bea
   & & \ddot \phi_{(i)} + 3 H \dot \phi_{(i)} + V_{,\phi_{(i)}}  
   \nonumber \\
   & & \quad
       + \delta \ddot \phi_{(i)} + 3 H \delta \dot \phi_{(i)}
       - {\Delta \over a^2} \delta \phi_{(i)} 
       + \sum_k V_{,\phi_{(i)}\phi_{(k)}} \delta \phi_{(k)} 
       - 2 A \ddot \phi_{(i)} - \dot \phi_{(i)} \left( \dot A + 6 H A 
       - {1 \over a} B^\alpha_{\;\; |\alpha} - \dot C^\alpha_\alpha \right) 
   \nonumber \\
   & & \quad \quad
       = 2 A \left[ \delta \ddot \phi_{(i)} 
       + 3 H \delta \dot \phi_{(i)}
       - 2 A \ddot \phi_{(i)}
       - \dot \phi_{(i)} \left( 2 \dot A + 6 H A
       - {1 \over a} B^\alpha_{\;\;|\alpha}
       - \dot C^\alpha_\alpha \right) \right]
       + \delta \dot \phi_{(i)} 
       \left( \dot A - {1 \over a} B^\alpha_{\;\;|\alpha} 
       - \dot C^\alpha_\alpha \right) 
       - 2 {1 \over a} B^\alpha \delta \dot \phi_{(i),\alpha} 
   \nonumber \\
   & & \quad \quad \quad
       + {1 \over a} \delta \phi_{(i),\alpha} 
       \left( {1 \over a} A^{,\alpha} - \dot B^\alpha 
       - 2 H B^\alpha - 2 {1 \over a} C^{\alpha\beta}_{\;\;\;\; |\beta}
       + {1 \over a} C^{\beta|\alpha}_{\beta} \right) 
       - 2 {1 \over a^2} \delta \phi_{(i),\alpha|\beta} C^{\alpha\beta} 
       - {1 \over 2} \sum_{k,l} V_{,\phi_{(i)}\phi_{(k)}\phi_{(l)}} 
       \delta \phi_{(k)} \delta \phi_{(l)}
   \nonumber \\
   & & \quad \quad \quad
       + \ddot \phi_{(i)} B_\alpha B^\alpha
       + \dot \phi_{(i)} \left[ 
       {1 \over a} A_{,\alpha} B^\alpha 
       + B_\alpha \dot B^\alpha 
       + 3 H B^\alpha B_\alpha 
       + {1 \over a} B_\beta \left( 2 C^{\alpha\beta}_{\;\;\;\; |\alpha} 
       - C^{\alpha|\beta}_\alpha \right) 
       + 2 C^{\alpha\beta} \left( {1 \over a} B_{\alpha|\beta} 
       + \dot C_{\alpha\beta} \right) \right]
   \nonumber \\
   & & \quad \quad
       \equiv N_{\phi_{(i)}}.
   \label{MSFs-EOM-pert}
\eea
The energy-momentum tensor gives: 
\bea
   \tilde T^{(\phi)}_{00}
   &=& {1 \over 2} \sum_k \phi_{(k)}^{\prime 2} + a^2 V
       + \sum_k \left( \phi_{(k)}^\prime \delta \phi_{(k)}^\prime 
       + a^2 V_{,\phi_{(k)}} \delta \phi_{(k)} \right)
       + 2 a^2 V A
       + \sum_{k} \Bigg( {1 \over 2} \delta \phi_{(k)}^{\prime 2}
       + {1 \over 2} \delta \phi_{(k),\alpha} 
       \delta \phi_{(k)}^{\;\;\;\; ,\alpha}
   \nonumber \\
   & & 
       + {1 \over 2} a^2 \sum_l V_{,\phi_{(k)} \phi_{(l)}} 
       \delta \phi_{(k)} \delta \phi_{(l)}
       + 2 a^2 V_{,\phi_{(k)}} \delta \phi_{(k)} A
       - \phi_{(k)}^\prime \delta \phi_{(k),\alpha} B^\alpha
       + {1 \over 2} \phi_{(k)}^{\prime 2} B_\alpha B^\alpha \Bigg),
   \nonumber \\
   \tilde T^{(\phi)}_{0\alpha}
   &=& \sum_k \phi^\prime_{(k)} \delta \phi_{(k),\alpha}
       - \left( {1 \over 2} \sum_k \phi_{(k)}^{\prime 2} 
       - a^2 V \right) B_\alpha
       + \sum_k \left[ \delta \phi_{(k)}^\prime \delta \phi_{(k),\alpha}
       + \left( - \phi_{(k)}^\prime \delta \phi_{(k)}^\prime 
       + a^2 V_{,\phi_{(k)}} \delta \phi_{(k)} 
       + \phi_{(k)}^{\prime 2} A \right) B_\alpha \right],
   \nonumber \\
   \tilde T^{(\phi)}_{\alpha\beta}
   &=& g^{(3)}_{\alpha\beta} \left( {1 \over 2} \sum_k \phi_{(k)}^{\prime 2} 
       - a^2 V \right)
       + g^{(3)}_{\alpha\beta} \sum_k \left(
       \phi_{(k)}^\prime \delta \phi_{(k)}^\prime
       - a^2 V_{,\phi_{(k)}} \delta \phi_{(k)}
       - \phi_{(k)}^{\prime 2} A \right)
       + \left( \sum_k \phi_{(k)}^{\prime 2} - 2 a^2 V \right) C_{\alpha\beta}
   \nonumber \\
   & & + \sum_k \Bigg\{ \delta \phi_{(k),\alpha} \delta \phi_{(k),\beta}
       + 2 \left[ \phi_{(k)}^\prime \delta \phi_{(k)}^\prime
       - a^2 V_{,\phi_{(k)}} \delta \phi_{(k)}
       - \phi_{(k)}^{\prime 2} A \right] C_{\alpha\beta}
       - {1 \over 2} g^{(3)}_{\alpha\beta} \Bigg[
       - \delta \phi_{(k)}^{\prime 2}
       + \delta \phi_{(k),\gamma} \delta \phi_{(k)}^{\;\;\;\; ,\gamma}
   \nonumber \\
   & & 
       + a^2 \sum_l V_{,\phi_{(k)}\phi_{(l)}} 
       \delta \phi_{(k)} \delta \phi_{(l)}
       + 4 \phi_{(k)}^\prime \delta \phi_{(k)}^\prime A
       - 2 \phi_{(k)}^\prime \delta \phi_{(k),\gamma} B^\gamma
       + \phi_{(k)}^{\prime 2} \left( - 4 A^2 
       + B_\gamma B^\gamma \right) \Bigg] \Bigg\}.
   \label{MSFs-Tab-pert}
\eea
{}Fluid quantities can be read from eq. (\ref{Tab-fluid}) as:
\bea
   \mu^{(\phi)} + \delta \mu^{(\phi)} 
   &=& {1 \over 2} \sum_k \dot \phi_{(k)}^2 + V
       + \sum_k \left( \dot \phi_{(k)} \delta \dot \phi_{(k)} 
       - \dot \phi_{(k)}^2 A 
       + V_{,\phi_{(k)}} \delta \phi_{(k)} \right) 
       + \sum_k \Bigg[
       {1 \over 2} \delta \dot \phi_{(k)}^2 
       + {1 \over 2 a^2} \delta \phi_{(k),\alpha} 
       \delta \phi_{(k)}^{\;\;\;\; ,\alpha} 
   \nonumber \\
   & & 
       + {1 \over 2} \sum_l V_{,\phi_{(k)}\phi_{(l)}} 
       \delta \phi_{(k)} \delta \phi_{(l)}
       - 2 \dot \phi_{(k)} \delta \dot \phi_{(k)} A 
       + {1 \over a} \dot \phi_{(k)} \delta \phi_{(k),\alpha} B^\alpha
       + \left( 2 A^2 - {1 \over 2} B^\alpha B_\alpha \right)
       \dot \phi_{(k)}^2 \Bigg], 
   \nonumber \\
   p^{(\phi)} + \delta p^{(\phi)} 
   &=& \sum_k {1 \over 2} \dot \phi_{(k)}^2 - V
       + \sum_k \left( \dot \phi_{(k)} \delta \dot \phi_{(k)} 
       - \dot \phi_{(k)}^2 A
       - V_{,\phi_{(k)}} \delta \phi_{(k)} \right) 
       + \sum_k \Bigg[ 
       {1 \over 2} \delta \dot \phi_{(k)}^2
       - {1 \over 6 a^2} \delta \phi_{(k),\alpha} 
       \delta \phi_{(k)}^{\;\;\;\; ,\alpha} 
   \nonumber \\
   & & 
       - {1 \over 2} \sum_l V_{,\phi_{(k)}\phi_{(l)}} 
       \delta \phi_{(k)} \delta \phi_{(l)}
       - 2 \dot \phi_{(k)} \delta \dot \phi_{(k)} A
       + {1 \over a} \dot \phi_{(k)} \delta \phi_{(k),\alpha} B^\alpha 
       + \left( 2 A^2 - {1 \over 2} B^\alpha B_\alpha \right)
       \dot \phi_{(k)}^2 \Bigg],
   \nonumber \\
   Q^{(\phi)}_\alpha 
   &=& - {1 \over a} \sum_k \left[ \dot \phi_{(k)} \delta \phi_{(k),\alpha} 
       + \left( \delta \dot \phi_{(k)} - \dot \phi_{(k)} A \right)
       \delta \phi_{(k),\alpha} \right],
   \nonumber \\
   \Pi^{(\phi)}_{\alpha\beta} 
   &=& {1 \over a^2} \sum_k \left( \delta \phi_{(k),\alpha} 
       \delta \phi_{(k),\beta} 
       - {1 \over 3} g^{(3)}_{\alpha\beta} 
       \delta \phi_{(k),\gamma} \delta \phi_{(k)}^{\;\;\;\; ,\gamma} \right). 
   \label{fluid-MSFs}
\eea
We indicate the quadratic parts as $\delta \mu^{(q)}$, $\delta p^{(q)}$, 
$Q^{(q)}_\alpha$, and $\Pi^{(q)}_{\alpha\beta}$.

\subsection{Generalized gravity theories}
                                                   \label{sec:GGT}

\subsubsection{Covariant equations}

As the action for a class of generalized gravity theories
we consider
\bea
   & & S = \int \sqrt{- \tilde g}
       \left[ {1 \over 2} \tilde f (\tilde \phi^K, \tilde R)
       - {1 \over 2} \tilde g_{IJ} (\tilde \phi^K) \tilde \phi^{I,c} 
       \tilde \phi^J_{\;\;,c}
       - \tilde V (\tilde \phi^K)
       + \tilde L_m \right] d^4 x.
   \label{GGT-action}
\eea
$\tilde \phi^I$ is the $I$-th component of $N$ scalar fields.
The capital indices $I, J, K, \dots = 1,2,3,$ $\dots, N$ indicate the
scalar fields, and the summation convention is used for repeated indices.
$\tilde f(\tilde \phi^K, \tilde R)$ is a general algebraic function of 
$\tilde R$ and the scalar fields $\tilde \phi^I$, and 
$\tilde g_{IJ} (\tilde \phi^K)$ and $\tilde V(\tilde \phi^K)$ are general
algebraic functions of the scalar fields.
We include a nonlinear sigma type kinetic term where
the kinetic matrix $\tilde g_{IJ}$ is considered as a Riemannian metric on the
manifold with the coordinates $\tilde \phi^I$.
The matter part Lagrangian $\tilde L_m$ includes the fluids,
the kinetic components, and the interaction with the fields, as well.
We have introduced the general action in eq. (\ref{GGT-action}) 
in \cite{Hwang-GGT-1990,Hwang-1991} as a toy model which 
allows quite general handling of various different generalized gravity 
theories in a unified manner, see \cite{Hwang-Noh-GGT}. 
Our generalized gravity includes as subset: 
$f(R)$ gravity which includes $R^2$ gravity,
the scalar-tensor theory which includes the Jordan-Brans-Dicke theory,
the non-minimally coupled scalar field, the induced gravity,
the low-energy effective action of string theory, etc,
and various combinations of such gravity theories with additional multiple
fields and fluids.
It does not, however, include higher-derivative theories with
terms like $R^{ab} R_{ab}$, see \cite{NH-Rab} for its role.

The gravitational field equation and the equation of motion become:
\bea
   & & \tilde G_{ab} = {1 \over \tilde F} \Bigg[ \tilde T_{ab}
       + \tilde g_{IJ} \left( \tilde \phi^I_{\;\;,a} \tilde \phi^J_{\;\;,b}
       - {1 \over 2} \tilde g_{ab} \tilde \phi^{I,c} \tilde \phi^J_{\;\;,c} 
       \right)
       + {1 \over 2} \left( \tilde f - \tilde R \tilde F - 2 \tilde V \right) 
       \tilde g_{ab} + \tilde F_{,a;b} 
       - \tilde g_{ab} \tilde F^{;c}_{\;\;\; c} \Bigg]
       \equiv 8 \pi G \tilde T_{ab}^{({\rm eff})},
   \label{GFE} \\
   & & \tilde \phi^{I;c}_{\;\;\;\;\; c} 
       + {1 \over 2} \left( \tilde f - 2 \tilde V \right)^{,I}
       + \tilde \Gamma^I_{JK} \tilde \phi^{J,c} \tilde \phi^K_{\;\;,c}
       = - \tilde L_m^{\;\;\; ,I} \equiv \tilde \Gamma^I,
   \label{EOM} \\
   & & \tilde T_{a;b}^{b} = \tilde L_{m,J} \tilde \phi^J_{\;\; ,a},
   \label{fluid-conservation}
\eea
where $\tilde F \equiv \partial \tilde f / \partial \tilde R$;
$\tilde g^{IJ}$ is the inverse metric of $\tilde g_{IJ}$,
$\tilde \Gamma^I_{JK} \equiv {1 \over 2} \tilde g^{IL}
\left( \tilde g_{LJ,K} + \tilde g_{LK,J} - \tilde g_{JK,L} \right)$,
and $\tilde V_{,\tilde I} \equiv \partial \tilde V/(\partial \tilde \phi^I)$.
Introduction of the effective energy-momentum tensor 
$\tilde T_{ab}^{({\rm eff})}$ provides a useful trick to derive and handle the
perturbed set of equations \cite{Hwang-GGT-1990}.
It allows the equations derived in Einstein gravity remain valid
with the energy-momentum parts replaced by the effective ones.

We note that the gravity theory in eq. (\ref{GGT-action}) 
can be transformed to Einstein's gravity through a conformal rescaling 
of the metric and rescaling of one of the fields.
As the result we have Einstein's gravity sector with only complications
appearing in the modified form of the field potential; the nonlinear
sigma type couplings in the kinetic part also remain.
We have studied the conformal transformation properties to the
linear order perturbation in \cite{Hwang-GGT-1990,Hwang-CT-1997},
and in most general form in the Appendix A of \cite{Hwang-Noh-CMB-2002}.
Extention to the second-order perturbation is trivial.

\subsubsection{Perturbed equations}

The perturbed set of equations can be derived similarly as in
the previous sections on the scalar fields.
We set
\bea
   & & \tilde F \equiv F + \delta F, \quad
       \tilde \Gamma^I \equiv \Gamma^I + \delta \Gamma^I.
\eea
Thus,
\bea
   & & \delta F = F_{,I} \delta \phi^I + F_{,R} \delta R
       + {1 \over 2} F_{,IJ} \delta \phi^I \delta \phi^J
       + F_{,IR} \delta \phi^I \delta R
       + {1 \over 2} F_{,RR} \delta R^2.
\eea
The equation of motion in eq. (\ref{EOM}) gives
\bea
   & & \ddot \phi^I + 3 H \dot \phi^I
       - {1 \over 2} g^{IJ} \left( f_{,J} - 2 V_{,J} \right)
       + \Gamma^I_{JK} \dot \phi^J \dot \phi^K + \Gamma^I
   \nonumber \\
   & & \quad
       + \delta \ddot \phi^I + 3 H \delta \dot \phi^I
       - {\Delta \over a^2} \delta \phi^I
       - 2 A \ddot \phi^I
       + \dot \phi^I \left( - \dot A - 6 H A
       + {1 \over a} B^\alpha_{\;\;|\alpha} + \dot C^\alpha_\alpha \right)
       - {1 \over 2} g^{IJ} \left[ F_{,J} \delta R
       + \left( f_{,LJ} - 2 V_{,LJ} \right) \delta \phi^L \right]
   \nonumber \\
   & & \quad
       - {1 \over 2} g^{IJ}_{\;\;\;\;,L} \delta \phi^L
       \left( f_{,J} - 2 V_{,J} \right)
       + 2 \Gamma^I_{JK} \left( \dot \phi^J \delta \dot \phi^K
       - A \dot \phi^J \dot \phi^K \right)
       + \Gamma^I_{JK,L} \delta \phi^L \dot \phi^J \dot \phi^K
       + \delta \Gamma^I
   \nonumber \\
   & & \quad \quad
       = 2 A \left[ \delta \ddot \phi^I
       + 3 H \delta \dot \phi^I
       - 2 A \ddot \phi^I 
       - \dot \phi^I \left( 2 \dot A + 6 H A 
       - {1 \over a} B^\alpha_{\;\;|\alpha}
       - \dot C^\alpha_\alpha \right)
       + 2 \Gamma^I_{JK} \left( \dot \phi^J \delta \dot \phi^K
       - A \dot \phi^J \dot \phi^K \right)
       + \Gamma^I_{JK,L} \delta \phi^L \dot \phi^J \dot \phi^K \right]
   \nonumber \\
   & & \quad \quad \quad
       + B^\alpha B_\alpha \left( \ddot \phi^I + 3 H \dot \phi^I
       + \Gamma^I_{JK} \dot \phi^J \dot \phi^K \right)
       + \left( \dot A - {1 \over a} B^\alpha_{\;\;|\alpha}
       - \dot C^\alpha_\alpha \right) \delta \dot \phi^I
       - 2 {1 \over a} B^\alpha \delta \dot \phi^I_{\;\;,\alpha}
       - 2 {1 \over a^2} C^{\alpha\beta} \delta \phi^I_{\;\;,\alpha|\beta}
   \nonumber \\
   & & \quad \quad \quad
       + {1 \over a} \delta \phi^I_{\;\;,\alpha}
       \left( {1 \over a} A^{,\alpha} - \dot B^\alpha - 2 H B^\alpha
       - 2 {1 \over a} C^{\alpha\beta}_{\;\;\;\;|\beta}
       + {1 \over a} C^{\beta|\alpha}_\beta \right)
   \nonumber \\
   & & \quad \quad \quad
       + \dot \phi^I \left[ { 1\over a} A_{,\alpha} B^\alpha
       + B^\alpha \dot B_\alpha
       + {1 \over a} B_\alpha \left( 2 C^{\alpha\beta}_{\;\;\;\;|\beta}
       - C^{\beta|\alpha}_\beta \right)
       + 2 C^{\alpha\beta} \left( {1 \over a} B_{\alpha|\beta}
       + \dot C_{\alpha\beta} \right) \right]
   \nonumber \\
   & & \quad \quad \quad
       + {1 \over 4} g^{IJ} \left[ F_{,RJ} \delta R^2
       + 2 F_{,LJ} \delta R \delta \phi^L
       + \left( f_{,LMJ} - 2 V_{,LMJ} \right)
       \delta \phi^L \delta \phi^M \right]
   \nonumber \\
   & & \quad \quad \quad
       + {1 \over 2} g^{IJ}_{\;\;\;\;,L} \delta \phi^L
       \left[ F_{,J} \delta R + \left( f_{,LJ} - 2 V_{,LJ} \right)
       \delta \phi^L \right]
       + {1 \over 4} g^{IJ}_{\;\;\;\;,LM} \delta \phi^L \delta \phi^M
       \left( f_{,J} - 2 V_{,J} \right)
   \nonumber \\
   & & \quad \quad \quad
       + \Gamma^I_{JK} \left( - \delta \dot \phi^J \delta \dot \phi^K
       - 2 {1 \over a} \dot \phi^J \delta \phi^K_{\;\;,\alpha} B^\alpha
       + {1 \over a^2} \delta \phi^{J|\alpha} \delta \phi^K_{\;\;,\alpha}
       \right)
       - 2 \Gamma^I_{JK,L} \delta \phi^L \dot \phi^J \delta \dot \phi^K
       - {1 \over 2} \Gamma^I_{JK,LM} \delta \phi^L \delta \phi^M
       \dot \phi^J \dot \phi^K
   \nonumber \\
   & & \quad \quad
       \equiv N_g.
   \label{GGT-EOM-pert}
\eea
$\delta R$ can be read from eq. (\ref{R}).
{}From eq. (\ref{GFE}) the effective energy momentum tensor gives:
\bea
   \tilde T_{00}^{({\rm eff})}
   &=&
       {1 \over 8 \pi G \tilde F} \Bigg\{ \tilde T_{00}
       + {1 \over 2} g_{IJ} \phi^{I\prime} \phi^{J\prime}
       - {1 \over 2} a^2 \left( f - RF - 2 V \right)
       - 3 {a^\prime \over a} F^\prime
   \nonumber \\
   & &
       + g_{IJ} \phi^{I\prime} \delta \phi^{J\prime}
       + {1 \over 2} g_{IJ,L} \delta \phi^L \phi^{I\prime} \phi^{J\prime}
       - {1 \over 2} a^2 \left[ \left( f_{,L} - 2 V_{,L} \right) \delta \phi^L
       - R \delta F \right]
   \nonumber \\
   & &
       - a^2 A \left( f - RF - 2 V \right)
       - 3 {a^\prime \over a} \delta F^\prime
       + \Delta \delta F
       - \left( B^\alpha_{\;\;|\alpha}
       + C^{\alpha\prime}_\alpha \right) F^\prime
   \nonumber \\
   & &
       + g_{IJ} \left( {1 \over 2} \delta \phi^{I\prime} \delta \phi^{J\prime}
       + {1 \over 2} \delta \phi^{I,\alpha} \delta \phi^J_{\;\;,\alpha}
       + {1 \over 2} B^\alpha B_\alpha \phi^{I\prime} \phi^{J\prime}
       - B^\alpha \phi^{I\prime} \delta \phi^J_{\;\;,\alpha} \right)
   \nonumber \\
   & &
       + g_{IJ,L} \delta \phi^L \phi^{I\prime} \delta \phi^{J\prime}
       + {1 \over 4} g_{IJ,LM} \delta \phi^L \delta \phi^M
       \phi^{I\prime} \phi^{J\prime}
   \nonumber \\
   & &
       + {1 \over 4} a^2 \left[ F_{,R} \delta R^2
       - \left( f_{,LM} - 2 V_{,LM} \right)
       \delta \phi^L \delta \phi^M \right]
       - a^2 A \left[ \left( f_{,L} - 2 V_{,L} \right) \delta \phi^L
       - R \delta F \right]
   \nonumber \\
   & &
       - \left( B^\alpha_{\;\;|\alpha}
       + C^{\alpha\prime}_\alpha \right) \delta F^\prime
       - 2 B^\alpha \delta F^\prime_{,\alpha}
       - \left( {a^\prime \over a} B^\alpha
       + 2 C^{\alpha\beta}_{\;\;\;\;|\beta}
       - C^{\beta|\alpha}_\beta \right) \delta F_{,\alpha}
       + 2 A \Delta \delta F
       - 2 C^{\alpha\beta} \delta F_{,\alpha|\beta}
   \nonumber \\
   & &
       + B^\alpha B_\alpha \left( F^{\prime\prime}
       + {a^\prime \over a} F^\prime \right)
       + \left[ 2 A_{,\alpha} B^\alpha
       + B_\alpha \left( 2 C^{\alpha\beta}_{\;\;\;\;|\beta}
       - C^{\beta|\alpha}_\beta \right)
       + 2 C^{\alpha\beta} \left( B_{\alpha|\beta} + C_{\alpha\beta}^\prime
       \right) \right] F^\prime \Bigg\},
   \nonumber \\
   \tilde T_{0\alpha}^{({\rm eff})}
   &=& {1 \over 8 \pi G \tilde F} \Bigg\{
       \tilde T_{0\alpha} + g_{IJ} \phi^{I\prime} \delta \phi^J_{\;\;,\alpha}
       - {1 \over 2} B_\alpha \left[ g_{IJ} \phi^{I\prime} \phi^{J\prime}
       + a^2 \left( f - RF - 2 V \right) \right]
       + \delta F^\prime_{,\alpha} - {a^\prime \over a} \delta F_{,\alpha}
   \nonumber \\
   & &
       - B_\alpha F^{\prime\prime}
       - \left( A_{,\alpha} + {a^\prime \over a} B_\alpha \right) F^\prime
   \nonumber \\
   & &
       + g_{IJ} \left[ \delta \phi^{I\prime} \delta \phi^J_{\;\;,\alpha}
       + B_\alpha \left( - \phi^{I\prime} \delta \phi^{J\prime}
       + A \phi^{I\prime} \phi^{J\prime} \right) \right]
       + g_{IJ,L} \delta \phi^L
       \left( \phi^{I\prime} \delta \phi^J_{\;\;,\alpha}
       - {1 \over 2} B_\alpha \phi^{I\prime} \phi^{J\prime} \right)
   \nonumber \\
   & &
       - {1 \over 2} a^2 B_\alpha \left[ \left( f_{,L} - 2 V_{,L}
       \right) \delta \phi^L - R \delta F \right]
       - B_\alpha \delta F^{\prime\prime}
       - \left( A_{,\alpha} + {a^\prime \over a} B_\alpha \right)
       \delta F^\prime
       - \left( {1 \over 2} B_\alpha^{\;\;|\beta}
       - {1 \over 2} B^\beta_{\;\;|\alpha}
       + C^{\beta\prime}_\alpha \right) \delta F_{,\beta}
   \nonumber \\
   & &
       + B_\alpha \Delta \delta F
       + 2 A B_\alpha F^{\prime\prime}
       + \left[ 2 A A_{,\alpha} + 2 {a^\prime \over a} A B_\alpha
       + B_\beta C^{\beta\prime}_\alpha
       - B^\beta B_{[\beta|\alpha]}
       + B_\alpha \left( A^\prime - B^\beta_{\;\;|\beta}
       - C^{\beta\prime}_\beta \right) \right] F^\prime \Bigg\},
   \nonumber \\
   \tilde T_{\alpha\beta}^{({\rm eff})}
   &=&
       {1 \over 8 \pi G \tilde F} \Bigg\{
       \tilde T_{\alpha\beta}
       + {1 \over 2} g^{(3)}_{\alpha\beta} \left[
       g_{IJ} \phi^{I\prime} \phi^{J\prime}
       + a^2 \left( f - R F - 2 V \right)
       + 2 F^{\prime\prime} + 2 {a^\prime \over a} F^\prime \right]
   \nonumber \\
   & &
       + C_{\alpha\beta} \left[ g_{IJ} \phi^{I\prime} \phi^{J\prime}
       + a^2 \left( f - R F - 2 V \right)
       + 2 F^{\prime\prime} + 2 {a^\prime \over a} F^\prime \right]
       + \delta F_{,\alpha|\beta}
       - \left( B_{(\alpha|\beta)} + C^\prime_{\alpha\beta} \right) F^\prime
   \nonumber \\
   & &
       + g^{(3)}_{\alpha\beta} \Bigg[
       g_{IJ} \left( \phi^{I\prime} \delta \phi^{J\prime}
       - A \phi^{I\prime} \phi^{J\prime} \right)
       + {1 \over 2} g_{IJ,L} \delta \phi^L \phi^{I\prime} \phi^{J\prime}
       + {1 \over 2} a^2 \left[ \left( f_{,L} - 2 V_{,L} \right)
       \delta \phi^L - R \delta F \right]
   \nonumber \\
   & &
       + \delta F^{\prime\prime} + {a^\prime \over a} \delta F^\prime
       - \Delta \delta F
       - 2 A F^{\prime\prime}
       - \left( A^\prime + 2 {a^\prime \over a} A
       - B^\gamma_{\;\;|\gamma} - C^{\gamma\prime}_\gamma
       \right) F^\prime \Bigg]
       + g_{IJ} \delta \phi^I_{\;\;,\alpha} \delta \phi^J_{\;\;,\beta}
   \nonumber \\
   & &
       + C_{\alpha\beta} \Bigg[
       2 g_{IJ} \left( \phi^{I\prime} \delta \phi^{J\prime}
       - A \phi^{I\prime} \phi^{J\prime} \right)
       + g_{IJ,L} \delta \phi^L \phi^{I\prime} \phi^{J\prime}
       + a^2 \left[ \left( f_{,L} - 2 V_{,L} \right) \delta \phi^L
       - R \delta F \right]
   \nonumber \\
   & &
       + 2 \left[ \delta F^{\prime\prime} + {a^\prime \over a} \delta F^\prime
       - \Delta \delta F
       - 2 A F^{\prime\prime}
       - \left( A^\prime + 2 {a^\prime \over a} A
       - B^\gamma_{\;\;|\gamma} - C^{\gamma\prime}_\gamma
       \right) F^\prime \right]
       \Bigg]
   \nonumber \\
   & &
       - \left( B_{(\alpha|\beta)} + C^\prime_{\alpha\beta} \right)
       \left( \delta F^\prime - 2 A F^\prime \right)
       - \left( 2 C^\gamma_{(\alpha|\beta)}
       - C^{\;\;\;\;|\gamma}_{\alpha\beta} \right)
       \left( \delta F_{,\gamma} - B_\gamma F^\prime \right)
   \nonumber \\
   & &
       + g^{(3)}_{\alpha\beta} \Bigg[
       {1 \over 2} g_{IJ} \left[ \delta \phi^{I\prime} \delta \phi^{J\prime}
       - 4 A \phi^{I\prime} \delta \phi^{J\prime}
       + \left( 4 A^2 - B^\gamma B_\gamma \right) \phi^{I\prime} \phi^{J\prime}
       + 2 B^\gamma \phi^{I\prime} \delta \phi^J_{\;\;,\gamma}
       - \delta \phi^{I,\gamma} \delta \phi^J_{\;\;,\gamma} \right]
   \nonumber \\
   & &
       + g_{IJ,L} \delta \phi^L \left( \phi^{I\prime} \delta \phi^{J\prime}
       - A \phi^{I\prime} \phi^{J\prime} \right)
       + {1 \over 4} g_{IJ,LM} \delta \phi^L \delta \phi^M \phi^{I\prime}
       \phi^{J\prime}
       + {1 \over 4} a^2 \left[ - F_{,R} \delta R^2
       + \left( f_{,LM} - 2 V_{,LM} \right) \delta \phi^L
       \delta \phi^M \right]
   \nonumber \\
   & &
       - 2 A \delta F^{\prime\prime}
       - \left( A^\prime + 2 {a^\prime \over a} A
       - B^\gamma_{\;\;|\gamma} - C^{\gamma\prime}_\gamma \right)
       \delta F^\prime
       + 2 B^\gamma \delta F^\prime_{,\gamma}
   \nonumber \\
   & &
       + \left( - A^{,\gamma}
       + B^{\gamma\prime} + {a^\prime \over a} B^\gamma
       + 2 C^{\gamma\delta}_{\;\;\;\;|\delta}
       - C^{\delta|\gamma}_\delta \right) \delta F_{,\gamma}
       + 2 C^{\gamma\delta} \delta F_{,\gamma|\delta}
       + \left( 4 A^2 - B^\gamma B_\gamma \right)
       \left( F^{\prime\prime} + {a^\prime \over a} F^\prime \right)
   \nonumber \\
   & &
       + \left[ 4 A A^\prime
       - A_{,\gamma} B^\gamma
       - B^\gamma B_\gamma^\prime
       - 2 A \left( B^\gamma_{\;\;|\gamma} + C^{\gamma\prime}_\gamma \right)
       - B_\gamma \left( 2 C^{\gamma\delta}_{\;\;\;\;|\delta}
       - C^{\delta|\gamma}_\delta \right)
       - 2 C^{\gamma\delta} \left( B_{\gamma|\delta}
       + C_{\gamma\delta}^\prime \right) \right] F^\prime
       \Bigg] \Bigg\}.
\eea
The fluid quantities follow from eq. (\ref{Tab-fluid}):
\bea
   \mu^{({\rm eff})} + \delta \mu^{({\rm eff})}
   &=&
       {1 \over 8 \pi G \tilde F} \Bigg\{
       \mu
       + {1 \over 2} g_{IJ} \dot \phi^I \dot \phi^J
       - {1 \over 2} \left( f - RF - 2V \right)
       - 3 H \dot F
   \nonumber \\
   & &
       + \delta \mu
       + g_{IJ} \left( \dot \phi^I \delta \dot \phi^J
       - A \dot \phi^I \dot \phi^J \right)
       + {1 \over 2} g_{IJ,L} \delta \phi^L \dot \phi^I \dot \phi^J
       - {1 \over 2} \left( f_{,L} - 2 V_{,L} \right) \delta \phi^L
   \nonumber \\
   & &
       - 3 H \delta \dot F
       + \left( {1 \over 2} R + {\Delta \over a^2} \right) \delta F
       - \left( - 6 H A + {1 \over a} B^\alpha_{\;\;|\alpha}
       + \dot C^\alpha_\alpha \right) \dot F
   \nonumber \\
   & &
       + {1 \over 2} g_{IJ} \left[ \delta \dot \phi^I \delta \dot \phi^J
       + {1 \over a^2} \delta \phi^{I,\alpha} \delta \phi^J_{\;\;,\alpha}
       - 4 A \dot \phi^I \delta \dot \phi^J
       + 2 {1 \over a} B^\alpha \dot \phi^I \delta \phi^J_{\;\;,\alpha}
       + \left( 4 A^2 - B^\alpha B_\alpha \right)
       \dot \phi^I \dot \phi^J \right]
   \nonumber \\
   & &
       + g_{IJ,L} \delta \phi^L \left( \dot \phi^I \delta \dot \phi^J
       - A \dot \phi^I \dot \phi^J \right)
       + {1 \over 4} g_{IJ,LM} \delta \phi^L \delta \phi^M
       \dot \phi^I \dot \phi^J
   \nonumber \\
   & &
       + {1 \over 4} \left[ F_{,R} \delta R^2
       - \left( f_{,LM} - 2 V_{,LM} \right)
       \delta \phi^L \delta \phi^M \right]
       - \left( - 6 H A
       + {1 \over a} B^\alpha_{\;\;|\alpha}
       + \dot C^\alpha_\alpha \right)
       \left( \delta \dot F - 2 A \dot F \right)
   \nonumber \\
   & &
       - \left[ 3 H B^\alpha
       + {1 \over a} \left( 2 C^{\alpha\beta}_{\;\;\;\;|\beta}
       - C^{\beta|\alpha}_\beta \right) \right]
       \left( {1 \over a} \delta F_{,\alpha} - B_\alpha \dot F \right)
       - 2 {1 \over a^2} C^{\alpha\beta} \delta F_{,\alpha|\beta}
       + 2 \dot F C^{\alpha\beta} \left( {1 \over a} B_{\alpha|\beta}
       + \dot C_{\alpha\beta} \right)
       \Bigg\},
   \nonumber \\
   p^{({\rm eff})} + \delta p^{({\rm eff})}
   &=&
       {1 \over 8 \pi G \tilde F} \Bigg\{
       p
       + { 1\over 2} g_{IJ} \dot \phi^I \dot \phi^J
       + {1 \over 2} \left( f - RF - 2V \right)
       + \ddot F + 2 H \dot F
   \nonumber \\
   & &
       + \delta p
       + g_{IJ} \left( \dot \phi^I \delta \dot \phi^J
       - A \dot \phi^I \dot \phi^J \right)
       + {1 \over 2} g_{IJ,L} \delta \phi^L \dot \phi^I \dot \phi^J
       + {1 \over 2} \left( f_{,L} - 2 V_{,L} \right) \delta \phi^L
   \nonumber \\
   & &
       + \delta \ddot F + 2 H \delta \dot F
       - \left( {1 \over 2} R 
       + {2 \over 3} {\Delta \over a^2} \right) \delta F
       - 2 A \ddot F
       - \left[ \dot A + 4 H A
       - {2 \over 3} \left( {1 \over a} B^\alpha_{\;\;|\alpha}
       + \dot C^\alpha_\alpha \right) \right] \dot F
   \nonumber \\
   & &
       + g_{IJ} \left[ {1 \over 2} \delta \phi^I \delta \dot \phi^J
       - 2 A \dot \phi^I \delta \dot \phi^J
       + {1 \over 2} \left( 4 A^2 - B^\alpha B_\alpha \right)
       \dot \phi^I \dot \phi^J
       + {1 \over a} B^\alpha \dot \phi^I \delta \phi^J_{\;\;,\alpha}
       - {1 \over 6 a^2} \delta \phi^{I,\alpha} \delta \phi^J_{\;\;,\alpha}
       \right]
   \nonumber \\
   & &
       + g_{IJ,L} \delta \phi^L \left( \dot \phi^I \delta \dot \phi^J
       - A \dot \phi^I \dot \phi^J \right)
       + {1 \over 4} g_{IJ,LM} \delta \phi^L \delta \phi^M
       \dot \phi^I \dot \phi^J
       - {1 \over 4} \left[ F_{,R} \delta R^2
       - \left( f_{,LM} - 2 V_{,LM} \right)
       \delta \phi^L \delta \phi^M \right]
   \nonumber \\
   & &
       - 2 A \delta \ddot F
       - \left[ \dot A + 4 H A
       - {2 \over 3} \left( {1 \over a} B^\alpha_{\;\;|\alpha}
       + \dot C^\alpha_\alpha \right) \right]
       \left( \delta \dot F - 2 A \dot F \right)
       + 2 { 1\over a} B^\alpha \delta \dot F_{,\alpha}
   \nonumber \\
   & &
       + \left[ - {1 \over a} A^{,\alpha}
       + \dot B^\alpha + H B^\alpha
       + {2 \over 3a} \left( 2 C^{\alpha\beta}_{\;\;\;\;|\beta}
       - C^{\beta|\alpha}_\beta \right) \right]
       \left( {1 \over a} \delta F_{,\alpha} - B_\alpha \dot F \right)
       + {4 \over 3 a^2} C^{\alpha\beta} \delta F_{,\alpha|\beta}
   \nonumber \\
   & &
       + \left( 4 A^2 - B^\alpha B_\alpha \right) \ddot F
       + \left[ 2 A \dot A - {2 \over a} A_{,\alpha} B^\alpha
       - H B^\alpha B_\alpha
       - {4 \over 3} C^{\alpha\beta}
       \left( {1 \over a} B_{\alpha|\beta} + \dot C_{\alpha\beta} \right)
       \right] \dot F
       \Bigg\},
   \nonumber \\
   Q_\alpha^{({\rm eff})}
   &=&
       {1 \over 8 \pi G \tilde F} \Bigg\{
       Q_\alpha
       - {1 \over a} g_{IJ} \dot \phi^I \delta \phi^J_{\;\;,\alpha}
       + {1 \over a} \left( - \delta \dot F_{,\alpha}
       + H \delta F_{,\alpha} \right)
       + {1 \over a} A_{,\alpha} \dot F
   \nonumber \\
   & &
       - {1 \over a} g_{IJ} \left( \delta \dot \phi^I 
       - A \dot \phi^I \right) \delta \phi^J_{\;\;,\alpha}
       - {1 \over a} g_{IJ,L} \delta \phi^L \dot \phi^I
       \delta \phi^J_{\;\;,\alpha}
       + A \left[ -  3 {1 \over a} \dot F A_{,\alpha}
       + {1 \over a} \left( \delta \dot F_{,\alpha}
       - H \delta F_{,\alpha} \right) \right]
   \nonumber \\
   & &
       + {1 \over a} A_{,\alpha} \delta \dot F
       + {1 \over a} \left[{1 \over 2a} \left( B^{\;\;|\beta}_\alpha
       - B^\beta_{\;\;|\alpha} \right)
       + \dot C^\beta_\alpha \right] \delta F_{,\beta}
       - {1 \over a^2} B^\beta \delta F_{,\alpha|\beta}
       + {1 \over a} B^\beta B_{\alpha|\beta} \dot F
       \Bigg\},
   \nonumber \\
   \Pi_{\alpha\beta}^{({\rm eff})}
   &=&
       {1 \over 8 \pi G \tilde F} \Bigg\{
       \Pi_{\alpha\beta}
       + { 1\over a^2} \delta F_{,\alpha|\beta}
       - \left( {1 \over a} B_{(\alpha|\beta)}
       + \dot C_{\alpha\beta} \right) \dot F
       - {1 \over 3} g^{(3)}_{\alpha\beta} \left[
       {\Delta \over a^2} \delta F
       - \left( { 1\over a} B^\gamma_{\;\;|\gamma}
       + \dot C^\gamma_\gamma \right) \dot F \right]
   \nonumber \\
   & &
       + {1 \over a^2} g_{IJ} \delta \phi^I_{\;\;,\alpha}
       \delta \phi^J_{\;\;,\beta}
       - {2 \over 3} C_{\alpha\beta} \left[ {\Delta \over a^2} \delta F
       - \left( {1 \over a} B^\gamma_{\;\;|\gamma}
       + \dot C^\gamma_\gamma \right) \dot F \right]
   \nonumber \\
   & &
       - \left( {1 \over a} B_{(\alpha|\beta)}
       + \dot C_{\alpha\beta} \right)
       \left( \delta \dot F - 2 A \dot F \right)
       - { 1\over a} \left( 2 C^\gamma_{(\alpha|\beta)}
       - C_{\alpha\beta}^{\;\;\;\;|\gamma} \right)
       \left( {1 \over a} \delta F_{,\gamma} - B_\gamma \dot F \right)
   \nonumber \\
   & &
       - {1 \over 3} g^{(3)}_{\alpha\beta} \Bigg[
       {1 \over a^2} g_{IJ} \delta \phi^{I,\gamma} \delta \phi^J_{\;\;,\gamma}
       + 2 C^{\gamma\delta} \left( {1 \over a} B_{\gamma|\delta}
       + \dot C_{\gamma\delta} \right) \dot F
       - {2 \over a^2} C^{\gamma\delta} \delta F_{,\gamma|\delta}
   \nonumber \\
   & &
       - \left( { 1\over a} B^\gamma_{\;\;|\gamma}
       + \dot C^\gamma_\gamma \right) \left( \delta \dot F - 2 A \dot F \right)
       - {1 \over a} \left( 2 C^{\gamma\delta}_{\;\;\;\;|\delta}
       - C^{\delta|\gamma}_\delta \right) \left( {1 \over a} \delta F_{,\gamma}
       - B_\gamma \dot F \right)
       \Bigg] \Bigg\}.
   \label{GGT-fluid-pert}
\eea
We have used eq. (\ref{Tab-fluid-1}) for the energy-momentum tensor.
We indicate the quadratic parts as $\delta \mu^{({\rm eff},q)}$,
$\delta p^{({\rm eff},q)}$, $Q_\alpha^{({\rm eff},q)}$ 
and $\Pi_{\alpha\beta}^{({\rm eff},q)}$.
We note again that in this generalized gravity the basic equations in 
\S \ref{sec:Basic-equations} {\it remain valid} with the fluid quantities 
replaced by the effective ones.

\subsection{Electromagnetic field}

\subsubsection{Covariant equations}

The Lagrangian of electromagnetic field is given as
\bea
   & & {\cal L}_{\rm em} 
       = - {1 \over 4} \sqrt{- \tilde g} \tilde F^{ab} \tilde F_{ab},
\eea
where $\tilde F_{ab} \equiv \tilde A_{a,b} - \tilde A_{b,a}$.
The energy-momentum tensor becomes
\bea
   & & \tilde T^{({\rm em})}_{ab} = \tilde F_{ac} \tilde F_b^{\;\;c} 
       - {1 \over 4} \tilde g_{ab} \tilde F_{cd} \tilde F^{cd}.
\eea
We introduce \cite{Ellis-1971}
\bea
   & & \tilde F_{ab} = \tilde u_a \tilde E_b - \tilde u_b \tilde E_a 
       - \tilde \eta_{abcd} \tilde u^c \tilde H^d,
   \nonumber \\
   & & \tilde E_a \equiv \tilde F_{ab} \tilde u^b, \quad
       \tilde H_a \equiv {1 \over 2} \tilde \eta_{abcd} 
       \tilde u^b \tilde F^{cd}, \quad
       \tilde E^2 \equiv \tilde E^a \tilde E_a, \quad 
       \tilde H^2 \equiv \tilde H^a \tilde H_a, 
   \nonumber \\
   & & \tilde q \equiv - \tilde j^a \tilde u_a, \quad 
       \tilde J^a \equiv \tilde h^a_b \tilde j^b.
   \label{EM-decompose}
\eea
Then we have
\bea
   & & \tilde T^{({\rm em})}_{ab} = {1 \over 2} \tilde u_a \tilde u_b 
       \left( \tilde E^2 + \tilde H^2 \right)
       + 2 \tilde u_{(a} \tilde \eta_{b)cgd} \tilde u^c \tilde E^g \tilde H^d
       - \tilde E_a \tilde E_b - \tilde H_a \tilde H_b 
       + {1 \over 2} \tilde h_{ab} \left( \tilde E^2 + \tilde H^2 \right),
   \nonumber \\
   & & \tilde \mu^{({\rm em})} = {1 \over 2} 
       \left( \tilde E^2 + \tilde H^2 \right), \quad
       \tilde p^{({\rm em})} 
       = {1 \over 6} \left( \tilde E^2 + \tilde H^2 \right),
   \nonumber \\
   & & \tilde q^{({\rm em})}_a = \tilde \eta_{acgd} \tilde u^c 
       \tilde E^g \tilde H^d, \quad
       \tilde \pi^{({\rm em})}_{ab} 
       = - \tilde E_a \tilde E_b - \tilde H_a \tilde H_b
       + {1 \over 3} \tilde h_{ab} \left( \tilde E^2 + \tilde H^2 \right).
   \label{Maxwell-fluid}
\eea
{}From the Maxwell equations and the conservation equations
\bea
   & & \tilde F^{ab}_{\;\;\;\; ;b} = \tilde j^a, \quad 
       \tilde F_{[ab;c]} = 0, \quad
       \tilde j^a_{\;\; ;a} = 0,
\eea
we can derive the covariant forms of relativistic Maxwell's equations
\cite{Ellis-1971}:
\bea
   & & \tilde E^a_{\;\; ;b} \tilde h^b_a + 2 \tilde H_a \tilde \omega^a 
       = \tilde q, 
   \label{Maxwell-1} \\
   & & \tilde H^a_{\;\; ;b} \tilde h^b_a - 2 \tilde E_a \tilde \omega^a = 0,
   \label{Maxwell-2} \\
   & & \tilde h^a_b \tilde E^b_{\;\; ;c} \tilde u^c 
       = \tilde E^b \left( \tilde \omega^a_{\;\; b} + \tilde \sigma^a_{\;\; b}
       - {2 \over 3} \tilde \theta \tilde h^a_b \right) 
       + \tilde \eta^{abcd} u_b
       \left( \tilde a_c \tilde H_d - \tilde H_{c;d} \right) - \tilde J^a,
   \label{Maxwell-3} \\
   & & \tilde h^a_b \tilde H^b_{\;\; ;c} \tilde u^c 
       = \tilde H^b \left( \tilde \omega^a_{\;\; b} + \tilde \sigma^a_{\;\; b}
       - {2 \over 3} \tilde \theta \tilde h^a_b \right) 
       + \tilde \eta^{abcd} \tilde u_b
       \left( \tilde a_c \tilde E_d - \tilde E_{c;d} \right) - \tilde J^a,
   \label{Maxwell-4} \\
   & & \tilde q_{,a} \tilde u^a + \tilde \theta \tilde q 
       + \tilde h^a_b \tilde J^b_{\;\; ;a} + \tilde J^a \tilde a_a = 0.
   \label{Maxwell-5}
\eea

\subsubsection{Perturbations}

We take the normal-frame, thus $\tilde u_a = \tilde n_a$,
thus $\tilde \omega_{ab} = 0$.
Due to the high symmetry the Friedmann background does not support 
the electric or magnetic field.
Thus, $\tilde E_a$ and $\tilde H_a$ are already perturbed order.
We set 
\bea
   & & \tilde E_\alpha \equiv E_\alpha, \quad
       \tilde H_\alpha \equiv H_\alpha, 
\eea
where $E_\alpha$ and $H_\alpha$ are based on $g^{(3)}_{\alpha\beta}$.
Thus, $\tilde E_0 = - E_\alpha B^\alpha$ (which follows from
$\tilde E_a \tilde n^a = 0$), etc.
{}For $\eta^{abcd}$ see eq. (\ref{eta-relation}).
Equations (\ref{Maxwell-1}-\ref{Maxwell-5}) become:
\bea
   & & E^\alpha_{\;\;|\alpha} - a^2 \delta q 
       = 2 \left( C^{\alpha\beta} E_\beta \right)_{|\alpha}
       - C^\alpha_{\alpha|\beta} E^\beta,
   \label{Maxwell-1-pert} \\
   & & H^\alpha_{\;\;|\alpha} 
       = 2 \left( C^{\alpha\beta} H_\beta \right)_{|\alpha}
       - C^\alpha_{\alpha|\beta} H^\beta,
   \label{Maxwell-2-pert} \\
   & & \dot E^\alpha + H E^\alpha 
       + {1 \over a} \eta^{\alpha\beta\gamma} H_{\beta|\gamma}
       + J^\alpha
       = A \left( \dot E^\alpha + H E^\alpha \right)
       - {1 \over a} E^\alpha_{\;\;|\beta} B^\beta
       + E^\beta \left( {1 \over a} B^\alpha_{\;\;|\beta}
       + 2 \dot C^\alpha_\beta \right)
   \nonumber \\
   & & \qquad
       - E^\alpha \left( {1 \over a} B^\beta_{\;\;|\beta} 
       + \dot C^\beta_\beta \right)
       + {1 \over a} \eta^{\alpha\beta\gamma} \left(
       H_{\gamma} A_{,\beta} - H_{\beta|\gamma} C^\delta_\delta \right) 
       - {2 \over a} \eta^{\beta\gamma\delta} C^\alpha_\beta H_{\gamma|\delta},
   \label{Maxwell-3-pert} \\
   & & ( H_\alpha \Leftrightarrow E_\alpha ),
   \label{Maxwell-4-pert} \\
   & & \delta \dot q + 3 H \delta q 
       = - 3 H A \delta q - {1 \over a} \delta q_{,\alpha} B^\alpha
       + \delta K \delta q
       - {1 \over a^2} \left[ \left( J^\alpha - 2 C^{\alpha\beta} J_\beta
       \right)_{|\alpha}
       + J^\alpha \left( C^\beta_{\beta|\alpha} + A_{,\alpha} \right) \right],
   \label{Maxwell-5-pert} 
\eea
where we set
\bea
   & & \tilde q \equiv q + \delta q, \quad
       \tilde J_\alpha \equiv J_\alpha,
\eea
with $J_\alpha$ based on $g^{(3)}_{\alpha\beta}$;
$J_\alpha$ in this subsection differs from the flux term in ADM notation
used in the other sections.
We have $q = 0$.

\noindent
The energy-momentum tensor becomes:
\bea
   \tilde T^{({\rm em})}_{00}
   &=& {1 \over 2} \left( E^\alpha E_\alpha + H^\alpha H_\alpha \right),
   \nonumber \\
   \tilde T^{({\rm em})}_{0\alpha} 
   &=& - \eta_{\alpha\beta\gamma} E^\beta H^\gamma,
   \nonumber \\
   \tilde T^{({\rm em})}_{\alpha\beta} 
   &=& - E_\alpha E_\beta - H_\alpha H_\beta
       + {1 \over 2} g^{(3)}_{\alpha\beta} 
       \left( E^\gamma E_\gamma + H^\gamma H_\gamma \right).
   \label{EM-Tab}
\eea
{}Fluid quantities can be read from eq. (\ref{Tab-fluid}) as:
\bea
   & & \delta \mu^{({\rm em})} = 3 \delta p^{({\rm em})} 
       = {1 \over 2 a^2} \left( E^\alpha E_\alpha + H^\alpha H_\alpha \right),
   \nonumber \\
   & & Q^{({\rm em})}_\alpha = {1 \over a^2} \eta_{\alpha\beta\gamma} 
       E^\beta H^\gamma,
   \nonumber \\
   & & \Pi^{({\rm em})}_{\alpha\beta} = - {1 \over a^2} 
       \left[ E_\alpha E_\beta + H_\alpha H_\beta
       - {1 \over 3} g^{(3)}_{\alpha\beta} 
       \left( E^\gamma E_\gamma + H^\gamma H_\gamma \right) \right].
   \label{fluid-EM}
\eea
We have $\mu^{({\rm em})} = 0 = p^{({\rm em})}$.

\subsection{Null-geodesic and temperature anisotropy}

We introduce the photon four-velocity as:
\bea
   & & \tilde k^0 \equiv {1 \over a} \left( \nu + \delta \nu \right), \quad
       \tilde k^\alpha \equiv - {\nu \over a} 
       \left( e^\alpha + \delta e^\alpha \right);
   \nonumber \\
   & & \tilde k_0 = - a \nu \left( 1 + {\delta \nu \over \nu} 
       + 2 A - B_\alpha e^\alpha 
       + 2 A {\delta \nu \over \nu} - B_\alpha \delta e^\alpha \right),
   \nonumber \\
   & & \tilde k_\alpha = - a \nu \left( e_\alpha + \delta e_\alpha + B_\alpha
       + 2 C_{\alpha\beta} e^\beta
       + B_\alpha {\delta \nu \over \nu} + 2 C_{\alpha\beta} \delta e^\beta
       \right),
   \label{k-def}
\eea
where $e^\alpha$ and $\delta e^\alpha$ are based on $g^{(3)}_{\alpha\beta}$,
and $\nu$ and $e^\alpha$ are assumed to be the background order.
We have
\bea
   & & {d \over d \lambda} = {\partial x^a \over \partial \lambda}
       {\partial \over \partial x^a} = \tilde k^a \partial_a
       = {\nu \over a} \left( \partial_0 - e^\alpha \partial_\alpha
       + {\delta \nu \over \nu} \partial_0 - \delta e^\alpha \partial_\alpha
       \right).
\eea
Thus,
\bea
   & & {d \over d y} \equiv \partial_0 - e^\alpha \partial_\alpha,
   \label{y-def}
\eea
can be considered as a derivative along the background photon four-velocity.
The null equation, $\tilde k^a \tilde k_a = 0$, gives 
\bea
   \tilde k^a \tilde k_a 
   &=& \nu^2 \Bigg[ e^\alpha e_\alpha - 1
       + 2 \left( e^\alpha \delta e_\alpha - {\delta \nu \over \nu} - A
       + B_\alpha e^\alpha + C_{\alpha\beta} e^\alpha e^\beta \right)
   \nonumber \\
   & & + \delta e^\alpha \delta e_\alpha - {\delta \nu^2 \over \nu^2}
       - 2 {\delta \nu \over \nu} \left( 2 A - B_\alpha e^\alpha \right)
       + 2 \left( B_\alpha + 2 C_{\alpha\beta} e^\beta \right) \delta e^\alpha
       \Bigg]
       = 0.
   \label{null-eq}
\eea 
The geodesic equation, $\tilde k^a_{\;\;;b} \tilde k^b = 0$, 
using eq. (\ref{connections}), gives
\bea
   \tilde k^0_{\;\; ;b} \tilde k^b 
   &=& 
       {\nu^2 \over a^2} \Bigg\{ {(a \nu)^\prime \over a \nu} 
       + \left( {\delta \nu \over \nu} \right)^\prime
       + 2 {\nu^\prime \over \nu} {\delta \nu \over \nu}
       - {\delta \nu_{,\alpha} \over \nu} e^\alpha
       + 2 {a^\prime \over a} e^\alpha \delta e_\alpha
       + A^\prime - 2 {a^\prime \over a} A
   \nonumber \\
   & & 
       + \left( B_{\alpha|\beta} + C_{\alpha\beta}^\prime 
       + 2 {a^\prime \over a} C_{\alpha\beta} \right) e^\alpha e^\beta
       - 2 \left( A_{,\alpha} - {a^\prime \over a} B_\alpha \right) e^\alpha
   \nonumber \\
   & & 
       + {\delta \nu \over \nu} {\delta \nu^\prime \over \nu}
       - {\delta \nu_{,\alpha} \over \nu} \delta e^\alpha
       + 2 {\delta \nu \over \nu} A^\prime
       - 2 {\delta \nu \over \nu} \left( A_{,\alpha} 
       - {a^\prime \over a} B_\alpha \right) e^\alpha
       + {a^\prime \over a} \delta e^\alpha \delta e_\alpha
   \nonumber \\
   & & 
       - 2 \delta e^\alpha \left( A_{,\alpha} 
       - {a^\prime \over a} B_\alpha \right)
       - 4 {a^\prime \over a} e^\alpha \delta e_\alpha A
       + 2 e^\alpha \delta e^\beta \left( C_{\alpha\beta}^\prime
       + 2 {a^\prime \over a} C_{\alpha\beta}
       + B_{(\alpha|\beta)} \right)
   \nonumber \\
   & & 
       - \left[ A \left( 2 B_{\alpha|\beta} + 2 C_{\alpha\beta}^\prime 
       + 4 {a^\prime \over a} C_{\alpha\beta} \right)
       + B_\gamma \left( 2 C^\gamma_{\alpha|\beta} 
       - C_{\alpha\beta}^{\;\;\;\; |\gamma} \right) \right] e^\alpha e^\beta
       + {a^\prime \over a} \left( 4 A^2 - B^\alpha B_\alpha \right)
   \nonumber \\
   & &
       + 2 \left( 2 A A_{,\alpha} - 2 {a^\prime \over a} A B_\alpha
       + B_\beta C_\alpha^{\beta \prime} 
       + B^\beta B_{[\alpha|\beta]} \right) e^\alpha
       - 2 A A^\prime - A_{,\alpha} B^\alpha
       + B^\alpha \left( B_\alpha^\prime + {a^\prime \over a} B_\alpha 
       \right) \Bigg\} 
       = 0,
   \label{geodesic-eq-time} \\
   \tilde k^\alpha_{\;\; ;b} \tilde k^b 
   &=& 
       {\nu^2 \over a^2} \Bigg\{
       - e^{\alpha\prime} + e^\beta e^\alpha_{\;\;|\beta}
       - \delta e^{\alpha^\prime}
       - {\delta \nu \over \nu} e^{\alpha\prime}
       + \delta e^\alpha_{\;\; |\beta} e^\beta 
       + \delta e^\beta e^\alpha_{\;\; |\beta}
   \nonumber \\
   & & 
       + \left( 2 C^\alpha_{\beta|\gamma} - C_{\beta\gamma}^{\;\;\;\;|\alpha}
       \right) e^\beta e^\gamma
       - \left( B_\beta^{\;\;|\alpha} - B^\alpha_{\;\;|\beta}
       + 2 C^{\alpha\prime}_\beta \right) e^\beta
       + A^{,\alpha} - B^{\alpha\prime}
   \nonumber \\
   & &  
       - {\delta \nu \over \nu} \delta e^{\alpha\prime}
       + 2 {\delta \nu \over \nu} \left( A^{,\alpha} - B^{\alpha\prime} \right)
       - \left( \delta e^\beta + {\delta \nu \over \nu} e^\beta \right)
       \left( B_\beta^{\;\;|\alpha} - B^\alpha_{\;\;|\beta}
       + 2 C^{\alpha\prime}_\beta \right) 
       + \delta e^\beta \delta e^\alpha_{\;\;\;|\beta}
   \nonumber \\
   & & 
       + 2 e^\beta \delta e^\gamma \left( 2 C^\alpha_{\beta|\gamma}
       - C_{\beta\gamma}^{\;\;\;\;|\alpha} \right) + A^\prime B^\alpha
       - 2 A_{,\beta} C^{\alpha\beta}
       + 2 C^\alpha_\beta B^{\beta\prime} 
       - 2 B^\alpha A_{,\beta} e^\beta
       + 4 C^{\alpha\gamma} \left( B_{[\beta|\gamma]} 
       + C_{\beta\gamma}^\prime \right) e^\beta
   \nonumber \\
   & & 
       - 2 C^\alpha_\delta \left( 2 C^\delta_{\beta|\gamma}
       - C_{\beta\gamma}^{\;\;\;\; |\delta} \right) e^\beta e^\gamma
       + B^\alpha \left( B_{\beta|\gamma} + C^\prime_{\beta\gamma}
       \right) e^\beta e^\gamma
       \Bigg\}
       = 0,
   \label{geodesic-eq-space}
\eea
where we used the null equation in eq. (\ref{null-eq}).
To the background order eqs. (\ref{null-eq}-\ref{geodesic-eq-space}) give
\bea
   & & e^\alpha e_\alpha = 1, \quad
       \nu \propto a^{-1}, \quad
       e^{\alpha\prime} = e^\beta e^\alpha_{\;\;\;|\beta}.
   \label{null-geodesic-BG}
\eea
Using eqs. (\ref{y-def},\ref{null-eq}), eq. (\ref{geodesic-eq-time}) becomes
\bea
   & & {d \over dy} \left( {\delta \nu \over \nu} + A \right)
       - A_{,\alpha} e^\alpha 
       + \left( B_{\alpha|\beta} + C_{\alpha\beta}^\prime \right) 
       e^\alpha e^\beta
   \nonumber \\
   & & \qquad
       = - {\delta \nu \over \nu} {\delta \nu^\prime \over \nu}
       - {a^\prime \over a} {\delta \nu^2 \over \nu^2} 
       - 2 {\delta \nu \over \nu} 
       \left( A^\prime + 2 {a^\prime \over a} A \right)
       + {\delta \nu_{,\alpha} \over \nu} \delta e^\alpha
       + 2 {\delta \nu \over \nu} A_{,\alpha} e^\alpha
   \nonumber \\
   & & \qquad \qquad
       + 2 \delta e^\alpha A_{,\alpha}
       + 4 {a^\prime \over a} e^\alpha \delta e_\alpha A
       - 2 e^\alpha \delta e^\beta \left( B_{(\alpha|\beta)} 
       + C_{\alpha\beta}^\prime \right)
   \nonumber \\
   & & \qquad \qquad
       + \left[ 2 A \left( B_{\alpha|\beta} + C_{\alpha\beta}^\prime 
       + 2 {a^\prime \over a} C_{\alpha\beta} \right)
       + B_\gamma \left( 2 C^\gamma_{\alpha|\beta} 
       - C_{\alpha\beta}^{\;\;\;\; |\gamma} \right) \right] e^\alpha e^\beta
       - {a^\prime \over a} \left( 4 A^2 - B^\alpha B_\alpha \right)
   \nonumber \\
   & & \qquad \qquad
       - 2 \left( 2 A A_{,\alpha} - 2 {a^\prime \over a} A B_\alpha
       + B_\beta C_\alpha^{\beta\prime} 
       + B^\beta B_{[\alpha|\beta]} \right) e^\alpha
       + 2 A A^\prime + A_{,\alpha} B^\alpha
       - B^\alpha \left( B_\alpha^\prime + {a^\prime \over a} B_\alpha \right)
   \nonumber \\
   & & \qquad
      \equiv N_\nu. 
   \label{geodesic-eq-time2} 
\eea
Thus, we have 
\bea
   & & \left( {\delta \nu \over \nu} + A \right) \Bigg|^O_E
       = \int_E^O \left[ A_{,\alpha} e^\alpha 
       - \left( B_{\alpha|\beta} + C_{\alpha\beta}^\prime \right) 
       e^\alpha e^\beta 
       + N_\nu \right] d y,
   \label{nu-integ}
\eea
where the integral is along the ray's null-geodesic path from $E$
the emitted event at the intersection of the ray and the last 
scattering surface to $O$ the observed event here and now.

The temperatures of the CMB at two different points ($O$ and $E$) along
a single null-geodesic ray in a given observational direction are 
\cite{KS-1966,SW-1967},
\bea
   & & {\tilde T_O \over \tilde T_E}
       \equiv {1 \over 1 + \tilde z}
       \equiv {(\tilde k^a \tilde u_a)_O \over (\tilde k^b \tilde u_b)_E},
   \label{T-ku}
\eea
where $\tilde u_a$ at $O$ and $E$ are the local four-velocities of the observer
and the emitter, respectively.
Thus, $\tilde u_a$ should be considered as the one based on the energy-frame
which sets $\tilde q_a \equiv 0$; or equivalently in a general frame vector
which absorbs the flux term to the frame vector to the second-order.

Using eqs. (\ref{u-def},\ref{k-def}) we have
\bea
   & & \tilde k^a \tilde u_a
       = - \nu \Bigg[ 1 + {\delta \nu \over \nu} + A 
       + \left( V^E_\alpha - B_\alpha \right) e^\alpha
       + {\delta \nu \over \nu} A 
       + \delta e^\alpha \left( V^E_\alpha - B_\alpha \right)
       + \left( A B_\alpha + 2 C_{\alpha\beta} V^{E\beta} \right) e^\alpha
       + {1 \over 2} V^{E\alpha} V^E_\alpha - {1 \over 2} A^2
       \Bigg].
   \nonumber \\
   \label{ku}
\eea
We have denoted the energy-frame nature
by replacing $V_\alpha$ to $V^E_\alpha$; 
if we consider eq. (\ref{frame-inv-def-E}),
eq. (\ref{ku}) in this form is valid in the general frame.
Since the calculations in the rest of this paper are based on 
the normal-frame vector we use eq. (\ref{frame-E-N}) to derive the result
in the normal-frame.
We have
\bea
   \tilde k^a \tilde u_a
   &=& - \nu \Bigg[ 1 + {\delta \nu \over \nu} + A 
       + {1 \over \mu + p} Q_\alpha e^\alpha
       - {\left( \delta \mu + \delta p \right) Q_\alpha 
       + \Pi_{\alpha\beta} Q^\beta \over (\mu + p)^2} e^\alpha
   \nonumber \\
   & & 
       + {\delta \nu \over \nu} A 
       + \delta e^\alpha {Q_\alpha \over \mu + p} 
       + {1 \over 2} \left( B_\alpha + {Q_\alpha \over \mu + p} \right)
       \left( B^\alpha + {Q^\alpha \over \mu + p} \right)
       - {1 \over 2} A^2
       \Bigg].
\eea
Using eq. (\ref{T-ku}) we have
\bea
   & & \left( 1 - {T_E \over T_O} {\nu_O \over \nu_E} \right) 
       \left( 1 + {\delta T_E \over T_E} \right)
   \nonumber \\
   & & \qquad
       + {\delta T \over T} \Bigg|^O_E
       =
       {T_E \over T_O} {\nu_O \over \nu_E}
       \Bigg[ {\delta \nu \over \nu} + A
       + {1 \over \mu + p} Q_\alpha e^\alpha
       - {\left( \delta \mu + \delta p \right) Q_\alpha 
       + \Pi_{\alpha\beta} Q^\beta \over (\mu + p)^2} e^\alpha
       + {\delta \nu \over \nu} A
       + \delta e^\alpha {Q_\alpha \over \mu + p}
   \nonumber \\
   & & \qquad \qquad
       + {1 \over 2} \left( B_\alpha + {Q_\alpha \over \mu + p} \right)
       \left( B^\alpha + {Q^\alpha \over \mu + p} \right)
       - {1 \over 2} A^2
       \Bigg] \Bigg|^O_E
       \times 
       \left[ 1 - \left( {\delta \nu \over \nu} + A 
       + {1 \over \mu + p} Q_\gamma e^\gamma \right)
       + {\delta T \over T} \right] \Bigg|_E
   \nonumber \\
   & & \qquad
       \equiv {T_E \over T_O} {\nu_O \over \nu_E}
       \left( {\delta \nu \over \nu} + A
       + {1 \over \mu + p} Q_\alpha e^\alpha \right) \Bigg|_E^O 
       + N_T,
   \label{delta-T}
\eea
where 
${\delta T \over T}|^O_E \equiv {\delta T \over T}|_O - {\delta T \over T}|_E$
and ${\delta T \over T}|_E \equiv {\delta T \over T}$ at $E$.
Thus, if we take $T_O/T_E = \nu_O/\nu_E$,
eqs. (\ref{delta-T},\ref{nu-integ}) give
\bea
   & & {\delta T \over T} \Bigg|_O
       = {\delta T \over T} \Bigg|_E
       + {1 \over \mu + p} Q_\alpha e^\alpha \Big|^O_E
       + \int_E^O \left[ A_{,\alpha} e^\alpha 
       - \left( B_{\alpha|\beta} + C_{\alpha\beta}^\prime \right) 
       e^\alpha e^\beta + N_\nu \right] d y
       + N_T.
   \label{delta-T-2}
\eea

In the large angular scale we are considering
(larger than the horizon size at the last scattering era)
the detailed dynamics at last scattering is not important.
The physical processes of last scattering are important in the small
angular scale where we need to solve the Boltzmann equation for the
photon distribution function, see \S \ref{sec:Boltzmann}.

\subsection{Boltzmann equation}
                                                 \label{sec:Boltzmann}

\subsubsection{Covariant equations}

The relativistic Boltzmann equation is \cite{Lindquist-1966}
\bea
   & & {d \over d \lambda} \tilde f = {d x^a \over d \lambda}
       {\partial \tilde f \over \partial x^a} + {d \tilde p^a \over d \lambda}
       {\partial \tilde f \over \partial \tilde p^a}
       = \tilde p^a {\partial \tilde f \over \partial x^a}
       - \tilde \Gamma^a_{bc} \tilde p^b \tilde p^c 
       {\partial \tilde f \over \partial \tilde p^a}
       = \tilde C [\tilde f],
\eea
where $\tilde f(x^a, \tilde p^b)$ is a distribution function with the
phase space variables $x^a$ and $\tilde p^a \equiv d x^a/d\lambda$,
and $\tilde C [\tilde f]$ is the collision term.
The energy-momentum tensor of the collisionless (or collisional) component is
\bea
   & & \tilde T^{(c)}_{ab} 
       = \int 2 \theta (\tilde p^0) \delta (\tilde p^c \tilde p_c 
       + m^2 ) \tilde p_a \tilde p_b 
       \tilde f {d^4 \tilde p_{0123} \over \sqrt{-\tilde g} }
       = \int 2 \theta (\tilde p^0) \delta (\tilde p^c \tilde p_c + m^2 ) 
       \tilde p_a \tilde p_b \tilde f 
       \sqrt{- \tilde g} d^4 \tilde p^{0123},
\eea
where 
\bea
   & & \delta ( \tilde p^c \tilde p_c + m^2) 
       = \delta ( \tilde g_{00} \tilde p^0 \tilde p^0 
       + 2 \tilde g_{0\alpha} \tilde p^0 \tilde p^\alpha 
       + \tilde g_{\alpha\beta} \tilde p^\alpha \tilde p^\beta + m^2 )
       = { \delta ({\rm mass\; shell}) \over 
       | 2 \tilde g_{00} \tilde p^0 + 2 \tilde g_{0\alpha} \tilde p^\alpha | }
       = {\delta ( {\rm mass \; shell}) \over 2 |\tilde p_0| }.
\eea
Thus, after integrating over $\tilde p^0$, we have
\bea
   & & \tilde T^{(c)}_{ab} 
       = \int {\sqrt{-g} d^3 \tilde p^{123} \over |\tilde p_0|} 
       \tilde p_a \tilde p_b \tilde f,
\eea
with the mass-shell condition $\tilde p^a \tilde p_a + m^2 = 0$.

\subsubsection{Perturbed equations}

Under our metric, using $\tilde p^a$ as the phase space variable, we have
\bea
   & & \tilde p^0 \tilde f^\prime + \tilde p^\alpha \tilde f_{,\alpha} 
       - \Bigg\{ {a^\prime \over a} \left( \tilde p^0 \tilde p^0 
       + g^{(3)}_{\alpha\beta} \tilde p^\alpha \tilde p^\beta \right)
       + A^\prime \tilde p^0 \tilde p^0 
       + 2 \left( A_{,\alpha} - {a^\prime \over a} B_\alpha
       \right) \tilde p^0 \tilde p^\alpha
   \nonumber \\
   & & \qquad
       + \left( - 2 {a^\prime \over a} g^{(3)}_{\alpha\beta} A
       + B_{\alpha|\beta} + C^\prime_{\alpha\beta} 
       + 2 {a^\prime \over a} C_{\alpha\beta} \right) 
       \tilde p^\alpha \tilde p^\beta 
   \nonumber \\
   & & \qquad
       + \left( - 2 A A^\prime - A_{,\alpha} B^\alpha 
       + B^\alpha B_\alpha^\prime + {a^\prime \over a} B^\alpha B_\alpha
       \right) \tilde p^0 \tilde p^0
       + 2 \left( - 2 A A_{,\alpha} + 2 {a^\prime \over a} A B_\alpha
       - B_\beta C^{\beta \prime}_\alpha
       + B^\beta B_{[\beta|\alpha]} \right) \tilde p^0 \tilde p^\alpha
   \nonumber \\
   & & \qquad
       + \left[ {a^\prime \over a} g^{(3)}_{\alpha\beta} 
       \left( 4 A^2 - B^\gamma B_\gamma \right)
       - 2 A \left( B_{\alpha|\beta} + C^\prime_{\alpha\beta}
       + 2 {a^\prime \over a} C_{\alpha\beta} \right)
       - B_\gamma \left( 2 C^\gamma_{\alpha|\beta} 
       - C_{\alpha\beta}^{\;\;\;\;|\gamma} \right) \right] 
       \tilde p^\alpha \tilde p^\beta
       \Bigg\} {\partial \tilde f \over \partial \tilde p^0}
   \nonumber \\
   & & \qquad
       - \Bigg[ \left( 2 {a^\prime \over a} \tilde p^0 \tilde p^\alpha
       + \Gamma^{(3)\alpha}_{\;\;\;\;\;\beta\gamma} 
       \tilde p^\beta \tilde p^\gamma \right)
       + \left( A^{,\alpha} - B^{\alpha\prime} - {a^\prime \over a} B^\alpha
       \right) \tilde p^0 \tilde p^0
       + \left( B_\beta^{\;\;|\alpha} - B^\alpha_{\;\;|\beta}
       + 2 C^{\alpha\prime}_\beta \right) \tilde p^0 \tilde p^\beta
   \nonumber \\
   & & \qquad
       + \left( {a^\prime \over a} g^{(3)}_{\beta\gamma} B^\alpha
       + 2 C^\alpha_{\beta|\gamma} - C_{\beta\gamma}^{\;\;\;\;|\alpha}
       \right) \tilde p^\beta \tilde p^\gamma
       \Bigg] {\partial \tilde f \over \partial \tilde p^\alpha}
       = \tilde C[\tilde f].
   \label{f-eq-p}
\eea
As the phase space variable it is convenient to use ($q$, $\gamma^\alpha$)
introduced as
\bea
   q 
   &\equiv& a \sqrt{ a^2 ( \tilde p^0 )^2 - m^2 } \left[ 1 
       + {a^2 (\tilde p^0)^2 \over a^2 ( \tilde p^0 )^2 - m^2} \left( A
       + {1 \over 2} B^\alpha B_\alpha \right)
       - {1 \over 2} {a^4 (\tilde p^0)^4 
       \over [ a^2 ( \tilde p^0 )^2 - m^2 ]^2} A^2 \right],
   \nonumber \\
   \gamma^{\alpha} 
   &\equiv& {a \over \sqrt{ a^2 ( \tilde p^0 )^2 - m^2 }}
       \Bigg[ \tilde p^\alpha - {a^2 (\tilde p^0)^2 \over a^2 (\tilde p^0)^2 
       - m^2} A \tilde p^\alpha 
       - B^\alpha \tilde p^0 + C^\alpha_\beta \tilde p^\beta
       + {3 \over 2} {a^4 (\tilde p^0)^4 
       \over [a^2 ( \tilde p^0 )^2 - m^2]^2} A^2 \tilde p^\alpha 
   \nonumber \\
   & &
       + {a^2 (\tilde p^0)^2 \over a^2 ( \tilde p^0 )^2 - m^2} 
       \left( A B^\alpha \tilde p^0 - A C^\alpha_\beta \tilde p^\beta 
       - {1 \over 2} B^\beta B_\beta \tilde p^\alpha \right)
       + C^\alpha_\beta B^\beta \tilde p^0
       - {1 \over 2} C^\gamma_\beta C^\alpha_\gamma \tilde p^\beta 
       \Bigg],
   \nonumber \\
   \tilde p^0 
   &=& {1 \over a^2} \sqrt{q^2 + m^2 a^2}
       \left( 1 - A + {3 \over 2} A^2 - {1 \over 2} B^\alpha B_\alpha \right), 
   \nonumber \\
   \tilde p^\alpha 
   &=& {1 \over a^2} \left[ q \gamma^{\alpha}
       + \sqrt{q^2 + m^2 a^2} B^\alpha - q C^\alpha_\beta \gamma^\beta
       - \sqrt{q^2 + m^2 a^2} \left( A B^\alpha 
       + 2 C^\alpha_\beta B^\beta \right)
       + {3 \over 2} q C^\gamma_\beta C^\alpha_\gamma \gamma^\beta 
       \right],
   \label{q-def}
\eea
where $\gamma^\alpha$ is based on $g^{(3)}_{\alpha\beta}$.
The mass-shell condition gives
\bea
   & & ( \tilde p^0 )^2 - g^{(3)}_{\alpha\beta} \tilde p^\alpha \tilde p^\beta
       - {m^2/a^2} + 2 A ( \tilde p^0)^2 
       + 2 B_\alpha \tilde p^\alpha \tilde p^0
       - 2 C_{\alpha\beta} \tilde p^\alpha \tilde p^\beta = 0,
\eea
and we can show $\gamma^\alpha \gamma_\alpha = 1$.
The Boltzmann equation becomes
\bea
   & & \tilde f^\prime + {q \over \sqrt{q^2 + m^2 a^2} }
       \left( \gamma^\alpha \delta f_{,\alpha}
       - \Gamma^{(3)\alpha}_{\;\;\;\;\;\beta\gamma}
       \gamma^\beta \gamma^\gamma 
       {\partial \delta f \over \partial \gamma^\alpha} \right)
       - \left[ {\sqrt{q^2 + m^2 a^2} \over q}
       \gamma^\alpha A_{,\alpha} + \left( B_{\alpha|\beta}
       + C_{\alpha\beta}^\prime \right) \gamma^\alpha \gamma^\beta
       \right] q {\partial \tilde f \over \partial q}
   \nonumber \\
   & & \qquad
       = - {1 \over \sqrt{q^2 + m^2 a^2}} \left( q A \gamma^\alpha
       + \sqrt{q^2 + m^2 a^2} B^\alpha - q C^\alpha_\beta \gamma^\beta \right)
       \delta f_{,\alpha}
   \nonumber \\
   & & \qquad
       - \Bigg\{ {\sqrt{q^2 + m^2 a^2} \over q} \left[ A_{,\alpha}
       \left( A \gamma^\alpha + C^\alpha_\beta \gamma^\beta \right)
       - B^\beta B_{\beta|\alpha} \gamma^\alpha \right]
       + 2 C^\beta_\gamma \left( C^\prime_{\alpha\beta} + B_{(\alpha|\beta)}
       \right) \gamma^\alpha \gamma^\gamma
   \nonumber \\
   & & \qquad
       + B_\gamma \left( 2 C^\gamma_{\alpha|\beta} 
       - C_{\alpha\beta}^{\;\;\;\;|\gamma} \right) \gamma^\alpha \gamma^\beta 
       \Bigg\} q {\partial f \over \partial q}
   \nonumber \\
   & & \qquad
       + \Bigg[ \Gamma^{(3)\alpha}_{\;\;\;\;\beta\gamma}
       \left( {q \over \sqrt{q^2 + m^2 a^2}} A \gamma^\beta \gamma^\gamma
       + B^\beta \gamma^\gamma
       - {q \over \sqrt{q^2 + m^2 a^2}} C^\gamma_\delta \gamma^\beta 
       \gamma^\delta \right)
       + {\sqrt{q^2 + m^2 a^2} \over q} \left( A^{,\alpha} 
       - A_{,\beta} \gamma^\beta \gamma^\alpha \right)
   \nonumber \\
   & & \qquad
       + \left( B_\beta^{\;\;|\alpha} 
       + C^{\alpha\prime}_\beta \right) \gamma^\beta
       - \left( B_{\beta|\gamma} + C^\prime_{\beta\gamma} \right)
       \gamma^\beta \gamma^\gamma \gamma^\alpha
       + {q \over \sqrt{q^2 + m^2 a^2}} \left( C^\alpha_{\beta|\gamma}
       - C_{\beta\gamma}^{\;\;\;\;|\alpha} \right) \gamma^\beta \gamma^\gamma
       \Bigg]
       {\partial \delta f \over \partial \gamma^\alpha}
   \nonumber \\
   & & \qquad
       + {a^2 \over \sqrt{q^2 + m^2 a^2} } \left( 1 + A - {1 \over 2} A^2
       + {1 \over 2} B^\alpha B_\alpha \right) \tilde C [\tilde f]
   \nonumber \\
   & & \qquad
       \equiv N_c.
   \label{f-eq-q} 
\eea
{}For convenience we located the collision term in $N_c$.
The energy-momentum tensor becomes
\bea
   & & \tilde T^{(c)}_{ab}
       = {1 \over a^2} \int \tilde p_a \tilde p_b \tilde f {q^2 d q d \Omega_q
       \over \sqrt{ q^2 + m^2 a^2 } }.
\eea
Thus, using $\int \gamma_\alpha \gamma_\beta d \Omega_q 
= {1 \over 3} g^{(3)}_{\alpha\beta}$, we have
\bea
   \tilde T^{(c)}_{00}
   &=& {1 \over a^2} \int \sqrt{q^2 + m^2 a^2} q^2 dq d \Omega_q
       \Bigg[ f (1 + 2 A) + \delta f (1 + 2 A) + \left( 1 + {1 \over 3}
       {q^2 \over q^2 + m^2 a^2} \right) B^\alpha B_\alpha f
   \nonumber \\
   & &
       + {2 q \over \sqrt{q^2 + m^2 a^2}} B_\alpha \gamma^\alpha \delta f
       \Bigg],
   \nonumber \\
   \tilde T^{(c)}_{\alpha0}
   &=& - {1 \over a^2} \int {q^4 dq d \Omega_q \over \sqrt{q^2 + m^2 a^2}}
       \left\{ {1 \over 3} B_\alpha f + \left[ {\sqrt{q^2 + m^2 a^2} \over q}
       \left( \gamma_\alpha + A \gamma_\alpha 
       + C_{\alpha\beta} \gamma^\beta \right) 
       + B_\beta \gamma^\beta \gamma_\alpha \right] \delta f \right\},
   \nonumber \\
   \tilde T^{(c)}_{\alpha\beta}
   &=& {1 \over a^2} \int {q^4 dq d \Omega_q \over \sqrt{q^2 + m^2 a^2} }
       \left( {1 \over 3} g^{(3)}_{\alpha\beta} f
       + \delta f \gamma_\alpha \gamma_\beta
       + {2 \over 3} f C_{\alpha\beta}
       + 2 \delta f \gamma_{(\alpha} C_{\beta)\gamma} \gamma^\gamma \right).
\eea
{}From eq. (\ref{Tab-fluid}) the fluid quantities become:
\bea
   \mu^{(c)} + \delta \mu^{(c)}
   &=& {1 \over a^4} \int \sqrt{q^2 + m^2 a^2} q^2 dq d \Omega_q
       \left( f + \delta f \right),
   \nonumber \\
   p^{(c)} + \delta p^{(c)}
   &=& {1 \over 3 a^4} \int {q^4 dq d \Omega_q
       \over \sqrt{q^2 + m^2 a^2} } \left( f + \delta f \right),
   \nonumber \\
   Q^{(c)}_{\alpha}
   &=& {1 \over a^4} \int q^3 dq d \Omega_q \left( \gamma_\alpha
       + C_{\alpha\beta} \gamma^\beta \right) \delta f,
   \nonumber \\
   \Pi^{(c)}_{\alpha\beta}
   &=& {1 \over a^4} \int {q^4 dq d \Omega_q \over \sqrt{q^2 + m^2 a^2} }
       \left( \gamma_\alpha \gamma_\beta 
       - {1\over 3} g^{(3)}_{\alpha\beta}
       + 2 \gamma_{(\alpha} C_{\beta)\gamma} \gamma^\gamma
       - {2 \over 3} C_{\alpha\beta} \right) \delta f. 
   \label{fluid-f}
\eea
If we have multiple components each described by the Boltzmann equation, 
all equations in this subsection remain valid for any component
with $\tilde f$ replaced by $\tilde f_{(i)}$, etc.,
and the total (collective) fluid quantities as the sum of the individual ones.

\section{Decomposition}
                                             \label{sec:Decomposition}

\subsection{Three perturbation types}

We decompose the perturbation variables as follow:
\bea
   A 
   &\equiv& \alpha,
   \nonumber \\ 
   B_\alpha 
   &\equiv& \beta_{,\alpha} + B_\alpha^{({v})},
   \nonumber \\ 
   C_{\alpha\beta} 
   &\equiv& \varphi g^{(3)}_{\alpha\beta}
       + \gamma_{,\alpha|\beta}
       + C_{(\alpha|\beta)}^{({v})}
       + C_{\alpha\beta}^{({t})},
   \nonumber \\ 
   Q_\alpha 
   &\equiv& Q_{,\alpha} + Q_\alpha^{({v})}
       \equiv (\mu + p) \left( - v_{,\alpha} 
       + v_\alpha^{(v)} \right),
   \nonumber \\ 
   \Pi_{\alpha\beta} 
   &\equiv& {1 \over a^2} \left(
       \Pi_{,\alpha|\beta} 
       - {1 \over 3} g^{(3)}_{\alpha\beta} \Delta \Pi \right)
       + {1 \over a} \Pi_{(\alpha|\beta)}^{({v})}
       + \Pi_{\alpha\beta}^{({t})},
   \label{metric-decomp-def}
\eea
with the properties:
\bea
   & & B^{({v})\alpha}_{\;\;\;\;\;\;\;|\alpha} \equiv 0, \quad
       C^{({v})\alpha}_{\;\;\;\;\;\;\;|\alpha} \equiv 0, \quad
       v^{({v})\alpha}_{\;\;\;\;\;\;\;|\alpha} \equiv 0, \quad
       \Pi^{({v})\alpha}_{\;\;\;\;\;\;\;|\alpha} \equiv 0, 
   \nonumber \\ 
   & & C^{({t})\alpha}_{\;\;\;\;\;\alpha} \equiv 0, \quad
       \Pi^{({t})\alpha}_{\;\;\;\;\;\alpha} \equiv 0, \quad
       C^{({t})\beta}_{\;\;\;\;\;\alpha|\beta} \equiv 0, \quad
       \Pi^{({t})\beta}_{\;\;\;\;\;\alpha|\beta} \equiv 0.
\eea
$\mu + p$ appearing in the decomposition of $Q_\alpha$ is
assumed to be the background order quantity.
The vector- and the tensor-type perturbations are denoted by 
superscripts $({v})$ and $({t})$, respectively.
We assume all these variables are based on $g^{(3)}_{\alpha\beta}$.
The decomposed variables also can be expressed in terms of 
the original variables.
{}For example, we have
$\beta = \Delta^{-1} \nabla^\alpha B_\alpha$ and
$B^{(v)}_\alpha = B_\alpha - \nabla_\alpha \Delta^{-1} \nabla^\beta B_\beta$,
etc, where $\nabla_\alpha$ means $\nabla^{(3)}_\alpha$.
{}For the fluid quantities we have
\bea
   Q
   &=& \Delta^{-1} \nabla^\alpha Q_\alpha,
   \nonumber \\
   Q^{(v)}_\alpha
   &=& Q_\alpha
       - \nabla_\alpha \Delta^{-1} \nabla^\beta Q_\beta,
   \nonumber \\
   \Pi
   &=& {3 \over 2} a^2 \left( \Delta + {1 \over 2} R^{(3)} \right)^{-1}
       \Delta^{-1} \nabla^\alpha \nabla^\beta \Pi_{\alpha\beta},
   \nonumber \\
   \Pi^{(v)}_\alpha
   &=& 2 a \left( \Delta + {1 \over 3} R^{(3)} \right)^{-1} \left(
       \nabla^\beta \Pi_{\alpha\beta}
       - \nabla_\alpha \Delta^{-1} \nabla^\beta \nabla^\gamma 
       \Pi_{\beta\gamma} \right),
   \nonumber \\
   \Pi^{(t)}_{\alpha\beta}
   &=& \Pi_{\alpha\beta}
       - {3 \over 2} \left( \nabla_\alpha \nabla_\beta 
       - {1 \over 3} g^{(3)}_{\alpha\beta} \Delta \right)
       \left( \Delta + {1 \over 2} R^{(3)} \right)^{-1} 
       \Delta^{-1} \nabla^\gamma \nabla^\delta \Pi_{\gamma\delta}
   \nonumber \\
   & & - 2 \nabla_{(\alpha} \left( \Delta + {1 \over 3} R^{(3)} \right)^{-1} 
       \left( \nabla^\gamma \Pi_{\beta)\gamma}
       - \nabla_{\beta)} \Delta^{-1} \nabla^\gamma \nabla^\delta 
       \Pi_{\gamma\delta} \right).
   \label{fluid-decomp}
\eea
We introduce
\bea
   & & \chi \equiv a \left( \beta + a \dot \gamma \right), \quad
       \Psi^{(v)}_\alpha \equiv B^{(v)}_\alpha + a \dot C^{(v)}_\alpha,
   \label{chi-def}
\eea
and let
\bea
   & & \kappa \equiv \delta K.
   \label{deltaK-def}
\eea
In the multi-component situation we have eq. (\ref{fluid-sum}). 
{}For the individual component we have
\bea
   & & Q_{(i)\alpha} 
       \equiv (\mu_{(i)} + p_{(i)}) \left( - v_{(i),\alpha}
       + v_{(i)\alpha}^{(v)} \right),
   \nonumber \\
   & & \Pi_{(i)\alpha\beta} \equiv {1 \over a^2} \left(
       \Pi_{(i),\alpha|\beta} 
       - {1 \over 3} g^{(3)}_{\alpha\beta} \Delta \Pi_{(i)} \right)
       + {1 \over a} \Pi_{(i)(\alpha|\beta)}^{({v})}
       + \Pi_{(i)\alpha\beta}^{({t})},
\eea
with
\bea
   & & v^{(v)\alpha}_{(i)\;\;\;|\alpha} \equiv 0, \quad
       \Pi^{(v)\alpha}_{(i)\;\;\;|\alpha} \equiv 0, \quad
       \Pi^{(t)\alpha}_{(i)\alpha} \equiv 0, \quad
       \Pi^{(t)\beta}_{(i)\alpha|\beta} \equiv 0,
\eea
and
\bea
   & & \delta I_{(i)\alpha} \equiv \delta I_{(i),\alpha}
       + \delta I^{(v)}_{(i)\alpha}, \quad
       \delta I^{(v)\alpha}_{(i)\;\;\; |\alpha} \equiv 0.
\eea
The definitions for the scalar-type perturbation variables are 
introduced to match with our notation used in the linear analysis 
\cite{Bardeen-1988,Hwang-1991,Hwang-Noh-CMB-2002};
compared with our previous definitions in the linear theory 
our $v$ and $v_{(i)}$ correspond to $v/k$ 
and $v_{(i)}/k$ in \cite{Hwang-Noh-CMB-2002} where $k$ is the wave number.
These are the notations introduced by Bardeen in 1988 \cite{Bardeen-1988}.
A complete set of equations written separately for the three perturbation
types will be presented in equations (\ref{scalar-0})-(\ref{tensor-4})
where the quadratic combinations of the linear-order variables contribute
to the second-order perturbations.
Thus, to the second-order the three perturbation types couple each other 
through quadratic combinations of the linear-order terms.
If needed we may decompose the perturbed order quantities explicitly as
in eq. (\ref{metric-expansion})
\bea
   & & \alpha \equiv \alpha^{(1)} + \alpha^{(2)}, \quad
       \varphi \equiv \varphi^{(1)} + \varphi^{(2)}, 
   \label{pert-decompose}
\eea
etc.

\subsection{Background equations}

To the background order, eqs. 
(\ref{pert-eq1},\ref{pert-eq3},\ref{pert-eq5},\ref{MSF-EOM-pert}) give
\bea
   & & H^2 = {8 \pi G \over 3} \mu - {K \over a^2} + {\Lambda \over 3},
   \label{BG1} \\
   & & {\ddot a \over a} = - {4 \pi G \over 3} \left( \mu + 3 p \right)
       + {\Lambda \over 3},
   \label{BG2} \\
   & & \dot \mu + 3 H \left( \mu + p \right) = 0,
   \label{BG3} \\
   & & \ddot \phi + 3 H \dot \phi + V_{,\phi} = 0.
   \label{BG4}
\eea
In the multi-component situations from eqs. 
(\ref{pert-eq5i},\ref{MSFs-EOM-pert}) we have
\bea
   & & \dot \mu_{(i)} + 3 H \left( \mu_{(i)} + p_{(i)} \right) 
       = - {1 \over a} I_{(i)0},
   \\
   & & \ddot \phi_{(i)} + 3 H \dot \phi_{(i)} + V_{,\phi_{(i)}} = 0,
\eea
with
\bea
   & & \mu_{(\phi)} = {1 \over 2} \sum_k \dot \phi_{(k)}^2 + V, \quad
       p_{(\phi)} = {1 \over 2} \sum_k \dot \phi_{(k)}^2 - V,
   \label{fluid-MSF}
\eea
which follow from eq. (\ref{fluid-MSFs}).

In the generalized gravity considered in \S \ref{sec:GGT},
eqs. (\ref{BG1},\ref{BG2}) remain valid by replacing the 
fluid quantities to the effective one in eq. (\ref{GGT-fluid-pert}):
\bea
   \mu^{({\rm eff})}
   &=&
       {1 \over 8 \pi G F}
       \left[ \mu + {1 \over 2} g_{IJ} \dot \phi^I \dot \phi^J
       - {1 \over 2} \left( f - R F - 2 V \right)
       - 3 H \dot F \right],
   \nonumber \\
   p^{({\rm eff})}
   &=&
       {1 \over 8 \pi G F}
       \left[ p + {1 \over 2} g_{IJ} \dot \phi^I \dot \phi^J
       + {1 \over 2} \left( f - R F - 2 V \right)
       + \ddot F + 2 H \dot F \right].
   \label{GGT-BG-fluid-decomp}
\eea
{}For the equation of motion, eq. (\ref{GGT-EOM-pert}) gives
\bea
   & & \ddot \phi^I + 3 H \dot \phi^I
       - {1 \over 2} g^{IJ} \left( f_{,J} - 2 V_{,J} \right)
       + \Gamma^I_{JK} \dot \phi^J \dot \phi^K + \Gamma^I
       = 0.
\eea

The null-geodesic equations are presented in eq. (\ref{null-geodesic-BG}).
The Boltzmann equation in eq. (\ref{f-eq-q}) gives
\bea
   & & f^\prime = {a^2 \over \sqrt{q^2 + m^2 a^2}} C [f],
\eea
and from eq. (\ref{fluid-f}) we have
\bea
   & & \mu^{(c)}
       = {1 \over a^4} \int \sqrt{q^2 + m^2 a^2} q^2 dq d \Omega_q f, \quad
       p^{(c)} = {1 \over 3 a^4} \int {q^4 dq d \Omega_q
       \over \sqrt{q^2 + m^2 a^2} } f.
\eea

\subsection{Decomposed equations}
                                              \label{sec:decomposed-eqs}

We summarize a complete set of equations necessary to analyse each 
perturbation type.
We decompose the perturbation variables according to equation 
(\ref{metric-decomp-def}).
Algebraic manipulations are made which can be recognized by examining the 
right hand side of the following equations.

The scalar-type perturbation: 
\bea
   & & \kappa - 3 H \alpha + 3 \dot \varphi + {\Delta \over a^2} \chi 
       = N_0,
   \label{scalar-0} \\
   & & 4 \pi G \delta \mu + H \kappa + {\Delta + 3K \over a^2} \varphi 
       = {1\over 4} N_1,
   \label{scalar-1} \\
   & & \kappa 
       + {\Delta + 3K \over a^2} \chi - 12 \pi G (\mu + p) a v
       = {3 \over 2} \Delta^{-1} \nabla^{\alpha} N_{2\alpha}
       \equiv N_2^{({\rm s})},
   \label{scalar-2} \\
   & & \dot \kappa + 2 H \kappa 
       - 4 \pi G \left( \delta \mu + 3 \delta p \right)
       + \left( 3 \dot H + {\Delta \over a^2} \right) \alpha
       = N_3,
   \label{scalar-3} \\
   & & \dot \chi + H \chi - \varphi - \alpha - 8 \pi G \Pi 
       = {3 \over 2} a^2 
       \left( \Delta + 3K \right)^{-1}
       \Delta^{-1} \nabla^{\alpha} \nabla_\beta
       N_{4\alpha}^{\;\;\beta}
       \equiv N_4^{({\rm s})},
   \label{scalar-4} \\
   & & \delta \dot \mu + 3 H \left( \delta \mu + \delta p \right)
       - \left( \mu + p \right) \left( \kappa - 3 H \alpha 
       + {1 \over a} \Delta v \right)
       = N_5,
   \label{scalar-5} \\
   & & {[a^4 (\mu + p) v]^\cdot \over a^4(\mu + p)}
       - {1 \over a} \alpha
       - {1 \over a (\mu + p)} \left( \delta p 
       + {2 \over 3} {\Delta + 3K \over a^2} \Pi \right)
       = - {1 \over \mu + p} \Delta^{-1} \nabla^\alpha N_{6\alpha}
       \equiv N_6^{({\rm s})},
   \label{scalar-6} \\
   & & \delta \dot \mu_{(i)} 
       + 3 H \left( \delta \mu_{(i)} + \delta p_{(i)} \right)
       - \left( \mu_{(i)} + p_{(i)} \right) \left( \kappa - 3 H \alpha
       + {1 \over a} \Delta v_{(i)} \right)
       + {1 \over a} \delta I_{(i)0}
       = N_{5(i)},
   \label{scalar-5-i} \\
   & & {[a^4 (\mu_{(i)} + p_{(i)}) v_{(i)}]^\cdot 
       \over a^4(\mu_{(i)} + p_{(i)})} 
       - {1 \over a} \alpha
       - {1 \over a (\mu_{(i)} + p_{(i)})} \left( \delta p_{(i)}
       + {2 \over 3} {\Delta + 3K \over a^2} \Pi_{(i)}
       - \delta I_{(i)} \right)
   \nonumber \\
   & & \qquad
       = - {1 \over \mu_{(i)} + p_{(i)}} \Delta^{-1} \nabla^\alpha 
       N_{6(i)\alpha}
       \equiv N_{6(i)}^{({\rm s})},
   \label{scalar-6-i} \\
   & & \delta \ddot \phi + 3 H \delta \dot \phi
       - {\Delta \over a^2} \delta \phi + V_{,\phi\phi} \delta \phi 
       - \dot \phi \left( \kappa + \dot \alpha \right)
       - \left( 2 \ddot \phi + 3 H \dot \phi \right) \alpha
       = N_\phi - \dot \phi N_0,
   \label{scalar-phi} \\
   & & \delta \ddot \phi_{(i)} + 3 H \delta \dot \phi_{(i)} 
       - {\Delta \over a^2} \delta \phi_{(i)} 
       + \sum_k V_{,\phi_{(i)}\phi_{(k)}} \delta \phi_{(k)}
       - \dot \phi_{(i)} \left( \kappa + \dot \alpha \right)
       - \left( 2 \ddot \phi_{(i)} + 3 H \dot \phi_{(i)} \right) \alpha
       = N_{\phi_{(i)}} - \dot \phi_{(i)} N_0.
   \label{scalar-phi-i}
\eea
The vector-type perturbation:
\bea
   & & {\Delta + 2 K \over 2 a^2} \Psi^{(v)}_\alpha 
       + 8 \pi G (\mu + p) v^{(v)}_\alpha
       = {1 \over a} \left( N_{2\alpha}
       - \nabla_\alpha \Delta^{-1} \nabla^{\beta} N_{2\beta} \right)
       \equiv N_{2\alpha}^{({v})},
   \label{vector-2} \\
   & & \dot \Psi^{(v)}_\alpha + 2 H \Psi^{(v)}_\alpha
       - 8 \pi G \Pi^{(v)}_\alpha
       = 2 a \left( \Delta + 2 K \right)^{-1} \left(
       \nabla_\beta N_{4\alpha}^{\;\;\beta}
       - \nabla_\alpha \Delta^{-1} 
       \nabla^{\gamma} \nabla_\beta N_{4\gamma}^{\;\;\beta} \right)
       \equiv N_{4\alpha}^{(v)},
   \label{vector-4} \\
   & & {[a^4 (\mu + p) v^{(v)}_\alpha]^\cdot \over a^4 (\mu + p)}
       + {\Delta + 2K \over 2 a^2} {\Pi^{(v)}_\alpha \over \mu + p}
       = {1 \over \mu + p} \left( N_{6\alpha} 
       - \nabla_\alpha \Delta^{-1} \nabla^{\beta} N_{6\beta} \right)
       \equiv N_{6\alpha}^{({v})},
   \label{vector-6} \\
   & & {[a^4 (\mu_{(i)} + p_{(i)}) v^{(v)}_{(i)\alpha}]^\cdot 
       \over a^4 (\mu_{(i)} + p_{(i)})} 
       + {\Delta + 2K \over 2 a^2} 
       {\Pi^{(v)}_{(i)\alpha} \over \mu_{(i)} + p_{(i)}}
       - {1 \over a} {\delta I^{(v)}_{(i)\alpha} \over \mu_{(i)} + p_{(i)}}
   \nonumber \\
   & & \qquad
       = {1 \over \mu_{(i)} + p_{(i)}} \left( N_{6(i)\alpha}
       - \nabla_\alpha \Delta^{-1} \nabla^{\beta} N_{6(i)\beta} \right)
       \equiv N_{6(i)\alpha}^{({v})}.
   \label{vector-6-i} 
\eea
The tensor-type perturbation
\bea
   & & \ddot C^{({t})}_{\alpha\beta} 
       + 3 H \dot C^{({t})}_{\alpha\beta}
       - {\Delta - 2K \over a^2} C^{({t})}_{\alpha\beta} 
       - 8 \pi G \Pi^{({t})}_{\alpha\beta}
   \nonumber \\ 
   & & \qquad 
       = N_{4\alpha\beta}
       - {3 \over 2} \left( \nabla_{\alpha} \nabla_\beta
       - {1\over 3} g^{(3)}_{\alpha\beta} \Delta \right)
       \left( \Delta + 3K \right)^{-1}
       \Delta^{-1} \nabla^{\gamma} \nabla_\delta
       N_{4\gamma}^{\;\;\delta}
   \nonumber \\ 
   & & \qquad \qquad
       - 2 \nabla_{(\alpha} \left( \Delta + 2K \right)^{-1} 
       \left( \nabla^{\gamma} N_{4\beta)\gamma}
       - \nabla_{\beta)} \Delta^{-1} \nabla^{\gamma}
       \nabla_\delta N_{4\gamma}^{\;\;\delta} \right)
       \equiv N_{4\alpha\beta}^{(t)}.
   \label{tensor-4} 
\eea
          In order to derive eqs. (\ref{scalar-4},\ref{vector-4},\ref{tensor-4})
          it is convenient to show
          \bea
             & & {1 \over a^2} \left( \nabla_\alpha \nabla_\beta 
                 - {1 \over 3} g^{(3)}_{\alpha\beta} \Delta \right) 
                 \left( \dot \chi + H \chi - \varphi - \alpha 
                 - 8 \pi G \Pi \right)
                 + {1 \over a^3} 
                 \left( a^2 \Psi^{(v)}_{(\alpha|\beta)} \right)^\cdot
                 - 8 \pi G {1 \over a} \Pi^{(v)}_{(\alpha|\beta)}
             \nonumber \\
             & & \qquad
                 + \ddot C^{({t})}_{\alpha\beta} 
                 + 3 H \dot C^{({t})}_{\alpha\beta}
                 - {\Delta - 2 K \over a^2} C^{({t})}_{\alpha\beta} 
                 - 8 \pi G \Pi^{({t})}_{\alpha\beta}
                = N_{4\alpha\beta},
          \eea
          which follows from eq. (\ref{pert-eq4}).
In our perturbative approach, the second-order perturbations
are sourced by the quadratic combinations of all three-types of
linear-order terms.

{}For the scalar field, from eq. (\ref{fluid-MSFs}), we have:
\bea
   \delta \mu^{(\phi)} 
   &=& \sum_k \left( \dot \phi_{(k)} \delta \dot \phi_{(k)} 
       - \dot \phi_{(k)}^2 \alpha
       + V_{,\phi_{(k)}} \delta \phi_{(k)} \right) 
       + \delta \mu^{(q)},
   \nonumber \\
   \delta p^{(\phi)} 
   &=& \sum_k \left( \dot \phi_{(k)} \delta \dot \phi_{(k)}
       - \dot \phi_{(k)}^2 \alpha
       - V_{,\phi_{(k)}} \delta \phi_{(k)} \right) 
       + \delta p^{(q)},
   \nonumber \\
   Q^{(\phi)}
   &=& - (\mu^{(\phi)} + p^{(\phi)}) v^{(\phi)}
       = - {1 \over a} \sum_k \dot \phi_{(k)} \delta \phi_{(k)}
       + \Delta^{-1} \nabla^\alpha Q^{(q)}_\alpha,
   \nonumber \\
   Q^{(\phi,v)}_\alpha 
   &=& (\mu^{(\phi)} + p^{(\phi)}) v^{(\phi,v)}_\alpha
       = Q^{(q)}_\alpha - \nabla_\alpha \Delta^{-1} \nabla^\beta Q^{(q)}_\beta.
   \label{fluid-MSFs-decomp}
\eea
The anisotropic pressure follows from 
eqs. (\ref{fluid-MSFs},\ref{fluid-decomp}).

In the generalized gravity theory in \S \ref{sec:GGT}, 
eq. (\ref{GGT-EOM-pert}) gives
\bea
   & & \delta \ddot \phi^I + 3 H \delta \dot \phi^I
       - {\Delta \over a^2} \delta \phi^I
       - 2 \alpha \ddot \phi^I
       + \dot \phi^I \left( 3 \dot \varphi - \dot \alpha - 6 H \alpha
       + {\Delta \over a^2} \chi \right)
   \nonumber \\
   & & \qquad
       - {1 \over 2} g^{IJ} \left[ F_{,J} \delta R
       + \left( f_{,LJ} - 2 V_{,LJ} \right) \delta \phi^L \right]
       - {1 \over 2} g^{IJ}_{\;\;\;\;,L} \delta \phi^L
       \left( f_{,J} - 2 V_{,J} \right)
   \nonumber \\
   & & \qquad
       + 2 \Gamma^I_{JK} \left( \dot \phi^J \delta \dot \phi^K
       - A \dot \phi^J \dot \phi^K \right)
       + \Gamma^I_{JK,L} \delta \phi^L \dot \phi^J \dot \phi^K
       + \delta \Gamma^I
       = N_g.
\eea
Equations 
(\ref{scalar-0}-\ref{scalar-6},\ref{vector-2}-\ref{vector-6},\ref{tensor-4}) 
remain valid even in the generalized gravity by replacing
the fluid quantities to the effective ones.
The decomposed effective fluid quantities follow from 
eqs. (\ref{GGT-fluid-pert},\ref{fluid-decomp}) as:
\bea
   \delta \mu^{({\rm eff})}
   &=&
       {1 \over 8 \pi G F}
       \Bigg[ \delta \mu
       + g_{IJ} \left( \dot \phi^I \delta \dot \phi^J
       - \alpha \dot \phi^I \dot \phi^J \right)
       + {1 \over 2} g_{IJ,L} \delta \phi^L \dot \phi^I \dot \phi^J
       - {1 \over 2} \left( f_{,L} - 2 V_{,L} \right) \delta \phi^L
   \nonumber \\
   & &
       - 3 H \delta \dot F
       + \left( {1 \over 2} R + {\Delta \over a^2} \right) \delta F
       + \left( 6 H \alpha
       - {\Delta \over a^2} \chi
       - 3 \dot \varphi \right) \dot F
       - 8 \pi G \mu^{({\rm eff})} \delta F
       \Bigg]
       + \delta \mu^{({\rm eff},q)},
   \nonumber \\
   \delta p^{({\rm eff})}
   &=&
       {1 \over 8 \pi G F}
       \Bigg[ \delta p
       + g_{IJ} \left( \dot \phi^I \delta \dot \phi^J
       - \alpha \dot \phi^I \dot \phi^J \right)
       + {1 \over 2} g_{IJ,L} \delta \phi^L \dot \phi^I \dot \phi^J
       + {1 \over 2} \left( f_{,L} - 2 V_{,L} \right) \delta \phi^L
       + \delta \ddot F
       + 2 H \delta \dot F
   \nonumber \\
   & &
       - \left( {1 \over 2} R + {2 \over 3} {\Delta \over a^2} \right) \delta F
       - 2 \alpha \ddot F
       - \left( \dot \alpha + 4 H \alpha
       - {2 \over 3} {\Delta \over a^2} \chi
       - 2 \dot \varphi \right) \dot F
       - 8 \pi G p^{({\rm eff})} \delta F
       \Bigg]
       + \delta p^{({\rm eff},q)},
   \nonumber \\
   Q^{({\rm eff})}
   &=&
       {1 \over 8 \pi G F}
       \left[ Q - { 1\over a} g_{IJ} \dot \phi^I \delta \phi^J
       + {1 \over a} \left( - \delta \dot F + H \delta F \right)
       + {1 \over a} A \dot F \right]
       + Q^{({\rm eff},q)},
   \nonumber \\
   Q_\alpha^{({\rm eff},v)}
   &=&
       {1 \over 8 \pi G F} Q^{(v)}_\alpha
       + Q^{({\rm eff},v,q)}_\alpha,
   \nonumber \\
   \Pi^{({\rm eff})}
   &=&
       {1 \over 8 \pi G F} \left( \Pi + \delta F - \chi \dot F \right)
       + \Pi^{({\rm eff},q)},
   \nonumber \\
   \Pi_\alpha^{({\rm eff},v)}
   &=&
       {1 \over 8 \pi G F} \left( \Pi^{(v)}_\alpha
       - \Psi^{(v)}_\alpha \dot F \right)
       + \Pi^{({\rm eff},v,q)}_\alpha,
   \nonumber \\
   \Pi_{\alpha\beta}^{({\rm eff},t)}
   &=&
       {1 \over 8 \pi G F} \left( \Pi^{(t)}_{\alpha\beta}
       - \dot C^{(t)}_{\alpha\beta} \dot F \right)
       + \Pi^{({\rm eff},t,q)}_{\alpha\beta},
   \label{GGT-pert-fluid-decomp}
\eea
where the quadratic parts follow from eq. (\ref{fluid-decomp}).
As an example, from eqs. (\ref{tensor-4},\ref{GGT-pert-fluid-decomp}),
the gravitational wave equation in generalized gravity becomes
\bea
   & & \ddot C^{({t})}_{\alpha\beta} 
       + \left( 3 H + {\dot F \over F} \right) \dot C^{({t})}_{\alpha\beta}
       - {\Delta - 2K \over a^2} C^{({t})}_{\alpha\beta} 
       = {1 \over F} \Pi^{({t})}_{\alpha\beta}
       + 8 \pi G \Pi^{({\rm eff},t,q)}_{\alpha\beta}
       + N_{4\alpha\beta}^{(t)}.
   \label{tensor-GGT} 
\eea

{}For the electromagnetic field we can decompose:
\bea
   & & E_\alpha \equiv E^{({\rm em})}_{,\alpha} + E^{(v)}_\alpha, \quad
       H_\alpha \equiv H^{({\rm em})}_{,\alpha} + H^{(v)}_\alpha; \quad
       E^{(v)|\alpha}_\alpha \equiv 0 \equiv H^{(v)|\alpha}_\alpha.
   \label{E-decomp}
\eea
The decomposed forms of fluid quantities
can be read from eqs. (\ref{fluid-EM},\ref{fluid-decomp}).
Similarly, for the null-geodesic equations we decompose
\bea
   & & \delta e_\alpha \equiv \delta e_{,\alpha} + \delta e_\alpha^{(v)}; \quad
       \delta e^{(v)|\alpha}_\alpha \equiv 0.
   \label{delta-e-decomp}
\eea
{}For the temperature anisotropy, eq. (\ref{delta-T-2}) gives
\bea
   {\delta T \over T} \Big|_O
   &=& 
       {\delta T \over T} \Big|_E
       - v_{,\alpha} e^\alpha \Big|^O_E
       + \int^O_E \Big( - \varphi^\prime + \alpha_{,\alpha} e^\alpha
       - {1 \over a} \chi_{,\alpha|\beta} e^\alpha e^\beta \Big) dy
   \nonumber \\
   & & + v^{(v)}_\alpha e^\alpha \Big|^O_E
       - \int^O_E \Psi^{(v)}_{\alpha|\beta} e^\alpha e^\beta dy
       - \int^O_E C^{(t)\prime}_{\alpha\beta} e^\alpha e^\beta dy
       + \int_E^O N_\nu dy + N_T.
   \label{SW-general}
\eea
To the linear-order this result was first presented by
Sachs and Wolfe \cite{SW-1967};
for further analyses using our notation, see \cite{Hwang-SW}.

{}For the Boltzmann equation, eq. (\ref{f-eq-q}) becomes
\bea
   & & \tilde f^\prime + {q \over \sqrt{q^2 + m^2 a^2} }
       \left( \gamma^\alpha \delta f_{,\alpha}
       - \Gamma^{(3)\alpha}_{\;\;\;\;\;\beta\gamma}
       \gamma^\beta \gamma^\gamma 
       {\partial \delta f \over \partial \gamma^\alpha} \right)
   \nonumber \\
   & & \qquad
       - \left[ \varphi^\prime + {\sqrt{q^2 + m^2 a^2} \over q}
       \gamma^\alpha \alpha_{,\alpha} 
       + \left( { 1\over a} \chi_{,\alpha|\beta}
       + \Psi^{(v)}_{\alpha|\beta}
       + C^{(t)\prime}_{\alpha\beta} \right) \gamma^\alpha \gamma^\beta
       \right] q {\partial \tilde f \over \partial q}
       = N_c.
   \label{Boltzmann-decomp}
\eea
The fluid quantities can be read from eqs. (\ref{fluid-f},\ref{fluid-decomp}).

We emphasize that all the equations up to this point are
presented without fixing the gauge conditions.
In order to solve the equations in a given situation,
we can choose {\it any} allowed gauge conditions suitable
for the situation.
In this sense, the equations are presented in a {\it gauge-ready} form.

\section{Gauge Issue}
                                                \label{sec:Gauges}

\subsection{Gauge transformation}

We consider the following transformation between two coordinates
$x^a$ and $\hat x^a$ 
\bea
   & & \hat  x^a 
       \equiv x^a + \tilde \xi^a (x^e)
       \equiv x^a + \tilde \zeta^a 
       + {1 \over 2} \tilde \zeta^a_{\;\; ,b} \tilde \zeta^b.
   \label{GT-x}
\eea
The variables $\tilde \xi^a$ and $\tilde \zeta^a$ are perturbed order quantities.
To the second-order we may have
\bea
   & & \tilde \xi^a \equiv \tilde \xi^{(1)a} + \tilde \xi^{(2)a},
\eea
and similarly for $\tilde \zeta^a$.
{}For any tensor quantity we use the tensor transformation property
between $x^a$ and $\hat x^a$ spacetimes.
\bea
   & & \tilde \phi (x^e)
       = \hat {\tilde \phi} (\hat x^e), \quad
       \tilde v_a (x^e)
       = {\partial \hat x^b \over \partial x^a} \hat {\tilde v}_b (\hat x^e),
       \quad
       \tilde t_{ab} (x^e)
       = {\partial \hat x^c \over \partial x^a}
       {\partial \hat x^d \over \partial x^b} 
       \hat {\tilde t}_{cd} (\hat x^e).
   \label{coord-tr}
\eea
Comparing the tensor quantities at the same spacetime point, $x^a$,
we can derive the gauge transformation property of the tensor quantity.
We can show that a tensor quantity ${\bf t}$ transforms as \cite{tensor-GT}
\bea
   & & \hat {\bf t} (x^c) = {\bf t} (x^c) 
       - { \pounds_{\tilde \zeta} {\bf t} }
       + {1 \over 2} \pounds_{\tilde \zeta}^2 {\bf t},
\eea
where $\pounds_{\tilde \zeta}$ is a Lie derivative along $\tilde \zeta^a$.
We have
\bea
   \hat {\tilde \phi} (x^e) 
   &=& \tilde \phi (x^e) - \tilde \phi_{,c} \tilde \xi^c
       + \tilde \phi_{,b} \tilde \xi^b_{\;\; ,c} \tilde \xi^c
       + {1 \over 2} \tilde \phi_{,bc} \tilde \xi^b \tilde \xi^c, 
   \label{GT-scalar} \\ 
   \hat {\tilde v}_a (x^e)
   &=& \tilde v_a (x^e) - \tilde v_{a,b} \tilde \xi^b
       - \tilde v_b \tilde \xi^b_{\;\;,a}
       + {1 \over 2} \tilde v_{a,bc} \tilde \xi^b \tilde \xi^c
       + \tilde v_{a,b} \tilde \xi^b_{\;\;,c} \tilde \xi^c
       + \tilde v_{b,c} \tilde \xi^b_{\;\;,a} \tilde \xi^c 
       + \tilde v_b \tilde \xi^b_{\;\;,ac} \tilde \xi^c
       + \tilde v_c \tilde \xi^c_{\;\;,b} \tilde \xi^b_{\;\;,a},
   \label{GT-vector} \\ 
   {\hat {\tilde t}}_{ab} (x^e)
   &=& \tilde t_{ab} (x^e) - 2 \tilde t_{c(a} \tilde \xi^c_{\;\;,b)} 
       - \tilde t_{ab,c} \tilde \xi^c
   \nonumber \\ 
   & & + 2 \tilde t_{c(a} \tilde \xi^d_{\;\;,b)} \tilde \xi^c_{\;\;,d} 
       + \tilde t_{cd} \tilde \xi^c_{\;\;,a} \tilde \xi^d_{\;\;,b} 
       + \tilde \xi^d \left( 
       2 \tilde \xi^c_{\;\;,(b} \tilde t_{a)c,d}
       + 2 \tilde t_{c(a} \tilde \xi^c_{\;\;,b)d} 
       + {1 \over 2} \tilde t_{ab,cd} \tilde \xi^c 
       + \tilde t_{ab,c} \tilde \xi^c_{\;\;,d} 
       \right). 
   \label{GT-tensor}
\eea

We define
\bea
   & & \tilde \xi^0 \equiv \xi^0, \quad
       \tilde \xi^\alpha \equiv \xi^\alpha, 
   \label{GT-0,alpha}
\eea
where $\xi^\alpha$ is based on $g^{(3)}_{\alpha\beta}$.
In terms of $\tilde \zeta^a$ we set $\tilde \zeta^0 \equiv \zeta^0$ and 
$\tilde \zeta^\alpha \equiv \zeta^\alpha$
where $\zeta^\alpha$ is based on $g^{(3)}_{\alpha\beta}$.
Thus, we have
\bea
   & & \xi^0 = \zeta^0 + {1 \over 2} \zeta^{0\prime} \zeta^0
       + {1 \over 2} \zeta^0_{\;\;,\alpha} \zeta^\alpha, \quad
       \xi^\alpha = \zeta^\alpha 
       + {1 \over 2} \zeta^{\alpha\prime} \zeta^0
       + {1 \over 2} \zeta^\alpha_{\;\;,\beta} \zeta^\beta.
\eea

{}From the gauge transformation property of $\tilde g_{ab}$ 
and the definitions of our perturbation variables we can derive:
\bea
   \hat A 
   &=& A 
       - \left( \xi^{0\prime} 
       + {a^\prime \over a} \xi^0 \right)
       - A^\prime \xi^0 
       - 2 A \left( \xi^{0\prime} + {a^\prime \over a} \xi^0 \right) 
       - A_{,\alpha} \xi^\alpha - B_\alpha \xi^{\alpha\prime} 
   \nonumber \\
   & & 
       + {3 \over 2} \xi^{0\prime} \xi^{0\prime} 
       + \xi^0_{\;\;,\alpha} \xi^{\alpha\prime} 
       + \xi^\alpha \left( \xi^{0\prime}_{\;\;,\alpha} 
       + {a^\prime \over a} \xi^0_{\;\;,\alpha} \right) 
       + \xi^0 \left[ \xi^{0\prime\prime} 
       + 3 {a^\prime \over a} \xi^{0\prime}  
       + {1 \over 2} \left( {a^{\prime\prime} \over a} 
       + {a^{\prime 2} \over a^2 } \right) \xi^0 \right]
       - {1\over 2} \xi^{\alpha\prime} \xi^\prime_{\alpha}, 
   \label{GT-A} \\
   \hat B_\alpha 
   &=& B_\alpha 
       - \xi^0_{\;\;,\alpha} 
       + \xi^\prime_\alpha - 2 A \xi^0_{\;\;,\alpha} 
       - \left( B^\prime_\alpha 
       + 2 {a^\prime \over a} B_\alpha \right) \xi^0 
       - B_\alpha \xi^{0\prime} 
       - B_{\alpha,\beta} \xi^\beta 
       - B_\beta \xi^\beta_{\;\;,\alpha} 
       + 2 C_{\alpha\beta} \xi^{\beta\prime} 
   \nonumber \\
   & & 
       - \xi^\prime_\alpha \xi^{0\prime} 
       + 2 \xi^{0\prime} \xi^0_{\;\;,\alpha}
       + \xi^0_{\;\;,\beta} \xi^\beta_{\;\;,\alpha} 
       + \xi^\gamma \xi^0_{\;\;,\alpha\gamma} 
       - \xi^0 \left( \xi^{\prime\prime}_\alpha 
       + 2 {a^\prime \over a} \xi^\prime_\alpha 
       - \xi^{0\prime}_{\;\;,\alpha} 
       - 2 {a^\prime \over a} \xi^0_{\;\;,\alpha} \right)
   \nonumber \\
   & & 
       - \xi^{\beta}_{\;\;,\alpha} \xi^\prime_\beta 
       - g^{(3)}_{\alpha\beta} \xi^\beta_{\;\;,\gamma} 
       \xi^{\gamma\prime}
       - \xi^\gamma \left( g^{(3)}_{\alpha\beta,\gamma} 
       \xi^{\beta\prime}
       + g^{(3)}_{\alpha\beta} \xi^{\beta\prime}_{\;\;,\gamma} \right),
   \label{GT-B} \\
   \hat C_{\alpha\beta} 
   &=& C_{\alpha\beta} 
       - {a^\prime \over a} \xi^0 g^{(3)}_{\alpha\beta}
       - {1\over 2} g^{(3)}_{\alpha\beta,\gamma} \xi^\gamma 
       - g^{(3)}_{\gamma(\alpha} \xi^\gamma_{\;\;,\beta)}
   \nonumber \\
   & & 
       + B_{(\alpha} \xi^0_{\;\;,\beta)} 
       - \left( C^\prime_{\alpha\beta} 
       + 2 {a^\prime \over a} C_{\alpha\beta} \right) \xi^0 
       - C_{\alpha\beta,\gamma} \xi^\gamma 
       - 2 C_{\gamma(\alpha} \xi^\gamma_{\;\;,\beta)} 
       + \xi^\prime_{(\alpha} \xi^0_{\;\;,\beta)} 
       - {1\over 2} \xi^0_{\;\;,\alpha} \xi^0_{\;\;,\beta} 
       + {a^\prime \over a} g^{(3)}_{\alpha\beta} \xi^\gamma \xi^0_{\;\;,\gamma}
   \nonumber \\
   & & 
       + \xi^0 \Bigg[ 
       {a^\prime \over a} g^{(3)}_{\alpha\beta} \xi^{0\prime}
       + {1 \over 2} \left( {a^{\prime\prime} \over a} 
       + {a^{\prime 2} \over a^2} \right) g^{(3)}_{\alpha\beta} \xi^0
       + \left( {1\over 2} \xi^{\gamma\prime} 
       + {a^\prime \over a} \xi^\gamma \right) 
       g^{(3)}_{\alpha\beta,\gamma} 
       + 2 {a^\prime \over a} g^{(3)}_{\gamma(\alpha} 
       \xi^\gamma_{\;\;,\beta)}
       + g^{(3)}_{\gamma(\alpha} \xi^{\gamma\prime}_{\;\;,\beta)} 
       \Bigg]
   \nonumber \\
   & & 
       + \xi^\delta_{\;\;,(\beta} 
       g^{(3)}_{\alpha)\gamma} \xi^\gamma_{\;\;,\delta} 
       + {1\over 2} g^{(3)}_{\gamma\delta} \xi^\gamma_{\;\;,\alpha} 
       \xi^\delta_{\;\;,\beta}
       + \xi^\delta \left( {1\over 2} g^{(3)}_{\alpha\beta,\gamma} 
       \xi^\gamma_{\;\;,\delta}
       + \xi^\gamma_{\;\;,(\beta} g^{(3)}_{\alpha)\gamma,\delta} 
       + {1\over 4} g^{(3)}_{\alpha\beta,\gamma\delta} \xi^\gamma
       + g^{(3)}_{\gamma(\alpha} \xi^\gamma_{\;\;,\beta)\delta} 
       \right).
   \label{GT-C} 
\eea   
{}From the gauge transformation property of $\tilde T_{ab}$ 
and using the definitions of our perturbed fluid
variables in the normal-frame we can derive:
\bea
   \delta \hat \mu 
   &=& 
       \delta \mu 
       - \left(\mu^\prime + \delta \mu^\prime \right) \xi^0
       - \delta \mu_{,\alpha} \xi^\alpha 
       + \mu^\prime \left( \xi^0 \xi^{0\prime} 
       + \xi^0_{\;\;,\alpha} \xi^\alpha \right) 
       + {1\over 2} \mu^{\prime\prime} \xi^0 \xi^0 
       + \left[ 2 Q^\alpha + ( \mu + p) \xi^{0,\alpha} \right] 
       \xi^0_{\;\;,\alpha}, 
   \label{GT-delta-mu} \\ 
   \delta \hat p 
   &=& 
       \delta p 
       - \left( p^\prime + \delta p^\prime \right) \xi^0
       - \delta p_{,\alpha} \xi^\alpha 
       + p^\prime \left( \xi^0 \xi^{0\prime} 
       + \xi^0_{\;\;,\alpha} \xi^\alpha \right) 
       + {1\over 2} p^{\prime\prime} \xi^0 \xi^0
       + {1 \over 3} \left[ 2 Q^\alpha 
       + (\mu + p) \xi^{0,\alpha} \right] \xi^0_{\;\;,\alpha}, 
   \label{GT-delta-p} \\ 
   \hat Q_\alpha 
   &=& 
       Q_\alpha + (\mu + p) \xi^0_{\;\; ,\alpha}
       - Q_\beta \xi^\beta_{\;\;,\alpha}
       - Q_{\alpha,\beta} \xi^\beta - \left( Q^\prime_\alpha 
       + {a^\prime \over a} Q_\alpha \right) \xi^0 
       + (\mu + p) A \xi^0_{\;\;,\alpha} 
       + \left[ \delta \mu + \delta p - \left( \mu^\prime 
       + p^\prime \right) \xi^0 \right] \xi^0_{\;\;,\alpha} 
   \nonumber \\
   & &
       + \Pi^\beta_\alpha \xi^0_{\;\;,\beta} 
       - (\mu + p) \left[ \xi^0_{\;\;,\alpha} \xi^{0\prime}
       + \xi^\beta_{\;\;,\alpha} \xi^0_{\;\;,\beta} 
       + \xi^0 \left( \xi^{0\prime}_{\;\;,\alpha} 
       + {a^\prime \over a} \xi^0_{\;\;,\alpha} \right) 
       + \xi^\beta \xi^0_{\;\;,\alpha\beta} \right],
   \label{GT-Q} \\
   \hat \Pi_{\alpha\beta} 
   &=& 
       \Pi_{\alpha\beta} - 2 \Pi_{\gamma(\alpha}
       \xi^\gamma_{\;\;,\beta)} - \left( \Pi^\prime_{\alpha\beta} 
       + 2 {a^\prime \over a} \Pi_{\alpha\beta} \right) \xi^0 
       - \Pi_{\alpha\beta,\gamma} \xi^\gamma 
   \nonumber \\
   & & 
       + \left[ 2 Q_{(\alpha} + ( \mu + p) \xi^0_{\;\;,(\alpha} \right] 
       \xi^0_{\;\;,\beta)} - {1\over 3} g^{(3)}_{\alpha\beta}
       \left[ 2 Q^\gamma + ( \mu + p) \xi^{0,\gamma} \right] 
       \xi^0_{\;\;,\gamma}. 
   \label{GT-Pi} 
\eea
Under the gauge transformation the individual fluid quantities 
$\delta \mu_{(i)}$, $\delta p_{(i)}$, $Q_{(i)\alpha}$, 
$\Pi_{(i)\alpha\beta}$, and $\delta \phi_{(i)}$ 
transform just like the corresponding collective fluid quantities
in eqs. (\ref{GT-delta-mu}-\ref{GT-phi}) with all the fluid quantities
changed into the ones for the individual one.
Using the vector nature of $\tilde I_{(i)a}$ we have
\bea
   \delta \hat I_{(i)0}
   &=& \delta I_{(i)0} - \left( I_{(i)0} \xi^0 \right)^\prime
       - \left( \delta I_{(i)0} \xi^0 \right)^\prime
       - \delta I_{(i)0,\alpha} \xi^\alpha
       + \delta I_{(i)\alpha} \xi^{\alpha\prime}
       + I_{(i)0}^\prime \xi^{0\prime} \xi^0
       + \left[ I_{(i)0} \left( \xi^{0\prime} \xi^0
       + \xi^0_{\;\;,\alpha} \xi^\alpha \right) \right]^\prime,
   \label{GT-I_0} \\
   \delta \hat I_{(i)\alpha} 
   &=& \delta I_{(i)\alpha} - I_{(i)0} \xi^0_{\;\;,\alpha}
       - \delta I_{(i)0} \xi^0_{\;\;,\alpha}
       - \delta I_{(i)\alpha}^\prime \xi^0
       - \delta I_{(i)\alpha,\beta} \xi^\beta
       - \delta I_{(i)\beta} \xi^\beta_{\;\;,\alpha}
       + I_{(i)0}^\prime \xi^0_{\;\;,\alpha} \xi^0
   \nonumber \\
   & &
       + I_{(i)0} \left[ \left( \xi^0_{\;\;,\alpha} \xi^0 \right)^\prime
       + \left( \xi^0_{\;\;,\beta} \xi^\beta \right)_{,\alpha} \right].
   \label{GT-I_alpha}
\eea

The fluid quantities we use in this work are based on the 
normal-frame four-vector where 
$\tilde n_\alpha = 0$, see eq. (\ref{frames}).
It is convenient to have the gauge-transformation properties
of the fluid quantities in the energy-frame where we set $Q_\alpha = 0$.
These can be derived either by applying the frame-transformation rule
presented in eqs. (\ref{frame-N-E},\ref{frame-E-N}) or directly from the
gauge transformation property of the energy-momentum tensor
in eq. (\ref{Tab-pert-general}) with $Q_\alpha = 0$ in the energy-frame.
We have
\bea
   \delta \hat \mu^E
   &=& 
       \delta \mu^E - \left( \mu^\prime + \delta \mu^{E\prime} \right) \xi^0
       - \delta \mu^E_{\;\;,\alpha} \xi^\alpha
       + {1 \over 2} \mu^{\prime\prime} \xi^0 \xi^0
       + \mu^\prime \left( \xi^0 \xi^{0\prime}
       + \xi^\alpha \xi^0_{\;\;,\alpha} \right),  
   \nonumber \\
   \delta \hat p^E
   &=& 
       \delta p^E - \left( p^\prime + \delta p^{E\prime} \right) \xi^0
       - \delta p^E_{\;\;,\alpha} \xi^\alpha
       + {1 \over 2} p^{\prime\prime} \xi^0 \xi^0
       + p^\prime \left( \xi^0 \xi^{0\prime}
       + \xi^\alpha \xi^0_{\;\;,\alpha} \right),  
   \nonumber \\
   \hat V^E_\alpha - \hat B_\alpha
   &=& 
       V^E_\alpha - B_\alpha + \xi^0_{\;\;,\alpha}
       - \left( V^E_\alpha - B_\alpha \right)^\prime \xi^0 
       + {a^\prime \over a} \left( V^E_\alpha - B_\alpha \right) \xi^0 
       - \left( V^E_\beta - B_\beta \right) \xi^\beta_{\;\;,\alpha} 
       - \left( V^E_\alpha - B_\alpha \right)_{,\beta} \xi^\beta
   \nonumber \\
   & & 
       + \left( V^{E\beta} + \xi^{\beta\prime} \right) 
       \left( g^{(3)}_{\alpha\beta,\gamma} \xi^\gamma
       + 2 g^{(3)}_{\gamma(\alpha} \xi^\gamma_{\;\;,\beta)} \right)
       + \left( A - \xi^{0\prime} - {a^\prime \over a} \xi^0 \right)
       \left( 2 \xi^0_{\;\;,\alpha} - \xi^\prime_\alpha \right)
       + B_\alpha \left( \xi^{0\prime} + 3 {a^\prime \over a} \xi^0 \right)
       - 2 C_{\alpha\beta} \xi^{\beta\prime}
   \nonumber \\
   & & 
       - \xi^\beta_{\;\;,\alpha} \xi^0_{\;\;,\beta}
       - \xi^0 \xi^{0\prime}_{\;\;,\alpha}
       - \xi^\beta \xi^0_{\;\;,\alpha\beta}
       + 2 {a^\prime \over a} \xi^0 \xi_\alpha^\prime,
   \nonumber \\
   \hat \Pi^E_{\alpha\beta}
   &=& 
       \Pi^E_{\alpha\beta} 
       - \left( \Pi^{E\prime}_{\alpha\beta}
       + 2 {a^\prime \over a} \Pi^E_{\alpha\beta} \right) \xi^0
       - \Pi^E_{\alpha\beta,\gamma} \xi^\gamma
       - 2 \Pi^E_{\gamma(\alpha} \xi^\gamma_{\;\;,\beta)}.
   \label{GT-E-frame}
\eea

{}From the gauge transformation of $\tilde \phi$ we have
\bea
   & & \delta \hat \phi = \delta \phi 
       - \left( \phi^\prime + \delta \phi^\prime \right) \xi^0 
       - \delta \phi_{,\alpha} \xi^\alpha
       + \phi^\prime \left( \xi^{0\prime} \xi^0 
       + \xi^0_{\;\;,\alpha} \xi^\alpha \right)
       + {1 \over 2} \phi^{\prime\prime} \xi^0 \xi^0.
   \label{GT-phi} 
\eea

Using the gauge-transformation property of a vector quantity 
$\tilde k^a$ similar to eqs. (\ref{coord-tr},\ref{GT-vector}),
and using the definition of $\tilde k^a$ in eq. (\ref{k-def}) 
we can derive
\bea
   {\delta \hat \nu \over \nu}
   &=&
       {\delta \nu \over \nu}
       + \xi^{0\prime} + 2 {a^\prime \over a} \xi^0
       - \xi^0_{\;\;,\alpha} e^\alpha
       - {\delta \nu^\prime \over \nu} \xi^0
       + {\delta \nu \over \nu} \left( \xi^{0\prime}
       + {a^\prime \over a} \xi^0 \right)
       - {\delta \nu_{,\alpha} \over \nu} \xi^\alpha
       - \xi^0_{\;\;,\alpha} \delta e^\alpha
   \nonumber \\
   & &
       + \xi^0 \left[ - \xi^{0\prime\prime}
       - \left( {a^{\prime\prime} \over a}
       - 3 {a^{\prime 2} \over a^2} \right) \xi^0
       + \xi^{0\prime}_{\;\;,\alpha} e^\alpha
       + \xi^0_{\;\;,\alpha} \left( e^{\alpha\prime}
       - 2 {a^\prime \over a} e^\alpha \right) \right]
   \nonumber \\
   & &
       + \xi^\alpha \left( - \xi^{0\prime}_{\;\;,\alpha}
       - 2 {a^\prime \over a} \xi^0_{\;\;,\alpha}
       + \xi^0_{\;\;,\alpha\beta} e^\beta
       + \xi^0_{\;\;,\beta} e^\beta_{\;\;,\alpha} \right),
   \label{GT-delta-nu} \\
   \delta \hat e^\alpha
   &=&
       \delta e^\alpha
       - \xi^0 \left( e^{\alpha\prime} - 2 {a^\prime \over a} e^\alpha \right)
       - \xi^{\alpha\prime}
       + \xi^\alpha_{\;\;,\beta} e^\beta
       - e^\alpha_{\;\;,\beta} \xi^\beta
       - \xi^0 \left( \delta e^{\alpha\prime}
       - 2 {a^\prime \over a} \delta e^\alpha \right)
       - \xi^{\alpha\prime} {\delta \nu \over \nu}
       + \xi^\alpha_{\;\;,\beta} \delta e^\beta
       - \delta e^\alpha_{\;\;,\beta} \xi^\beta
   \nonumber \\
   & &
       + \xi^0 \Bigg\{
       \xi^{0\prime} \left( e^{\alpha\prime}
       - 2 {a^\prime \over a} e^\alpha \right)
       + \xi^0 \left[ {1 \over 2} e^{\alpha\prime\prime}
       - 2 {a^\prime \over a} e^{\alpha\prime}
       - \left( {a^{\prime\prime} \over a} - 3 {a^{\prime 2} \over a^2} \right)
       e^\alpha \right]
   \nonumber \\
   & & + \xi^{\alpha\prime\prime}
       - 2 {a^\prime \over a} \xi^{\alpha\prime}
       - \xi^{\alpha\prime}_{\;\;,\beta} e^\beta
       + e^\alpha_{\;\;,\beta} \xi^{\beta\prime}
       + \xi^\beta \left( e^{\alpha\prime}_{\;\;,\beta}
       - 2 {a^\prime \over a} e^\alpha_{\;\;,\beta} \right)
       - \xi^\alpha_{\;\;,\beta} \left( e^{\beta\prime}
       - 2 {a^\prime \over a} e^\beta \right) \Bigg\}
   \nonumber \\
   & & + \xi^\beta \left[ {1 \over 2} e^\alpha_{\;\;,\beta\gamma} \xi^\gamma
       - \xi^\alpha_{\;\;,\gamma} e^\gamma_{\;\;,\beta}
       + e^\alpha_{\;\;,\gamma} \xi^\gamma_{\;\;,\beta}
       + \xi^{\alpha\prime}_{\;\;,\beta}
       - \xi^\alpha_{\;\;,\beta\gamma} e^\gamma
       + \xi^0_{\;\;,\beta} \left( e^{\alpha\prime}
       - 2 {a^\prime \over a} e^\alpha \right) \right].
   \label{GT-delta-e}
\eea
Using the scalar nature of the temperature $\tilde T$ and
eq. (\ref{GT-scalar}) we can show
\bea
   & & \delta \hat T (x^e)
       = \delta T (x^e) - \left( T^\prime + \delta T^\prime \right) \xi^0
       - \delta T_{,\alpha} \xi^\alpha
       + T^\prime \left( \xi^{0\prime} \xi^0
       + \xi^0_{\;\;,\alpha} \xi^\alpha \right)
       + {1 \over 2} T^{\prime\prime} \xi^0 \xi^0.
   \label{GT-T}
\eea
Using the vector nature of the electric and magnetic vectors and
eq. (\ref{GT-vector}) we can show
\bea
   & & \hat E_\alpha (x^e)
       = E_\alpha (x^e)
       - E_\alpha^\prime \xi^0
       - E_{\alpha,\beta} \xi^\beta
       - E_\beta \xi^\beta_{\;\;,\alpha},
\eea
and similarly for $H_\alpha$.
Thus, $E_\alpha$ and $H_\alpha$ are gauge-invariant to the linear-order.

Since $\tilde p^a \equiv d x^a/ d \lambda$, under the gauge
transformation we have 
${\hat {\tilde p}}^a = \tilde p^a + \tilde \xi^a_{\;\;,b} \tilde p^b$.
Using the definitions of $q$ and $\gamma^\alpha$ in Eq. (\ref{q-def}) 
we can derive
\bea
   \hat q 
   &=& q \Bigg\{ 1 + {a^\prime \over a} \xi^0 
       + {\sqrt{q^2 + m^2 a^2} \over q} \xi^0_{\;\;,\alpha} \gamma^\alpha
       + {\sqrt{q^2 + m^2 a^2} \over q} \xi^0_{\;\;,\alpha} \left[
       \gamma^\alpha \left( A - \xi^{0\prime} + {a^\prime \over a} \xi^0 \right)
       - C^\alpha_\beta \gamma^\beta \right]
   \nonumber \\
   & & 
       + {q^2 + m^2 a^2 \over q^2} \Bigg[
       - A^\prime \xi^0 - A_{,\alpha} \xi^\alpha
       + {1 \over 2} \xi^0_{\;\;,\alpha} \xi^{0,\alpha} 
       + \xi^\alpha \left( \xi^{0\prime}_{\;\;,\alpha} 
       + {a^\prime \over a} \xi^0_{\;\;,\alpha} \right)
       + \xi^0 \xi^{0\prime\prime}
       + {a^\prime \over a} \xi^0 \xi^{0\prime}
   \nonumber \\
   & & 
       + \left( {1 \over 2} {a^{\prime\prime} \over a} 
       {3 q^2 + 2 m^2 a^2 \over q^2 + m^2 a^2}
       - {a^{\prime 2} \over a^2} \right) \xi^0 \xi^0 \Bigg]
       - {1 \over 2} {m^2 a^2 \over q^2} \xi^0_{\;\;,\alpha} \xi^0_{\;\;,\beta}
       \gamma^\alpha \gamma^\beta \Bigg\},
   \label{q-GT} \\
   \hat \gamma^\alpha 
   &=& \gamma^\alpha + {\sqrt{q^2 + m^2 a^2} \over q} \left( \xi^{0,\alpha}
       - \xi^0_{\;\;,\beta} \gamma^\beta \gamma^\alpha \right)
       + \left( \xi^\alpha_{\;\;,\beta} - {1 \over 2} \xi^\alpha_{\;\;|\beta}
       - {1 \over 2} \xi^{\;\;|\alpha}_\beta \right) \gamma^\beta.
   \label{gamma-GT}
\eea
As $\hat \gamma^\alpha$ always appears together with perturbed order 
terms multiplied, it is evaluated only to the linear-order.
{}From the scalar nature of $\tilde f$ we have
\bea
   & & \tilde f (x^e, q, \gamma^\epsilon)
       = \hat {\tilde f} (\hat x^e, \hat q, \hat \gamma^\epsilon)
       = \hat {\tilde f} (x^e + \tilde \xi^e, q + \delta q, 
       \gamma^\epsilon + \delta \gamma^\epsilon).
\eea
At the same momentum space and the spacetime point, we have
\bea
   & & \delta \hat f 
       = \delta f - \tilde f^\prime \xi^0 - \tilde f_{,q} \delta q
       + f^\prime \xi^0 \xi^{0\prime} 
       + {1 \over 2} f^{\prime\prime} \xi^0 \xi^0
       + 2 f^\prime_{,q} \xi^0 \delta q
       + f_{,q} \delta q_{,q} \delta q
       + {1 \over 2} f_{,qq} \delta q^2
       - \delta f_{,\alpha} \xi^\alpha
       - \delta f_{,\gamma^\alpha} \delta \gamma^\alpha.
\eea
Using eqs. (\ref{q-GT},\ref{gamma-GT},\ref{GT-A}) we have
\bea
   \delta \hat f
   &=& \delta f - \tilde f^\prime \xi^0
       - q \tilde f_{,q} \left( {a^\prime \over a} \xi^0
       + {\sqrt{q^2 + m^2 a^2} \over q} \xi^0_{\;\;,\alpha} \gamma^\alpha
       \right)
   \nonumber \\
   & & 
       - q f_{,q} \Bigg\{ 
       {\sqrt{q^2 + m^2 a^2} \over q} \xi^0_{\;\;,\alpha} 
       \left[ \gamma^\alpha \left( A - \xi^{0\prime} \right) 
       - C^\alpha_\beta \gamma^\beta \right]
       + {q^2 + m^2 a^2 \over q^2} \Bigg[ 
       - A^\prime \xi^0 - A_{,\alpha} \xi^\alpha
   \nonumber \\
   & &
       + \xi^\alpha \left( \xi^{0\prime}_{\;\;,\alpha} 
       + {a^\prime \over a} \xi^0_{\;\;,\alpha} \right)
       + \xi^0 \xi^{0\prime\prime}
       + {1 \over 2} {a^{\prime\prime} \over a} 
       {3 q^2 + 2 m^2 a^2 \over q^2 + m^2 a^2} \xi^0 \xi^0
       + {a^\prime \over a} \xi^0 \xi^{0\prime}
       + {1 \over 2} \xi^0_{\;\;,\alpha} \xi^{0,\alpha} \Bigg]
   \nonumber \\
   & &
       - {1 \over 2} {2 q^2 + m^2 a^2 \over q^2} \xi^0_{\;\;,\alpha}
       \xi^0_{\;\;,\beta} \gamma^\alpha \gamma^\beta
       - {2 q^2 + m^2 a^2 \over q^2} {a^{\prime 2} \over a^2} \xi^0 \xi^0
       - {q \over \sqrt{q^2 + m^2 a^2}} {a^\prime \over a} 
       \xi^0 \xi^0_{\;\;,\alpha} \gamma^\alpha 
       \Bigg\}
   \nonumber \\
   & & 
       + {1 \over 2} q^2 f_{,qq} \left( 
       {a^{\prime 2} \over a^2} \xi^0 \xi^0
       + 2 {\sqrt{q^2 + m^2 a^2} \over q} {a^\prime \over a}
       \xi^0 \xi^0_{\;\;,\alpha} \gamma^\alpha
       + {q^2 + m^2 a^2 \over q^2} \xi^0_{\;\;,\alpha} \xi^0_{\;\;,\beta}
       \gamma^\alpha \gamma^\beta \right)
   \nonumber \\
   & & 
       + 2 q f^\prime_{,q} \xi^0 \left(
       {a^\prime \over a} \xi^0
       + {\sqrt{q^2 + m^2 a^2} \over q} \xi^0_{\;\;,\alpha} \gamma^\alpha
       \right)
       + f^\prime \xi^0 \xi^{0\prime}
       + {1 \over 2} f^{\prime\prime} \xi^0 \xi^0
   \nonumber \\
   & & 
       - \delta f_{,\alpha} \xi^\alpha
       - \delta f_{,\gamma^\alpha} \left[
       {\sqrt{q^2 + m^2 a^2} \over q} \left( \xi^{0,\alpha}
       - \xi^0_{\;\;,\beta} \gamma^\beta \gamma^\alpha \right)
       + \left( \xi^\alpha_{\;\;,\beta}
       - {1 \over 2} \xi^\alpha_{\;\;|\beta} 
       - {1 \over 2} \xi^{\;\;|\alpha}_\beta \right) \gamma^\beta 
       \right].
   \label{f-GT}
\eea
Notice that with our phase space variables introduced in Eq. (\ref{q-def})
the distribution function $f$ is spatially gauge-invariant to the linear-order.
We can check that the gauge transformation property of $\delta f$
in Eq. (\ref{f-GT}) is consistent with the gauge transformation properties
of the fluid quantities identified in Eq. (\ref{fluid-f}).

We further decompose $\xi_\alpha$ (and similarly for $\zeta_\alpha$)
into the scalar- and vector-types as
\bea
   & & \xi_\alpha \equiv {1 \over a} \xi_{,\alpha} + \xi_\alpha^{({v})}, 
   \label{xi-decomp}
\eea
with $\xi^{({v})\alpha}_{\;\;\;\;\;\;\;|\alpha} \equiv 0$.
In order to fix the gauge we can impose three conditions on three 
variables such that 
these conditions can fix $\xi^0$, $\xi$ and $\xi_\alpha^{({v})}$.
We call these conditions fixing $\xi^0$, $\xi$ and $\xi_\alpha^{({v})}$
as the temporal, the spatial and the rotational gauge fixing conditions,
respectively.

The decomposed variables in eq. (\ref{metric-decomp-def}) and others 
transform as
\bea
   \hat \alpha 
   &=& \alpha 
       - {1 \over a} \left( a \xi^0 \right)^\prime
       + A_\xi,
   \nonumber \\
   \hat \beta 
   &=& \beta - \xi^0 
       + \left( {1 \over a} \xi \right)^\prime
       + \Delta^{-1} \nabla^\alpha B_{\xi\alpha},
   \nonumber \\
   \hat B_\alpha^{(v)} 
   &=& B_\alpha^{(v)} 
       + \xi_\alpha^{(v)\prime}
       + B_{\xi \alpha} 
       - \nabla_\alpha \Delta^{-1} \nabla^\beta B_{\xi \beta},
   \nonumber \\
   \hat \gamma 
   &=& \gamma 
       - {1 \over a} \xi
       + {1 \over 2} \left( \Delta + {1 \over 2} R^{(3)} \right)^{-1}
       \left( 3 \Delta^{-1} \nabla^\alpha \nabla^\beta C_{\xi \alpha\beta}
       - C_{\xi\alpha}^{\;\;\alpha} \right),
   \nonumber \\
   \hat \varphi 
   &=& \varphi 
       - {a^\prime \over a} \xi^0
       + {1 \over 3} C_{\xi\alpha}^{\;\;\alpha}
       - {1 \over 6} \Delta \left( \Delta + {1\over 2} R^{(3)} \right)^{-1}
       \left( 3 \Delta^{-1} \nabla^\alpha \nabla^\beta C_{\xi \alpha\beta}
       - C_{\xi \alpha}^{\;\;\alpha} \right),
   \nonumber \\
   \hat C_\alpha^{(v)} 
   &=& C_\alpha^{(v)} 
       - \xi_\alpha^{(v)}
       + 2 \left( \Delta + {1 \over 3} R^{(3)} \right)^{-1}
       \left( \nabla^\beta C_{\xi \alpha\beta}
       - \nabla_\alpha \Delta^{-1} \nabla^\gamma \nabla^\beta C_{\xi\gamma\beta}
       \right),
   \nonumber \\
   \hat C_{\alpha\beta}^{(t)} 
   &=& C_{\alpha\beta}^{(t)} 
       - C_{\xi\alpha\beta} 
       - {1 \over 3} C_{\xi\gamma}^{\;\;\gamma} g^{(3)}_{\alpha\beta}
       - {1 \over 2} \left( \nabla_\alpha \nabla_\beta 
       - {1 \over 3} g^{(3)}_{\alpha\beta} \Delta \right)
       \left( \Delta + {1 \over 2} R^{(3)} \right)^{-1}
       \left( 3 \Delta^{-1} \nabla^\gamma \nabla^\delta C_{\xi\gamma\delta}
       - C_{\xi\gamma}^{\;\;\gamma} \right)
   \nonumber \\
   & & 
       - \nabla_{(\alpha} \left( \Delta + {1 \over 3} R^{(3)} \right)^{-1}
       \left( \nabla^\gamma C_{\xi \beta)\gamma}
       - \nabla_{\beta)} \Delta^{-1} \nabla^\gamma \nabla^\delta
       C_{\xi\gamma\delta} \right),
   \nonumber \\
   \delta \hat \mu 
   &=& \delta \mu 
       - \mu^\prime \xi^0 + \delta \mu_\xi,
   \nonumber \\
   \delta \hat p 
   &=& \delta p - 
       p^\prime \xi^0 + \delta p_\xi,
   \nonumber \\
   \hat v 
   &=& v 
       - \xi^0 
       - {1 \over \mu + p} \Delta^{-1} \nabla^\alpha Q_{\xi\alpha},
   \nonumber \\
   \hat v_\alpha^{(v)} 
   &=& v_\alpha^{(v)}
       + {1 \over \mu + p} \left( Q_{\xi\alpha} 
       - \nabla_\alpha \Delta^{-1} \nabla^\beta Q_{\xi\beta} \right),
   \nonumber \\
   \hat \Pi 
   &=& \Pi 
       + {3 \over 2} a^2 \left( \Delta + {1 \over 2} R^{(3)} \right)^{-1}
       \Delta^{-1} \nabla^\alpha \nabla^\beta \Pi_{\xi\alpha\beta},
   \nonumber \\
   \hat \Pi^{(v)}_\alpha 
   &=& \Pi^{(v)}_\alpha
       + 2 a \left( \Delta + {1 \over 3} R^{(3)} \right)^{-1}
       \left( \nabla^\beta \Pi_{\xi\alpha\beta} 
       - \nabla_\alpha \Delta^{-1} \nabla^\beta \nabla^\gamma 
       \Pi_{\xi\beta\gamma} \right),
   \nonumber \\
   \hat \Pi_{\alpha\beta}^{(t)} 
   &=& \Pi_{\alpha\beta}^{(t)} 
       - \Pi_{\xi\alpha\beta} 
       - {3 \over 2} \left( \nabla_\alpha \nabla_\beta 
       - {1 \over 3} g^{(3)}_{\alpha\beta} \Delta \right)
       \left( \Delta + {1 \over 2} R^{(3)} \right)^{-1}
       \Delta^{-1} \nabla^\gamma \nabla^\delta \Pi_{\xi\gamma\delta}
   \nonumber \\
   & & 
       - \nabla_{(\alpha} \left( \Delta + {1 \over 3} R^{(3)} \right)^{-1}
       \left( \nabla^\gamma \Pi_{\xi \beta)\gamma}
       - \nabla_{\beta)} \Delta^{-1} \nabla^\gamma \nabla^\delta
       \Pi_{\xi\gamma\delta} \right),
   \nonumber \\
   \delta \hat \phi 
   &=& \delta \phi 
       - \phi^\prime \xi^0 + \delta \phi_\xi,
   \nonumber \\
   {\delta \hat \nu \over \nu} 
   &=& {\delta \nu \over \nu} 
       + \xi^{0\prime} + 2 {a^\prime \over a} \xi^0
       - \xi^0_{\;\;,\alpha} e^\alpha 
       + {\delta \nu_\xi \over \nu},
   \nonumber \\
   \delta \hat e
   &=& \delta e
       - {1 \over a} \xi^\prime + {a^\prime \over a^2} \xi
       - \Delta^{-1} \nabla_\alpha \left[ \xi^0
       \left( e^{\alpha \prime} - 2 {a^\prime \over a} e^\alpha \right) \right]
   \nonumber \\
   & &
       + \Delta^{-1} \left[ {1 \over a} \Delta (\xi_{,\alpha}) e^\alpha
       + 2 K \xi^{(v)}_\alpha e^\alpha
       - e^\alpha_{\;\;|\beta\alpha} 
       \left( {1 \over a} \xi^{,\beta} + \xi^{(v)\beta} \right)
       + \nabla_\alpha \delta e^\alpha_\xi \right],
   \nonumber \\
   \delta \hat e^{(v)}_\alpha
   &=& \delta e_\alpha 
       - \delta e_{,\alpha},
   \nonumber \\
   \delta \hat T
   &=& \delta T 
       - T^\prime \xi^0 + \delta T_\xi,
   \nonumber \\
   \hat E^{({\rm em})}
   &=& E^{({\rm em})} 
       + \Delta^{-1} \nabla^\alpha E_{\alpha\xi}, 
   \nonumber \\
   \hat E^{(v)}_\alpha
   &=& E^{(v)}_\alpha 
       + E_{\alpha\xi}
       - \nabla_\alpha \Delta^{-1} \nabla^\beta E_{\beta\xi},
   \nonumber \\
   \delta \tilde f 
   &=& \delta f - q {\partial f \over \partial q}
       \left( {a^\prime \over a} \xi^0
       + {\sqrt{q^2 + m^2 a^2} \over q} 
       \xi^0_{,\alpha} \gamma^\alpha \right)
       + \delta f_\xi,
   \label{GT-decomposed}
\eea
where $A_\xi$ indicates the quadratic parts of eq. (\ref{GT-A})
and similarly for other variables.
{}For $\delta f$ we have used $f^\prime = 0$ which follows
from eq. (\ref{f-eq-q}) for $C[f] = 0$ to the background order.

Using $t$ instead of $\eta$ (indicated as $0$) as the time variable,
from the definition $dt \equiv a d \eta$ we can show
\bea
   & & \xi^0 = {1 \over a} \xi^t \left( 1 - {1 \over 2} H \xi^t \right).
   \label{xi^t-def}
\eea

\subsection{Linear-order}

{}From eq. (\ref{GT-decomposed})
we find that the decomposed metric and matter variables 
transform to the linear-order as:
\bea
   & & \hat \alpha = \alpha - \dot \xi^t, \quad
       \hat \beta = \beta - {1 \over a} \xi^t 
       + a \left( {\xi \over a} \right)^\cdot, \quad
       \hat \gamma = \gamma - {1 \over a} \xi, \quad
       \hat \varphi = \varphi - H \xi^t, \quad
       \hat \chi = \chi - \xi^t, \quad
       \hat \kappa = \kappa 
       + \left( 3 \dot H + {\Delta \over a^2} \right) \xi^t,
   \nonumber \\
   & & \delta \hat \mu = \delta \mu - \dot \mu \xi^t, \quad
       \delta \hat p = \delta p - \dot p \xi^t, \quad
       \hat v = v - {1 \over a} \xi^t, \quad
       \hat \Pi = \Pi, \quad
       \delta \hat \phi = \delta \phi - \dot \phi \xi^t,
   \nonumber \\
   & & \hat B_\alpha^{({v})} = B_\alpha^{({v})} + a \dot \xi_\alpha^{({v})},
       \quad
       \hat C_\alpha^{({v})} = C_\alpha^{({v})} - \xi_\alpha^{({v})}, \quad
       \hat \Psi_\alpha^{({v})} = \Psi_\alpha^{({v})}, \quad
       \hat v^{(v)}_\alpha = v^{(v)}_\alpha, \quad
       \hat \Pi_\alpha^{({v})} = \Pi_\alpha^{({v})}, 
   \nonumber \\
   & & \hat C_{\alpha\beta}^{({t})} = C_{\alpha\beta}^{({t})}, \quad
       \hat \Pi_{\alpha\beta}^{({t})} = \Pi_{\alpha\beta}^{({t})}.
   \label{GT-decomposed-linear}
\eea

\subsubsection{Temporal gauge conditions}

Temporal gauge fixing condition, fixing $\xi^t$,
applies only to the scalar-type perturbation.
To the linear-order, we can impose any one of the following temporal 
gauge conditions to be valid at any spacetime point:
\bea
   & & {\rm synchronous \; gauge} : \hskip 1.8cm \alpha \equiv 0 
       \quad \rightarrow \quad \xi^t ({\bf x}),
   \nonumber \\
   & & {\rm comoving \; gauge} : \hskip 2.2cm v \equiv 0 \quad 
       \rightarrow \quad \xi^t = 0,
   \nonumber \\
   & & {\rm zero\!-\!shear \; gauge} : \hskip 1.9cm \chi \equiv 0 \quad 
       \rightarrow \quad \xi^t = 0,
   \nonumber \\
   & & {\rm uniform\!-\!expansion \; gauge} : \quad \hskip .3cm \kappa \equiv 0 
       \quad \rightarrow \quad \xi^t = 0,
   \nonumber \\
   & & {\rm uniform\!-\!curvature \; gauge} : \hskip .7cm \varphi \equiv 0 
       \quad \rightarrow \quad \xi^t = 0,
   \nonumber \\
   & & {\rm uniform\!-\!density \; gauge} : \hskip .8cm \delta \mu \equiv 0 
       \quad \rightarrow \quad \xi^t = 0,
   \nonumber \\
   & & {\rm uniform\!-\!pressure \; gauge} : \hskip .7cm \delta p \equiv 0 
       \quad \rightarrow \quad \xi^t = 0,
   \nonumber \\
   & & {\rm uniform\!-\!field \; gauge} : \hskip 1.3cm \delta \phi \equiv 0 
       \quad \rightarrow \quad \xi^t = 0.
   \label{temporal-gauges-1}
\eea
Except for the synchronous gauge condition, each of the other 
temporal gauge fixing conditions completely removes the temporal gauge mode.
In the multi-component situations in addition we can choose
one of the following conditions as the proper temporal gauge condition
which also removes the temporal gauge-mode completely:
\bea
   & & \delta \mu_{(i)} \equiv 0, \quad
       \delta p_{(i)} \equiv 0, \quad
       v_{(i)} \equiv 0, \quad
       \delta \phi_{(i)} \equiv 0.
   \label{temporal-gauges-2}
\eea
All these variables which can be used to fix the temporal gauge
freedom in fact do not depend on the spatial gauge transformation, $\xi$,
thus are naturally spatially gauge-invariant.

The followings are some examples of combinations of variables 
which are temporally gauge-invariant:
\bea
   & & \delta \mu_v \equiv \delta \mu - \dot \mu a v, \quad
       \varphi_\chi \equiv \varphi - H \chi, \quad
       v_\chi \equiv v - {1 \over a} \chi, \quad
       \varphi_v \equiv \varphi - a H v, \quad
       \varphi_{\delta \phi} \equiv \varphi - {H \over \dot \phi} \delta \phi
       \equiv - {H \over \dot \phi} \delta \phi_\varphi.
   \label{GI}
\eea
These are completely (i.e., both spatially and temporally) gauge-invariant
to the linear-order.
Any variable under any gauge condition in 
eqs. (\ref{temporal-gauges-1},\ref{temporal-gauges-2}) 
(except for the synchronous gauge), has a unique equivalent 
gauge-invariant combination.
{}For example, we have
\bea
   & & \varphi_\chi = \varphi|_{\chi \equiv 0}.
\eea
Thus, $\varphi_\chi$ is {\it the same} as $\varphi$ variable in 
the zero-shear gauge where we set $\chi \equiv 0$.

All the equations in \S \ref{sec:Equations} and \ref{sec:Decomposition} 
are presented without imposing any gauge condition.
The equations are arranged using the above variables in
eqs. (\ref{temporal-gauges-1},\ref{temporal-gauges-2}) 
which can be used in fixing the temporal gauge condition.
This allows us to use the various temporal gauge conditions optimally
depending on the situation,
thus the equations are presented in a sort of {\it gauge-ready} manner.
Usually we do not know the most suitable gauge condition {\it a priori}.
In order to take the advantage of gauge choice in the most optimal way
it is desirable to use the gauge-ready form equations presented in this paper.
Our set of equations is arranged so that we can easily impose various
fundamental gauge conditions in 
eqs. (\ref{temporal-gauges-1},\ref{temporal-gauges-2}),
and their suitable combinations as well.
As we have so many different ways of fixing the temporal gauge conditions
it is convenient to denote the gauge condition,
or equivalently, gauge-invariant combination, we are using.
Our notation for gauge-invariant combinations proposed in eq. (\ref{GI})
is convenient for such a purpose in the spirit of our gauge strategy
\cite{Hwang-1991,Hwang-Noh-CMB-2002}.
The notation is also practically convenient for connecting solutions 
in different gauge conditions as well as tracing the associated 
gauge conditions easily.
Compared with the notations for gauge-invariant variables which were
introduced by Bardeen \cite{Bardeen-1980,Bardeen-1988} we have
\bea
   & & \epsilon_m = \delta_v \equiv \delta \mu_v/\mu, \quad
       \Psi_H = \varphi_\chi, \quad
       v^{(0)}_s = k v_\chi, \quad
       p \pi_L^{(0)} = \delta p, \quad
       p \pi_T^{(0)} = - {\Delta \over a^2} \Pi, \quad
       \zeta \equiv \varphi_\delta, 
   \label{Bardeen-notation}
\eea
etc.; we ignored the harmonic functions used in \cite{Bardeen-1980}.
The perturbed curvature variable in the comoving gauge
${\cal R}$ often used in the literature is the same as our 
$\varphi_v$ which is the same as $\varphi_{\delta \phi}$ in the scalar field.

\subsubsection{Spatial gauge conditions}
                                                    \label{sec:GT-spatial}

The spatial gauge transformations $\xi$ and $\xi^{({v})}_\alpha$
affect the scalar- and the vector-type perturbations, respectively.
Due to spatial homogeneity of the background we have natural
spatial gauge fixing conditions to choose \cite{Bardeen-1988}.
We have two natural spatial gauge fixing conditions.
{}From eq. (\ref{GT-decomposed-linear}) we can see that:
\bea
   & & {\rm B\!-\!gauge}: \quad 
       \beta \equiv 0 , \quad B_\alpha^{({v})} \equiv 0 
       \quad \rightarrow \quad \xi ({\bf x},t) \propto a, \quad
       \xi_\alpha^{({v})} ({\bf x}),
   \label{B-gauge} \\
   & & {\rm C\!-\!gauge}: \quad 
       \gamma \equiv 0, \quad  C_\alpha^{({v})} \equiv 0 
       \quad \rightarrow \quad \xi = 0, \quad
       \xi_\alpha^{({v})} = 0.
   \label{C-gauge}
\eea
{}For $\beta$ we have considered a situation where the temporal 
gauge condition already completely removed $\xi^t$.
We call the spatial gauge fixing conditions in 
eqs. (\ref{B-gauge},\ref{C-gauge}) the B-gauge and the C-gauge, respectively
\cite{Bianchi-pert}.
These gauge conditions are imposed so that we have
\bea
   & & {\rm B\!-\!gauge}: \quad B_\alpha \equiv 0,
   \nonumber \\
   & & {\rm C\!-\!gauge}: \quad 
        C_{\alpha\beta} \equiv \varphi g^{(3)}_{\alpha\beta}
        + C_{\alpha\beta}^{({t})}.
\eea
Apparently, the $B$-gauge conditions fail to fix the spatial and the rotational
gauge modes completely, thus, even after imposing the gauge conditions
we still have the remaining gauge modes.
Whereas, the $C$-gauge conditions successfully remove the gauge modes.
To the linear-order, the variables $\chi$ 
and $\Psi^{(v)}_\alpha$ introduced in eq. (\ref{chi-def})
are natural and unique spatially gauge-invariant combinations.
Notice that in the $C$-gauge $\chi$ is the same as $a \beta$,
and $\Psi^{(v)}$ is the same as $B^{(v)}_\alpha$.
Thus, $\beta$ and $B^{(v)}_\alpha$ variables in the $C$-gauge
conditions are equivalent to the corresponding (spatially and rotationally)
gauge-invariant combinations $\chi/a$ and $\Psi^{(v)}_\alpha$, respectively.

\subsection{Second-order}

\subsubsection{Gauge conditions}

If we use any of the gauge conditions which completely fixes both the temporal 
and the spatial gauge modes to the linear-order,
the gauge transformation properties of the second-order variables,
say $\varphi^{(2)}$ in eq. (\ref{pert-decompose}), 
follow {\it exactly the same forms} as their linear counterparts.
Using the transformation of $\delta \phi$ in eq. (\ref{GT-phi}) as an example,
to the linear order we have
\bea
   & & \delta \hat \phi^{(1)} = \delta \phi^{(1)} - \phi^\prime \xi^{0(1)}.
   \label{GT-phi-linear}
\eea
If we take gauge conditions which remove (fix) $\xi^0$ and $\xi^\alpha$
completely to the linear order we have $\xi^{0(1)} = 0 = \xi^{\alpha(1)}$.
Thus, from eq. (\ref{GT-phi}) we have
\bea
   & & \delta \hat \phi^{(2)} = \delta \phi^{(2)} - \phi^\prime \xi^{0(2)},
\eea
which shows exactly the same form as in eq. (\ref{GT-phi-linear}).
Thus, the gauge conditions in eqs. 
(\ref{GT-decomposed-linear},\ref{temporal-gauges-2}) 
apply to second-order perturbation variables as well, and
we can impose similar gauge conditions even to the second-order.
{}For example, in the zero-shear gauge we impose $\chi = 0$ as the
gauge condition to the second-order, thus $\chi^{(1)} = 0 = \chi^{(2)}$;
otherwise mentioned, we always take the $C$-gauge for the spatial 
and rotational ones.
In this gauge condition the gauge transforamtion properties are completely 
fixed, and the gauge-modes do not appear. 
Thus, we anticipate that each variable in that gauge condition
has unique corresponding gauge-invariant combination of variables.
Thus, using $\varphi$, we have $\varphi |_{\chi = 0, C-{\rm gauge}}$
is free of the gauge-modes.
We denote the corresponding gauge-invariant combination as 
\bea
   & & \varphi_\chi
\eea
with the $C$-gauge condition assumed always.
To the linear-order we have $\varphi_\chi \equiv \varphi - H \chi$,
but to the second-order we need correction terms to make $\varphi_\chi$
gauge-invariant.
Construction of such a gauge-invariant combination will be shown below. 

{}From eqs. (\ref{Extrinsic-curvature},\ref{Intrinsic-curvature}), 
assuming pure scalar mode in the $C$-gauge we can show
\bea
   \bar K_{\alpha\beta}
   &=& - \left( 1 - \alpha \right) \chi_{,\alpha|\beta}
       + 2 \chi_{,(\alpha} \varphi_{,\beta)}
       - {1 \over 3} g^{(3)}_{\alpha\beta} 
       \left[ - \left( 1 - \alpha \right) \Delta \chi
       + 2 \chi^{,\gamma} \varphi_{,\gamma} \right],
   \\
   R^{(h)}
   &=& {1 \over a^2} \left[ R^{(3)} 
       - 4 \left( \Delta + {1 \over 2} R^{(3)} \right) \varphi
       + 16 \varphi \left( \Delta + {1 \over 4} R^{(3)} \right) \varphi
       + 6 \varphi^{,\alpha} \varphi_{,\alpha} \right].
\eea 
Thus, the gauge condition $\chi \equiv 0$ implies $\bar K_{\alpha\beta} = 0$,
justifying its name as the zero-shear gauge to the second-order.
Similarly, the gauge condition $\varphi \equiv 0$ implies 
$R^{(h)} = {1 \over a^2} R^{(3)}$ (we also have 
$R^{(h)}_{\alpha\beta} = {1 \over 3} R^{(3)} g^{(3)}_{\alpha\beta}$),
justifying its name as the uniform-curvature gauge to the second-order.
We can show that the names of gauge conditions in 
eq. (\ref{temporal-gauges-1})
remain valid to the second-order. 

In the perturbative approach, apparently, this method
can be similarly applied to any higher-order perturbations. 
As long as we work in any of these gauge conditions, the gauge modes are
completely removed and the behavior of all the variables is 
equivalently gauge-invariant.
As the variables are free of gauge mode, these can be considered as 
physically important ones in that particular gauge conditions we choose.
We can also choose different gauge conditions in the second-order compared 
with the ones imposed to the linear-order.
Examples will be shown below.

\subsubsection{Constructing gauge-invariant combinations}

Let us explain a method to derive the gauge-invariant combinations 
using an example.
Since the gauge transformation properties of $\delta \mu$ and
$\delta \phi$ are available in a convenient forms in 
eqs. (\ref{GT-delta-mu},\ref{GT-phi}) we consider the gauge-invariant 
combinations involving these two variables to the second-order.
Thus, we consider the case with a scalar field.
To the linear-order we can construct various gauge invariant combinations
involving $\delta \mu$, and as examples we consider two cases
\bea
   & & \delta \mu_{\delta \phi} 
       \equiv \delta \mu - {\mu^\prime \over \phi^\prime} \delta \phi, \quad
       \delta \mu_\varphi
       \equiv \delta \mu - \phi^\prime {a \over a^\prime} \varphi.
   \label{GI-eg}
\eea
Apparently, the combinations in eqs. (\ref{GI-eg}) are not gauge-invariant 
to the second-order.
In order to construct the gauge-invariant combination in the
gauge with $\delta \phi^{(2)} = 0$ we construct 
$\delta \hat \mu - {\mu^\prime \over \phi^\prime} \delta \hat \phi$
using eqs. (\ref{GT-delta-mu},\ref{GT-phi}).
Then, in the RHS we have quadratic combination of linear-order terms
involving $\xi^0$ and $\xi^\alpha$.
As the spatial and rotational gauge we consider the
$C$-gauge conditions which remove the corresponding gauge modes completely.
This can be achieved by taking
\bea
   & & \xi^\alpha 
       = - \left( \hat \gamma^{,\alpha} + \hat C^{(v)\alpha} \right)
       + \gamma^{,\alpha} + C^{(v)\alpha},
\eea
which follows from eqs. 
(\ref{GT-decomposed-linear},\ref{xi-decomp}),
and moving terms with overhat to the LHS. 
Now, coming to the temporal gauge freedom, if we want to consider
the uniform-field gauge we take
\bea
   & & \xi^0 = - {1 \over \phi^\prime} 
       \left( \delta \hat \phi - \delta \phi \right),
\eea
which follows from eq. (\ref{GT-decomposed-linear}), 
and move terms with overhat to the LHS. 
Then we have a gauge-invariant combination
\bea
   \delta \mu_{\delta \phi}
   &\equiv&
       \delta \mu - {\mu^\prime \over \phi^\prime} \delta \phi
       - \left( \delta \mu - {\mu^\prime \over \phi^\prime} \delta \phi 
       \right)_{,\alpha} \left( \gamma^{,\alpha} + C^{(v)\alpha} \right)
       - {1 \over \phi^\prime} \left( \delta \mu 
       - {\mu^\prime \over \phi^\prime} \delta \phi \right)^\prime \delta \phi
       - {1 \over 2} \left( {\mu^\prime \over \phi^\prime} \right)^\prime
       {1 \over \phi^\prime} \delta \phi^2
       - {1 \over a^2} \delta \phi^{,\alpha} \delta \phi_{,\alpha}
   \nonumber \\
   &\equiv& 
       \delta \mu - {\mu^\prime \over \phi^\prime} \delta \phi
       + \delta \mu^{(q)}_{\delta \phi}.
   \label{GI-second1}
\eea
We have $\delta \mu_{\delta \phi} = \delta \mu 
\big|_{\delta \phi^{(1)} = 0 = \delta \phi^{(2)},C-{\rm gauge}}$,
thus $\delta \mu_{\delta \phi}$ is the same as
$\delta \mu$ under the gauge conditions
$\delta \phi^{(1)} = 0 = \delta \phi^{(2)}$ and the $C$-gauges.
If we want to take the uniform-curvature gauge to the linear-order
we take
\bea
   & & \xi^0 = - {a \over a^\prime} \left( \hat \varphi - \varphi \right)
\eea
which follows from eq. (\ref{GT-decomposed-linear}), 
and move terms with overhat to the LHS. 
Then we can identify a gauge-invariant combination
\bea
   & &
       \delta \mu
       - {\mu^\prime \over \phi^\prime} \delta \phi
       - \left( \delta \mu - {\mu^\prime \over \phi^\prime} \delta \phi 
       \right)_{,\alpha} \left( \gamma^{,\alpha} + C^{(v)\alpha} \right)
       - {a \over a^\prime} \left( \delta \mu 
       - {\mu^\prime \over \phi^\prime} \delta \phi \right)^\prime \varphi
   \nonumber \\
   & & \qquad
       - \left( {\mu^\prime \over \phi^\prime} \right)^\prime
       {a \over a^\prime} \left( \delta \phi
       - {1 \over 2} \phi^\prime {a \over a^\prime} \varphi \right) \varphi
       - 2 {1 \over a^2} \phi^\prime {a \over a^\prime}
       \left( \delta \phi 
       - {1 \over 2} \phi^\prime {a \over a^\prime} \varphi \right)^{,\alpha} 
       \varphi_{,\alpha}.
   \label{GI-second2}
\eea
This combination is equivalent to $\delta \mu$
in the following gauge conditions:
$\delta \phi = 0$ in the linear and the pure second-order part
[i.e., $\delta \mu^{(2)} - (\mu^\prime/\phi^\prime) \delta \phi^{(2)}$],
and $\varphi = 0$ in the quadratic parts, and the $C$-gauges.
By replacing the linear-order part of eq. (\ref{GI-second2})
with $\delta \mu^{(1)} - \phi^\prime (a / a^\prime) \varphi^{(1)}$
we can make another gauge-invariant combination
\bea
   \delta \mu_{\varphi^{(1)},\delta \phi^{(2)}}
   &\equiv&
       \delta \mu^{(1)} 
       - \phi^\prime {a \over a^\prime} \varphi^{(1)}
       + \delta \mu^{(2)} 
       - {\mu^\prime \over \phi^\prime} \delta \phi^{(2)}
       - \left( \delta \mu - {\mu^\prime \over \phi^\prime} \delta \phi 
       \right)_{,\alpha} \left( \gamma^{,\alpha} + C^{(v)\alpha} \right)
       - {a \over a^\prime} \left( \delta \mu 
       - {\mu^\prime \over \phi^\prime} \delta \phi \right)^\prime \varphi
   \nonumber \\
   & &
       - \left( {\mu^\prime \over \phi^\prime} \right)^\prime
       {a \over a^\prime} \left( \delta \phi
       - {1 \over 2} \phi^\prime {a \over a^\prime} \varphi \right) \varphi
       - 2 {1 \over a^2} \phi^\prime {a \over a^\prime}
       \left( \delta \phi 
       - {1 \over 2} \phi^\prime {a \over a^\prime} \varphi \right)^{,\alpha} 
       \varphi_{,\alpha},
   \label{GI-second3}
\eea
which is the same as
$\delta \mu \big|_{\varphi^{(1)} = 0 = \delta \phi^{(2)},C-{\rm gauge}}$.
The calculation becomes simpler if we take the $C$-gauge condition:
this sets $\gamma \equiv 0 \equiv C^{(v)}_\alpha$,
thus we can simply set $\xi_\alpha \equiv 0$ 
($\xi \equiv 0 \equiv \xi^{(v)}_\alpha$).
Similarly we can construct diverse combinations of the
gauge-invariant variables: several useful gauge-invariant combinations
will be presented in the next subsection.

In the following, as in eq. (\ref{GI-second1}), a
gauge-invariant notation, say, $\varphi_v$ indicates
a combination which is equivalent to $\varphi$ in the
comoving gauge ($v = 0$) to all orders, thus, $v^{(1)} = 0 = v^{(2)}$,
and in the $C$-gauge.
In order to denote gauge-invariant combinations valid to
the second-order, we introduce the following notation
\bea
   & & \varphi_v \equiv \varphi - a H v + \varphi^{(q)}_v, \quad
       \varphi_\chi \equiv \varphi - H \chi + \varphi_\chi^{(q)}, \quad
       \delta \mu_v \equiv \delta \mu - \dot \mu a v + \delta \mu^{(q)}_v, \quad
       v_\chi \equiv v - {1 \over a} \chi + v_\chi^{(q)},
\eea
etc., where the upper $(q)$ index indicates the quadratic combinations
of the linear-order terms.
In the following we always take the spatial $C$-gauge.
We note that $\delta \mu_v^{(q)}$ is the quadratic correction terms
to make $\delta \mu_v$ a gauge-invariant combination to the second-order;
thus it differs from, say, $\delta \mu^{(q)} - \dot \mu a v^{(q)}$.
As $\varphi_v$ is the same as $\varphi$ in the $v=0$ gauge, we have
$\varphi_v^{(q)}$ vanishes under the $v = 0$ gauge, i.e., 
\bea
   & & \varphi_v^{(q)} |_v = \varphi_v^{(q)} |_{v = 0} = 0.
   \label{GI-correction}
\eea
Using the definition of our gauge-invariant combinations we can show,
for example,
\bea
   \varphi_v = \varphi - a H v + \varphi^{(q)}_v
       = \varphi_\chi - a H v_\chi + \varphi^{(q)}_v|_\chi,
\eea 
where in the second step we have evaluated the first step in
the zero-shear gauge.
Thus
\bea
   & & \varphi^{(q)}_v|_\chi
       = \varphi_v - \left( \varphi_\chi - a H v_\chi \right)
       = \varphi^{(q)}_v - \left( \varphi_\chi^{(q)} - a H v_\chi^{(q)} \right),
\eea
and similarly for other correction terms.

\subsubsection{Gauge-invariant variables}

Now, we present several useful gauge-invariant combinations explicitly.
We assume $R^{(3)} = 0$ and pure scalar-type perturbations.
We take the spatial $C$-gauge, $\gamma \equiv 0$.
As long as we take the temporal gauge which fixes $\xi^0$ completely,
we can set $\xi_\alpha \equiv 0$.
The metric becomes
\bea
   A = \alpha, \quad
       B_\alpha = {1 \over a} \chi_{,\alpha}, \quad
       C_{\alpha\beta} = \varphi g^{(3)}_{\alpha\beta}.
\eea
{}From eqs. (\ref{GT-A}-\ref{GT-Pi},\ref{GT-phi}) we have
\bea
   \hat \alpha
   &=& \alpha
       - {1 \over a} \left( a \xi^0 \right)^\prime
       - \alpha^\prime \xi^0
       - 2 \alpha \left( \xi^{0\prime} + {a^\prime \over a} \xi^0 \right)
       + {3 \over 2} \xi^{0\prime} \xi^{0\prime}
       + \xi^0 \left[ \xi^{0 \prime \prime}
       + 3 {a^\prime \over a} \xi^{0\prime}
       + {1 \over 2} \left( {a^{\prime \prime} \over a}
       + {a^{\prime 2} \over a^2} \right) \xi^0 \right],
   \nonumber \\
   \hat \varphi
   &=& \varphi
       - {a^\prime \over a} \xi^0
       + \xi^0 \left[ - \varphi^\prime - 2 {a^\prime \over a} \varphi
       + {a^\prime \over a} \xi^{0\prime}
       + {1 \over 2} \left( {a^{\prime\prime} \over a}
       + {a^{\prime 2} \over a^2} \right) \xi^0 \right]
   \nonumber \\
   & &
       + {1 \over 2} \left( {1 \over a} \chi^{,\alpha} \xi^0_{\;\;,\alpha}
       - {1 \over 2} \xi^{0,\alpha} \xi^0_{\;\;,\alpha} \right)
       - {1 \over 2} \Delta^{-1} \nabla^\alpha \nabla^\beta
       \left( {1 \over a} \chi_{,\alpha} \xi^0_{\;\;,\beta}
       - {1 \over 2} \xi^0_{\;\;,\alpha} \xi^0_{\;\;,\beta} \right),
   \nonumber \\
   \hat \chi
   &=& \chi
       - a \xi^0
       + a \xi^0 \left( \xi^{0\prime} + {a^\prime \over a} \xi^0 \right)
       + a \Delta^{-1} \nabla^\alpha \left[
       - 2 \alpha \xi^0_{\;\;,\alpha}
       - {1 \over a} \left( \chi^\prime
       + {a^\prime \over a} \chi \right)_{,\alpha} \xi^0
       - {1 \over a} \chi_{,\alpha} \xi^{0\prime}
       + \xi^{0\prime} \xi^0_{\;\;,\alpha} \right]
   \nonumber \\
   & &
       - {a \over 2} \Delta^{-1} \left[
       {1 \over a} \chi^{,\alpha} \xi^0_{\;\;,\alpha}
       - {1 \over 2} \xi^{0,\alpha} \xi^0_{\;\;,\alpha}
       - 3 \Delta^{-1} \nabla^\alpha \nabla^\beta \left(
       {1 \over a} \chi_{,\alpha} \xi^0_{\;\;,\beta}
       - {1 \over 2} \xi^0_{\;\;,\alpha} \xi^0_{\;\;,\beta} \right)
       \right]^\prime,
   \nonumber \\
   \hat \kappa
   &=& \kappa
       + \left( 3 {a^{\prime\prime} \over a^2} - 6 {a^{\prime 2} \over a^3}
       + {\Delta \over a} \right) \xi^0
       + {\rm quadratic \; terms},
   \nonumber \\
   \delta \hat \mu
   &=& \delta \mu
       - \mu^\prime \xi^0 - \delta \mu^\prime \xi^0
       + \mu^\prime \xi^{0\prime} \xi^0
       + {1 \over 2} \mu^{\prime\prime} \xi^0 \xi^0
       + \left( \mu + p \right) \left( - 2 v + \xi^0 \right)^{,\alpha}
       \xi^0_{\;\;,\alpha},
   \nonumber \\
   \hat v
   &=& v
       - \xi^0 + \xi^0 \xi^{0\prime}
       + {1 \over 2} \left( {a^\prime \over a}
       + {\mu^\prime + p^\prime \over \mu + p} \right) \xi^0 \xi^0
       - \left[ v^\prime + \left( 4 {a^\prime \over a}
       + {\mu^\prime + p^\prime \over \mu + p} \right) v \right] \xi^0
   \nonumber \\
   & &
       - \Delta^{-1} \nabla^\alpha \left[
       - 3 {a^\prime \over a} v_{,\alpha} \xi^0
       + {\delta \mu \over \mu + p} \xi^0_{\;\;,\alpha}
       + {1 \over a^2} {1 \over \mu + p} \left( \Pi^{,\beta}_{\;\;\;\alpha}
       - \delta^\beta_\alpha \Delta \Pi \right) \xi^0_{\;\;,\beta} \right],
   \nonumber \\
   \delta \hat \phi
   &=& \delta \phi
       - \phi^\prime \xi^0 - \delta \phi^\prime \xi^0
       + \phi^\prime \xi^{0\prime} \xi^0
       + {1 \over 2} \phi^{\prime\prime} \xi^0 \xi^0.
   \label{GT-temporal}
\eea
In the transformation of $v$ we have used eq. (\ref{scalar-6}).
We have ignored writing the quadratic terms in the transformation of $\kappa$;
this can be read from the definition of $\kappa$
\bea
   \kappa
      \equiv - {1 \over a} \left[
       3 \varphi^\prime
       - 3 {a^\prime \over a} \alpha
       + {\Delta \chi \over a}
       - \left( \alpha + 2 \varphi \right)
       \left( 3 \varphi^\prime + {\Delta \chi \over a} \right)
       + {3 a^\prime \over 2 a} \left( 3 \alpha^2
       - {1 \over a^2} \chi^{,\alpha} \chi_{,\alpha} \right)
       - {1 \over a} \chi^{,\alpha} \varphi_{,\alpha} \right],
\eea
which follows from eqs. (\ref{deltaK-def},\ref{Extrinsic-curvature}).

{}Following the prescription in the previous subsection, 
from eq. (\ref{GT-temporal})
we can construct the following gauge-invariant combinations
\bea
   \varphi_\chi
   &\equiv&
       \varphi - H \chi
       - \left( \dot \varphi_\chi + 2 H \varphi_\chi \right) \chi
       - {1 \over 2} \left( \dot H + H^2 \right) \chi^2
       + {1 \over 4 a^2} \left[ \chi^{,\alpha} \chi_{,\alpha}
       - \Delta^{-1} \nabla^\alpha \nabla^\beta
       \left( \chi_{,\alpha} \chi_{,\beta} \right) \right]
   \nonumber \\
   & &
        + H \Delta^{-1} \nabla^\alpha \left[ 2 \alpha_\chi \chi_{,\alpha}
        + \left( \dot \chi - H \chi \right) \chi_{,\alpha} \right]
        + {1 \over 4} a^2 H \Delta^{-1} \left[
        {1 \over a^2} \chi^{,\alpha} \chi_{,\alpha}
        - 3 {1 \over a^2} \Delta^{-1} \nabla^\alpha \nabla^\beta
        \left( \chi_{,\alpha} \chi_{,\beta} \right) \right]^\cdot,
   \label{GI-varphi-chi} \\
   \varphi_v
   &\equiv&
       \varphi - a H v
       - \left( \dot \varphi_v + 2 H \varphi_v \right) a v
       - {1 \over 2} \left( \dot H + 2 H^2
       - H {\dot \mu + \dot p \over \mu + p} \right) a^2 v^2
       + {1 \over 4 a} \left( 2 \chi - a v \right)^{,\alpha} v_{,\alpha}
   \nonumber \\
   & &
       - {1 \over 4a} \Delta^{-1} \nabla^\alpha \nabla^\beta
       \left[ \left( 2 \chi - a v \right)_{,\alpha} v_{,\beta} \right]
       + a H \Delta^{-1} \nabla^\alpha \left[
       {\delta \mu_v \over \mu + p} v_{,\alpha}
       + {1 \over a^2} {1 \over \mu + p}
       \left( \Pi^{,\beta}_{\;\;\; \alpha} - \delta^\beta_\alpha \Delta \Pi
       \right) v_{,\beta} \right],
   \label{GI-varphi-v} \\
   \delta_v
   &\equiv&
       \delta - {\dot \mu \over \mu} a v
       - {\delta \dot \mu_v \over \mu} a v
       - {1 \over 2} {\dot \mu \over \mu} {\dot H \over H} a^2 v^2
       - {\mu + p \over \mu} v^{,\alpha} v_{,\alpha}
   \nonumber \\
   & &
       + {\dot \mu \over \mu} a \Delta^{-1} \nabla^\alpha \left[
       {\delta \mu_v \over \mu + p} v_{,\alpha}
        + {1 \over a^2} {1 \over \mu + p}
       \left( \Pi^{,\beta}_{\;\;\;\alpha} - \delta^\beta_\alpha \Delta \Pi
       \right) v_{,\beta} \right],
   \label{GI-delta-v} \\
   v_\chi
   &\equiv&
       v - {1 \over a} \chi
       - \left[ \dot v_\chi + \left( H + {\dot \mu + \dot p \over \mu + p}
       \right) v_\chi \right] \chi
       + {1 \over 2 a} \left( H - {\dot \mu + \dot p \over \mu + p} \right)
       \chi^2
   \nonumber \\
   & &
       + {1 \over a} \Delta^{-1} \nabla^\alpha \left[
       2 \alpha_\chi \chi_{,\alpha}
       + \chi_{,\alpha} \left( \dot \chi - H \chi \right)
       - {\delta \mu_v \over \mu + p} \chi_{,\alpha}
       - {1 \over a^2} {1 \over \mu + p} \left( \Pi^{,\beta}_{\;\;\;\alpha}
       - \delta^\beta_\alpha \Delta \Pi \right) \chi_{,\beta} \right]
   \nonumber \\
   & &
       + {a \over 4} \Delta^{-1} \left[
       {1 \over a^2} \chi^{,\alpha} \chi_{,\alpha}
       - {3 \over a^2} \Delta^{-1} \nabla^\alpha \nabla^\beta
       \left( \chi_{,\alpha} \chi_{,\beta} \right) \right]^\cdot,
   \label{GI-v-chi} \\
   \chi_v
   &\equiv&
       \chi - a v
       - {1 \over 2} a^2 \left( H - {\dot \mu + \dot p \over \mu + p} \right)
       v^2
   \nonumber \\
   & &
       + a \Delta^{-1} \nabla^\alpha \left[ - 2 \alpha_v v_{,\alpha}
       - \left( \dot \chi_v + H \chi_v \right)_{,\alpha} v
       - \chi_{,\alpha} \dot v
       +  {\delta \mu_v \over \mu + p} v_{,\alpha}
       + {1 \over a^2} {1 \over \mu + p} \left( \Pi^{,\beta}_{\;\;\;\alpha}
       - \delta^\beta_\alpha \Delta \Pi \right) v_{,\beta} \right]
   \nonumber \\
   & &
       + {1 \over 2} a^2 \Delta^{-1} \left[
       - \left( {1 \over a} \chi^{,\alpha} v_{,\alpha}
       - {1 \over 2} v^{,\alpha} v_{,\alpha} \right)
       + 3 \Delta^{-1} \nabla^\alpha \nabla^\beta
       \left( {1 \over a} \chi_{,\alpha} v_{,\beta}
       - {1 \over 2} v_{,\alpha} v_{,\beta} \right) \right]^\cdot.
   \label{GI-chi-v}
\eea
To the linear-order, $\delta_v$ (equivalently, $\delta$ in the comoving gauge)
behaves like the Newtonian density perturbation,
and $v_\chi$ and $- \varphi_\chi$ (equivalently, $v$ and $- \varphi$
in the zero-shear gauge) behave like the Newtonian 
velocity and the gravitational potential.
Also to the linear-order $\varphi_v$ is known to be the best conserved
quantity in the super-sound-horizon scale.
{}For extensions of these results to the second-order,
see \S \ref{sec:Pressureless} and \ref{sec:pure-scalar}, respectively.

In the case of a scalar field we have
\bea
   \varphi_{\delta \phi}
   &\equiv&
       \varphi - {H \over \dot \phi} \delta \phi
       - {1 \over a^2} \left( a^2 \varphi_{\delta \phi} \right)^\cdot
       {\delta \phi \over \dot \phi}
       - {1 \over 2 a^2} \left( {a^2 H \over \dot \phi} \right)^\cdot
       {\delta \phi^2 \over \dot \phi}
   \nonumber \\
   & &
       + {1 \over 2 a^2 \dot \phi} \left[
       \chi^{,\alpha} \delta \phi_{,\alpha}
       - {1 \over 2 \dot \phi} \delta \phi^{,\alpha} \delta \phi_{,\alpha}
       - \Delta^{-1} \nabla^\alpha \nabla^\beta \left(
       \chi_{,\alpha} \delta \phi_{,\beta}
       - {1 \over 2 \dot \phi} \delta \phi_{,\alpha} \delta \phi_{,\beta}
       \right) \right],
   \label{GI-varphi-delta-phi} \\
   \delta \phi_\varphi
   &\equiv&
       \delta \phi - {\dot \phi \over H} \varphi
       - {1 \over H} \delta \dot \phi_\varphi \varphi
       + {\dot \phi^2 \over 2 a^2 H^3} \left( {a^2 H \over \dot \phi}
       \right)^\cdot \varphi^2
   \nonumber \\
   & &
       - {\dot \phi \over 2 a^2 H^2} \left[
       \chi^{,\alpha} \varphi_{,\alpha}
       - {1 \over 2H} \varphi^{,\alpha} \varphi_{,\alpha}
       - \Delta^{-1} \nabla^\alpha \nabla^\beta \left(
       \chi_{,\alpha} \varphi_{,\beta}
       - {1 \over 2 H} \varphi_{,\alpha} \varphi_{,\beta} \right) \right].
   \label{GI-delta-phi-varphi}
\eea
Thus, to the second-order we have
$\varphi_{\delta \phi} \neq - {H \over \dot \phi} \delta \phi_\varphi$.
By evaluating the RHSs of
eqs. (\ref{GI-varphi-delta-phi},\ref{GI-delta-phi-varphi})
in the $\varphi = 0$ gauge and $\delta \phi = 0$ gauge, respectively, we have
the following relations between the two gauge-invariant variables
\bea
   \varphi_{\delta \phi}
   &=&
       - {H \over \dot \phi} \delta \phi_\varphi
       + {H \over \dot \phi^2} \delta \phi_\varphi \delta \dot \phi_\varphi
       + {1 \over 2 a^2 \dot \phi} \left( {a^2 H \over \dot \phi} \right)^\cdot
       \delta \phi_\varphi^2
   \nonumber \\
   & &
       + {1 \over 2 a^2 \dot \phi} \left[
       \chi_\varphi^{\;\;,\alpha} \delta \phi_{\varphi,\alpha}
       - {1 \over 2 \dot \phi} \delta \phi_\varphi^{\;\;,\alpha}
       \delta \phi_{\varphi,\alpha}
       - \Delta^{-1} \nabla^\alpha \nabla^\beta \left(
       \chi_{\varphi,\alpha} \delta \phi_{\varphi,\beta}
       - {1 \over 2 \dot \phi} \delta \phi_{\varphi,\alpha}
       \delta \phi_{\varphi,\beta} \right) \right],
   \label{GT-varphi-phi} \\
   \delta \phi_\varphi
   &=&
       - {\dot \phi \over H} \varphi_{\delta \phi}
       + {\dot \phi \over H^2} \varphi_{\delta \phi} \dot \varphi_{\delta \phi}
       + {1 \over 2 a^2 H} \left( {a^2 \dot \phi \over H}
       \right)^\cdot \varphi_{\delta \phi}^2
   \nonumber \\
   & &
       - {\dot \phi \over 2 a^2 H^2} \left[
       \chi_{\delta \phi}^{\;\;\;,\alpha} \varphi_{\delta \phi,\alpha}
       - {1 \over 2H} \varphi_{\delta \phi}^{\;\;\;,\alpha}
       \varphi_{\delta \phi,\alpha}
       - \Delta^{-1} \nabla^\alpha \nabla^\beta \left(
       \chi_{\delta \phi,\alpha} \varphi_{\delta \phi,\beta}
       - {1 \over 2 H} \varphi_{\delta \phi,\alpha} \varphi_{\delta \phi,\beta}
       \right) \right].
   \label{GT-phi-varphi}
\eea
To linear-order $\delta \phi_\varphi$ (equivalently, $\delta \phi$ in the
uniform-curvature gauge) most closely resembles the scalar-field
equation in the fixed cosmological background metric \cite{Hwang-QFT}.
Since $\delta \phi \equiv 0$ implies $Q_\alpha^{(\phi)} = 0$
(thus $v^{(\phi)} = 0$), we have $\varphi_{\delta \phi} = \varphi_v$;
from eq. (\ref{MSF-fluid}) we notice that this is valid to the second-order.

\subsubsection{Spatial gradient variable}

A covariant density gradient variable 
\bea
   & & \tilde \Delta_a \equiv {1 \over \tilde \mu} \tilde h^b_a \tilde \mu_{,b},
\eea
is gauge-invariant to the linear-order \cite{EB}.
In \cite{EB} the energy-frame is taken.
Using the $\tilde u_a$-frame in eq. (\ref{u-def}) we have
\bea
   \Delta_\alpha
   &=&
       {1 \over 1 + \delta} \left\{
       \delta^E_{,\alpha} 
       + {\mu^\prime \over \mu} \left( V^E_\alpha - B_\alpha 
       + A B_\alpha + 2 V^{E\beta} C_{\alpha\beta} \right)
       + \left[ \delta^\prime 
       + {\mu^\prime \over \mu} \left( \delta - A \right) \right]
       \left( V^E_\alpha - B_\alpha \right)
       \right\},
   \label{Delta-E}
\eea
where we set $\tilde \Delta_\alpha \equiv \Delta_\alpha$ with
$\Delta_\alpha$ based on $g^{(3)}_{\alpha\beta}$;
from $\tilde u^a \tilde \Delta_a = 0$ 
we have $\tilde \Delta_0 = - V^\alpha \Delta_\alpha$.
{}From eqs. (\ref{frame-inv-def-N},\ref{frame-inv-def-E}) 
we notice that $\delta$ is frame invariant to the linear-order, and
we ignore superscript $E$ in such cases.
Using the prescription in eq. (\ref{frame-E-N})
we can express $\Delta_\alpha$ in the normal-frame as
\bea
   \Delta_\alpha
   &=&
       {1 \over 1 + \delta} \Bigg\{
       \delta^N_{,\alpha} 
       + {\mu^\prime \over \mu} {Q^N_\alpha \over \mu + p}
       + \left[ \delta^\prime + {\mu^\prime \over \mu} 
       \left( \delta - A \right) \right] {Q^N_\alpha \over \mu + p}
       - {( Q^{N\beta} Q^N_\beta )_{,\alpha} \over \mu (\mu + p)}
       - {\mu^\prime \over \mu} {1 \over (\mu + p)^2} \left[
       \left( \delta \mu + \delta p \right) Q^N_\alpha
       + \Pi_{\alpha\beta} Q^{N\beta} \right] \Bigg\}.
   \nonumber \\
   \label{Delta-N}
\eea
Thus, under the condition $Q_\alpha = 0$ we have
$\Delta_\alpha = \delta_{,\alpha}/(1 + \delta)$;
$Q_\alpha = 0$ can be achieved by the comoving gauge condition
($Q \equiv 0$) and the irrotational condition ($Q^{(v)}_\alpha \equiv 0$);
see \cite{EBH} and the note added in proof of \cite{HV-1990}.
Under the gauge transformation, 
either using eqs. (\ref{GT-delta-mu}-\ref{GT-Pi}) for eq. (\ref{Delta-N})
or using eq. (\ref{GT-E-frame}) for eq. (\ref{Delta-E}) we can show
\bea
   & & \hat \Delta_\alpha
       = \Delta_\alpha - \Delta^\prime_\alpha \xi^0
       - \left( \Delta_\beta \xi^\beta \right)_{,\alpha}
       - 2 \Delta_{[\alpha,\beta]} \xi^\beta.  
\eea
Thus, $\Delta_\alpha$ is not gauge-invariant to the second-order.
To the linear-order, the scalar-type part becomes
$\Delta_\alpha = \delta_{v,\alpha}$ where the gauge-invariant combination
$\delta_v$ is the same as $\delta$ in the comoving gauge.

\section{Applications}
                                         \label{sec:Applications}

\subsection{Closed form equations}
                                         \label{sec:closed-forms}

{}From eqs. (\ref{scalar-1},\ref{scalar-2}),
eqs. (\ref{scalar-2},\ref{scalar-5},\ref{scalar-6}),
eqs. (\ref{scalar-4}), 
eqs. (\ref{scalar-4},\ref{scalar-6}) 
and
eqs. (\ref{scalar-0},\ref{scalar-2},\ref{scalar-4})
we can derive, respectively:
\bea
   & & {\Delta + 3 K \over a^2} \varphi_\chi + 4 \pi G \delta \mu_v
       = {\Delta + 3 K \over a^2} \varphi_\chi^{(q)} 
       + 4 \pi G \delta \mu_v^{(q)}
       + {1 \over 4} N_1 - H N^{(s)}_2,
   \label{eqs-1} \\
   & & \delta \dot \mu_v + 3 H \delta \mu_v
       - {\Delta + 3 K \over a^2} \left[ a (\mu + p) v_\chi + 2 H \Pi \right]
       = \delta \dot \mu_v^{(q)} + 3 H \delta \mu_v^{(q)}
       - {\Delta + 3 K \over a^2} a (\mu + p) v_\chi^{(q)}
   \nonumber \\
   & & \qquad
       + N_5 + ( \mu + p) \left( N^{(s)}_2 + 3 aH N^{(s)}_6 \right),
   \label{eqs-2} \\
   & & \varphi_\chi + \alpha_\chi + 8 \pi G \Pi
       = \varphi_\chi^{(q)} + \alpha_\chi^{(q)}
       - N^{(s)}_4, 
       \quad {\rm or} \quad
       \varphi_\chi + \alpha_\chi + 8 \pi G \Pi_\chi
       = - N^{(s)}_{4\chi},
   \label{eqs-3} \\ 
   & & \dot v_\chi + H v_\chi 
       - {1 \over a} \left( \alpha_\chi
       + {\delta p_v \over \mu + p} 
       + {2 \over 3} {\Delta + 3 K \over a^2} {\Pi \over \mu + p} \right)
       = \dot v_\chi^{(q)} + H v_\chi^{(q)} 
       - {1 \over a} \left( \alpha_\chi^{(q)}
       + {\delta p_v^{(q)} \over \mu + p} \right)
       + N^{(s)}_6,
   \label{eqs-4} \\
   & & \dot \varphi_\chi + H \varphi_\chi
       + 4 \pi G ( \mu + p) a v_\chi + 8 \pi G H \Pi
       = \dot \varphi_\chi^{(q)} + H \varphi_\chi^{(q)}
       + 4 \pi G ( \mu + p) a v_\chi^{(q)}
       + {1 \over 3} \left( N_0 - N^{(s)}_2 \right) - H N^{(s)}_4.
   \label{eqs-5} 
\eea
These equations are presented using mixed gauge-invariant variables.
We note that to the linear-order, $\delta \mu_v$, $- \varphi_\chi$,
and $k v_\chi (\sim \nabla v_\chi)$ closely resemble the Newtonian 
density perturbation, perturbed gravitational potential, and the 
perturbed velocity perturbation, respectively 
\cite{Harrison-1967,Nariai-1969,Hwang-Noh-Newtonian}.
To the linear-order these equations were presented by Bardeen in 1980
\cite{Bardeen-1980}; see eqs. (4.3,4.8,4.4,4.5,4.7) in \cite{Bardeen-1980},
and compared with our notation, see eq. (\ref{Bardeen-notation}).
Using eq. (\ref{eqs-5}) and eqs. (\ref{eqs-1},\ref{eqs-4},\ref{eqs-5}) 
we can show
\bea
   \Phi
   &\equiv& \varphi_v - {K / a^2 \over 4 \pi G ( \mu + p)} \varphi_\chi
   \nonumber \\
   &=& {H^2 \over 4 \pi G (\mu + p) a}  
       \left[ {a \over H} \left( \varphi_\chi - \varphi_\chi^{(q)} \right)
       \right]^\cdot
       + 2 H^2 {\Pi \over \mu + p} 
       + \Phi^{(q)} 
       + N_\Phi,
   \label{Phi-eq} \\
   \dot \Phi
   &=& {H c_s^2 \Delta \over 4 \pi G (\mu + p) a^2} 
       \left( \varphi_\chi - \varphi_\chi^{(q)} \right)
       - {H \over \mu + p} \left( e + {2 \over 3} {\Delta \over a^2} \Pi \right)
       + \dot \Phi^{(q)}
       + N_{\dot \Phi},
   \label{dot-Phi-eq}
\eea
where 
\bea
   \Phi^{(q)}
   &\equiv& \varphi_v^{(q)} 
       - {K / a^2 \over 4 \pi G ( \mu + p)} \varphi_\chi^{(q)},
   \nonumber \\
   N_\Phi 
   &\equiv& {H^2 \over 4 \pi G (\mu + p)} \left[  
       N^{(s)}_4 + {1 \over 3H} \left( N^{(s)}_2 - N_0 \right) \right],
   \nonumber \\
   N_{\dot \Phi} 
   &\equiv& {1 \over 3} \left( 1 - {K/a^2 \over 4 \pi G (\mu + p)} \right)
       \left( N_0 - N^{(s)}_2 \right)
       - {H c_s^2 \over 4 \pi G ( \mu + p)} \left( {1 \over 4} N_1
       - H N^{(s)}_2 \right)
       + {K/a^2 \over 4 \pi G (\mu + p)} H N^{(s)}_4
       - aH N^{(s)}_6.
   \label{N_dot-Phi}
\eea
We have introduced an entropic perturbation $e$ by
\bea
   & & \delta p \equiv c_s^2 \delta \mu + e, \quad
       c_s^2 \equiv {\dot p \over \dot \mu}.
   \label{e-def}
\eea
Defined in this way, $e$ is not necessarily gauge-invariant to the second-order.
To the linear order we have $e = \tilde \pi$ introduced above 
eq. (\ref{entropy}).
Combining eqs. (\ref{Phi-eq},\ref{dot-Phi-eq}) we can derive
\bea
   & & {H^2 c_s^2 \over (\mu + p) a^3} \left[ {(\mu + p) a^3 \over H^2 c_s^2}
       \dot \Phi \right]^\cdot
       - c_s^2 {\Delta \over a^2} \Phi
       = {H c_s \over a^3 \sqrt{\mu + p}} \left[ v^{\prime\prime}
       - \left( {z^{\prime\prime} \over z} + c_s^2 \Delta \right) v \right]
   \nonumber \\
   & & \qquad
       = {H^2 c_s^2 \over (\mu + p) a^3} \left\{ 
       {(\mu + p) a^3 \over H^2 c_s^2} \left[ 
       - {H \over \mu + p} 
       \left( e + {2 \over 3} {\Delta \over a^2} \Pi \right) 
       + \dot \Phi^{(q)}
       + N_{\dot \Phi} \right] \right\}^\cdot
      - c_s^2 {\Delta \over a^2} 
      \left( 2 H^2 {\Pi \over \mu + p} + \Phi^{(q)} + N_\Phi \right),
   \label{ddot-Phi-eq} \\
   & & {\mu + p \over H} \left[ {H^2 \over (\mu + p) a}
       \left( {a \over H} \varphi_\chi \right)^\cdot \right]^\cdot
       - c_s^2 {\Delta \over a^2} \varphi_\chi
       = {\sqrt{\mu + p} \over a^2} \left[ u^{\prime\prime}
       - \left( {(1/\bar z)^{\prime\prime} \over 1/\bar z} + c_s^2 \Delta 
       \right) u \right]
   \nonumber \\
   & & \qquad
       = {4 \pi G (\mu + p) \over H} \left[
       - {H \over \mu + p} \left( e + {2 \over 3} {\Delta \over a^2} \Pi \right)
       - 2 \left( H^2 {\Pi \over \mu + p} \right)^\cdot
       + N_{\dot \Phi} - \dot N_\Phi \right]
   \nonumber \\
   & & \qquad \qquad
       + {\mu + p \over H} \left[ {H^2 \over (\mu + p) a}
       \left( {a \over H} \varphi_\chi^{(q)} \right)^\cdot \right]^\cdot
       - c_s^2 {\Delta \over a^2} \varphi_\chi^{(q)},
   \label{ddot-varphi_chi-eq}
\eea
where we used
\bea
   & & v \equiv z \Phi, \quad
       u \equiv {1 \over \bar z} {a \over H} \varphi_\chi, \quad
       c_s z \equiv {a \sqrt{\mu + p} \over H} \equiv \bar z.
   \label{u-v-def}
\eea
[$v$ in eqs. (\ref{ddot-Phi-eq}-\ref{u-v-def}) differs from the
 perturbed velocity related variable used in the rest of this paper.]
The equation using $v$ in the linear theory was first derived
by Field and Shepley in 1968 \cite{Field-Shepley-1968},
see also \cite{action,HV-1990}.
Using eq. (\ref{eqs-1}), eq. (\ref{ddot-varphi_chi-eq}) gives
an equation for $\delta_v$.
Using eqs. (\ref{eqs-1}-\ref{eqs-4}) 
we can derive equation for $\delta_v$ in {\it another} form
\bea
   & & {\mu + p \over a^2 \mu H} \left[ {H^2 \over ( \mu + p) a}
       \left( {a^3 \mu \over H} \delta_v \right)^\cdot \right]^\cdot
       - c_s^2 {\Delta \over a^2} \delta_v
   \nonumber \\ 
   & & \qquad
       = {\Delta + 3 K \over a^2} \left[ {e \over \mu}
       + {2 \over 3} {\Delta \over a^2} {\Pi \over \mu}
       + 2 {\mu + p \over \mu H} 
       \left( {H^2 \over \mu + p} \Pi \right)^\cdot \right]   
       + {\mu + p \over a^2 \mu H} \left[ {H^2 \over ( \mu + p) a}
       \left( {a^3 \mu \over H} \delta_v^{(q)} \right)^\cdot \right]^\cdot
       - c_s^2 {\Delta \over a^2} \delta_v^{(q)}
   \nonumber \\ 
   & & \qquad \qquad
       + {\mu + p \over \mu} \Bigg\{ - {1 \over 4} N_1
       + H N^{(s)}_2
       + {\Delta + 3 K \over a^2} \left( a N^{(s)}_6 - N^{(s)}_4 \right)
       + {1 \over a^2} \left[ a^2 \left( N^{(s)}_2 
       + {N_5 \over \mu + p} + 3 a H N^{(s)}_6 \right) \right]^\cdot
       \Bigg\}.
   \label{delta_v-eq}
\eea

The above set of equations is valid for a general imperfect fluid.
A minimally coupled scalar field can be regarded as an imperfect
fluid with special fluid quantities. 
We additionally have an equation of motion of the field which is 
actually included in the energy and momentum conservation equations.
In fact, the above set of equations is valid even for multi-component 
fluids and fields.
In such cases, the fluid quantities become collective fluid quantities
and we additionally need the energy and momentum conservation equations
for individual fluid and the equations of motion for the individual field.

In the single scalar field case, from eqs. 
(\ref{fluid-MSFs-decomp},\ref{fluid-MSF},\ref{eqs-1},\ref{e-def}) we have
\bea
   & & e = - {1 - c_s^2 \over 4 \pi G} {\Delta + 3 K \over a^2} 
       \left( \varphi_\chi - \varphi_\chi^{(q)} \right)
       + N_e,
   \nonumber \\
   & & N_e \equiv
       {1 - c_s^2 \over 4 \pi G} \left( {1 \over 4} N_1 
       - H N^{(s)}_2 \right)
       + \delta p^{(q)} - \delta \mu^{(q)}
       + 3H ( 1- c_s^2) a \Delta^{-1} \nabla^\alpha Q^{(q)}_\alpha.
\eea
Equation (\ref{Phi-eq}) remains valid, and eq. (\ref{dot-Phi-eq}) becomes
\bea
   & & \dot \Phi = {H c_A^2 \Delta \over 4 \pi G (\mu + p) a^2} 
       \left( \varphi_\chi - \varphi_\chi^{(q)} \right)
       - {H \over \mu + p} \left( N_e 
       + {2 \over 3} {\Delta \over a^2} \Pi \right)
       + \dot \Phi^{(q)}
       + N_{\dot \Phi},
   \label{dot-Phi-MSF-eq}
\eea
where 
\bea
   & & c_A^2 \Delta \equiv \Delta + 3 ( 1 - c_s^2 ) K.
   \label{c_A-def}
\eea
Therefore, eqs. (\ref{ddot-Phi-eq},\ref{ddot-varphi_chi-eq},\ref{u-v-def})
remain valid with $c_s$ and $e$ replaced by $c_A$ and $N_e$;
in eq. (\ref{ddot-Phi-eq}) one can show that we can ignore 
the operator nature of $\Delta^{-1}$ in the $c_A^2$.

The rotational perturbation and the gravitational wave are
described by eqs. (\ref{vector-6},\ref{tensor-4}), respectively
\bea
   & & {[a^4 (\mu + p) v^{(v)}_\alpha]^\cdot \over a^4 (\mu + p)}
       = - {\Delta + 2K \over 2 a^2} {\Pi^{(v)}_\alpha \over \mu + p}
       + N_{6\alpha}^{({v})},
   \label{rotation-pert} \\
   & & \ddot C^{({t})}_{\alpha\beta} 
       + 3 H \dot C^{({t})}_{\alpha\beta}
       - {\Delta - 2K \over a^2}
       C^{({t})}_{\alpha\beta} 
       = {1 \over a^3} \left[ v^{({t})\prime\prime}_{\alpha\beta}
       - \left( {a^{\prime\prime} \over a} + \Delta - 2 K \right) 
       v^{({t})}_{\alpha\beta} \right]
       = 8 \pi G \Pi^{({t})}_{\alpha\beta}
       + N_{4\alpha\beta}^{(t)},
   \label{GW-pert} 
\eea
where 
$v^{({t})}_{\alpha\beta} \equiv a C^{({t})}_{\alpha\beta}$.
We note that all the equations above in this section are valid for 
{\it general} $K$.

\subsection{Solutions to the linear-order}
                                          \label{sec:linear-sols}

\subsubsection{Scalar-type}

We consider a single component ideal fluid.
Several known solutions in the literature are the followings.

\noindent 
(i)
In the large-scale limit (the super-sound-horizon scale), i.e., ignoring
$c_s^2 \Delta$ term compared with $z^{\prime\prime}/z$ and
$(1/\bar z)^{\prime\prime}/(1/\bar z)$ terms in 
eqs. (\ref{ddot-Phi-eq},\ref{ddot-varphi_chi-eq}), we have general solutions
\footnote{
       ${\bf k}$ is the wave vector with $k \equiv |{\bf k}|$.
       With the wave number $k$ appearing in the equation the variables
       can be regarded as the Fourier transformed ones.
       To the {\it linear-order} each Fourier mode decouple from the
       other modes and evolves independently.
       The same equations in configuration space
       remain valid in Fourier space as well.
       Thus, we ignore specific symbols distinguishing the variables
       in the two spaces.
       Only in this subsection concerning the linear theory
       we use the Fourier transformation.
       }
\bea
   \Phi (k, t) 
   &=& C (k) - d (k) {k^2 \over 4 \pi G} \int^t
       {c_s^2 H^2 \over a^3 (\mu + p)} dt,
   \label{Phi-sol} \\
   \varphi_\chi (k, t) 
   &=& 4 \pi G C (k) {H \over a} \int^t
       {a (\mu + p) \over H^2} dt + d (k) {H \over a}.
   \label{varphi_chi-sol}
\eea
$C(k)$ and $d(k)$ are the coefficients of the growing and decaying solutions
(in an expanding medium), respectively.
To the second-order in the large-scale expansion we have
\bea
   \Phi 
   &=& C \left\{ 1 + k^2 \left[ \int^\eta \bar z^2 \left( \int^\eta
       {d \eta \over z^2} \right) d \eta
       - \int^\eta \bar z^2 d \eta \int^\eta {d \eta \over z^2} \right]
       \right\}
       - d {k^2 \over 4 \pi G} \int^\eta {d \eta \over z^2},
   \label{Phi-sol-2} \\
   \varphi_\chi 
   &=& 4 \pi G C {H \over a} \int^\eta \bar z^2 d \eta
       + d {H \over a} \left\{ 1 + k^2 \left[ \int^\eta {1 \over z^2}
       \left( \int^\eta \bar z^2 d \eta \right) d \eta
       - \int^\eta \bar z^2 d \eta \int^\eta {d \eta \over z^2} \right]
       \right\}.
   \label{varphi_chi-sol-2}
\eea
We emphasize that these solutions are valid for {\it general} $K$ and $\Lambda$,
and time-varying equation of state.

\noindent 
(ii)
In the small-scale limits [$c_s^2 k^2 \gg z^{\prime\prime}/z, \;
(1/\bar z)^{\prime\prime}/(1/\bar z)$], if we further assume 
that $c_s$ is constant in time, 
eqs. (\ref{ddot-Phi-eq},\ref{ddot-varphi_chi-eq}) give
the general solutions:
\bea
   & & v = z \Phi \propto e^{\pm i c_s k \eta}, \quad
       u = {1 \over \sqrt{\mu + p}} \varphi_\chi \propto e^{\pm i c_s k \eta}.
\eea

\noindent
(iii)
{}For $K = 0 = \Lambda$ and $w \equiv p/\mu = {\rm constant}$ 
we have exact solutions \cite{Bardeen-1980,Hwang-Fluid}.
{}For the background, from eqs. (\ref{BG1},\ref{BG3}), we have
\bea
   & & a \propto t^{2 \over 3 (1 + w)} \propto \eta^{2 \over 1 + 3w}, \quad
       a H \eta = {2 \over 1 + 3 w}.
   \label{BG-sol}
\eea
Equation (\ref{u-v-def}) gives $z \propto \bar z \propto a$, thus
\bea
   {z^{\prime\prime} \over z} = {2 (1-3w) \over (1+3w)^2} {1 \over \eta^2},
       \quad
       {(1/\bar z)^{\prime\prime} \over (1/\bar z)}
       = {6 (1+w) \over (1+3w)^2} {1 \over \eta^2}.
\eea
In this case, eqs. (\ref{ddot-Phi-eq},\ref{ddot-varphi_chi-eq}) 
become Bessel's equations with solutions:
\bea
   & & v = z \Phi \propto \sqrt{\eta} \left( J_\nu (x), Y_\nu(x) \right), \quad
       x \equiv c_s k |\eta|, \quad
       \nu \equiv {3 (1-w) \over 2(1+3w)},
   \\
   & & u = {1 \over \sqrt{\mu + p}} \varphi_\chi
       \propto \sqrt{\eta} \left( J_{\bar \nu} (x), Y_{\bar \nu} (x) \right),
       \quad
       \bar \nu \equiv {5 + 3w \over 2(1+3w)}.
\eea
We have $\bar \nu = \nu + 1$.
Using eqs. (\ref{Phi-eq},\ref{dot-Phi-eq}) we can normalize the solutions as
\bea
   \Phi
   &\equiv& c_1 (k) {J_\nu (x) \over x^\nu} + c_2 (k) {Y_\nu (x) \over x^\nu},
   \\
   \varphi_\chi
   &=& {3 ( 1 + w) \over 1 + 3w}
       \left( c_1 (k) {J_{\bar \nu} (x) \over x^{\bar \nu}}
       + c_2 (k) {Y_{\bar \nu} (x) \over x^{\bar \nu}} \right).
\eea
Equation (\ref{eqs-1}) gives
\bea
   \delta_v = {(1+3w)^2 \over 6w} x^2 \varphi_\chi.
\eea
In the large-scale limit ($x \ll 1$) we have
\bea
   & & \Phi = {c_1 \over 2^\nu \Gamma(\nu + 1)}
       - 2^\nu {\Gamma(\nu) \over \pi} x^{-2\nu} c_2,
\eea
where for $\nu = 0$ we have additional $2 \ln{x}$ factor in the $c_2$-mode.
By matching with the general large-scale solution in eq. (\ref{Phi-sol})
we can identify
\bea
   c_1 = 2^\nu \Gamma (\nu + 1) C, \quad
       c_2 = - {1 \over 3 (1+w)} {\pi \over 2^\nu \Gamma (\bar \nu)}
       {x^{2 \bar \nu} \over a^2 \eta} d.
\eea

In the large-scale limit ($x \ll 1$) we have
\bea
   & & \Phi \quad \propto \quad C, \; d a^{-{3 \over 2} (1-w)},
   \label{Phi-LS-sol} \\
   & & \varphi_\chi \quad \propto \quad C, \; d a^{-{5 + 3w \over 2}},
   \label{varphi_chi-LS-sol} \\
   & & \delta_v \quad \propto \quad Ca^{1+3w}, \; d a^{-{3 \over 2} (1-w)}
       \quad \propto \quad C t^{2(1+3w) \over 3(1+w)}, \; d t^{-{1-w \over 1+w}}
       \quad \propto \quad C \eta^2, \; d \eta^{- {3 (1-w) \over 1+ 3w}}.
   \label{delta_v-LS-sol}
\eea
Equation (\ref{delta_v-LS-sol}) follows from eq. (\ref{eqs-1}) which gives
$\delta_v \propto a^{1+3w} \varphi_\chi \propto \eta^2 \varphi_\chi$
in general.
Equation (\ref{delta_v-LS-sol}) includes the well known
solutions in the matter ($w= 0$) and radiation ($w = {1 \over 3}$) eras
\cite{Lifshitz-1946}:
\bea
   & & {\rm mde}: \quad
       \delta_v \quad \propto \quad C a, \; d a^{-{3 \over 2}}
       \quad \propto \quad C t^{2 \over 3}, \; d t^{-1}
       \quad \propto \quad C \eta^2, \; d \eta^{-3},
   \nonumber \\
   & & {\rm rde}: \quad \;
       \delta_v \quad \propto \quad C a^2, \; d a^{-1}
       \quad \propto \quad C t, \; d t^{-{1 \over 2}}
       \quad \propto \quad C \eta^2, \; d \eta^{-1}.
\eea
If we consider only the $C$-mode which is the relatively growing-mode
in an expanding phase, we have
\bea
   & & \Phi ({\bf x}, t) = C ({\bf x}),
   \\
   & & \varphi_\chi ({\bf x}, t) = {3 + 3w \over 5 + 3w} C({\bf x}).
   \label{varphi_chi-LS-sol-C}
\eea
The nontransient mode of $\Phi$ remains constant in the
super-sound-horizon scale.
Whereas, the one of $\varphi_\chi$ jumps as the background
equation of state changes.
Still, it is $- \varphi_\chi$, not $- \varphi_v$, which closely resembles 
the perturbed Newtonian gravitational potential \cite{Hwang-Noh-Newtonian}.

The asymptotic solutions (i) and (ii) remain valid 
for the scalar field with $K = 0$; in this case we have 
$c_s^2$ replaced by $1$.
The background solutions for $w = {\rm constant}$ case considered in (iii),
eq. (\ref{BG-sol}), are valid for a scalar field with exponential potential
of the following form \cite{Lucchin-Matarrese-1985}
\bea
   & & V = {1 - w \over 12 \pi G ( 1 + w)^2} e^{-\sqrt{24 \pi G (1+w)} \phi},
       \quad
       \phi = \sqrt{1 \over 6 \pi G ( 1 + w )} \ln{t}.
\eea
{}For the perturbation, from eq. (\ref{c_A-def}) and the prescription 
below it, the same equations of the fluid remain valid with
the coefficient of Laplacian term replaced by $1$ (for the field) 
instead of $c_s^2$ (for the fluid) \cite{Hwang-QFT}.
Thus, the perturbation solutions in the fluid remain valid for such 
the scalar field with $x = k|\eta|$, instead of $x = c_s k |\eta|$
for the fluid case.

We emphasize that, if a solution is known in a given gauge condition
the rest of the variables in all gauge conditions can be derived
through simple algebra; such solutions are presented in tabular forms
for an ideal fluid and a scalar field in 
\cite{Hwang-Fluid,Hwang-MDE,Hwang-MSF}.
Solutions in the situations of generalized gravity theories 
considered in \S \ref{sec:GGT} can be found in \cite{Hwang-Noh-GGT}. 
The cases with multiple components of the fluids and fields
were analysed in \cite{KS2,HN-multiple}.

\subsubsection{Vector-type}

The rotational perturbation is described by eq. (\ref{rotation-pert}).
If we assume $\Pi_\alpha^{(v)} = 0$, we have a general solution
\bea
   & & a \cdot a^3 (\mu + p) \cdot v_\alpha^{(v)} ({\bf x}, t) 
       = L_\alpha^{(v)} ({\bf x}).
   \label{rotation-sol}
\eea
Thus, to the linear-order the rotational perturbation is described by 
the conservation of angular momentum, and is transient in expanding media.
We note that eq. (\ref{rotation-pert}) follows from the
$T^b_{\alpha;b} = 0$, thus {\it independent} of the
gravitational field equation.
Thus the presence of scalar fields or the generalized gravity theories
considered in \S \ref{sec:GGT} do not affect the vector-type
perturbation of the fluids \cite{Hwang-GGT-1990}.

\subsubsection{Tensor-type}
                                             \label{sec:tensor-sol-linear}

Now, we consider the gravitational wave
with $K = 0$ and $\Pi^{({t})}_{\alpha\beta} = 0$.
The basic equation is presented in eq. (\ref{GW-pert}).

\noindent 
(i)
The general large-scale ($k^2 \ll a^{\prime\prime}/a$) solution is
\bea
   & & C_{\alpha\beta}^{(t)} (k, t) 
       = c_{\alpha\beta}^{(t)} (k) 
       + d_{\alpha\beta}^{(t)} (k) \int^t {dt \over a^3}.
   \label{GW-sol-LS}
\eea
Thus, ignoring the transient mode ($d_{\alpha\beta}$) in an expanding phase
the tensor-type perturbation is characterized by its conserved amplitude
$c_{\alpha\beta}^{(t)} (k)$.

\noindent 
(ii)
In the small-scale limit ($k^2 \gg a^{\prime\prime}/a$) we have a
general solution 
\bea
   & & C_{\alpha\beta}^{(t)} (k, t) \propto {1 \over a} e^{\pm i k \eta}.
   \label{GW-sol-SS}
\eea
Thus the gravitational wave redshifts away.

\noindent 
(iii)
{}For $K = 0 = \Lambda$ and $w = {\rm constant}$, we have an exact solution:
\bea
   & & v_{\alpha\beta}^{(t)} 
       = a C_{\alpha\beta}^{(t)} \propto \sqrt{\eta} 
       \left( J_\nu (x), Y_\nu (x) \right), \quad
       \nu \equiv {3(1-w) \over 2(1+3w)}, \quad
       x \equiv k |\eta|.
\eea
Thus, we set
\bea
   & & C_{\alpha\beta}^{(t)} = c_{1\alpha\beta}^{(t)} {J_\nu (x) \over x^\nu}
       + c_{2\alpha\beta}^{(t)} {Y_\nu(x) \over x^\nu}.
   \label{GW-sol-w}
\eea
In the large-scale limit ($x \ll 1$) we have
\bea
   & & C_{\alpha\beta}^{(t)} 
       = {1 \over 2^\nu \Gamma(\nu + 1)} c^{(t)}_{1\alpha\beta} 
       - 2^\nu {\Gamma(\nu) \over \pi} x^{-2\nu} c_{2\alpha\beta}^{(t)},
\eea
where for $\nu = 0$ we have additional $2 \ln{x}$ factor in the
$c_{2\alpha\beta}^{(t)}$-mode.
By matching with the general large-scale solution in
eq. (\ref{GW-sol-LS}) we can identify
\bea
   & & c_{1\alpha\beta}^{(t)} 
       = 2^\nu \Gamma (\nu + 1) c_{\alpha\beta}^{(t)}, \quad
       c_{2\alpha\beta}^{(t)} = {\pi \over 2^{\nu+1} \Gamma (\nu+1)}
       {\eta x^{2 \nu} \over a^2} d_{\alpha\beta}^{(t)}.
\eea
Thus,
\bea
   & & C_{\alpha\beta}^{(t)} 
       \quad \propto \quad
       c_{\alpha\beta}^{(t)}, 
       \; d_{\alpha\beta}^{(t)} a^{-{3 \over 2}(1-w)}
       \quad \propto \quad
       c_{\alpha\beta}^{(t)}, 
       \; d_{\alpha\beta}^{(t)} t^{-{1-w \over 1+w}}
       \quad \propto \quad
       c_{\alpha\beta}^{(t)}, 
       \; d_{\alpha\beta}^{(t)} \eta^{-{3(1-w) \over 1+3w}}.
   \label{GW-LS-sol}
\eea

\subsection{Pressureless irrotational fluid}
                                               \label{sec:Pressureless}

The equation of $\delta_v$ was derived in eq. (\ref{delta_v-eq}) 
or can be derived from eqs. (\ref{ddot-varphi_chi-eq},\ref{eqs-4}).
In the pressureless case, a simpler route is to use
the basic equations in eqs. (\ref{scalar-0}-\ref{scalar-6}).

We {\it consider} a pressureless fluid, 
thus $\delta p = 0 = \Pi_{\alpha\beta}$,
and {\it ignore} the vector-type perturbation. 
{}For the spatial gauge we take $\gamma \equiv 0$, thus $\beta = \chi/a$.
If we take the temporal comoving gauge ($v \equiv 0$) we have $Q_\alpha = 0$.
Equation (\ref{scalar-6}) gives 
$\alpha_v = - a N_{6v}^{(s)} 
 = - {1 \over 2} \beta_{v,\alpha} \beta_v^{\;\;,\alpha}$, 
thus $\alpha$ vanishes to the linear-order;
in the pressureless medium, to the linear-order,
the comoving particle follows geodesic, thus $v= 0$ implies $\alpha = 0$.
{}From eqs. (\ref{scalar-5},\ref{scalar-3}), first evaluating these in the 
comoving gauge, we can derive
\bea
   & & \dot \delta_v = \kappa_v 
       + {1 \over a} \nabla \cdot \left( \delta_v \nabla v_\chi \right),
   \label{dot-delta_v-eq} \\
   & & \dot \kappa_v + 2 H \kappa_v - 4 \pi G \mu \delta_v
       = {1 \over a^2} \left[ \left( \nabla v_\chi \right) \cdot
       \left( \nabla v_\chi \right)^{,\alpha} \right]_{|\alpha} 
       + \dot C^{(t)}_{\alpha\beta} \left( \dot C^{(t)\alpha\beta} 
       + {2 \over a^2} \chi_v^{\;,\alpha|\beta} \right),
   \label{dot-kappa_v-eq} 
\eea
where we have used $\chi_v = - a v_\chi$ following from eq. (\ref{GI}), and
$\kappa_v = {1 \over a} \Delta v_\chi$ following
from eq. (\ref{scalar-2}) with $K = 0$, both to the linear-order. 

In order to compare with Newtonian analysis we introduce 
${\bf u} = - \nabla v_\chi$ to the linear-order. 
By combining eqs. (\ref{dot-delta_v-eq},\ref{dot-kappa_v-eq}) we have
\bea
   & & \ddot \delta_v + 2 H \dot \delta_v - 4 \pi G \bar \mu \delta_v
       = - {1 \over a^2} \left[ a \nabla \cdot \left( \delta_v {\bf u}
       \right) \right]^\cdot
       + {1 \over a^2} \nabla \cdot \left( {\bf u} \cdot
       \nabla {\bf u} \right)
       + \dot C^{(t)}_{\alpha\beta} 
       \left( {2 \over a} \nabla^\alpha u^{\beta} 
       + \dot C^{(t)\alpha\beta} \right).
   \label{ddot-delta-eq-second}
\eea
We note that, to the linear-order, the growing solution of the gravitational 
wave remains constant in time in the super-horizon scale, whereas it redshifts
away ($C^{(t)}_{\alpha\beta} \propto a^{-1}$) in the sub-horizon scale;
see \S \ref{sec:tensor-sol-linear}.
Ignoring the gravitational wave we reproduce correctly
the corresponding Newtonian equation 
\bea
   & & \ddot \delta + 2 H \dot \delta - 4 \pi G \bar \varrho \delta
       = - {1 \over a^2} \left[ a \nabla \cdot \left( \delta {\bf u}
       \right) \right]^\cdot
       + {1 \over a^2} \nabla \cdot \left( {\bf u} \cdot
       \nabla {\bf u} \right).
   \label{ddot-delta-eq-N}
\eea
We note that our eq. (\ref{ddot-delta-eq-second}) is valid in the
super-sound-horizon (Jeans scale) which is negligible in the
pressureless medium, thus valid even in the super-(visual)-horizon.
In the Newtonian context eq. (\ref{ddot-delta-eq-N})
is {\it valid to all orders} in perturbation, and
follows from the mass conservation, the momentum conservation, and 
the Poissons' equation given as \cite{Peebles-1980}
\bea
   & & \dot \delta + {1 \over a} \nabla \cdot {\bf u}
       = - {1 \over a} \nabla \cdot \left( \delta {\bf u} \right),
   \label{dot-delta-eq-N} \\
   & & \dot {\bf u} + H {\bf u} + {1 \over a} \nabla \delta \Phi
       = - {1 \over a} {\bf u} \cdot \nabla {\bf u},
   \label{dot-delta-v-eq-N} \\
   & & {1\over a^2} \nabla^2 \delta \Phi 
       = 4 \pi G \bar \varrho \delta.
   \label{perturbed-Poisson-eq-N}
\eea
To the linear-order these equations can be compared with
the relativistic version in eqs. (\ref{eqs-2},\ref{eqs-4},\ref{eqs-1})
with eq. (\ref{eqs-3}).
To the second-order, however, 
although the final result in eq. (\ref{ddot-delta-eq-second}) coincides with 
the Newtonian one in eq. (\ref{ddot-delta-eq-N}),
we notice some difference between 
eqs. (\ref{dot-delta_v-eq},\ref{dot-kappa_v-eq},\ref{eqs-2}-\ref{eqs-4})
with eq. (\ref{eqs-3})
and eqs. (\ref{dot-delta-eq-N}-\ref{perturbed-Poisson-eq-N}).
{}From eq. (\ref{scalar-2}), to the second-order, we have
\bea
   & & \kappa_v - {\Delta \over a} v_\chi
       = N_{2v}^{(s)}
       - {\Delta \over a^2} \left( \chi_v^{(q)} + a v_\chi^{(q)} \right).
\eea
Since the RHSs of this equation and of eq. (\ref{eqs-1}) do not
vanish we cannot directly relate $-\varphi_\chi$ and
$-\nabla v_\chi$ (or $a \Delta^{-1} \nabla \kappa_v$) to the Newtonian 
counterparts $\delta \Phi$ and ${\bf u}$, respectively.
Still, we emphasize that the final equation identified
in eq. (\ref{ddot-delta-eq-second}) coincides exactly with 
the Newtonian one in eq. (\ref{ddot-delta-eq-N}).
{}For a similar conlusion in the relativistic situation, 
see \cite{Kofman-Pogosyan-1995}.
{}For analyses of eq. (\ref{ddot-delta-eq-N})
in the Newtonian context, see \cite{quasilinear}.

Using $\beta = 0$ as the spatial gauge condition Kasai \cite{Kasai-1992}
has derived a different equation compared with ours in 
eq. (\ref{ddot-delta-eq-second}).
Kasai \cite{Kasai-1992} took both the comoving gauge 
$v \equiv 0$ and the original synchronous gauge which takes 
$\alpha = 0 = \beta$.  
As we have shown above eq. (\ref{dot-delta_v-eq}) 
in a pressureless medium, the comoving gauge $v = 0$ implies   
$\alpha = - {1 \over 2} \beta_{,\alpha} \beta^{,\alpha}$,
thus vanishes if we take $\beta = 0$ as the spatial gauge condition.
However, in that gauge condition (the spatial $B$-gauge) the spatial 
gauge-mode is incompletely fixed.
Thus, comparison with the Newtonian analyses is not tranparent 
in that gauge condition.

\subsubsection{Nonlinear equation based on $3+1$ formulation}

The general equation of the pressureless and irrotational ideal fluid
can be derived from eqs. 
(\ref{Trace-prop},\ref{E-conservation},\ref{Mom-conservation}).
The pressureless ideal fluid implies $S_{\alpha\beta} = 0$.
We take the temporal comoving gauge condition, $v = 0$.
Together with the irrotational condition we have $Q_\alpha = 0$,
thus $J_\alpha = 0$.
Equation (\ref{Mom-conservation}) gives $N_{,\alpha} = 0$; if we use
the normalization in  eq. (\ref{ADM-metric-pert}), we have $N = a$.
Equations (\ref{Trace-prop},\ref{E-conservation}) give
\bea
   & & K = {\dot E \over E} - {1 \over N} {E_{,\alpha} \over E} N^\alpha,
   \label{K-eq} \\
   & & \dot K - {1 \over N} K_{,\beta} N^\beta - {1 \over 3} K^2
       = 4 \pi G E - \Lambda + \bar K^{\alpha\beta} \bar K_{\alpha\beta}.
   \label{dot-K-eq}
\eea
Apparently, the spatial $B$-gauge condition, $B_\alpha \equiv 0$,
leads to $N_\alpha = 0$, thus simplifying the equations.
However, such a gauge condition leaves
the spatial gauge-mode removed incompletely. 
We prefer to take the spatial $C$-gauge condition
which fixes the spatial gauge modes completely, in this way 
the analyses can be equivalently considered as spatially gauge-invariant ones.
{}From eqs. (\ref{K-eq},\ref{dot-K-eq}) we can derive 
eq. (\ref{ddot-delta-eq-second}) to the second-order.
We notice that, in contrast with the Newtonian case, in general we 
anticipate to have infinite perturbation series expansion, 
and eq. (\ref{ddot-delta-eq-second}) looks valid only to the second-order.
If we have the higher-order terms nonvanishing, these can be
regarded as purely relativistic effects.

\subsubsection{Nonlinear equation based on $1+3$ formulation}

Assuming pressureless condition, eqs. 
(\ref{cov-E-conserv},\ref{cov-Mom-conserv},\ref{Raychaudhury-eq})
in the energy-frame ($\tilde q_a = 0$) become:
\bea
   & & {\tilde {\dot {\tilde \mu}}} + \tilde \theta \tilde \mu = 0,
   \label{E-conservation-cov} \\
   & & \tilde a_a = 0,
   \label{Mom-conservation-cov} \\
   & & {\tilde {\dot {\tilde \theta}}}
       + {1 \over 3} {\tilde \theta}^2 + 4 \pi G \tilde \mu - \Lambda
       + 2 \left( \tilde \sigma^2 - \tilde \omega^2 \right) = 0.
   \label{Raychaudhuri}
\eea
In the energy-frame the frame-vector follows the possible energy flux,
thus, the energy flux term $\tilde q_a$ vanishes.
Equations (\ref{E-conservation-cov}) and (\ref{Raychaudhuri}) can be combined
to give
\bea
   & & \left( {{\tilde {\dot {\tilde \mu}}} \over \tilde \mu} 
       \right)^{\tilde \cdot} - {1 \over 3}
       \left( {{\tilde {\dot {\tilde \mu}}} \over \tilde \mu} \right)^2 
       - 4 \pi G \tilde \mu + \Lambda 
       - 2 \left( \tilde \sigma^2 - \tilde \omega^2 \right) = 0.
   \label{ddot-mu}
\eea
If we set $\tilde \mu \equiv \mu ( 1 + \delta )$, where $\mu$ is
the background energy density, equation (\ref{ddot-mu}) becomes
\bea
   & & {\tilde {\ddot \delta}} 
       - {2 \over 3} {{\tilde {\dot \mu}} \over \mu} 
       {\tilde {\dot \delta}}
       - 4 \pi G \mu \left( 1 + \delta \right) \delta
       - {4 \over 3} {{\tilde {\dot \delta}}^2 \over 1 + \delta}
       - 2 \left( \tilde \sigma^2 - \tilde \omega^2 \right) 
       \left( 1+ \delta \right)
       + \left( 1 + \delta \right) \left[
       \left( {{\tilde {\dot \mu}} \over \mu} \right)^\cdot
       - {1 \over 3} \left( {\tilde {\dot \mu} \over \mu} \right)^2
       - 4 \pi G \mu + \Lambda \right] = 0.
   \label{ddot-delta-1}
\eea
This is a completely nonlinear equation.

Now, we {\it assume} an irrotational fluid.
In the energy-frame, the comoving gauge condition leads to 
$\tilde u_\alpha \equiv 0$; this is equivalent to taking the normal-frame 
with vanishing energy flux.
The momentum conservation equation in eq. (\ref{Mom-conservation-cov})
implies that our frame-vector follows geodesic path.
In the comoving gauge eqs. (\ref{Mom-conservation-cov},\ref{a-normal}) 
lead to $A = - {1 \over 2} B^\alpha B_\alpha$ to the second-order.
Thus, to the linear-order we have $A = 0$ which coincides with taking
the synchronous gauge.
We have 
\bea
   & & {\tilde {\dot {\tilde \mu}}}
       = \tilde \mu_{,a} \tilde u^a
       = \partial_t \tilde \mu + {1\over a} \tilde \mu_{,\alpha} B^\alpha; \quad
       {\tilde {\dot \mu}} = \partial_t \mu, \quad
       {\tilde {\dot \delta}} = \partial_t \delta 
       + {1\over a} \delta_{,\alpha} B^\alpha.
   \label{dot-mu}
\eea
Only in the comoving gauge condition the covariant derivative 
along $\tilde u^a$ simplifies to the second-order as in eq. (\ref{dot-mu}).
In this derivation the pressureless condition is used essentially.
Thus, in this comoving gauge equation (\ref{ddot-delta-1}) becomes 
\bea
   \ddot \delta + 2 H \dot \delta - 4 \pi G \mu \delta
       = {1\over a^2} \left( a^2 \delta \dot \delta \right)^\cdot
       + {1\over 3} \dot \delta^2
       - {1\over a^2} \left( a \delta_{,\alpha} B^\alpha \right)^\cdot
       - {1\over a} \dot \delta_{,\alpha} B^\alpha
       + 2 \tilde \sigma^2, 
   \label{ddot-delta-2}
\eea
to the second-order.
{}From this we can derive eq. (\ref{ddot-delta-eq-second}) where
$\tilde \sigma^2$ follows from eqs. 
(\ref{kinematic-cov-3},\ref{Extrinsic-curvature},\ref{shear-pert-normal}).

\subsubsection{Gravitational wave as a source}

{}From eq. (\ref{ddot-delta-eq-second}) considering the
pure gravitational wave as the source for density perturbation we have
\bea
   & & \ddot \delta_v + 2 H \dot \delta_v - 4 \pi G \mu \delta_v
       = \dot C^{(t)}_{\alpha\beta} \dot C^{(t)\alpha\beta} \equiv S.
   \label{GW-source}
\eea
In the matter dominated era an exact solution is given in
eq. (\ref{GW-sol-w}) with $\nu = {3 \over 2}$;
in the large-scale limit, considering the relatively growing mode,
$S$ vanishes, whereas, in the small scale limit it decays proportional to
$1/(at)^2 \propto a^{-5}$, see eq. (\ref{GW-sol-SS}). 
If $\delta_g$ and $\delta_d$ denote two linear-order solutions,
the general solution can be written as
\bea
   & & \delta_v ({\bf x},t)
       = \delta_g ({\bf x},t) + \delta_d ({\bf x},t) +
       \int^t S ({\bf x}, t^\prime)
       { \delta_g ({\bf x}, t^\prime) \delta_d ({\bf x}, t)
       - \delta_d ({\bf x}, t^\prime) \delta_g ({\bf x}, t) \over
       \delta_g ({\bf x}, t^\prime) \dot \delta_d ({\bf x}, t^\prime)
       - \delta_d ({\bf x}, t^\prime) \dot \delta_g ({\bf x}, t^\prime) }
       d t^\prime.
   \label{general-sol-delta}
\eea
The particular solution is proportional to $a^{-5} t^2 \propto a^{-2}$,  
and decays more rapidly even compared with the decaying mode in the 
linear theory which behaves as $t^{-1}$.

\subsection{Pure scalar-type perturbation}
                                                   \label{sec:pure-scalar}

Equation (\ref{ddot-Phi-eq}) can be written as
\bea
   & & {H^2 c_s^2 \over (\mu + p) a^3} \left\{ {(\mu + p) a^3 \over H^2 c_s^2}
       \left[ \dot \Phi - \dot \Phi^{(q)}
       - N_{\dot \Phi} 
       + {H \over \mu + p} \left( e 
       + {2 \over 3} {\Delta \over a^2} \Pi \right) 
       \right] \right\}^\cdot
       = c_s^2 {\Delta \over a^2} \left( \Phi - \Phi^{(q)}
       - 2 H^2 {\Pi \over \mu + p} - N_\Phi \right).
   \label{ddot-Phi-eq2} 
\eea
In the large-scale limit, if we could ignore the second-order 
spatial derivative terms, we have
\bea
   & & \dot \Phi - \dot \Phi^{(q)} - N_{\dot \Phi} + {H \over \mu + p} e
       \propto {H^2 c_s^2 \over (\mu + p) a^3}.
   \label{LS-sol2}
\eea
Now, we consider $K = 0$ and ideal fluids, thus $e = 0$.
{}From eq. (\ref{Phi-eq}) we have $\Phi = \varphi_v$.
Ignoring the second-order spatial derivatives, from eqs. 
(\ref{N_dot-Phi},\ref{scalar-0}-\ref{scalar-6},\ref{pert-eq0}-\ref{pert-eq6}) 
we can show that $N_{\dot \Phi} = ( \varphi_v^2 )^\cdot$. 
In the comoving gauge we have $\Phi^{(q)} |_v = \varphi_v^{(q)} |_v = 0$,
see eq. (\ref{GI-correction}).
Thus, we have
\bea
   & & \varphi_v - \varphi_v^2 
       = C ({\bf x}) + d ({\bf x}) \int^t {H^2 c_s^2 \over (\mu + p) a^3} dt. 
   \label{LS-sol3}
\eea
Therefore, ignoring the transient mode in the expanding phase, we have
\bea
   & & \varphi_v - \varphi_v^2 = C ({\bf x}),
   \label{LS-sol4}
\eea
which remains constant even to the second-order in perturbations
\cite{Salopek-Bond-1990,Maldacena-2002}. 
Thus, $\varphi_v$ is conserved to the second-order 
in the large-scale (super-sound-horizon) limit.

Equation (\ref{ddot-Phi-eq2}) is valid for $p \neq 0$.
{}For $p = 0$ we have a simpler form in eq. (\ref{dot-Phi-eq}) which gives
\bea
   & & \dot \Phi - \dot \Phi^{(q)} - N_{\dot \Phi}
       + {H \over \mu} \left( e 
       + {2 \over 3} {\Delta \over a^2} \Pi \right) = 0.
   \label{dot-Phi-eq2}
\eea
Equation (\ref{LS-sol2}) includes this as a case in the large-scale 
(super-sound-horizon) limit.
Thus, the above results in eqs. (\ref{LS-sol2}-\ref{LS-sol4}) 
remain valid for general $p$.

{}From eq. (\ref{MSF-fluid}) we notice that for a minimally coupled scalar 
field $\delta \phi = 0$ implies $v = 0$ to the second-order, thus
\bea
   & & \varphi_v = \varphi_{\delta \phi},
   \label{CG-UFG}
\eea
and the uniform-field gauge coincides with the comoving gauge.
Thus, the above analyses are valid even for a minimally coupled scalar field.

\subsection{Pure rotation}
                                                   \label{sec:pure-vector}

In the case of pure rotation eqs. (\ref{vector-6},\ref{pert-eq6})
provide a complete set for a single component fluid; 
for the multi-component case see eqs. (\ref{vector-6-i},\ref{pert-eq6i}).
Assuming $\Pi^{(v)}_\alpha = 0$ we have
\bea
   & & {[a^4 (\mu + p) v^{(v)}_\alpha]^\cdot \over a^4 (\mu + p)}
       = N_{6\alpha}^{({v})},
   \nonumber \\
   & & N_{6\alpha}^{({v})} 
       = - {1 \over a} \left[ v^{(v)}_{\alpha|\beta} B^{(v)\beta}
       + v^{(v)}_\beta B^{(v)\beta}_{\;\;\;\;\;\;|\alpha}
       - \nabla_\alpha \Delta^{-1} \nabla^\beta \left(
       v^{(v)}_{\beta|\gamma} B^{(v)\gamma}
       + v^{(v)}_\gamma B^{(v)\gamma}_{\;\;\;\;\;\;\;|\beta} \right) \right].
   \label{pure-vector} 
\eea
As a simple exercise, using eq. (\ref{GT-decomposed}),
one can check the gauge transformation properties of both sides.
In the $C$-gauge condition ($C^{(v)}_\alpha \equiv 0$) we have
$B^{(v)}_\alpha = \Psi^{(v)}_\alpha$ which is gauge-invariant.
{}From eqs. (\ref{rotation-sol},\ref{vector-2}), to the linear-order
we have 
$\Psi_\alpha^{(v)} \propto a^2 (\mu + p) v^{(v)}_\alpha \propto a^{-2}$,
thus from eq. (\ref{pure-vector}) we have 
\bea
   & & \left[ a^4 ( \mu + p) v^{(v)}_\alpha \right]^\cdot 
       = a^{-3} \left[ a^3 \cdot a^4 \left( \mu + p \right) 
       N_{6\alpha}^{(v)} \right]
       \propto a^{-3}.
   \label{AM-2nd-eq}
\eea
Thus, the additional second-order perturbation sourced by
the RHS of eq. (\ref{pure-vector}) behaves as
\bea
   & & a^4 (\mu + p) v^{(v)}_\alpha 
       = L_\alpha^{(v)} ({\bf x}) 
       + \left[ a^3 \cdot a^4 \left( \mu + p \right) 
       N_{6\alpha}^{(v)} \right] \int^t {dt \over a^3}.
   \label{AM-2nd-sol}
\eea
The time dependent nonlinear solution is proportional
to $\int^t dt/a^3$; for $w = {\rm constant}$ it is proportional to
$a^{-3 (1 - w)/2}$, thus it always decays (in expanding phase) for $w<1$. 
The lower bound of integration which could give a temporally constant 
nonlinear solution can be absorbed to $L_\alpha^{(v)}$.

As we explained in \S \ref{sec:GT-spatial}, to the linear-order, 
the $C$-gauge condition removes the rotational gauge-mode completely,
whereas the $B$-gauge condition fails to fix it completely.
That is, even after imposing the gauge condition we have some modes
which are coordinate effect;
under the $B$-gauge, from eq. (\ref{GT-decomposed-linear}) we have 
$\xi_\alpha^{(v)} = \xi_\alpha^{(v)} ({\bf x})$.
Then, in eq. (\ref{pure-vector}) we notice an ironic situation 
where the $B$-gauge condition gives vanishing quadratic terms, whereas 
these terms do not vanish in the $C$-gauge condition.
That is, although we anticipate the nonlinear solution in the $C$-gauge
in eq. (\ref{AM-2nd-sol}) is physical, in the $B$-gauge condition the 
RHS of eq. (\ref{AM-2nd-eq}) vanishes, and we do not have the 
nonlinear solution in eq. (\ref{AM-2nd-sol}).
We can check this situation by using the gauge transformation
property of $v_\alpha^{(v)}$ variable in the two gauge conditions.

Considering pure vector-type perturbation, from 
eqs. (\ref{metric-decomp-def},\ref{fluid-decomp},\ref{GT-Q},\ref{xi-decomp}) 
we have
\bea
   & & \hat v_\alpha^{(v)}
       = v_\alpha^{(v)}
       - v_\beta^{(v)} \xi^{(v)\beta}_{\;\;\;\;\;\;\;,\alpha}
       - v_{\alpha,\beta}^{(v)} \xi^{(v)\beta}
       + \nabla_\alpha \Delta^{-1} \nabla^\beta \left(
       v_\gamma^{(v)} \xi^{(v)\gamma}_{\;\;\;\;\;\;\;,\beta}
       + v_{\beta,\gamma}^{(v)} \xi^{(v)\gamma} \right).
   \label{GT-v-vector}
\eea
Now, let the variables with hat and no-hat correspond
to the ones in the $B$- and the $C$-gauge conditions, respectively.
As the $\xi^{(v)}_\alpha$s appear in quadratic combination,
we need it only to the linear-order.
{}From eq. (\ref{GT-decomposed-linear}), we have
$\hat B_\alpha^{({v})} = B_\alpha^{({v})} + a \dot \xi_\alpha^{({v})}$.
Since the hat indicates the $B$ gauge, we have
$a \dot \xi_\alpha^{({v})} = - B_\alpha^{({v})}$,
thus $\xi_\alpha^{({v})} = - \int^t (B_\alpha^{({v})}/a) dt$.
Thus, eq. (\ref{GT-v-vector}) gives
\bea
   & & a^4 (\mu + p) \hat v^{(v)}_\alpha \Big|_{B-{\rm gauge}}
       = \left\{ a^4 (\mu + p) v^{(v)}_\alpha 
       - \left[ a^3 \cdot a^4 \left( \mu + p \right) 
       N_{6\alpha}^{(v)} \right] \int^t {dt \over a^3} \right\} 
       \Bigg|_{C-{\rm gauge}}.
\eea
Therefore, in the $B$-gauge the nonlinear solution in
the $C$-gauge in eq. (\ref{AM-2nd-sol}) disappears exactly.
We note, however, that the solution in eq. (\ref{AM-2nd-sol})
is physical (gauge-invariant) one in the $C$-gauge.

\subsection{Pure gravitational wave}
                                                   \label{sec:pure-tensor}

In the case of pure gravitational wave
eqs. (\ref{tensor-4},\ref{pert-eq4}) provide a complete set.
In the large-scale limit, thus ignoring second-order spatial derivative terms,
and assuming $K = 0$ and $\Pi^{(t)}_{\alpha\beta} = 0$, we have
\bea
   & & \ddot C^{({t})\alpha}_{\;\;\;\;\;\beta} 
       + 3 H \dot C^{({t})\alpha}_{\;\;\;\;\;\beta}
       = N_{4\;\;\;\beta}^{(t)\alpha},
   \nonumber \\ 
   & & N_{4\;\;\;\beta}^{(t)\alpha} 
       = N_{4\beta}^{\;\;\alpha}
       - \Delta^{-1} \left( \nabla^{\alpha} \nabla_\gamma
       N_{4\beta}^{\;\;\gamma}
       + \nabla_{\beta} \nabla^\gamma
       N_{4\gamma}^{\;\;\alpha} \right)
       + {1 \over 2} \left( \Delta^{-1} 
       \nabla^\alpha \nabla_\beta + \delta^\alpha_\beta \right)
       \Delta^{-1} \nabla^\gamma \nabla_\delta N_{4\gamma}^{\;\;\delta},
   \nonumber \\ 
   & & N_{4\beta}^{\;\;\alpha} = 2 \left( 
       \dot C^{({t})\alpha\gamma} \dot C^{({t})}_{\beta\gamma}
       - {1 \over 3} \delta^\alpha_\beta
       \dot C^{({t})\gamma\delta} \dot C^{({t})}_{\gamma\delta} \right).
   \label{pure-tensor} 
\eea
Notice that in this large-scale limit 
we have $C^{({t})}_{\alpha\beta} = {\rm constant}$
as the relatively growing solution (in expanding phase)
even to the second-order perturbation.
In this sense, ignoring the transient mode in expanding phase,
the amplitude of $C^{({t})}_{\alpha\beta}$ remains constant
even to the second-order in perturbations
\cite{Salopek-Bond-1990,Maldacena-2002}. 


\subsection{Action formulation}

We consider the action expanded to the second-order in perturbations 
which will give equations of motion to the linear-order in perturbation
\cite{Ford-Parker-1977,action,MFB-1992}.
We consider the action for a scalar field in eq. (\ref{action-MSF}).
The perturbed action can be derived by using 
eqs. (\ref{eta-relation},\ref{R}) and
the ADM quantities presented in \S \ref{sec:perturbed-quantities}.
To the background order, ignoring the surface terms, we have
\bea
   & & S_{\rm BG} 
       = {1 \over 16 \pi G} \int \sqrt{g^{(3)}} a^3
       \left[ - 6 \left( {\dot a \over a} \right)^2
       + {6K \over a^2} + 16 \pi G \left( {1 \over 2} \dot \phi^2 - V \right)
       \right] d t d^3 x. 
   \label{action-BG}
\eea
To the second-order perturbation, ignoring the surface terms, 
the pure gravitational wave part becomes
\bea
   & & S_{\rm GW} 
       = {1 \over 16 \pi G} \int \sqrt{g^{(3)}} a^3 
       \left( \dot C^{(t)\alpha\beta} \dot C^{(t)}_{\alpha\beta} 
       - {1 \over a^2} C^{(t)\alpha\beta|\gamma} C^{(t)}_{\alpha\beta|\gamma} 
        - {2K \over a^2} C^{(t)\alpha\beta} C^{(t)}_{\alpha\beta} \right)
        d t d^3 x. 
   \label{action-GW}
\eea 
This action is valid for arbitrary number of scalar fields
and fluids with vanishing tensor-type anisotropic stress.
Now, we consider the pure scalar-type perturbation.
We assume $K = 0$.
To the second-order perturbation, ignoring the surface terms, we have
\bea
   S_{\rm scalar} 
   &=& {1 \over 2} \int a^3 \left\{
       \delta \dot \phi_\varphi^2 
       - {1 \over a^2} \delta \phi_\varphi^{\;\;,\alpha} 
       \delta \phi_{\varphi,\alpha}
       + {H \over a^3 \dot \phi} \left[ a^3 \left( {\dot \phi \over H}
       \right)^\cdot \right]^\cdot \delta \phi_\varphi^2 \right\} dt d^3 x 
   \nonumber \\
   &=& {1 \over 2} \int a^3 {\dot \phi^2 \over H^2}
       \left(\dot \varphi_{\delta \phi}^2 
       - {1 \over a^2} \varphi_{\delta \phi}^{\;\;\;\;,\alpha} 
       \varphi_{\delta \phi,\alpha} \right) dt d^3 x. 
   \label{action-scalar}
\eea 
In this case we used the linear-order equations of motions, 
thus it is an on-shell action. 
In the second step we used eq. (\ref{GI}).

Maldacena has considered the perturbed action to the third-order
in perturbations which is needed to have equations of motion
valid to the second-order \cite{Maldacena-2002}. 
{}For the temporal gauge he used two gauge conditions,
the uniform-field gauge ($\delta \phi \equiv 0$)
and the uniform-curvature gauge ($\varphi \equiv 0$),
and the $C$-gauge for the spatial and rotational ones.
Compared with our notations we have
\bea
   & & 
       \zeta_{\rm Maldacena} = \varphi_{\delta \phi} - \varphi_{\delta \phi}^2, 
       \quad
       \varphi_{\rm Maldacena} = \delta \phi_\varphi. 
\eea
Thus, $\zeta_{\rm Maldacena}$ is conserved in the large-scale limit,
see eqs. (\ref{LS-sol4},\ref{CG-UFG}).
To the linear-order, from eq. (\ref{GI}) we have
$\varphi_{\delta \phi} = - (H/\dot \phi) \delta \phi_\varphi$, thus
$\zeta_{\rm Maldacena} = - (H/\dot \phi) \varphi_{\rm Maldacena}$.

\section{Discussions}
                                                   \label{sec:Discussions}

We have presented the basic equations to investigate
the second-order perturbation of the Friedmann world model.
In order to serve as a convenient reference for future studies and 
applications we have presented some useful relations and quantities needed
for the second-order perturbations.
The present study is, apparently, not entirely new in this rich field 
of cosmology and the large-scale structure formation.
In the late sixties Tomita has presented a series of work on the subject 
in the context of a fluid \cite{Tomita}.
Studies in the context of the ideal fluid or the minimally coupled field 
can be found in \cite{tensor-GT,second-order-pert,Maldacena-2002}.
In the case of a pressureless irrotational fluid, see
\cite{Kasai-1992,Kofman-Pogosyan-1995}.
The case with the null-geodesic equations was studied in \cite{null-geodesic}.
And, the case with Boltzmann equation was considered in \cite{Hu-etal-1994}.

Compared with the previous works, perhaps we could emphasize the following 
as the new points in our work:

(i)   We present the complete sets of perturbed equations in the gauge-ready 
      form, so that we could easily apply the equations to any gauge conditions
      which make the mathematical analyses of given problems simplest.

(ii)  We consider the most general Friedmann background with $K$ and $\Lambda$.
      Previous studies have considered the flat Friedmann background only.

(iii) We consider the most general imperfect fluid situation which includes 
      multiple imperfect fluids with general interactions among them.
      In addition we also include minimally coupled scalar fields, 
      a class of generalized gravity theories, the electromagnetic fields,
      the null geodesic, and the relativistic Boltzmann equation.

(iv)  In \S \ref{sec:closed-forms} we present closed form equations 
      which are simlar to the known ones in linear theory. 

(v)   In \S \ref{sec:Pressureless} we show that up to 
      the second-order in perturbations the relativistic pressureless fluid
      coincides exactly with the Newtonian one.
      We note that suitable choices and combinations of different gauges
      (thus gauge-invariant combinations) are important to show the equivalence.

(vi)  In \S \ref{sec:pure-scalar} we have derived the large-scale 
      (super-sound-horizon) conserved quantity to the second-order, 
      $\varphi_v$, directly from the differential equation governing its
      evolution.
      This consreved variable was first studied by Salopek and Bond
      \label{Salopek-Bond-1990}

Our equations are suitable to handle the nonlinear evolutions 
in the perturbative manner.
If we have the solutions to the linear-order (see \S \ref{sec:linear-sols} 
for some examples), the evolution of second-order perturbations
can be derived using the quadratic combination of the linear variables
as sources; our basic sets of equations in \S \ref{sec:decomposed-eqs}
and some closed forms in \S \ref{sec:closed-forms}
are presented with such a purpose.
As long as we take such a perturbative approach our formulation 
in this work can be trivially extended to any higher-order perturbation;
except for the fact that, of course, the needed algebra would be 
quite demanding.
We also have shown in \S \ref{sec:Gauges} that the gauge issue 
can be similarly handled even in such higher-order perturbation.

Our formulation can be applied using several different methods 
in the following. 

(i)   Quasilinear analyses using Fourier analyses as often used 
      in the Newtonian case \cite{quasilinear}.
      In this approach the quadratic combination of the linear-order terms
      will lead to the mode-mode coupling among different scales,
      as well as among different types of perturbations.

(ii)  Nonlinear backreaction.
      In our approach we have assumed the presence of ``fictitious''
      background metric which is spatially homogeneous and isotropic.
      As the basic equations of Einstein gravity are nonlinear the
      nonlinear fluctuations in the metric and the matter can affect
      the background world model.
      One anticipates to recover the background Friedmann world model
      through averaging the more realistic lumpy world model and finding
      best fit to the idealized world model \cite{Ellis-fitting}.

(iii) Fitting and averaging.
      Our basic equations in \S \ref{sec:Equations} are presented without
      separating the background order quantities from the perturbed-order ones.
      Thus, the equations are suitable for the operation of averaging.
      Using our formulation we could apply and check the various different
      averaging prescriptions suggested in the literature 
      \cite{Ellis-fitting,averaging,Kasai-1992}.

Our perturbative formulation would be a useful 
complement to the following formulations aiming to investigate the nonlinear 
evolutions of the cosmological structures.

(i)   The large-scale (long wavelength) approximation 
      or the spatial gradient expansion
      studied by Salopek, Tomita, Deruelle and others in
      \cite{Salopek-Bond-1990,LS-expansion}.

(ii)  Cosmological post-Newtonian formulation studied 
      by Futamase, Tomita and others in \cite{post-Newtonian}.

(iii) Relativistic Zel'dovich approximation studied in 
      \cite{Kasai-1992,Kofman-Pogosyan-1995,Zeldovich-approximation}. 

(iv)  General (spatially inhomogeneous and anisotropic) solutions
      near singularity where the large-scale conditions are well met;
      in such a situation it was shown that
      the spatially different points decouple and evolve separately.
      These were studied by Belinsky, Lifshitz, Khalatnikov, and others in
      \cite{near-singularity}.

Our general formulation can be used to study the following situations
anticipated during the evolution of our universe: 

(i)   We can check the limit of linear theory.
      Current cosmological observations can be successfully explained
      within the current standard theoretical paradigm.
      In that paradigm the linear perturbation theory plays significant roles 
      to explain the quantum generation
      stage in the early universe and the classical evolution processes
      in the large-scale and in the early era.
      The linear theory provides a self-consistent explanation
      of some important aspects of the origin and evolution of
      large-scale structures.
      However, the limit of the linear theory {\it cannot} be
      estimated within the linear theory.
      We expect the second-order perturbation theory could provide 
      a meaningful ways to investigate such limits.

(ii)  We can investigate the quasilinear process in the relativistic context.
      In the literature it is commonly assumed that the
      relativistic linear perturbation theory is sufficient to handle the
      large-scale structure, and the nonlinear processes occur only in the 
      Newtonian context which are often handled by the numerical simulations.
      The quasilinear evolution would be useful to investigate the
      transition regions between the linear and nonlinear evolutions.
      Our perturbative approach may have its own limit, because
      if we find the importance of second-order contributions 
      it may naturally follow that higher-order contributions would 
      become important immediately as well.
      Thus, we anticipate, if successful, the relativistic quasilinear
      analyses can be developed similarly as the Newtonian cosmological
      quasilinear analyses studied in \cite{quasilinear}.

(iii) {}Fate of fluctuations in the collapsing phase, 
      and possibly through a bounce.
      The fluctuations of single component medium and the gravitational wave are
      described by the second-order differential equations.
      In the linear stage and in the large-scale limit, we have general 
      solutions in eqs. (\ref{Phi-sol},\ref{varphi_chi-sol},\ref{GW-sol-LS}).
      In an expanding phase, the $C$-mode is relatively growing and
      the $d$-mode is decaying, thus transient.
      If the initial conditions (say, generated from the quantum fluctuations)
      are imposed in the early expanding phase, the $d$-mode parts
      disappear in a few e-folding time of the scale factor increase, 
      thus uninteresting.
      The relatively growing modes both for the scalar- and tensor-type 
      perturbations are characterized by the conserved amplitude of 
      certain gauge-invariant variable, $\varphi_v$ (and the curvature
      variables in other gauges) and $C^{(t)}_{\alpha\beta}$, 
      in the large-scale limit.
      In the collapsing phase, however, the roles of growing and decaying
      modes are switched.
      In the collapsing phase the $d$-mode, and the vector-mode as well,
      grows quite rapidly, see 
      eqs. (\ref{Phi-LS-sol},\ref{GW-LS-sol},\ref{rotation-sol});
      our solutions in \S \ref{sec:linear-sols} also cover the collapsing 
      phase by considering
      $t \rightarrow |t|$ with $t$ approaching $0$,
      see \cite{HN-bounce}.
      Thus, the linear perturbations grow rapidly and inevitably reach 
      the nonlinear stage \cite{Bardeen-1980,Lyth-collapse}.
      Such growths would cause the transition of our simple 
      (spatially homogeneous and isotropic) background world model to the 
      anisotropic and inhomogeneous ones studied in \cite{near-singularity}.
      Although we anticipate the perturbations would become quite nonlinear,
      we hope we could investigate the transition region based on our
      second-order perturbation formulation.
      One simplifying fact is that in the collapsing phase the
      local range covered by the dynamical time scale $\sim H^{-1}$ 
      shrinks relative to the comoving scale, thus effectively 
      the scales we are interested in satisfy the conditions of
      large-scale limit 
      \footnote{
                Professor James Bardeen and Ewan Stewart have suggested
                that the large-scale (long wavelength) expansion 
                or the spatial gradient expansion technique 
                \cite{Salopek-Bond-1990,LS-expansion}
                would be useful to investigate such situations.
                }.
      Such large-scale conditions are well met for a given comoving scale
      during the early evolution stage (near singularity)
      and as the background model approaches the singularity in
      the collapsing phase.
      Investigation of situations in the collapsing and subsequent bouncing 
      background is left for future study; 
      for evolutions under the linear assumption, see\cite{HN-bounce}.
      {}For the general cosmological investigation near singularity, 
      see \cite{near-singularity}.

(iv)  It is well known that the nonlinear effect (either in the quantum
      generation or in the classical evolution processes) could lead
      to non-Gaussian effects in the observed quantities of
      the CMB anisotropies and the large-scale galaxy distribution and motion.
      Maldacena has recently investigated such an effect on the CMB
      based on the second-order perturbation theory, 
      see\cite{Maldacena-2002}.
      The first year WMAP data shows no positive detection of non-Gaussian
      nature of the CMB sky maps under a couple of non-Gaussianity tests 
      \cite{WMAP-Gaussian}.

(v)   Evolutions in the super-horizon scale where the scale is
      larger than causal domain during the dynamic time scale.
      See \S \ref{sec:pure-scalar} and \ref{sec:pure-tensor} 
      for the conserved quantities to the second-order 
      which were found by Salopek and Bond in \cite{Salopek-Bond-1990}.

(vi)  In \S \ref{sec:Pressureless} we have shown that to the second-order
      a pressureless fluid with pure scalar-type perturbation reproduces
      Newtonian result.
      It is likely that the relativistic effect appears in the higher-order 
      perturbation which is left for future investigation.

\section*{Acknowledgments}

We thank Ewan Stewart for encouraging us to continue the project
and sharing his enthusiasm and opinions on the subject.
We also wish to thank Prof. George Efstathiou for inviting us to the
Institute of Astronomy where the project was reassumed.
JH wishes to thank Toshi Futamase and Masumi Kasai for their
hospitality and encouragement in the beginning stage of the work at Hirosaki.
JH and HN were supported by grant No. R02-2003-000-10051-0 and
No. R04-2003-10004-0, respectively, from the
Basic Research Program of the Korea Science and Engineering Foundation.



\begin{references}
\bibitem{Lifshitz-1946}
         E.M. Lifshitz, J. Phys. (USSR) {\bf 10}, 116 (1946);
         E.M. Lifshitz and I.M. Khalatnikov, Adv. Phys. {\bf 12}, 185 (1963).
\bibitem{Friedmann-1922}
         A.A. Friedmann, Zeitschrift f\"ur Physik {\bf 10}, 377 (1922); 
              translated in {\it Cosmological-constants: papers in modern
              cosmology}, edited by J. Bernstein and G. Feinberg
              (Columbia Univ. Press, New York, 1986), p49.
\bibitem{WMAP}
         C.L. Bennett, {\it et al.}, astro-ph/0302207;
         T.J. Pearson, {\it et al.}, astro-ph/0205388;
         C.L. Kuo, {\it et al.}, astro-ph/0212289.
\bibitem{Bond-Efstathiou-1987}
         J.R. Bond and G. Efstathiou, Mon. Not. R. Astron. Soc. {\bf 226},
            655 (1987).
\bibitem{Bianchi-pert}
         H. Noh and J. Hwang, Phys. Rev. D {\bf 52}, 1970, 5643 (1995);
         H. Noh, {\it ib id.} {\bf 53}, 690, 4311 (1996).
\bibitem{Bardeen-1980}
         J.M. Bardeen, Phys. Rev. D {\bf 22}, 1882 (1980).
\bibitem{Bardeen-1988}
         J.M. Bardeen, {\it Particle Physics and Cosmology}, edited by 
                       L. Fang and A. Zee (Gordon and Breach, London, 1988), p1.
\bibitem{Kodama-Sasaki-1984}
         H. Kodama and M. Sasaki, Prog. Theor. Phys. Suppl. {\bf 78}, 1 (1984).
\bibitem{MFB-1992}
         V.F. Mukhanov, H.A. Feldman, and R.H. Brandenberger,
                        Phys. Rep. {\bf 215}, 203 (1992).
\bibitem{Peebles-1980}
         P.J.E. Peebles, {\it The large-scale structure of the universe},
                (Princeton Univ. Press, Princeton, 1980);
         Ya.B. Zel'dovich and I.D. Novikov, {\it Relativistic astrophysics, 
                          Vol 2, The structure and evolution of the universe},
                          (Univ. Chicago Press, Chicago, 1983).
\bibitem{Liddle-Lyth-2000}
         T. Padmanabhan, {\it Structure formation in the universe},
                         (Cambridge University Press, Cambrige, 1993);
         P.J.E. Peebles, {\it Principles of physical cosmology},
                         (Princeton Univ. Press, Princeton, 1993);
         A.R. Liddle and D.H. Lyth,
                      {\it Cosmological inflation and large-scale structure}
                      (Cambridge Univ. Press, Cambridge, 2000).
\bibitem{Hwang-1991}
         J. Hwang, Astrophys. J. {\bf 375}, 443 (1991).
\bibitem{Hwang-Noh-Newtonian}
         J. Hwang and H. Noh, Gen. Rel. Grav. {\bf 31}, 1131 (1999).
\bibitem{Hwang-QFT}
         J. Hwang, Phys. Rev. D {\bf 48}, 3544 (1993);
                   Class. Quant. Grav. {\bf 11}, 2305 (1994).
\bibitem{Hwang-MSF}
         J. Hwang, Astrophys. J. {\bf 427}, 542 (1994).
\bibitem{Gauge-theory}
         R. Utiyama, Phys. Rev. {\bf 101}, 1597 (1958);
         T.W.B. Kibble, J. Math. Phys. {\bf 2}, 212 (1961).
\bibitem{SW-1967}
         R.K. Sachs and A.M. Wolfe, Astrophys. J. {\bf 147}, 73 (1967).
\bibitem{ADM}
         R. Arnowitt, S. Deser, and C.W. Misner, in {\it Gravitation: an
                      introduction to current research}, edited by  L. Witten
                      (Wiley, New York, 1962) p. 227.
\bibitem{Ehlers-1993}
         J. Ehlers, Proceedings of the mathematical-natural science of 
                    the Mainz academy of science and literature, 
                    Nr. {\bf 11}, 792 (1961),
                    translated in Gen. Rel. Grav. {\bf 25}, 1225 (1993).
\bibitem{Ellis-1971}
         G.F.R. Ellis, in {\it General relativity and cosmology, Proceedings of
                       the international summer school of physics Enrico 
                       Fermi course 47}, edited by R. K. Sachs (Academic 
                       Press, New York, 1971), 104;
                       in {\it Cargese Lectures in Physics}, edited by 
                       E. Schatzmann (Gorden and Breach, New York, 1973), 1.
\bibitem{HV-1990}
         J. Hwang and E.T. Vishniac, Astrophys. J.  {\bf 353}, 1 (1990).
\bibitem{Israel-1976}
         W. Israel, Ann. Phys. {\bf 100}, 310 (1976).
\bibitem{Hwang-GGT-1990}
         J. Hwang, Class. Quant. Grav. {\bf 7}, 1613 (1990).
\bibitem{Hwang-Noh-GGT}
         J. Hwang, Phys. Rev. D {\bf 53}, 762 (1996);
         J. Hwang and H. Noh, {\it ib id.} {\bf 54}, 1460 (1996);
                  Class. Quant. Grav. {\bf 15}, 1387, 1401 (1998).
\bibitem{NH-Rab}
         H. Noh and J. Hwang, Phys. Rev. D {\bf 55}, 5222 (1997);
                    {\bf 59}, 047501 (1999).
\bibitem{Hwang-CT-1997}
         J. Hwang, Class. Quant. Grav. {\bf 14}, 1981 (1997).
\bibitem{KS-1966}
         J. Kristian and R.K. Sachs, Astrophys. J. {\bf 143}, 379 (1966).
\bibitem{Lindquist-1966}
         R.W. Lindquist, Ann. Phys., {\bf 37}, 487 (1966);
         J. Ehlers, in {\it General relativity and cosmology, Proceedings of the
                    international summer school of physics Enrico Fermi course 
                    47}, edited by R.K. Sachs (Academic Press, New York, 1971),
                    p1.
\bibitem{Hwang-Noh-CMB-2002}
         J. Hwang and H. Noh, Phys. Rev. D {\bf 65}, 023512 (2002).
\bibitem{Hwang-SW}
         J. Hwang and H. Noh, Phys. Rev. D {\bf 59}, 067302 (1999);
         J. Hwang, T. Padmanabhan, O, Lahav, and H. Noh, {\it ib id.}
                   {\bf 65}, 043005 (2002).
\bibitem{tensor-GT}
         A.H. Taub, J. Math. Phys. {\bf 2}, 787 (1961);
         L.R.W. Abramo, R.H. Brandenberger, and V.F. Mukhanov, 
                        Phys. Rev. D {\bf 56}, 3248 (1997);
         S. Matarrese, S. Mollerach, and M. Bruni,
                       {\it ib id.} {\bf 58}, 043504 (1998).
\bibitem{EB}
         G.F.R. Ellis and M. Bruni, Phys. Rev. D {\bf 40}, 1804 (1989);
         A. Woszczyna and A. Ku{\l}ak, Class. Quant. Grav. {\bf 6}, 
                      1665 (1989).
\bibitem{EBH}
         G.F.R. Ellis, M. Bruni, and J. Hwang, Phys. Rev. D {\bf 42}, 
                       1035 (1990).
\bibitem{Harrison-1967}
         E.R. Harrison, Rev. Mod. Phys. {\bf 39}, 862 (1967).
\bibitem{Nariai-1969}
         H. Nariai, Prog. Theor. Phys. {\bf 41}, 686 (1969).
\bibitem{Field-Shepley-1968}
         G.B. Field and L.C. Shepley, Astrophys. Space. Sci.
                      {\bf 1}, 309 (1968).
\bibitem{action}
         V.N. Lukash, Sov. Phys. JETP Lett. {\bf 31}, 596 (1980);
                      Sov. Phys. JETP {\bf 52}, 807 (1980);
         G.V. Chibisov and V.F. Mukhanov, Mon. Not. R. Astron. Soc.
                       {\bf 200}, 535 (1982);
         V.F. Mukhanov, Sov. Phys. JETP {\bf 68}, 1297 (1988).
\bibitem{Hwang-Fluid}
         J. Hwang, Astrophys. J. {\bf 415}, 486 (1993).
\bibitem{Lucchin-Matarrese-1985}
         F. Lucchin and S. Matarrese, Phys. Rev. D {\bf 32}, 1316 (1985).
\bibitem{Hwang-MDE}
         J. Hwang, Astrophys. J. {\bf 427}, 533 (1994).
\bibitem{KS2}
         H. Kodama and M. Sasaki, Int. J. Mod. Phys. A {\bf 1}, 265 (1986);
                       {\bf 2}, 491 (1987).
\bibitem{HN-multiple}
         J. Hwang and H. Noh, Phys. Lett. B, {\bf 495}, 277 (2000);
                  Class. Quant. Grav. {\bf 19}, 527 (2002). 
\bibitem{Kofman-Pogosyan-1995}
         E. Bertschinger and A.J.S. Hamilton, Astrophys. J. {\bf 435}, 
                         1 (1994);
         L. Kofman and D. Pogosyan, {\it ib id.} {\bf 442}, 30 (1995);
         S. Matarrese and D. Terranova, Monthly Not. R. Astron. Soc. 
                       {\bf 283}, 400 (1996).
\bibitem{quasilinear}
         E.T. Vishniac, Monthly Not. R. Astron. Soc. {\bf 203}, 345 (1983);
         M.H. Goroff, B. Grinstein, S.-J. Rey, and M.B. Wise,
                      Astrophys. J., {\bf 311}, 6 (1986);
         N. Makino, M. Sasaki, and Y. Suto, Phys. Rev. D {\bf 46}, 585 (1992).
\bibitem{Kasai-1992}
         M. Kasai, Phys. Rev. Lett. {\bf 69}, 2330 (1992);
                   Phys. Rev. D {\bf 47}, 3214 (1993).
\bibitem{Salopek-Bond-1990}
         D.S. Salopek and J.R. Bond, Phys. Rev. D {\bf 42}, 3936 (1990).
\bibitem{Maldacena-2002}
         J. Maldacena, JHEP {\bf 0305}, 013 (2003).
\bibitem{Ford-Parker-1977}
         L.H. Ford and L. Parker, Phys. Rev. D {\bf 16}, 1601 (1977).
\bibitem{Tomita}
         K. Tomita, Prog. Theor. Phys. {\bf 37}, 831 (1967);
                    {\bf 45}, 1747 (1971);
                    {\bf 47}, 416 (1972).
\bibitem{second-order-pert}
         M. Bruni, S. Matarrese, S. Mollerach, and S. Sonego,
                   Class. Quant. Grav. {\bf 14}, 2585 (1997);
         W. Unruh, astro-ph/9802323;
         V. Acquaviva, N. Bartolo, S. Matarrese, and A. Riotto,
                       astro-ph/0209156;
         G. Rigopoulos, astro-ph/0212141.
\bibitem{null-geodesic}
         T. Pyne and S.M. Carroll, Phys. Rev. D {\bf 53}, 2920 (1996);
         S. Mollerach and S. Matarrese, {\it ib id.} {\bf 56}, 4494 (1997).
\bibitem{Hu-etal-1994}
         W. Hu, D. Scott, J. Silk, Phys. Rev. D {\bf 49}, 648 (1994);
         R. Maartens, T. Gebbie, and G.F.R. Ellis, {\it ib id.} {\bf 59},
                      083506 (1999). 
\bibitem{Ellis-fitting}
         G.F.R. Ellis, in {\it General Relativity and Gravitation}, edited by
                       B. Bertotti, {\it et al.} (Reidel, Dordrecht, 1984) 
                       p215;
         G.F.R. Ellis and W. Stoeger, Class. Quant. Grav. {\bf 4}, 1697 (1987).
\bibitem{averaging}
         M.F. Shirokov and I.Z. Fisher, Sov. Astron. {\bf 6}, 699 (1963);
         R.A. Isaacson, Phys. Rev. {\bf 166}, 1263, 1272 (1968);
         N.R. Sibgatullin, Sov. Phys. Doklady {\bf 16}, 697 (1972);
         A.V. Byalko, Sov. Phys. JETP {\bf 38}, 421 (1974);
         H. Nariai, Prog. Theor. Phys. {\bf 52}, 1539 (1974);
                    {\bf 53}, 656 (1975); {\bf 54}, 1356 (1975);
         L.S. Marochnik, N.V. Pelikhov, and G.M. Vereshkov,
                         Astrophys. Space Sci. {\bf 34}, 249, 281, (1975);
         G.M. Vereshkov and A.N. Poltavtsev, Sov. Phys. JETP {\bf 44}, 
                          1 (1976);
         A.M. Krymsky, L.S. Marochnik, P.D. Naselsky, and N.V. Pelikhov,
                       Astrophys. Space Sci. {\bf 55}, 325, (1978);
         L.S. Marochnik, P.D. Naselsky, and N.V. Pelikhov,
                         {\it ib id.} {\bf 67}, 261, (1980);
         L.S. Marochnik, Sov. Astron. {\bf 24}, 518, 651, (1980);
                         {\bf 25}, 8 (1981);
         T.W. Noonan, Gen. Rel. Grav. {\bf 16}, 1103 (1984);
         I.K. Rozgach\"eva, Sov. Astron. {\bf 31}, 14 (1987);
         T. Futamase, Phys. Rev. D {\bf 53}, 681 (1996);
         M. Carfora and K. Piotrkowska, {\it ib id.} {\bf 52}, 4393 (1995);
         J.P. Boersma, {\it ib id.} {\bf 57}, 798 (1998);
         W.R. Stoeger, A. Helmi, and D. Torres, gr-qc/9904020;
         Y. Nambu, Phys. Rev. D {\bf 62}, 104010 (2000);
                                {\bf 63}, 044013 (2001);
                                {\bf 65}, 104013 (2002);
         T. Buchert and M. Carfora, Class. Quant. Grav. {\bf 19}, 6109 (2002).
\bibitem{LS-expansion}
         K. Tomita, Prog. Theor. Phys. {\bf 48}, 1503 (1972);
                    {\bf 54}, 730 (1975);
         D.S. Salopek, Phys. Rev. D {\bf 43}, 3214 (1991);
         D.S. Salopek and J.M. Stewart, Class. Quant. Grav. {\bf 9}, 
                      1943 (1992);
                      Phys. Rev. D {\bf 48}, 719 (1993);
                      {\bf 47}, 3235 (1993);
         K. Tomita, {\it ib id.} {\bf 48}, 5634 (1993);
         G.L. Comer, N. Deruelle, D. Langlois, and J. Parry,
                     {\it ib id.} {\bf 49}, 2759 (1994);
         J. Parry, D.S. Salopek, and J.M. Stewart, {\it ib id.} {\bf 49},
                   2872 (1994);
         K. Tomita and N. Deruelle, {\it ib id.} {\bf 50}, 7216 (1994);
         N. Deruelle and D. Langlois, {\it ib id.} {\bf 52}, 2007 (1995);
         Y. Nambu and A. Taruya, Class. Quant. Grav. {\bf 13}, 705 (1996);
         G.L. Comer, {\it ib id.} {\bf 14}, 407 (1997);
         G.L. Comer, N. Deruelle, and D. Langlois, Phys. Rev. D {\bf 55}, 
                     3497 (1997);
         D.S. Salopek, {\it ib id.} {\bf 56}, 2057 (1997);
         Y. Nambu and Y. Yamaguchi, {\it ib id.} {\bf 60}, 104011 (1999);
         Y. Nambu and S. Ohokata, Class. Quant. Grav. {\bf 19}, 4263 (2002).
\bibitem{post-Newtonian}
         T. Futamase, Phys. Rev. Lett. {\bf 61}, 2175 (1988);
         K. Tomita, Prog. Theor. Phys. {\bf 79}, 258 (1988);
         T. Futamase, Mon. Not. R. Astron. Soc. {\bf 237}, 187 (1989);
         K. Tomita, Prog. Theor. Phys. {\bf 85}, 1041 (1991);
         T. Futamase, {\it ib id.} {\bf 86}, 389 (1991); 
                      {\bf 89}, 581 (1993);
         M. Shibata and H. Asada, {\it ib id.} {\bf 94}, 11 (1995);
         M. Takada and T. Futamase, Monthly Not. R. Astron. Soc.
                       {\bf 306}, 64 (1999).
\bibitem{Zeldovich-approximation}
         S. Matarrese, O. Pantano, and D. Saez, Phys. Rev. D {\bf 47},
                       1331 (1993); 
                       Monthly, Not. R. Astron. Soc. {\bf 271}, 513 (1994);
         K.M. Croudace, J. Parry, D.S. Salopek, and J.M. Stewart,
                        Astrophys. J. {\bf 423}, 22 (1994);
         D.S. Salopek, J.M. Stewart, and K.M. Croudace,
                       Monthly, Not. R. Astron. Soc. {\bf 271}, 1005 (1994);
         M. Kasai, Phys. Rev. D {\bf 52}, 5605 (1995);
         H. Russ, M. Morita, M. Kasai, and G. B\"orner,
                  {\it ib id.} {\bf 53}, 6881 (1996);
         M. Morita, K. Nakamura, and M. Kasai, {\it ib id.} {\bf 57}, 6094
                    (1998).
\bibitem{near-singularity}
         V.A. Belinskii, I.M. Khalatnikov, and E.M. Lifshitz, 
                         Adv. Phys. {\bf 19}, 525 (1970);
         I.M. Khalatnikov, and E.M. Lifshitz, 
                         Phys. Rev. Lett. {\bf 24}, 76 (1970);
         V.A. Belinskii, E.M. Lifshitz, and I.M. Khalatnikov, 
                         Sov. Phys. JETP {\bf 35}, 838 (1972);
         D. Eardley, E. Liang, and R. Sachs, J. Math. Phys. {\bf 13}, 99 (1972);
         V.A. Belinskii, I.M. Khalatnikov, and E.M. Lifshitz,
                         Adv. Phys. {\bf 31}, 639 (1982);
         L. Andersson and A.D. Rendall, Commun. Math. Phys. 
                       {\bf 218}, 479 (2001);
         B.K. Berger, Living Rev. Relativity 2002-1, gr-qc/0201056.
\bibitem{HN-bounce}
         J. Hwang and H. Noh, Phys. Rev. D {\bf 65}, 124010 (2002).
\bibitem{Lyth-collapse}
         D.H. Lyth, Phys. Lett. B {\bf 526} 171 (2002).
\bibitem{WMAP-Gaussian}
         E. Komatsu, {\it et al.}, Astrophys. J. Suppl. {\bf 148}, 119 (2003).
\end{references}
\end{document}